\documentclass{aa}
\usepackage{lmodern}
\usepackage{orcidlink}
\usepackage[varg]{txfonts}
\usepackage{svg}
\usepackage{hyperref}
\hypersetup{colorlinks, allcolors=blue}
\usepackage{graphicx,amsmath}
\usepackage{color}
\usepackage{xcolor}
\usepackage{txfonts}
\usepackage{lipsum}
\usepackage{subcaption}
\usepackage{lscape}
\usepackage{placeins}

\newcommand{\Msun}{{\rm\,M_\odot}}
\newcommand{\kpc}{{\rm\,kpc}}
\newcommand{\Gyr}{{\rm\,Gyr}}
\newcommand{\kms}{{\rm\,km\,s^{-1}}}

\begin{document} 

   \title{
   The SMUGGLE-Ring project: \\ Bar and bulge effects on nuclear disk and ring formation
}

    \author{SungWon Kwak       \inst{\ref{aip}}\orcidlink{0000-0003-0957-6201},
            Federico Marinacci \inst{\ref{bol1},\ref{bol2}}\orcidlink{0000-0003-3816-7028},
            Matthias Steinmetz \inst{\ref{aip},\ref{uniP}}\orcidlink{0000-0001-6516-7459},
            Ivan Minchev       \inst{\ref{aip}}\orcidlink{0000-0002-5627-0355},\\
            Cristina Chiappini \inst{\ref{aip}}\orcidlink{0000-0003-1269-7282},
            Mathias Schultheis \inst{\ref{nice}}\orcidlink{0000-0002-6590-1657},
            Woong-Tae Kim      \inst{\ref{snu1},\ref{snu2}}\orcidlink{0000-0003-4625-229X},
            Mark Vogelsberger  \inst{\ref{mit}}\orcidlink{0000-0001-8593-7692},\\ 
            Laura V. Sales     \inst{\ref{ucr}}\orcidlink{0000-0002-3790-720X},
            Hui Li             \inst{\ref{tsing}}\orcidlink{0000-0002-1253-2763},
            \and 
            Seungwon Baek      \inst{\ref{snu1}}\orcidlink{0009-0002-0251-9570}
            }

    \institute{Leibniz-Instit\"ut f\"ur Astrophysik Potsdam (AIP), An der Sternwarte 16, 14482, Potsdam, Germany \label{aip} 
    \email{skwak@aip.de}
    \and Dipartimento di Fisica e Astronomia ``Augusto Righi'', Universit\`a di Bologna, Via Piero Gobetti 93/2, I-40129 Bologna, Italy\label{bol1}
    \and
    INAF, Osservatorio di Astrofisica e Scienza dello Spazio di Bologna, Via Piero Gobetti 93/3, I-40129 Bologna, Italy\label{bol2}
    \and Universit\"at Potsdam, Institut f\"ur Physik und Astronomie, Karl-Liebknecht-Str. 24-25, 14476, Potsdam, Germany \label{uniP}
    \and Universit\'e C\^ote d'Azur, Observatoire de la C\^ote d'Azur, Laboratoire Lagrange, CNRS, Blvd de l'Observatoire, 06304 Nice, France\label{nice}
    \and Department of Physics \& Astronomy, Seoul National University, Seoul 08826, Republic of Korea\label{snu1}
    \and SNU Astronomy Research Center, Seoul National University, Seoul 08826, Republic of Korea\label{snu2}
    \and Department of Physics and Kavli Institute for Astrophysics and Space Research, Massachusetts Institute of Technology, Cambridge, MA 02139, USA \label{mit}
    \and University of California, Riverside, 900 University Ave., Riverside, CA 92521, USA \label{ucr}
    \and Department of Astronomy, Tsinghua University, Haidian DS 100084, Beijing, China \label{tsing}
    }            

   \date{\today}
 
\abstract{
We present the first results from the SMUGGLE-Ring project, a suite of simulations employing the SMUGGLE (Stars and MUltiphase Gas in GaLaxiEs) ISM and stellar feedback model to explore nuclear structures in Milky Way-mass galaxies. We discuss results from three simulations evolved for 5 Gyr in isolation, in which we vary the classical bulge mass, while keeping the disk and halo structures identical. Nuclear stellar disks and rings emerge exclusively in our bulge models, with more massive bulges associated with earlier formation and more extended initial gas reservoirs shortly after bar formation. After gas depletion via active star formation, the nuclear stellar disks bifurcate into pressure-supported nuclear star clusters (NSCs, $v_{\phi}/\sigma_R < 0.7$) and rotationally supported nuclear stellar rings (NSRs, $v_{\phi}/\sigma_R = 1.2$--1.7, radii 0.64--0.76 kpc). In our bulgeless model, stellar feedback more effectively disrupts and delays the buildup of stable nuclear gas disks. The enclosed stellar mass of NSCs ($\sim10^{9}\Msun$) dominates over that of NSRs ($\sim10^{8}\Msun$). The star formation rates decline over time due to gas depletion (NSCs 0.1--1 $\Msun$yr$^{-1}$, NSRs 0.01--$0.1 \Msun$yr$^{-1}$). Kinematics reveal outward-shifting rotation peaks with $\sigma$-drops in NSRs, while a fraction of stars in NSCs exhibits radial shift after 3 Gyr. These findings support inside-out NSD formation via secular bar evolution, with NSRs tracing the star-forming outer edge of the nuclear gas disk and NSCs forming the kinematically hotter inner component. The range of nuclear stellar disk sizes (0.25--0.76 kpc) falls within the observationally inferred ranges, but the existence of larger rings would require external gas flow and/or a longer period of evolution. Future SMUGGLE-Ring extensions will incorporate varying gas fractions, tidal/merger effects, and the circumgalactic medium to further elucidate nuclear diversity and outliers.
} 

   \titlerunning{SMUGGLE-Ring: Bulge Mass}
   \authorrunning{Kwak et al.}
   \maketitle
   \nolinenumbers

\section{Introduction}\label{sec:intro}
Stellar bars are a prevalent feature in disk galaxies throughout the local Universe, with observations indicating that approximately 60\% of nearby disk galaxies possess bars \citep{eskridge00,whyte02,menendez07,marinova07, wang25}. As rotating non-axisymmetric components, bars exert a significant influence on the secular evolution of their host galaxies by redistributing angular momentum between the disk and halo. This induces substantial gas inflows toward the central regions, shaping the morphological and dynamical characteristics of galactic centers \citep{combes85, athanassoula92, combes93, buta96, athanassoula13, seo19}. Through these mechanisms, bars drive the emergence of diverse nuclear structures, including nuclear stellar disks (NSDs), nuclear stellar rings (NSRs), and nuclear star clusters (NSCs), all of which prove essential for investigating the complex processes at play in the galactic center \citep{schultheis25}. NSDs consist of rotationally supported, kinematically cold stellar populations that remain distinct from classical bulges and the bars themselves, frequently serving as sites of ongoing star formation \citep{bittner20,gadotti20}. NSRs, which are primarily gaseous and actively star-forming, typically emerge at the outer peripheries of NSDs, whereas NSCs form highly dense, pressure-supported stellar components at the galactic core \citep{neumayer20, erwin24,gleis26}.
The exact formation of NSCs is still an active area of research, but two primary mechanisms are (1) the spiraling of star clusters and (2) in-situ formation via gas inflow, which is more common in more massive galaxies \citep{fahrion22,fahrion24}. For instance, \cite{nogueras23} suggested that nuclear structures in the Milky Way might have originated from a single progenitor structure, but the scaling relations of NSC and NSD by \cite{gadotti25} find no indication that these nuclear structures share the same assembly history.

Empirical evidence demonstrates that NSDs are common in barred galaxies \citep{kormendy82, gadotti19, gadotti20}, also appearing in early-type barred galaxies \citep{bosch98}.
Their sizes typically range from 0.1 to 1 kpc with stellar masses on the order of $10^9$ M$_{\odot}$ \citep{gadotti19, bittner20}. In the TIMER survey and other studies \citep{erwin24}, NSDs have been identified frequently in the barred galaxies examined, featuring comparatively young stellar populations, higher metallicities, and decreasing [$\alpha$/Fe] abundances with increasing radius, all of which support the notion of inside-out formation \citep{gadotti19, gadotti20, bittner20}. In the PHANGS-ALMA dataset, NSRs exhibit median radii around 0.4 kpc and star formation rates averaging approximately 0.1 $\Msun$yr$^{-1}$ \citep{gleis26}. 
While NSDs are found in a wide range of galaxy types, including S0 and early-type galaxies \citep{bosch98,erwin21,schultheis25}, they are notably absent in very late-type galaxies \citep{erwin24}. These patterns imply that the presence of a central spheroidal component may contribute to the formation and growth of nuclear structures, while environmental conditions, such as isolated fields versus dense clusters, may additionally shape their evolutionary trajectories \citep{erwin21}.

In the Milky Way, the NSD is particularly compact, with a radius of approximately 0.1 kpc \citep{schultheis25}, exhibiting a kinematic parameter $v_{\phi}/\sigma$ of roughly 1.4 \citep{sormani22, shahzamanian22}. The effective range of NSD encompasses a prominent NSC and maintains an intimate association with the Central Molecular Zone (CMZ - \citealt{zoccali24}). Comprehensive mappings of kinematic and metallicity gradients within the NSD \citep{nogueras23, schultheis25b, ryde25} reinforce the hypothesis of inside-out growth, as the NSD remains chemodynamically distinct from the surrounding bar and bulge stars \citep{nogueras24}. The observed characteristics in the Milky Way indicate that bar-driven gas accumulation serves as a dominant process of the CMZ formation, so understanding the chemodynamical properties and formation epoch of NSCs and NSDs would also provide insights into the origin of nuclear structures and subsequent bar evolutionary history of the Milky Way \citep{spitoni26}.

Recent observational work has also provided strong evidence for the existence of a pressure-supported spheroidal bulge component in the Milky Way. Using field stars, \cite{nepal26} robustly identified a chemically and kinematically distinct spheroidal population coexisting with the bar and the disk. Their analysis also indicated that the Galactic bar has dynamically influenced the spheroidal bulge, inducing a mild triaxiality and radial extension. In addition, they identified stars on x4 orbits that likely originate from the early spheroidal component, sharing similar chemical signatures with the bulge population. These findings provide observational support for the presence of an early-formed spheroidal bulge whose interaction with the later bar evolution may play an important role in shaping the nuclear structures of the Milky Way, and other galaxies.

Theoretical studies propose that bars direct gas flows inside the inner Lindblad resonances (ILRs), but not specifically at a certain resonance \citep{kim12c}, allowing the gas to settle into x2 orbits as rings or disks where gravitational torques equilibrate, potentially inside the radial extent of the minor axis of bars \citep{athanassoula92,combes93,athanassoula13}. The presence of a classical bulge alters the rotation curve in the inner region, provides a strong central gravitational potential, and creates dual ILRs \citep{combes93, pettitt18}. In fact, simulations of nuclear rings in barred galaxies without a classical bulge show that stellar feedback scatters central gas concentrations, potentially delaying the formation of nuclear structures \citep{seo19}. Such feedback could even destroy the nuclear bar \citep{li23}. However, it becomes puzzling that the bulge and gas in disk galaxy simulations are known to weaken the influence of bars while observations indicate that nuclear disk formation is directly related to bar evolution. Indeed, the existence of a massive central spheroidal component with high concentration is known to stabilize the disk against bar formation \citep{athanassoula02a, kwak17, kataria18, jang23}. Also, the inclusion of a gas component attenuates bar formation and growth by hindering angular momentum exchange \citep{seo19,beane23,bland24}. Given the complex interplay of these effects, the impact of classical bulge mass on bar and nuclear ring formation has not been thoroughly explored.

In this paper, we investigate the formation and evolution of the nuclear disks, NSRs, and NSCs in bar-forming disk galaxies by evolving galaxy models with the SMUGGLE model \citep[Stars and MUltiphase Gas in GaLaxiEs,][]{marinacci19}. By varying the mass of the classical bulge in identical the disk and halo density structures, we examine the effects of bulge mass on the size and kinematics of the nuclear disk and ring. Section \ref{sec:method} outlines the simulation techniques and initial configurations. The principal outcomes are detailed in Section \ref{sec:results}. The discussion and summary appear in Section \ref{sec:discussion}.

\section{Method}\label{sec:method}

\subsection{SMUGGLE-Ring Project}\label{sec:smuggle}
This study serves as the first stepping stone for the SMUGGLE-Ring project by demonstrating the role of a classical stellar bulge on the size of the nuclear disk and ring in three bar-forming disk galaxies. 
Throughout the SMUGGLE-Ring project, we plan to conduct an extensive set of hydrodynamics simulations of bar-forming galaxies with various initial conditions and mass resolutions. This project comprehensively examines nuclear structures, such as nuclear rings, nuclear disks, and NSC  \citep{bittner20,erwin24,fahrion22, fahrion24,gadotti19,gadotti20,gadotti25,nogueras23,nogueras24,schultheis21,schultheis25b,schultheis25,gleis26}. We aim to investigate the different evolutionary paths induced by multiple factors, including disk and bulge properties, gas fraction, the presence of a circumgalactic medium (CGM), and various merger scenarios. For better treatments of hydrodynamics and stellar feedback, we evolve those initial conditions using the interstellar medium (ISM) and stellar feedback model model SMUGGLE   \citep{marinacci19}, which demonstrates convergence with resolution (in the range $10^3-10^5\Msun$ for baryon mass resolution) in observed star formation and has been adopted to study various topics in galaxy formation and evolution \citep{kannan20,burger22,beane23,li20,li22,li23,smith22,sivasankaran22,sivasankaran25,sivasankaran26,tacchella22, narayanan23,barbani23,barbani25,li24,zhange24,zhangz24,zhangz25}.

The \textsc{SMUGGLE} model introduces a multiphase ISM with realistic physics, thereby enhancing the accuracy of galaxy evolution simulations \citep{marinacci19}. This framework overcomes limitations of traditional effective equation-of-state models, which artificially pressurize dense gas and produce unresolved disk structures and overly smooth ISM morphologies \citep[see, e.g.,][]{benitez18}. Stellar feedback in \textsc{SMUGGLE} simultaneously regulates star formation and generates the hot, warm, and cold phases of the ISM \citep{mckee77}. It achieves this through three key features: (i) local feedback (energy and momentum injection tied directly to star particles, with no decoupled wind particles), (ii) self-consistent generation of outflows, and (iii) explicit resolution of the multiphase structure of the ISM, in particular its dense phase at $  n>0.1\,\mathrm{cm}^{-3}  $ -- the density threshold at which sub-grid models impose an effective equation of state \citep[][see also \citealt{vogelsberger20} for a detailed review of the numerical methods]{springel03,vogelsberger14a,vogelsberger14b,nelson19,pillepich18,pillepich19,grand17}. These features are essential for realistic galaxy evolution and detailed substructure formation, while accurately predicting ISM turbulence and outflows.

SMUGGLE is implemented in the AREPO code, a moving-mesh framework that combines the strengths of Lagrangian and Eulerian methods for hydrodynamics and gravity \citep{weinberger20}. AREPO employs a finite-volume solver on an unstructured Voronoi mesh that adapts to gas flows, ensuring Galilean invariance and high accuracy in capturing shocks and instabilities (\citealt{springel10}; \citealt{pakmor16}). Gravity is handled via a standard oct-tree algorithm (\citealt{barnes86}), allowing efficient computation over large volumes. This setup is ideal for our goals, as it facilitates high-resolution simulations of Milky Way-like galaxies while maintaining numerical stability. Unlike fixed-grid codes, AREPO's mesh motion reduces advection errors, enabling detailed tracking of multiphase gas interactions without artificial diffusion, which is essential for modeling realistic ISM structures.

Cooling and heating processes in SMUGGLE are modeled to achieve low-temperature gas phases necessary for molecular cloud formation and star formation. Primordial cooling includes hydrogen and helium two-body processes, Compton cooling off the CMB, and photoionization from a uniform UV background (\citealt{katz96}; \citealt{ikeuchi86}; \citealt{faucher09}). High-temperature metal-line cooling is tabulated from CLOUDY calculations (\citealt{ferland98}), scaled by gas metallicity (\citealt{vogelsberger13}). For low temperatures ($T \lesssim 10^4$ K), fine-structure and molecular cooling are fitted from CLOUDY tables (\citealt{hopkins18b}), with self-shielding corrections suppressing UV impacts at high densities ($n \gtrsim 10^{-3}$ cm$^{-3}$; \citealt{rahmati13}). Additional heating from cosmic rays (\citealt{guo08}) and photoelectric effects on dust grains (\citealt{wolfire03}) stabilizes the cold and warm ISM phases (\citealt{field69}; \citealt{wolfire95}). This network allows gas to reach $\sim 10$ K, producing dense gas regions critical for realistic star formation.

Star formation in SMUGGLE follows a stochastic, probabilistic approach, converting gas cells into stellar particles based on a density threshold and gravitational boundedness. Eligible gas must exceed $n_{\rm th} = 100$ cm$^{-3}$, typical of giant molecular clouds (\citealt{ferriere01}), and have a virial parameter $\alpha < 1$ to ensure collapse overcomes thermal and kinetic support (\citealt{semenov17}; \citealt{hopkins18b}). The star formation rate is $\dot{M}_\star = \epsilon M_{\rm gas} / t_{\rm dyn}$, with efficiency $\epsilon = 0.01$ aligned with observations (\citealt{krumholz07}), and dynamical time $t_{\rm dyn} = \sqrt{3\pi / (32 G \rho_{\rm gas})}$. Optionally, star formation rate can be scaled with molecular fraction $f_{\rm H_2}$ of the gas cell (\citealt{mckee10}; \citealt{krumholz11}; \citealt{bigiel08}). Stellar particles represent simple stellar populations with a Chabrier IMF (\citealt{chabrier01}), formed probabilistically to match the local SFR, inheriting phase-space from parent gas cells (\citealt{springel03}; \citealt{vogelsberger13}).

Stellar feedback in SMUGGLE encompasses supernovae (SNe), radiative processes, and winds from OB and AGB stars, regulating ISM dynamics and outflows. SN feedback injects energy and momentum locally, boosting momentum to its terminal value during the unresolved Sedov-Taylor phase to capture blast wave expansion (\citealt{martizzi15}; \citealt{cioffi88}; \citealt{hopkins18a}). Type II and Ia SNe are modeled with Poisson-sampled discrete events, using IMF integration and a delay-time distribution, respectively \citep[][for more details]{vogelsberger13}. Momentum is distributed radially to neighboring cells via solid-angle weights, limited by a user-defined "superbubble" radius to avoid artificial over-coupling (\citealt{maclow88}; \citealt{weaver77}). Radiative feedback from young massive stars includes photoionization, imposing a $1.7 \times 10^4$ K temperature floor probabilistically in H II regions (\citealt{rybicki86}), and radiation pressure with IR optical depth for multiple scatterings in dense gas ($\kappa_{\rm IR} = 10 (Z/Z_\odot)$ cm$^2$ g$^{-1}$; \citealt{agertz13}; \citealt{hopkins18b}). 
OB winds preprocess the gas surrounding young stars before the onset of SNe \citep{matzner02, krumholz09}, with momentum injection and mass loss parameterized using metallicity-dependent fits \citep{hopkins18a, marinacci19}. AGB winds inject energy both before and after SNe \citep{agertz13}, and the stellar evolution model for AGB stars is parameterized as described in \cite{vogelsberger13}. These channels collectively drive turbulence and outflows, producing galactic winds and regulating star formation \citep{li17}, as is also shown in similar models \citep[e.g., TIGRESS,][]{kim18}.

\subsection{Initial Conditions}\label{sec:ic}
Our initial conditions consist of a disk and a halo that are comparable to those of a Milky Way-mass galaxy. The mass of a classical bulge is the only property that is varied to study its effects on bar formation and the ensuing nuclear ring formation. The disk includes a stellar and a gaseous disk. We first construct the collisionless disk-halo systems for our galaxy models using the \textsc{galic} code \citep{yurin14}, which is publicly available. \textsc{galic} iteratively modifies the velocities of particles to achieve equilibrium for a specified density structure. We then convert a fraction of the stellar particles into gas particles to construct gaseous disks of our initial conditions.

The density distribution of the entire disk component is given by
$$\rho_{d} (R, z) = \frac{M_{d}}{4\pi z_{d} R_{d}^2} \exp \left( -\frac{R}{R_{d}} \right) \text{sech}^2 \left(\frac{z}{z_{d}}\right),$$
where $M_{d}$ is the total mass of the disk, $z_{d}$ is the vertical scale height, and $R_{d}$ is the radial scale length, which is determined by the disk spin parameter introduced in \cite{mo98}.
Here, $R$ and $z$ denote the radial and vertical distances in cylindrical coordinates, respectively. For our galaxy models, we assign $R_d=3.0 \, \kpc$ with $R_d/z_d=10$  comparable to the Milky Way's size. We assign $M_{d}=6\times10^{10} \, \Msun$ and specify the velocity anisotropy parameter $f_R = \sigma_{R}^2 / \sigma_{z}^2$ in \textsc{galic}, fixing $f_R = 1.4$ for all models. This initial choice of $f_R$ is lower than the observed values of $f_R \sim 4$ for the Milky Way in the solar neighborhood \citep{sharma14, guiglion15}, as we aim to generate cold disks that form a bar in isolation. The bar formation leads to an increase in $\sigma_R$ \citep{kwak17, seo19}, gradually approaching the observed value in the Milky Way \citep{seo19}.

To construct the collisional gas disk, we convert $5 \times 10^{9} \, \Msun$ of the stellar disk particles into gas particles, setting the gas fraction $M_g / M_s \approx 9.1\%$ or $M_g / (M_s+M_g) \approx 8.3\%$ in all models, where $M_g$ and $M_s$ are the masses of the gas disk and stellar disk, respectively. This choice is motivated by the observed gas fractions in Milky Way-mass galaxies, which are typically $<10\%$ \citep{sage93,leroy09,saintonge11,papovich16}. Thus, the stellar disk has $M_s = 5.5 \times 10^{10} \, \Msun$, and the gas and stellar disks share the same density structure. To assess the impact of initial relaxation, we evolve the system without star formation for 0.1 Gyr, which is similar to the approach in \cite{seo19}, to allow it to gradually relax into a quasi-equilibrium state before enabling star formation. Note that \cite{seo19} reduced the scale height of the gas disk from 0.3 kpc to 0.1 kpc before performing the relaxation process, whereas our gas disk initially has the same scale height as the stellar disk, resulting in a small offset from equilibrium. Furthermore, the stellar feedback included in \textsc{SMUGGLE} generates galactic outflows and bubbles \citep{marinacci19,li24} shortly after star formation is activated. Combined, these factors render the effect of the quasi-equilibrium relaxation step, as in \cite{seo19}, negligible in our simulations.

The dark matter (DM) halo is spherically symmetric and follows a \cite{hernquist90} profile
\begin{equation}\label{eq:hernquist}
\rho_\text{DM} (r) = \frac{M_\text{DM}}{2\pi} \frac{a_h}{r(r+a_h)^3} ,
\end{equation}
where $a_h$ is the scale length of the halo and $M_\text{DM}$ is the total mass. The scale length of the DM halo depends on the concentration parameter $c$ as \begin{equation}\label{eq:cc}
a_h = \frac{r_{200}}{c} \left[2 \ln(1+c)-\frac{c}{1+c}\right]^{1/2},
\end{equation}
where $r_{200}$ is the virial radius \citep{springel05}. 
In all models, we fix the halo concentration and the total mass of the dark matter as  $c=14$ and $M_\text{DM} = 1.14  \times 10^{12} \Msun$. The choice of $c$ falls within the range inferred from the Aquarius and TNG simulations \citep{springel08, bose19}. 

For the stellar bulge, we also adopt a Hernquist profile given by
\begin{equation}\label{eq:hernquist}
\rho_b (r) = \frac{M_b}{2\pi} \frac{a_b}{r(r+a_b)^3} .
\end{equation}
The bulge scale length $  a_b  $ is fixed at $  0.40\,\mathrm{kpc}  $, which corresponds to a 3D half-mass radius of $  \sim 0.97\,\mathrm{kpc}  $ (the 2D projected effective radius is $\sim 0.73\,\mathrm{kpc} $). This bulge size is consistent with the scaling relations between classical bulge effective radius and galaxy stellar mass for Milky Way-mass disk galaxies \citep{gadotti09}. The bulge mass is the only variable in our models. As shown in Table \ref{table:model}, we include one bulgeless disk-halo model, r1c14b00, while r1c14b05 and r1c14b10 have bulge masses of $2.5\times10^9,\Msun$ and $5.0\times10^9,\Msun$, respectively. These masses correspond to bulge-to-disk ratios of B/D $\approx$ 0.045 and 0.091, which lie well within the observed range for late-type barred galaxies \citep{gadotti09} and are consistent with the mean classical bulge-to-total stellar mass ratio of $ \approx 0.06  $ reported for barred S0$-$Sb galaxies \citep{erwin15}.

All models share the same baryonic mass resolution of $10^4 \Msun$ and DM resolution of $10^5\Msun$.  Since the bulge mass is the only variable, all models have the same number of stellar disk, gas disk, and DM particles: $N_s = 5.5 \times 10^6$, $N_g = 0.5 \times 10^6$, and $N_{\rm DM} = 11.4 \times 10^6$. The number of particles in the stellar bulge $N_b$ is 0, $2.5 \times 10^5$, and $5.0 \times 10^5$ for models r1c14b00, r1c14b05, and r1c14b10, respectively. The mass ratio between baryon and DM particles is 10 in all models to prevent the gravitational heating triggered by massive DM particles \citep{kwak25, kwak26}. The gravitational softening length for stellar particles, including newly formed stars, is $0.03 \, \kpc$, while that for DM particles is $0.42 \, \kpc$ based on the mean particle separation within the scale radius. The minimum adaptive softening length for gas cells is $0.01 \, \kpc$. We refer to the stellar particles distributed in the initial conditions as initial stars and to the newly formed stars as new stars. All three models are evolved for 5 Gyr in isolation, without additional gas and external perturbations. The final values of properties such as bar pattern speed, corotation radius, and ring size are listed in Table \ref{table:model}.

\begin{table}[h!]
\caption{\label{ic} Galaxy Models}
\centering
\begin{tabular}{lcccccc}
\hline\hline
Model & $M_b$ & $B/D$ & $\Omega_{b}(t_{f})$ & $R_{\rm CR}(t_{f})$ & $R_{\rm ring}(t_{f})$ \\
 & [$10^{9}\Msun$] &  & [km s$^{-1}$kpc$^{-1}$] & [kpc] & [kpc] \\
\hline
r1c14b00 & 0   & 0     & 35.8 & 4.4 & no ring \\
r1c14b05 & 2.5 & 0.045 & 42.3 & 3.9 & 0.64 \\
r1c14b10 & 5.0 & 0.091 & 48.7 & 3.6 & 0.76 \\
\hline
\end{tabular}
\tablefoot{Column (1) is the model name, column (2) gives the bulge mass, and column (3) is the bulge-to-disk mass ratio ($B/D$). Columns (4) and (5) are the bar pattern speed and corotation radius at $t_f = 5$ Gyr. Column (6) is the radius of the nuclear ring calculated from its center of mass. The bulge mass is the only variable in this work. The time evolution of bar pattern speed is shown in Figure \ref{fig:barpattern}.}
\label{table:model}
\end{table}

\begin{figure}[htbp]
    \centering
    \includegraphics[width=0.5 \textwidth]{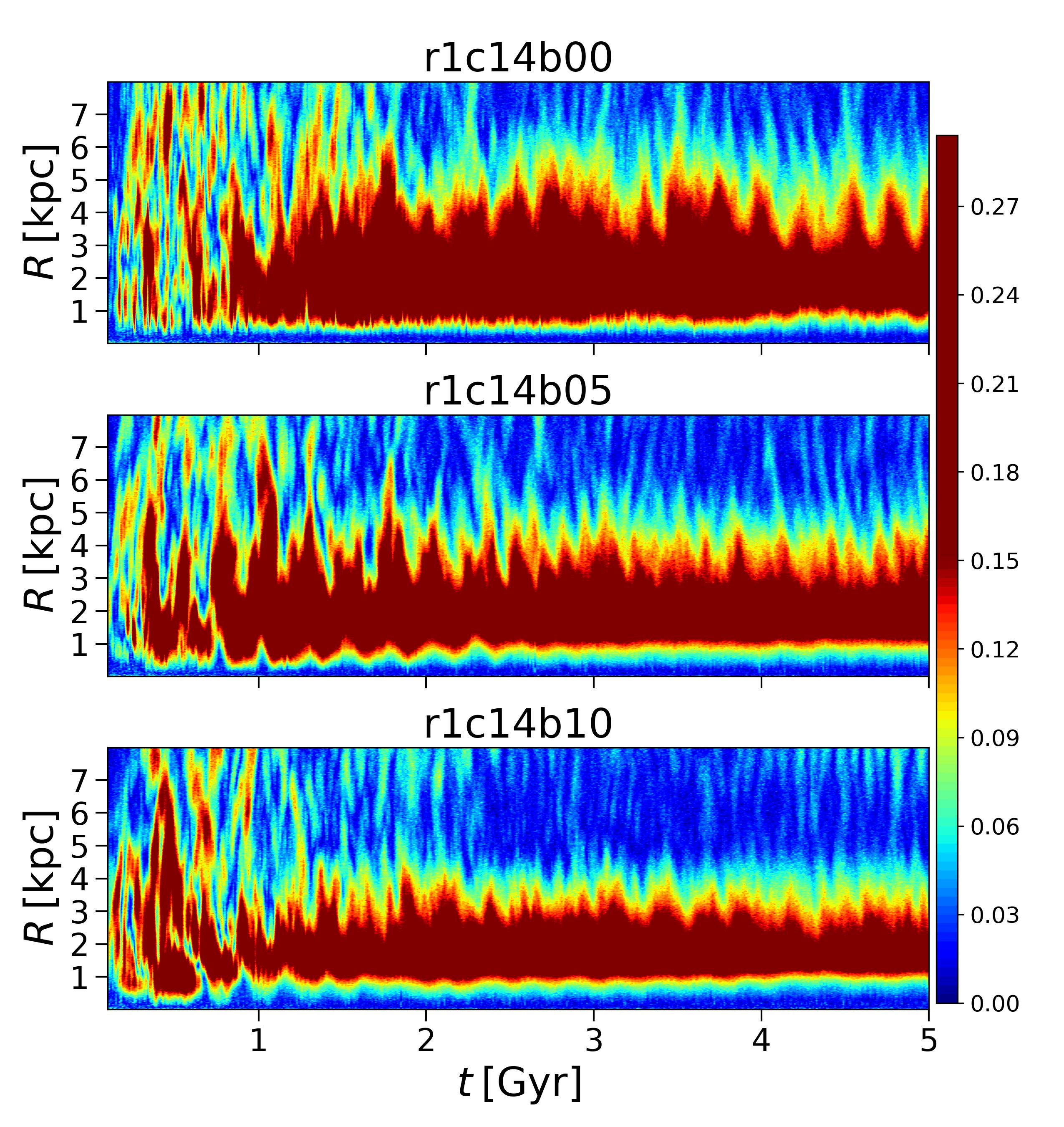}
    \caption{Distribution of the Fourier mode $m=2$ within 8 $\kpc$, illustrated by the time evolution of the radial Fourier distributions for each model. The time interval between snapshots is 0.01 Gyr. To ensure that the same color corresponds to the same value for ease of comparison, the color distribution in the color bar is fixed for each component across all models, so all values above 0.15 are shown in red. Only the initial stellar disk particles are selected to calculate the Fourier mode $m=2$, after excluding the initial classical bulge component. The radial profiles of the $m=2$ modes at representative times are presented in Figure \ref{fig:f2radial}.}
    \label{fig:f2map}
\end{figure}

\section{Results}\label{sec:results}

\begin{figure*}[htbp]
    \centering
    \includegraphics[width=0.33\textwidth]{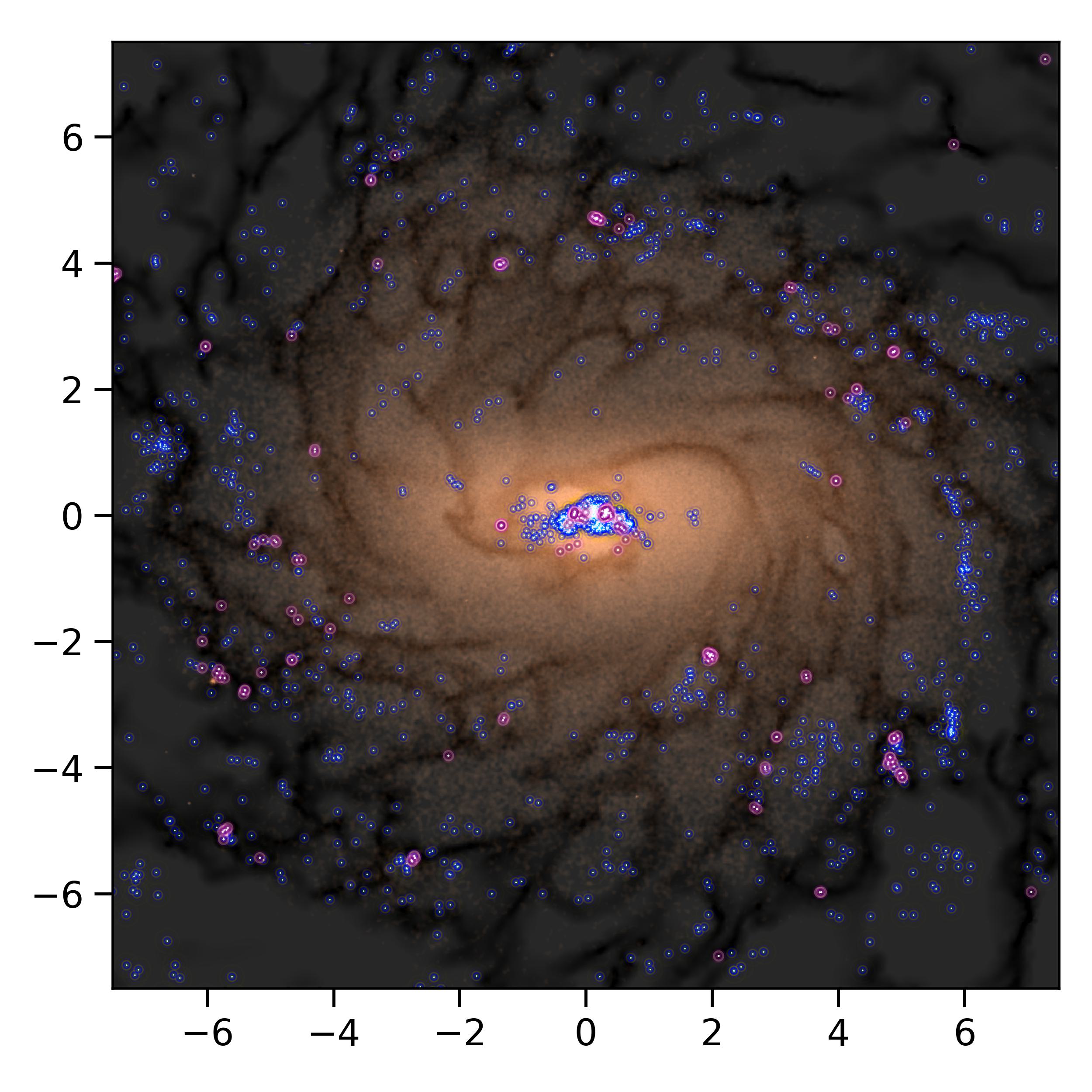}\includegraphics[width=0.33\textwidth]{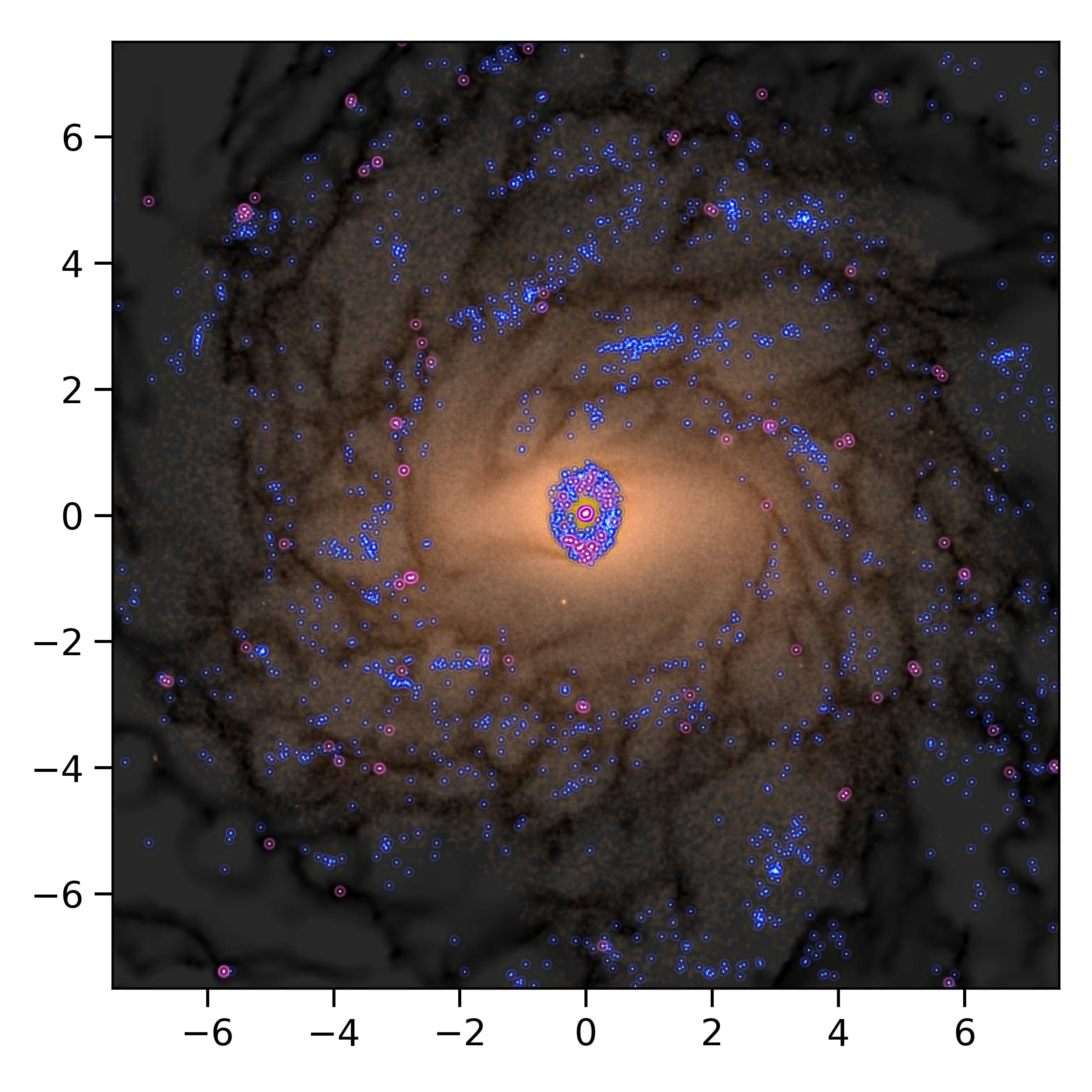}\includegraphics[width=0.33\textwidth]{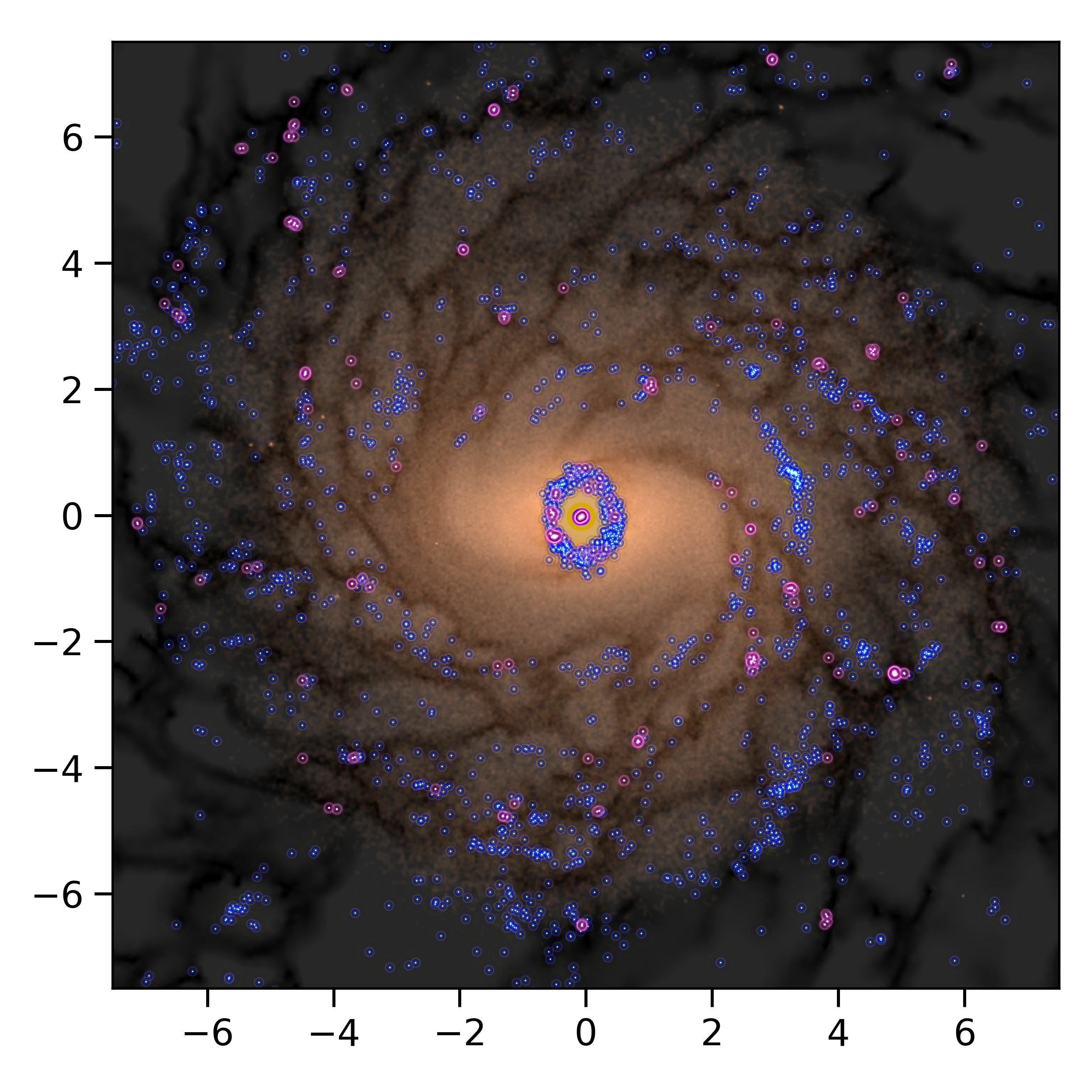}
    \caption{Face-on projections of the stacked surface density distributions of stars and gas in the X-Y plane within a $15 \times 15\,\kpc^{2}$ box at 2.5 Gyr for the models r1c14b00, r1c14b05, and r1c14b10 (from left to right). The yellowish-brown colors represent the stellar distribution, while the dark regions indicate the gas distribution. Overlaid blue and purple points mark young stars with ages $0.01 < t_\mathrm{age} < 0.1 \, \Gyr$ and $t_\mathrm{age} \leq 0.01 \, \Gyr$, respectively. The same projections at 1, 2, 3, 4, and 5 Gyr are shown in Fig.~\ref{fig:face_stack}. The separate surface density distributions of stars and gas are presented in Fig.~\ref{fig:face_stacking}.}
    \label{fig:face_250}
\end{figure*}

\begin{figure*}[t!]
    \centering
    \renewcommand{\arraystretch}{0}
    \begin{tabular}{@{}c@{}c@{}c@{}c@{}c@{}c@{}}
        & \textbf{r1c14b00} & \textbf{r1c14b05} & \textbf{r1c14b10} & \textbf{r1c14b10 (zoom-in)} \\
        \raisebox{2\height}{\rotatebox{90}{\textbf{1 Gyr}}} &
        \includegraphics[width=0.2\textwidth]{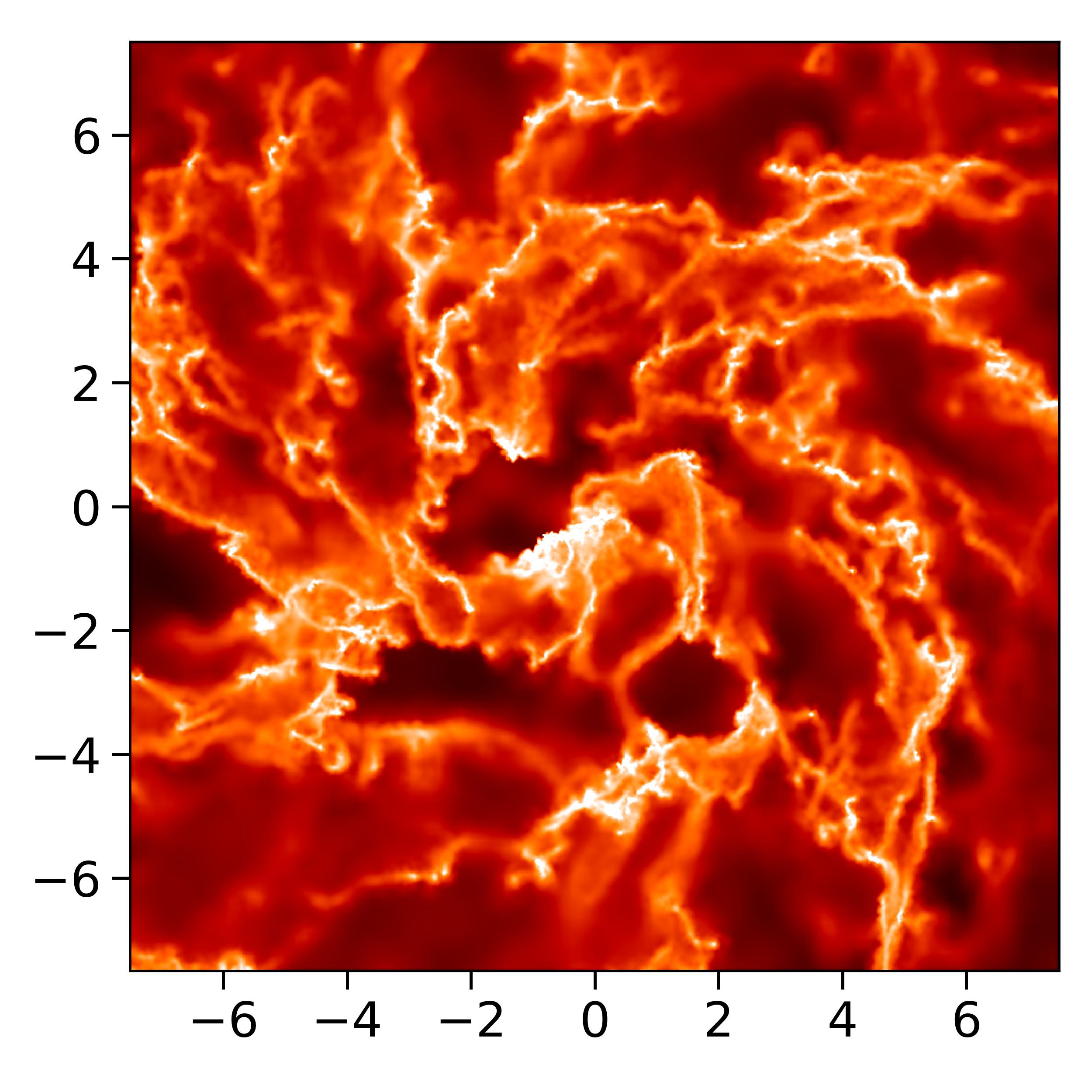} &
        \includegraphics[width=0.2\textwidth]{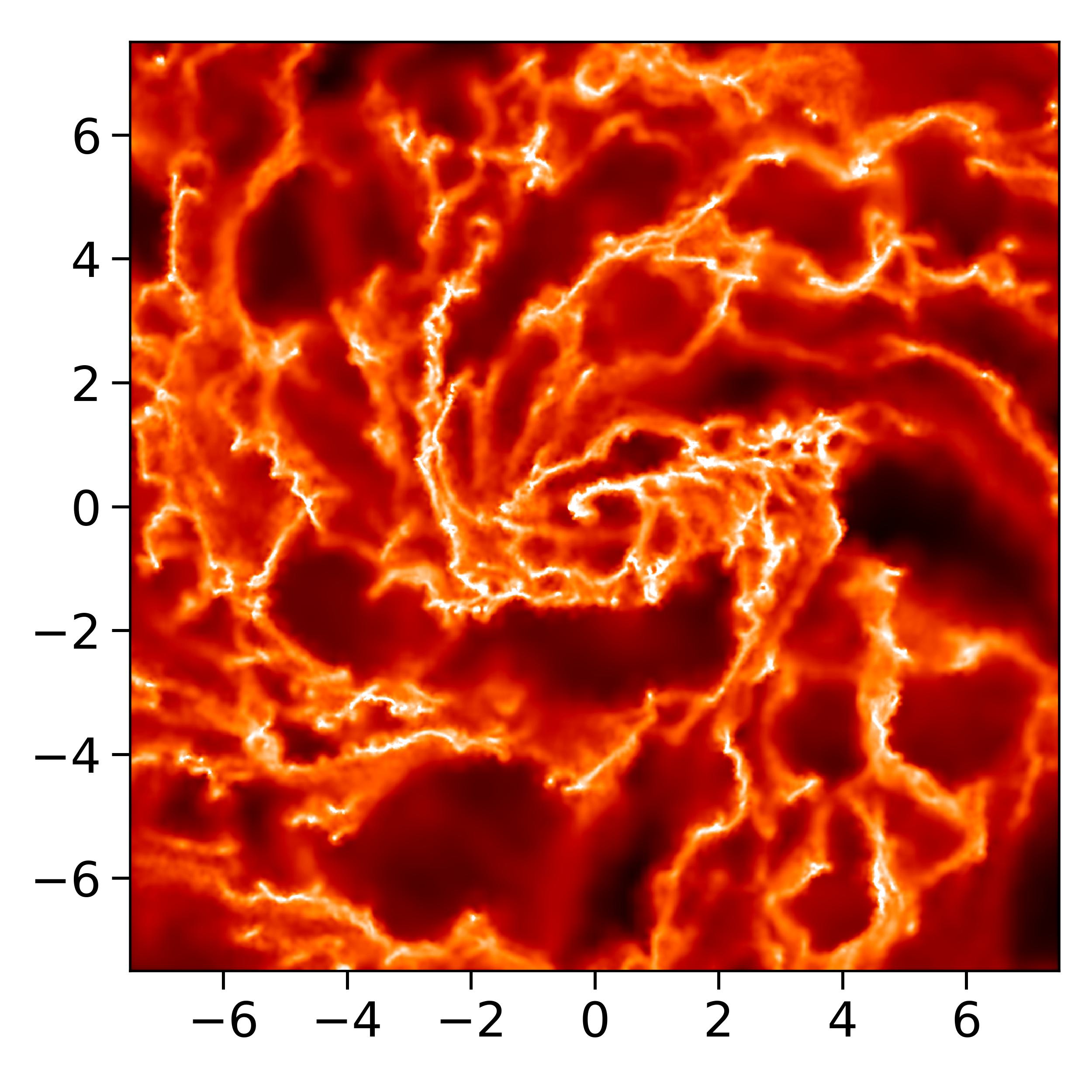} &
        \includegraphics[width=0.2\textwidth]{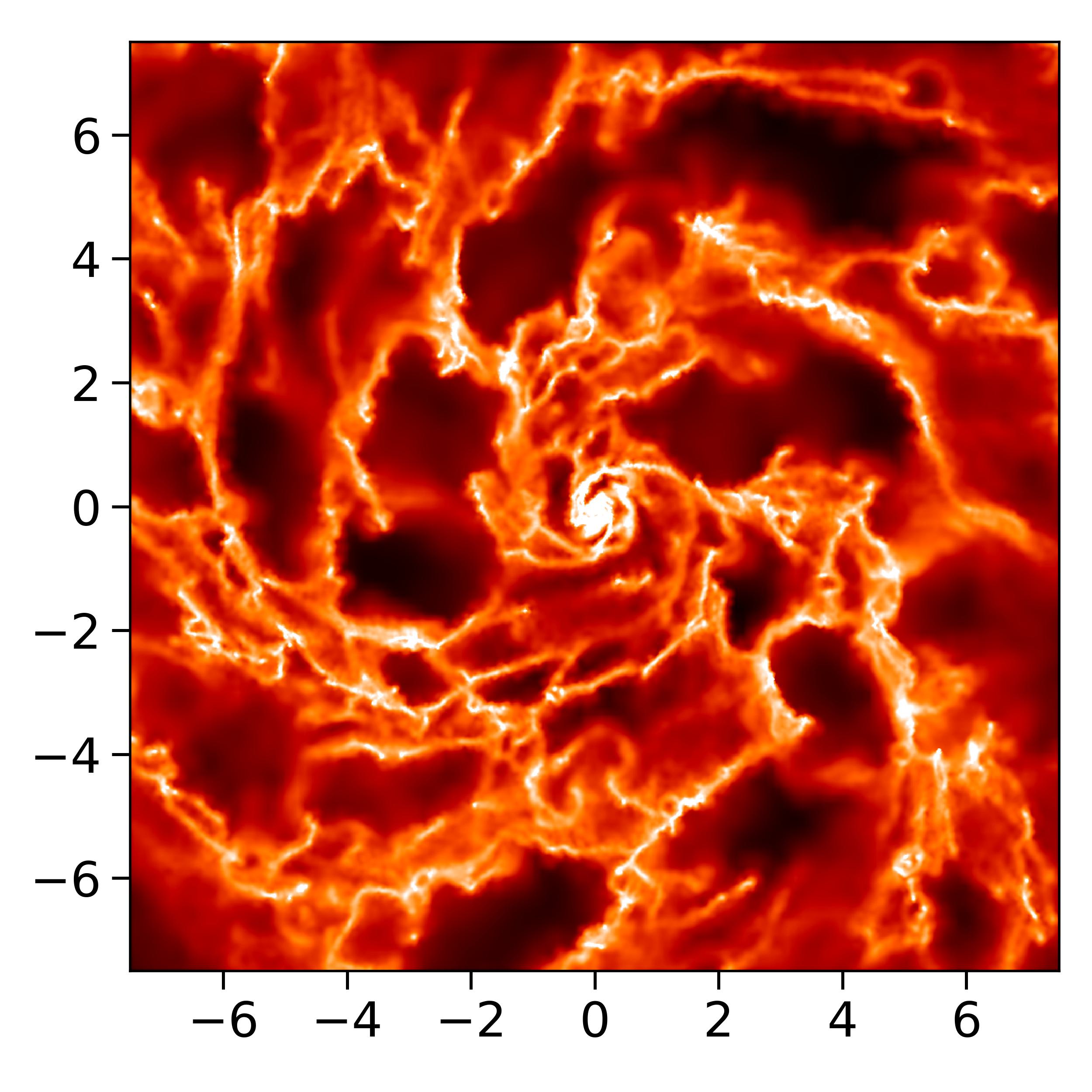} &
        \includegraphics[width=0.2\textwidth]{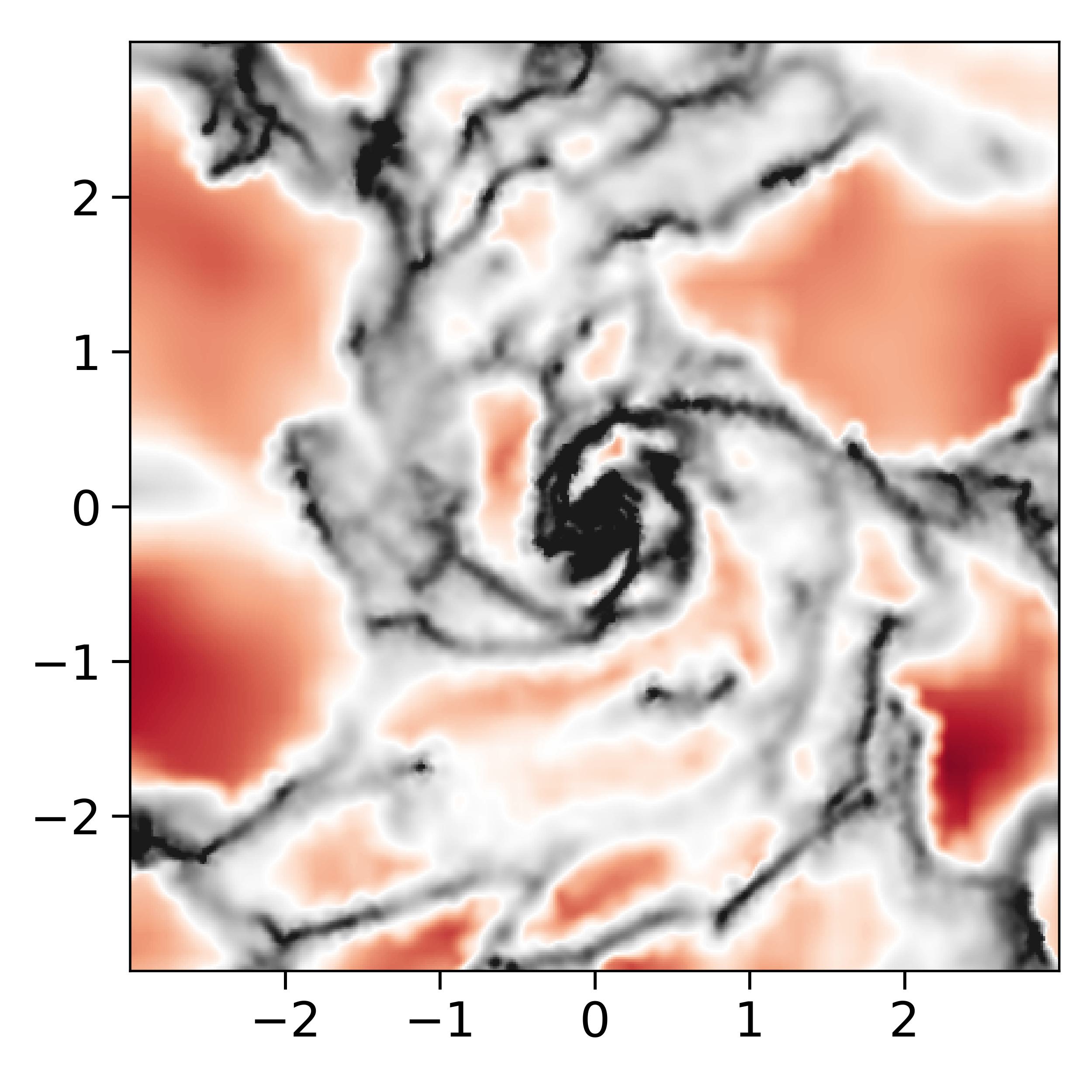}
        \\
        \raisebox{2\height}{\rotatebox{90}{\textbf{2 Gyr}}} &
        \includegraphics[width=0.2\textwidth]{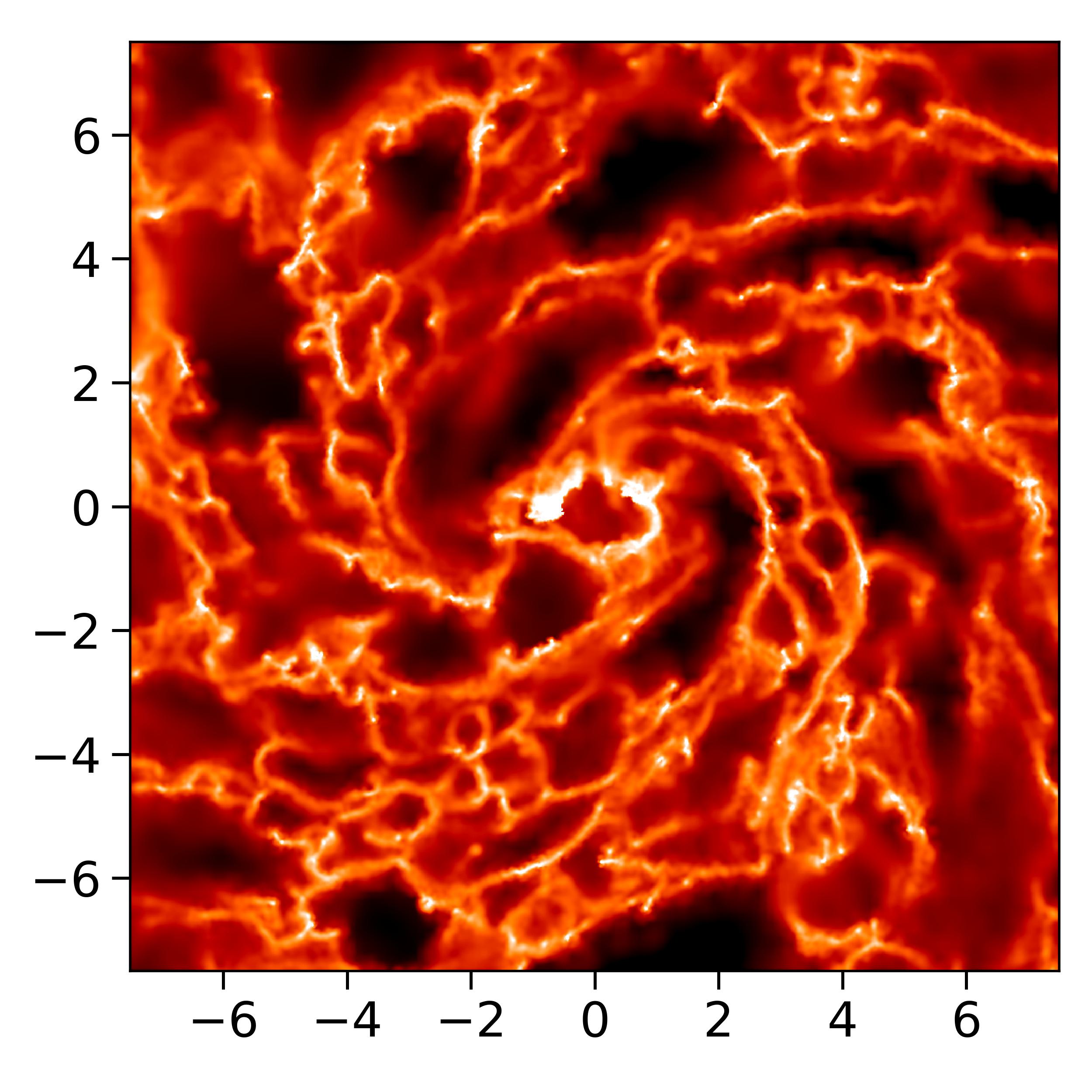} &
        \includegraphics[width=0.2\textwidth]{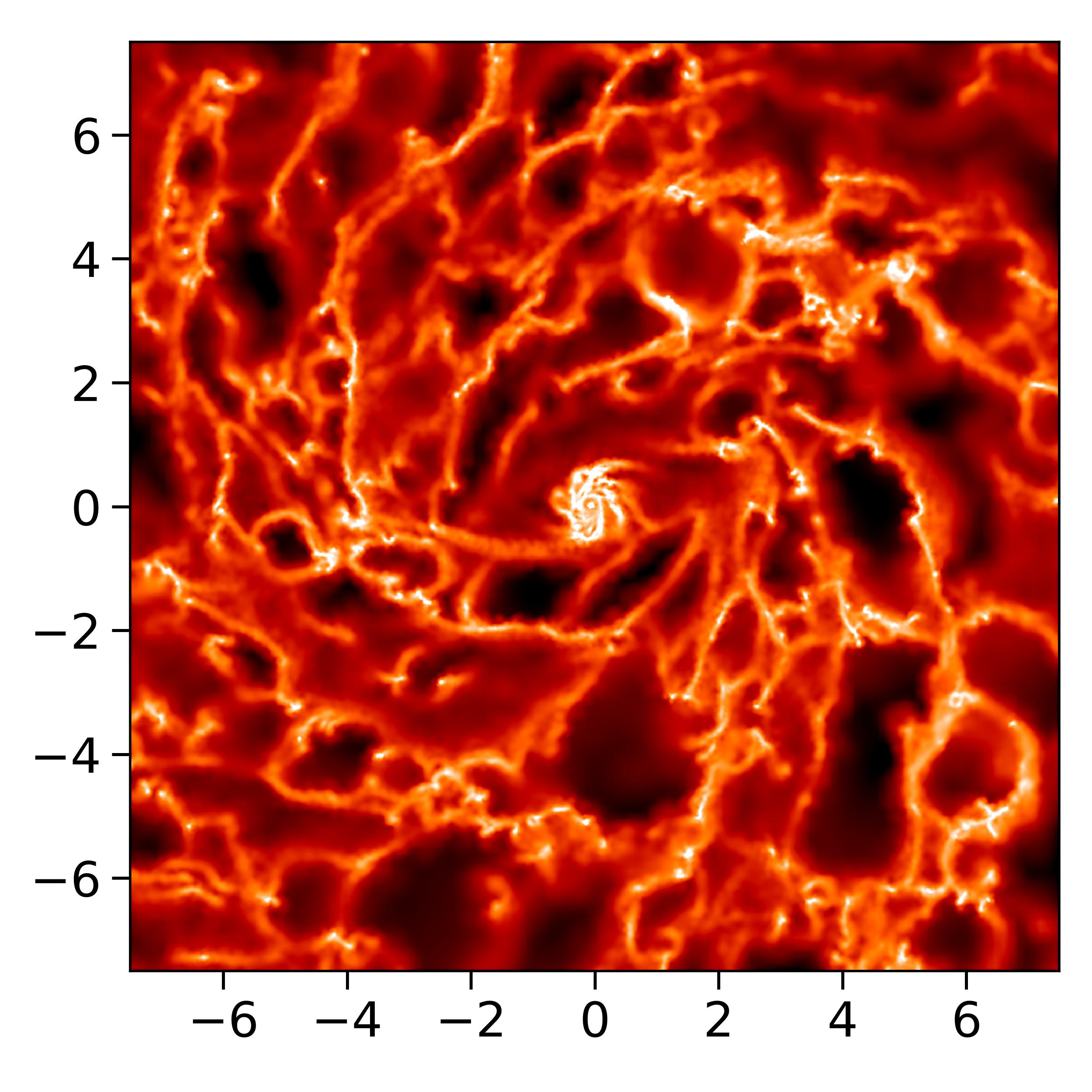} &
        \includegraphics[width=0.2\textwidth]{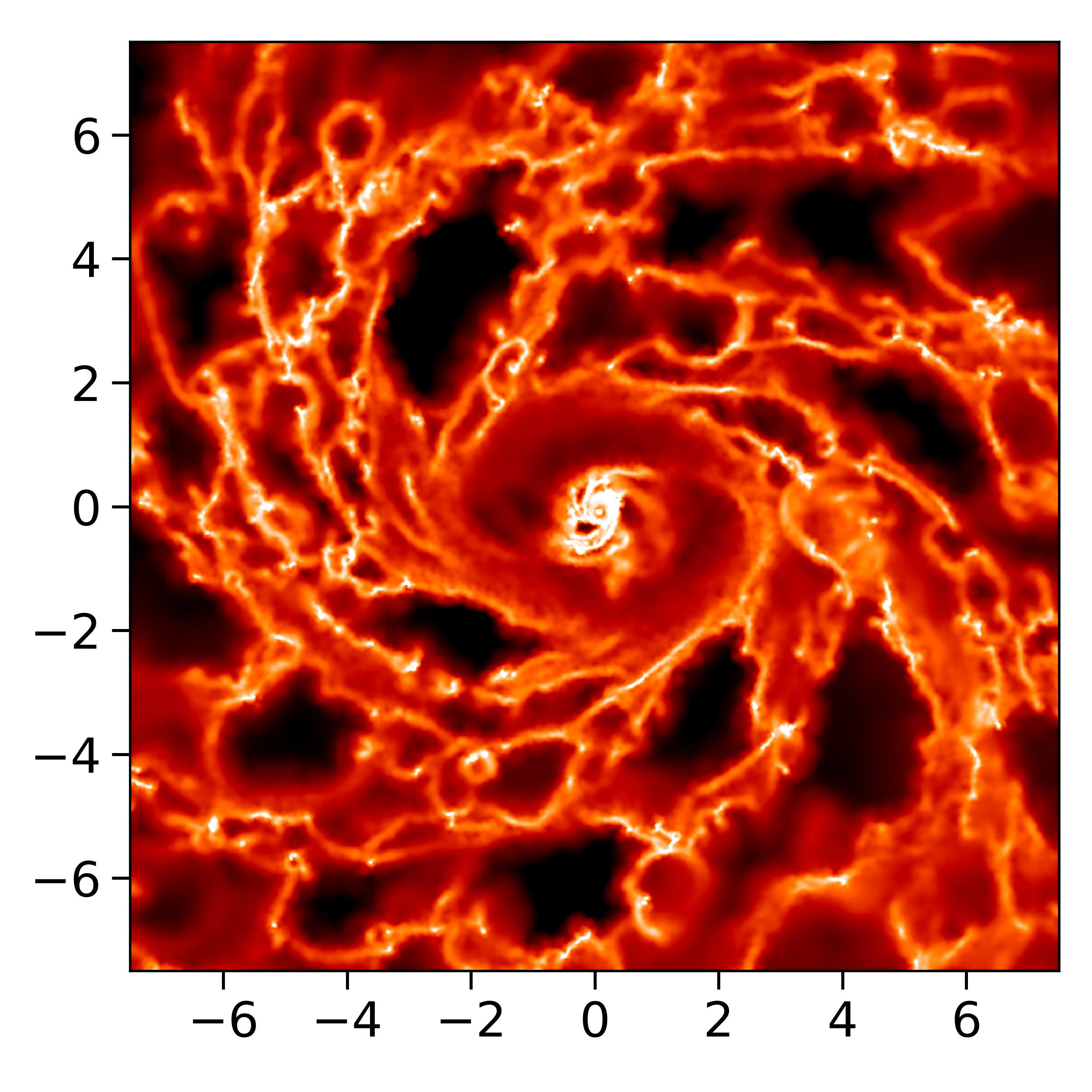} &
        \includegraphics[width=0.2\textwidth]{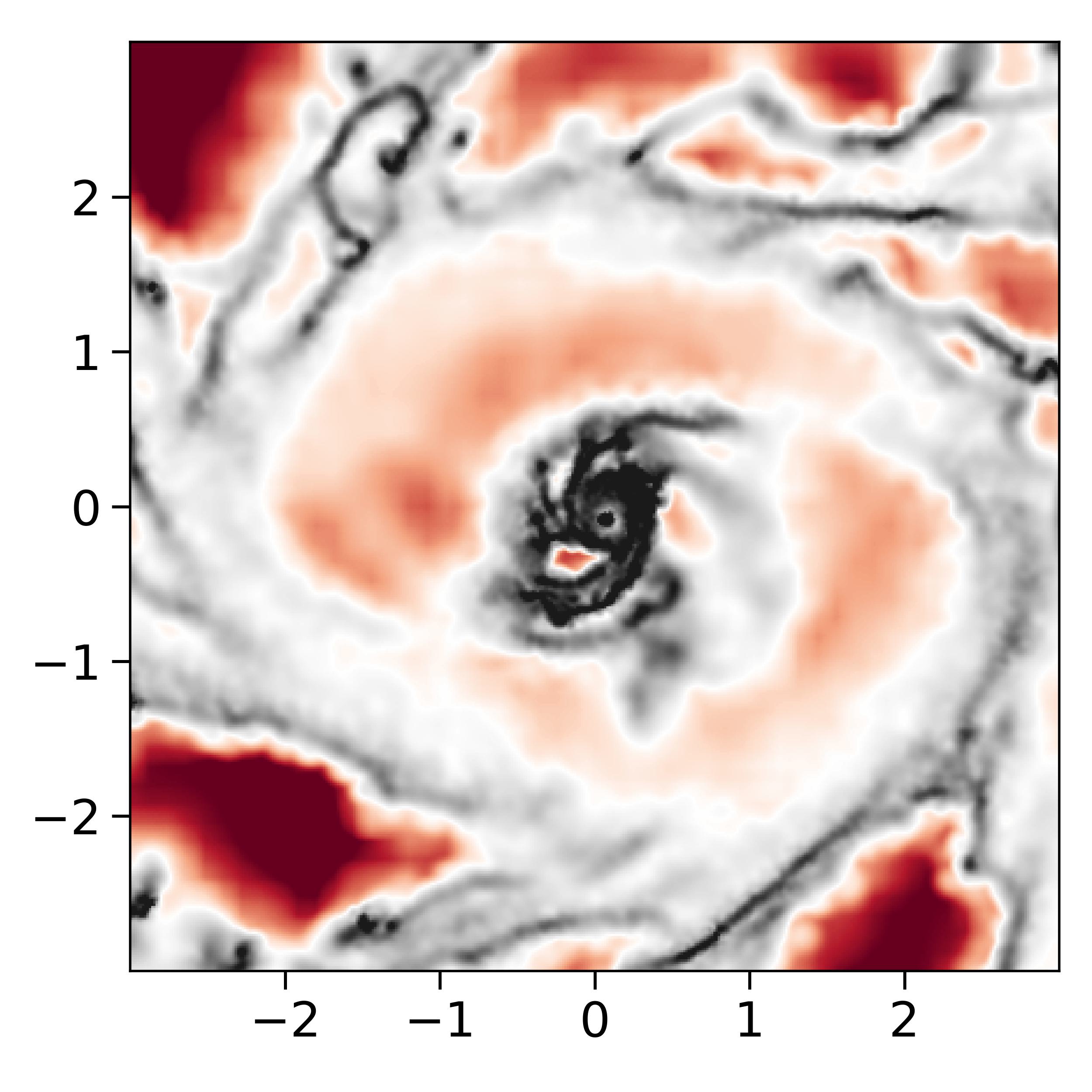}
        \\
        \raisebox{2\height}{\rotatebox{90}{\textbf{3 Gyr}}} &
        \includegraphics[width=0.2\textwidth]{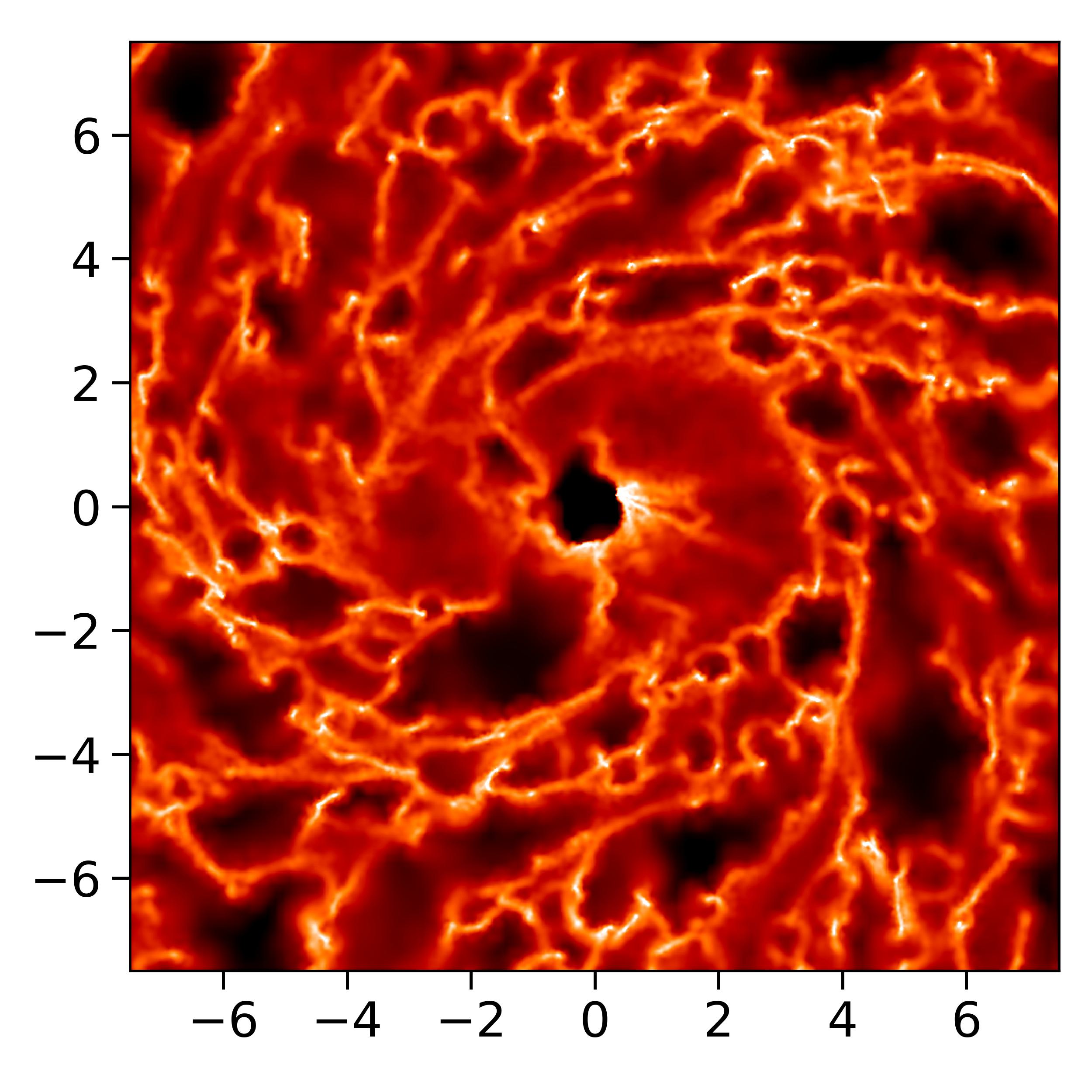} &
        \includegraphics[width=0.2\textwidth]{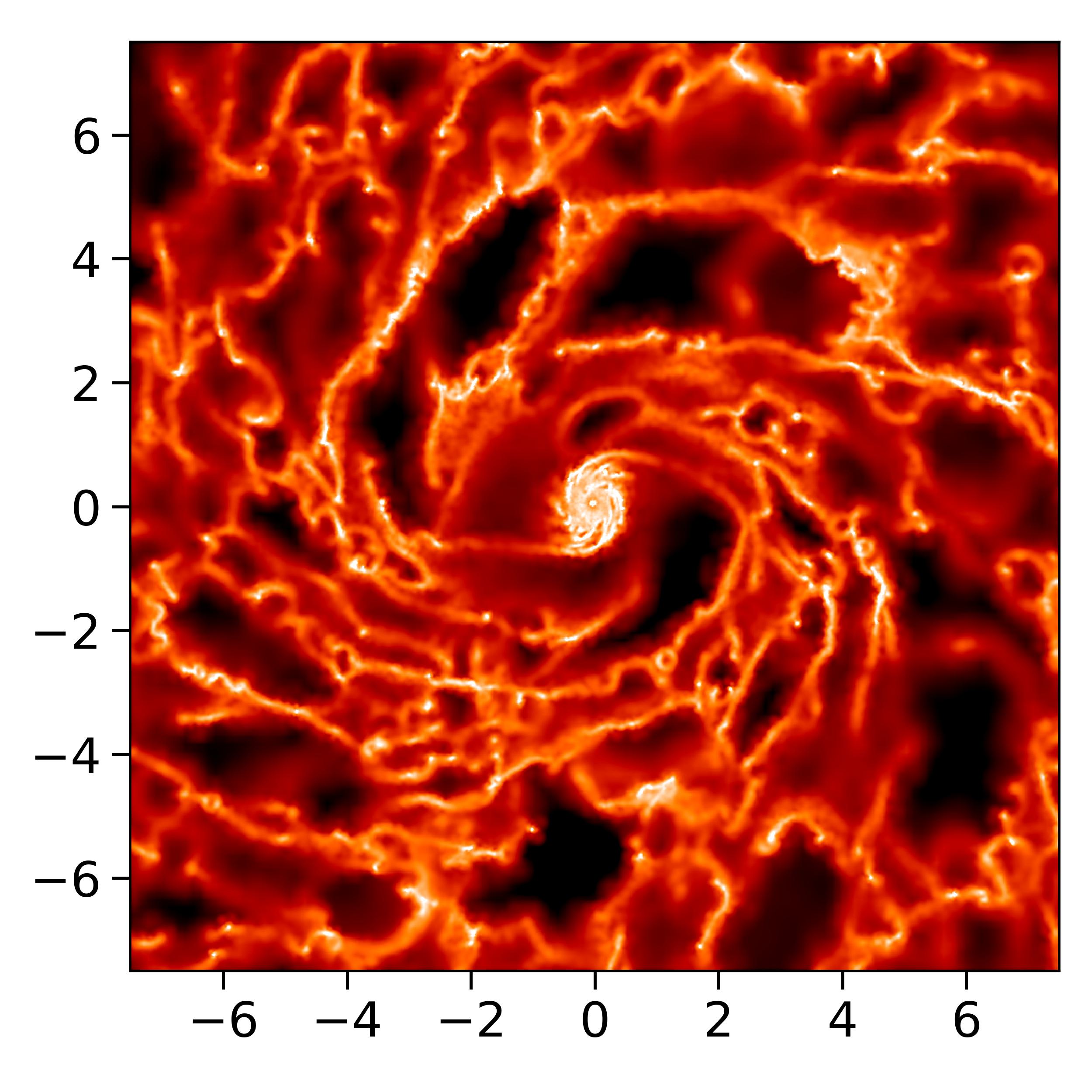} &
        \includegraphics[width=0.2\textwidth]{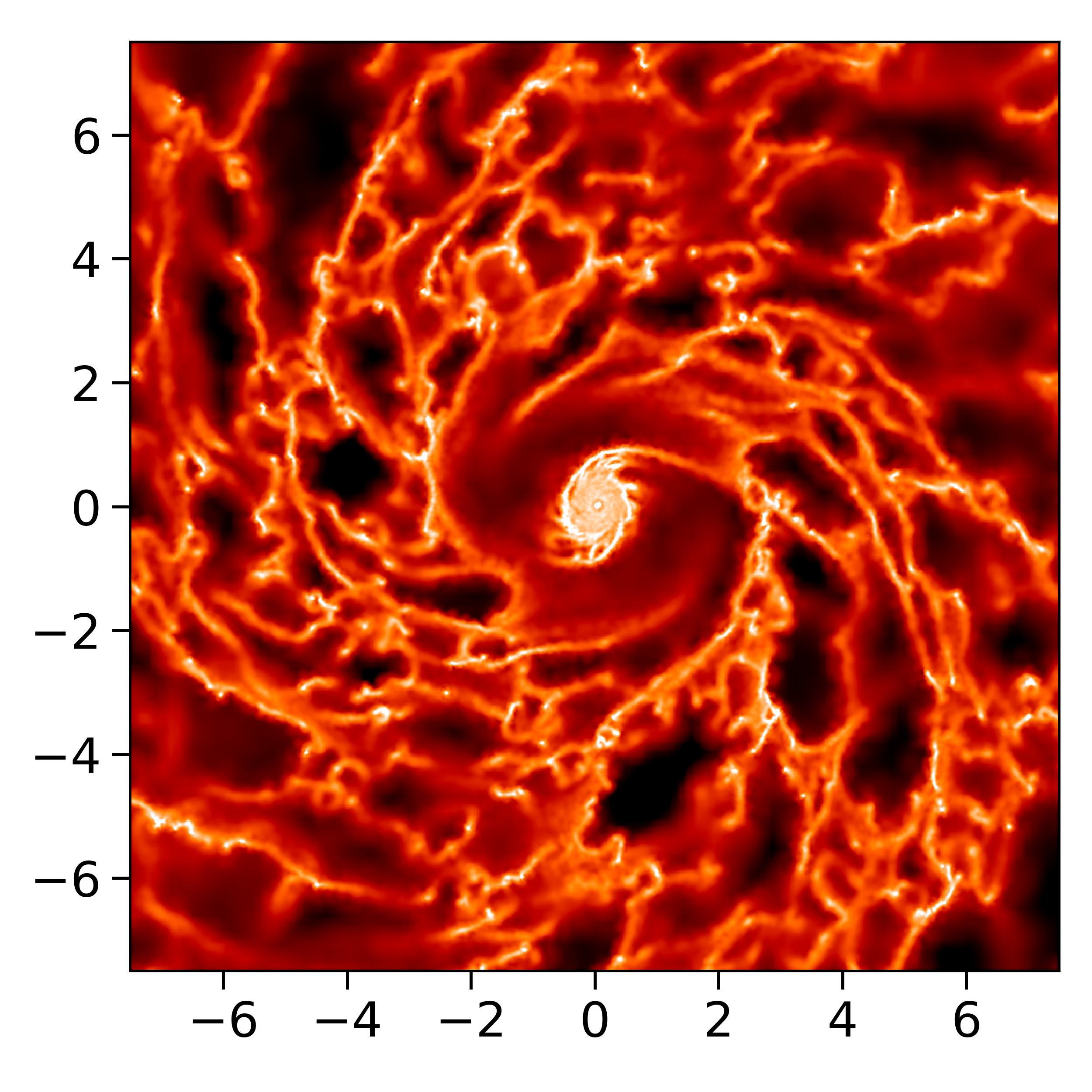} &
        \includegraphics[width=0.2\textwidth]{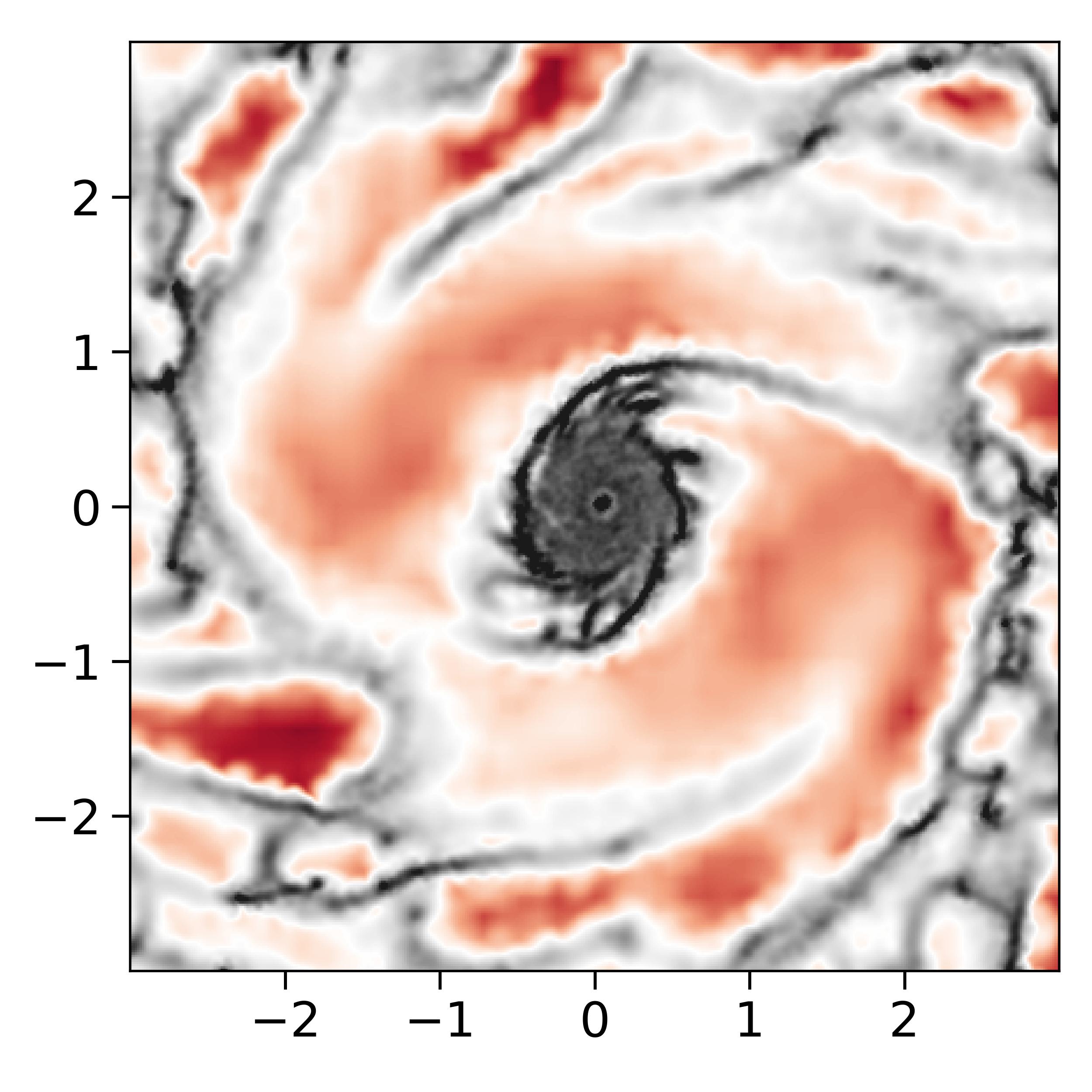}
        \\
        \raisebox{2\height}{\rotatebox{90}{\textbf{4 Gyr}}} &
        \includegraphics[width=0.2\textwidth]{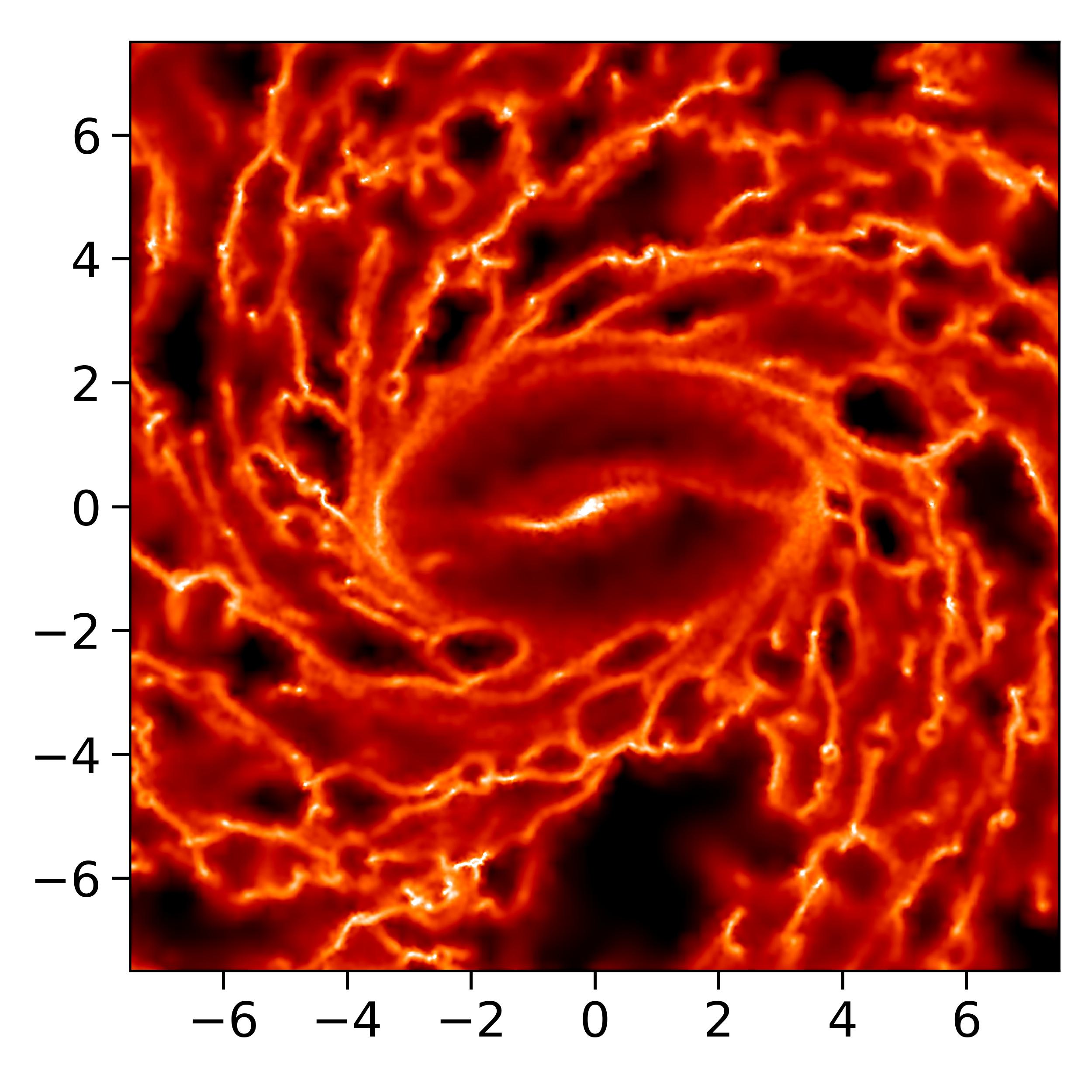} &
        \includegraphics[width=0.2\textwidth]{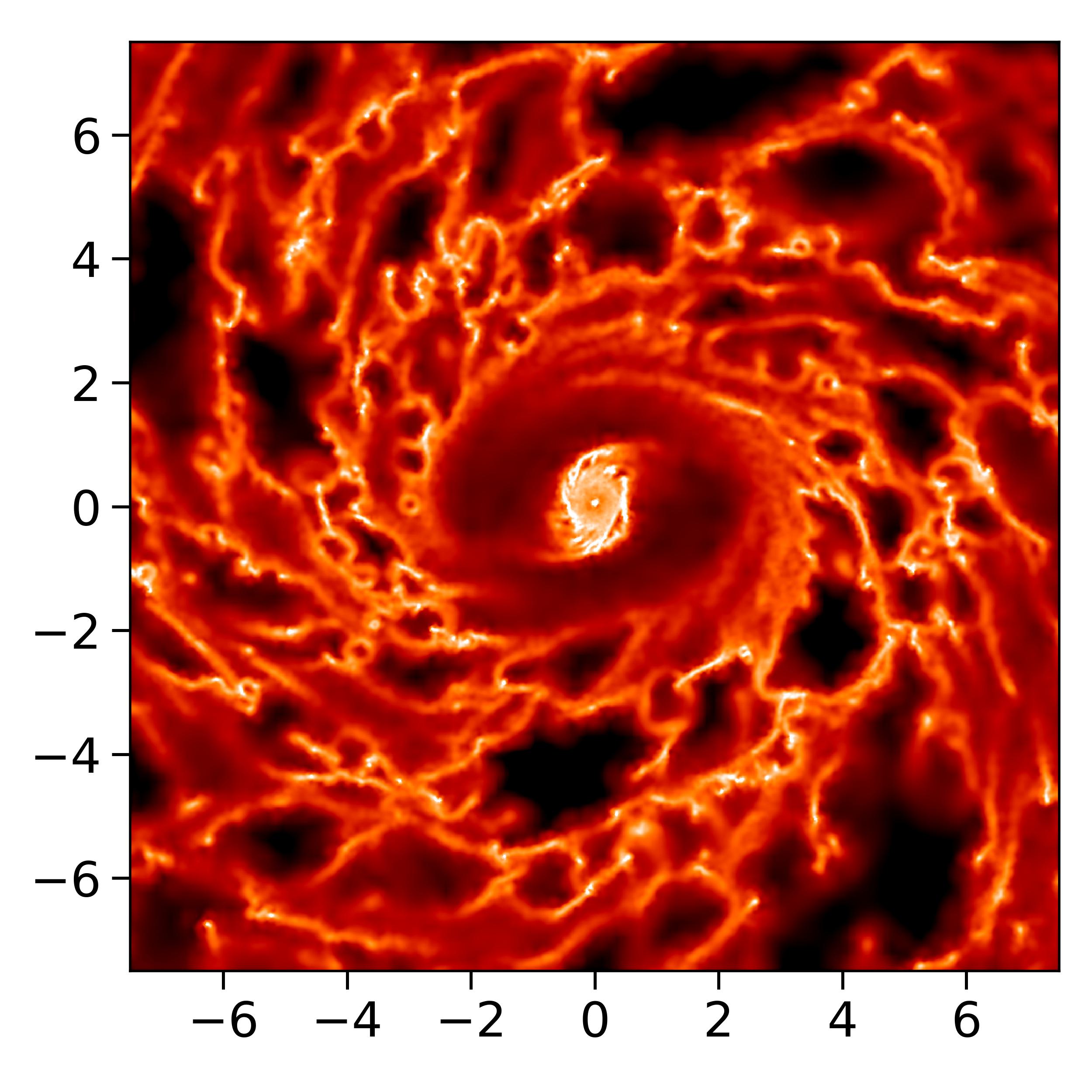} &
        \includegraphics[width=0.2\textwidth]{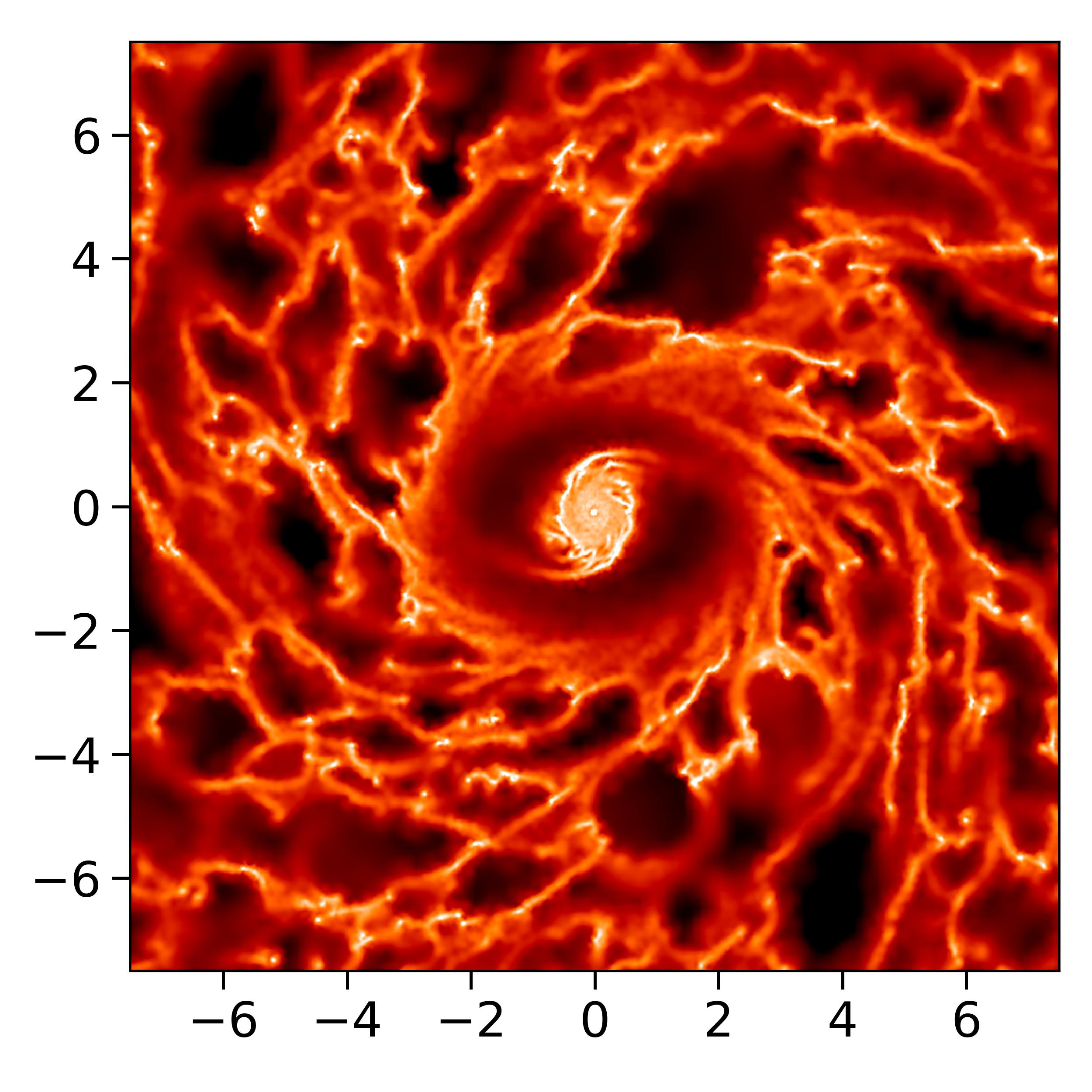} &
        \includegraphics[width=0.2\textwidth]{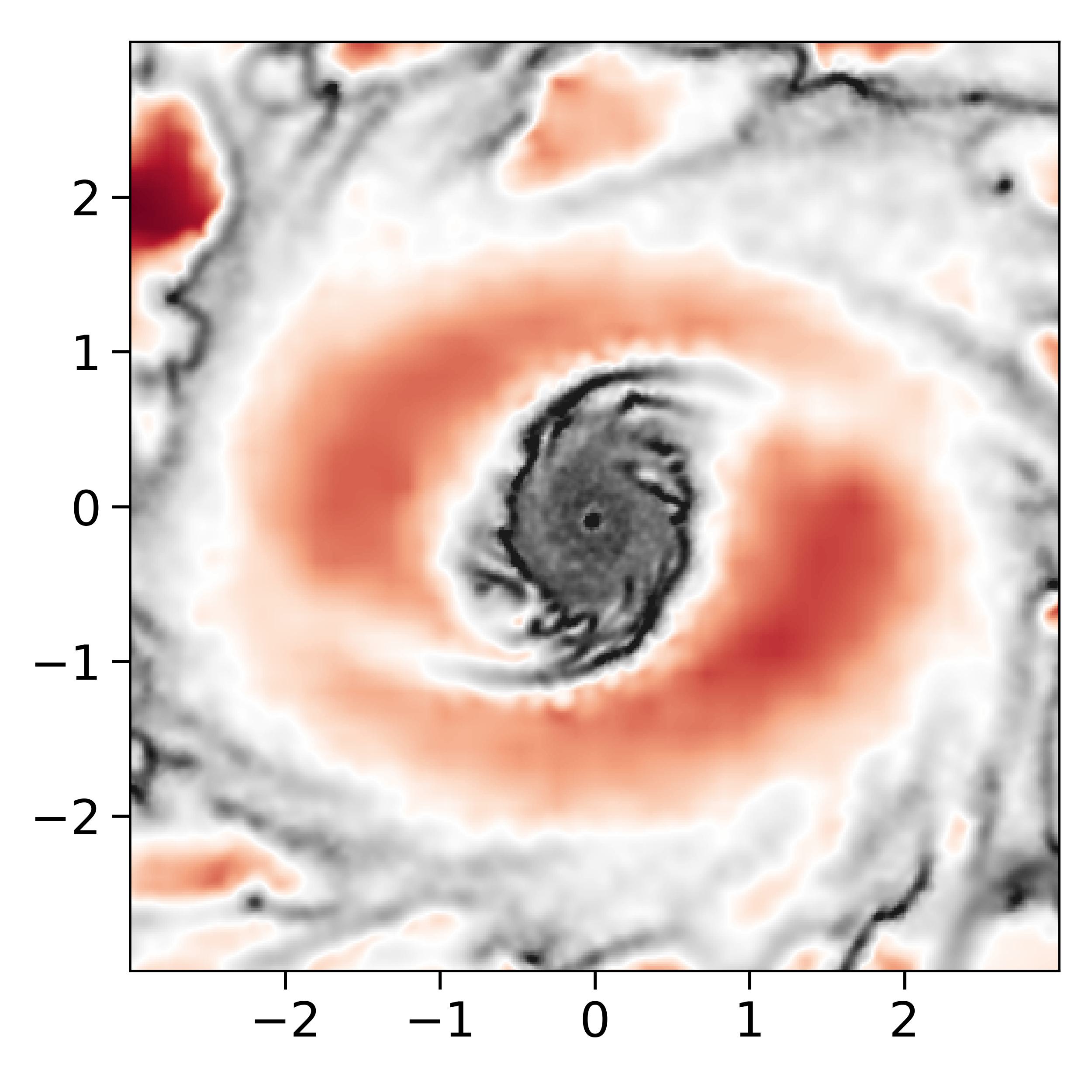}&
        \includegraphics[height=0.2\textwidth]{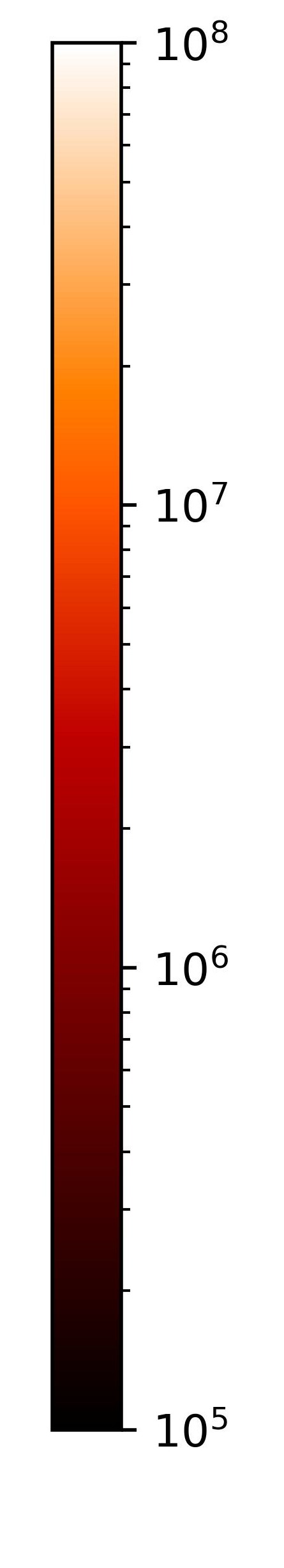}
        \\
        \raisebox{2\height}{\rotatebox{90}{\textbf{5 Gyr}}} &
        \includegraphics[width=0.2\textwidth]{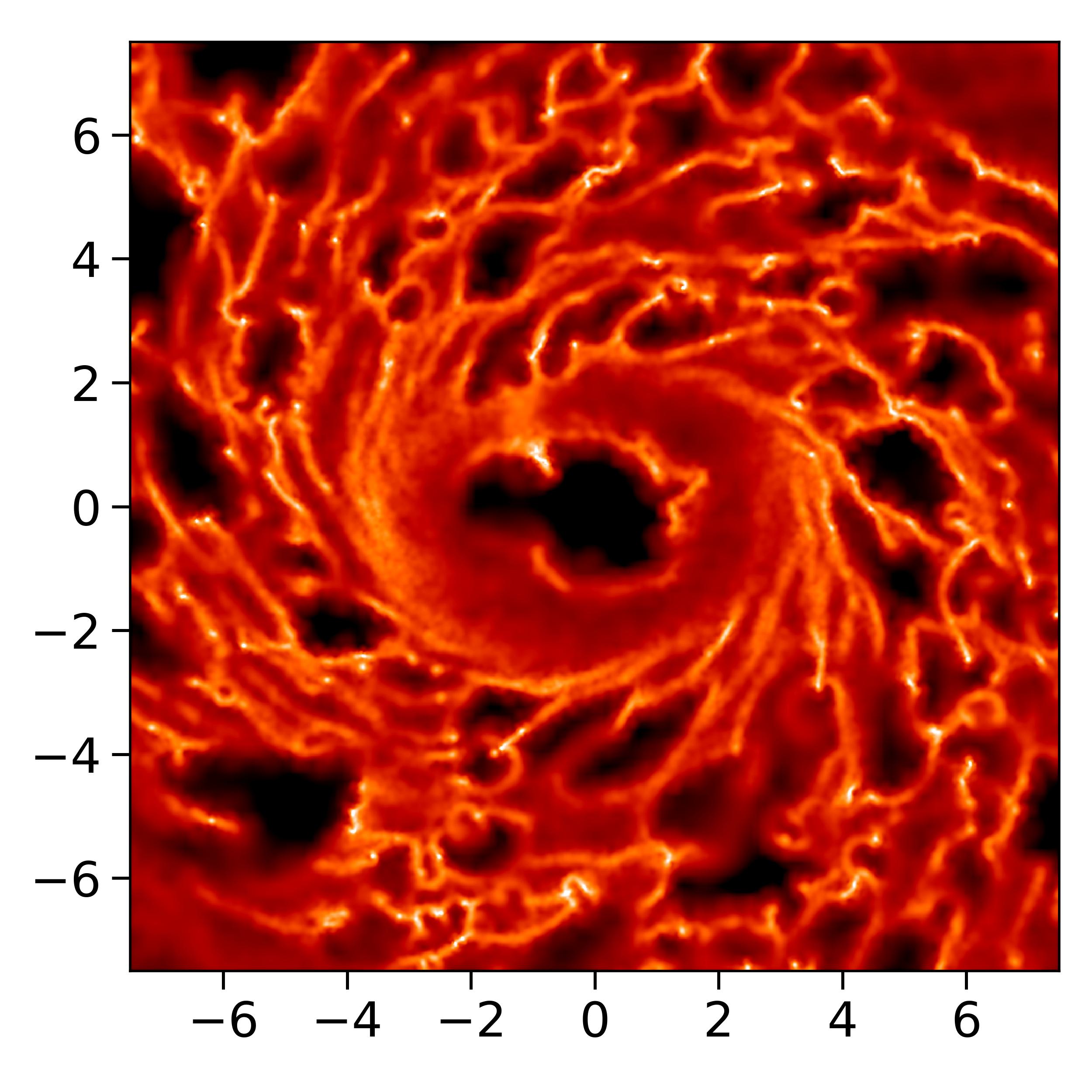} &
        \includegraphics[width=0.2\textwidth]{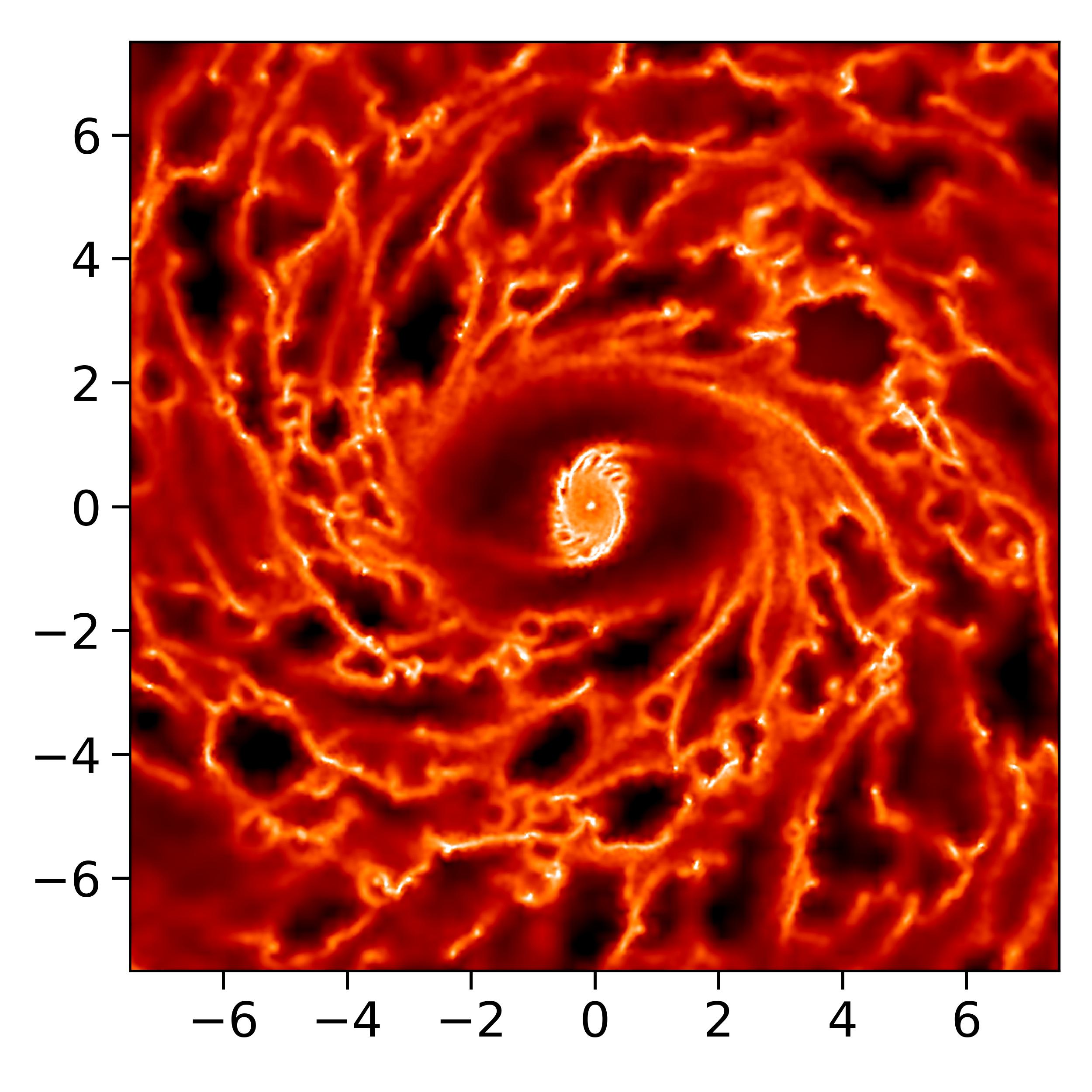} &
        \includegraphics[width=0.2\textwidth]{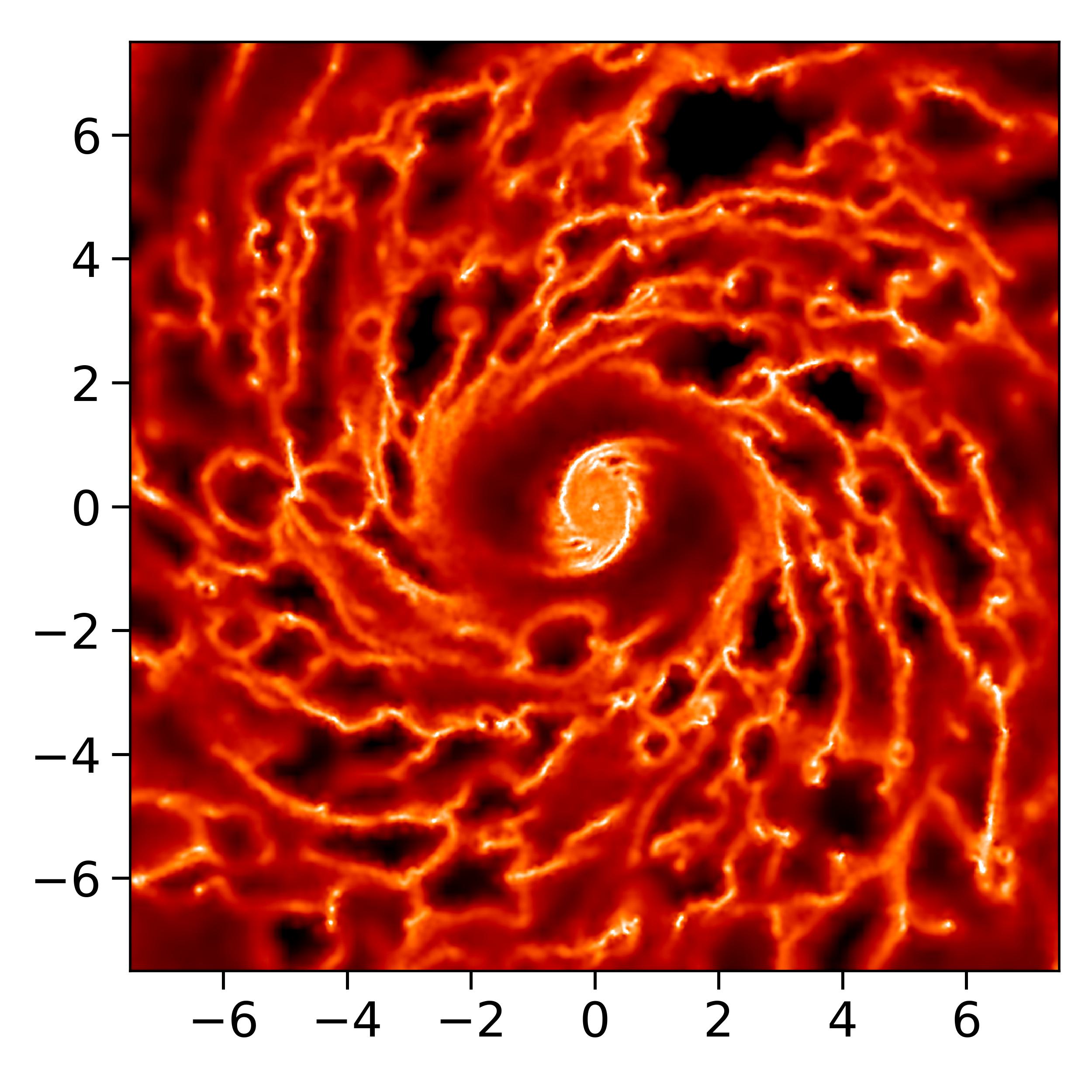} &
        \includegraphics[width=0.2\textwidth]{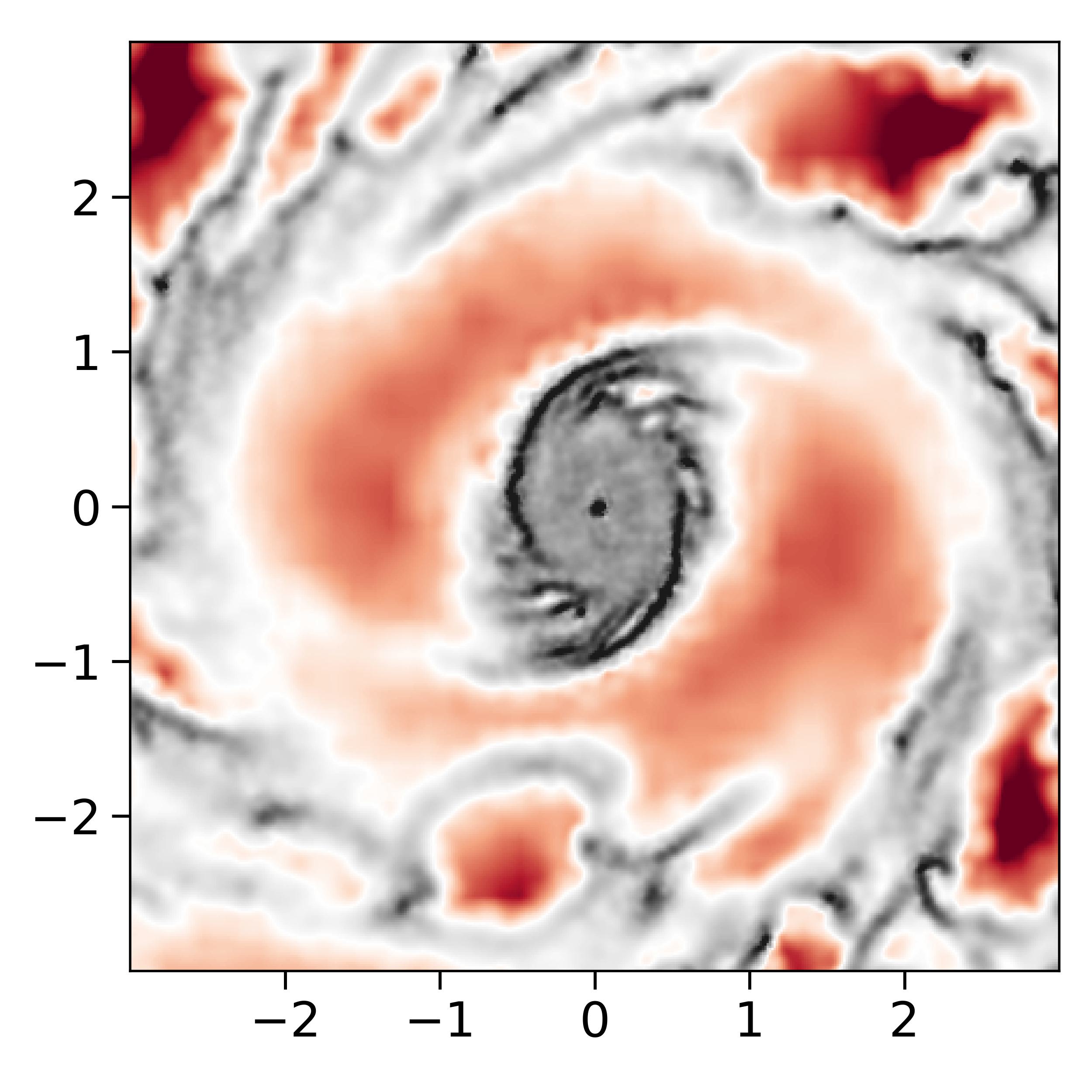} &
        \includegraphics[height=0.2\textwidth]{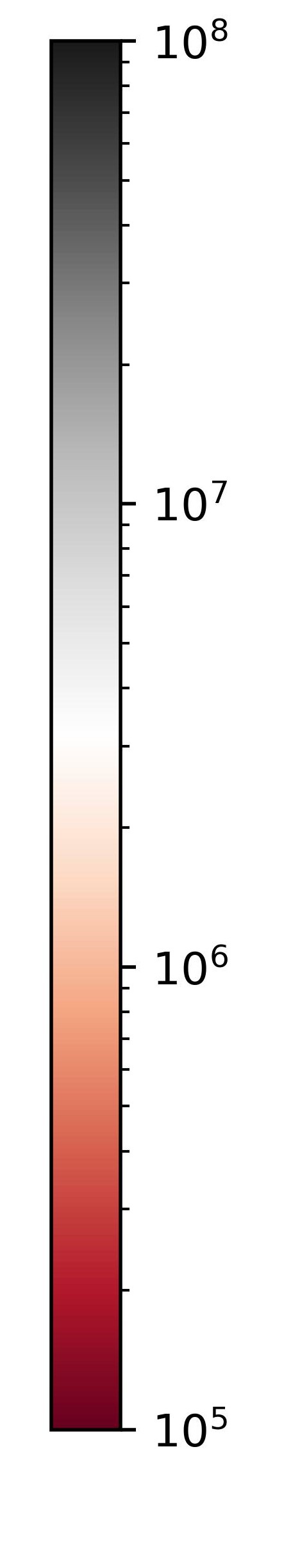}\\
    \end{tabular}
    \caption{Face-on projections of the gas surface density distributions in the X-Y plane within a $15 \times 15$ kpc box at 1, 2, 3, 4, and 5 Gyr (top to bottom rows) for models r1c14b00, r1c14b05, and r1c14b10 (left to right columns). The rightmost column shows zoom-in views of model r1c14b10 within a $6 \times 6$ kpc box at the corresponding times. The top and bottom color bars indicate units of $\Msun$ kpc$^{-2}$ and apply to the main gas density distributions and the zoom-in views, respectively.}
    \label{fig:face_gas}
\end{figure*}

\begin{figure*}[htbp]
    \centering
    \renewcommand{\arraystretch}{0}
    \begin{tabular}{@{}c@{}c@{}c@{}c@{}c@{}}
        & \textbf{r1c14b00} & \textbf{r1c14b05} & \textbf{r1c14b10}
        \\
        \raisebox{2\height}{\rotatebox{90}{\textbf{1 Gyr}}} &
        \includegraphics[width=0.22\textwidth]{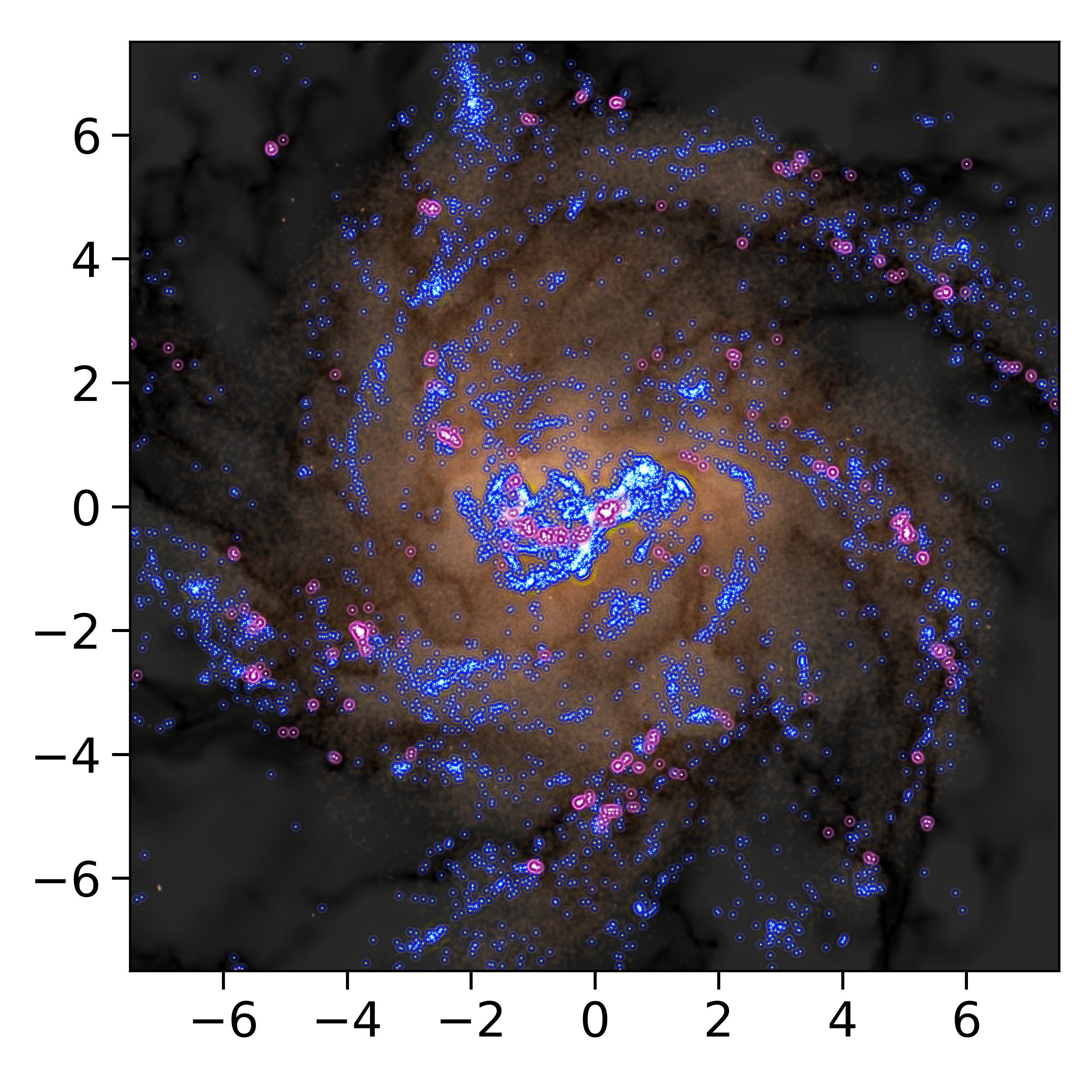} &
        \includegraphics[width=0.22\textwidth]{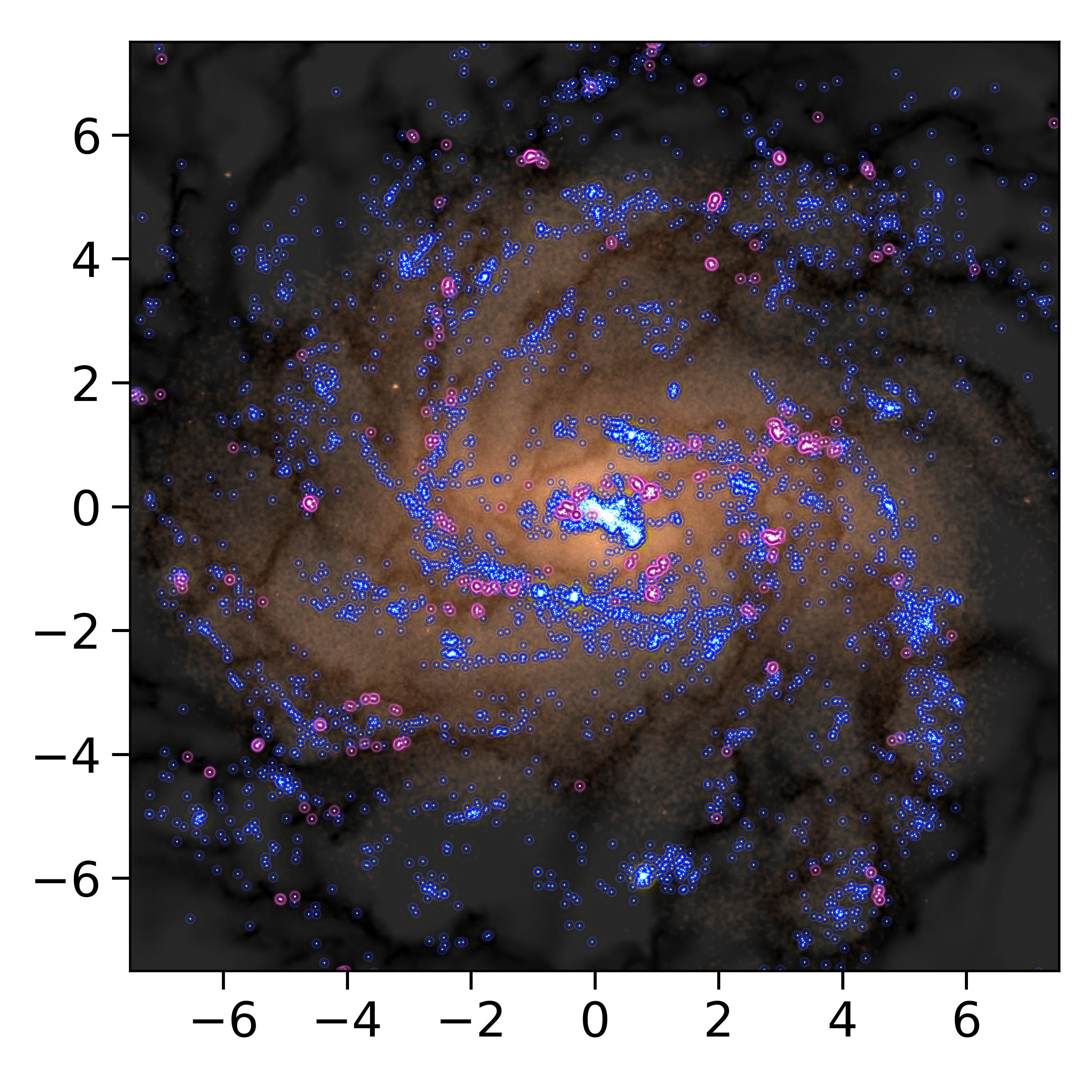} &
        \includegraphics[width=0.22\textwidth]{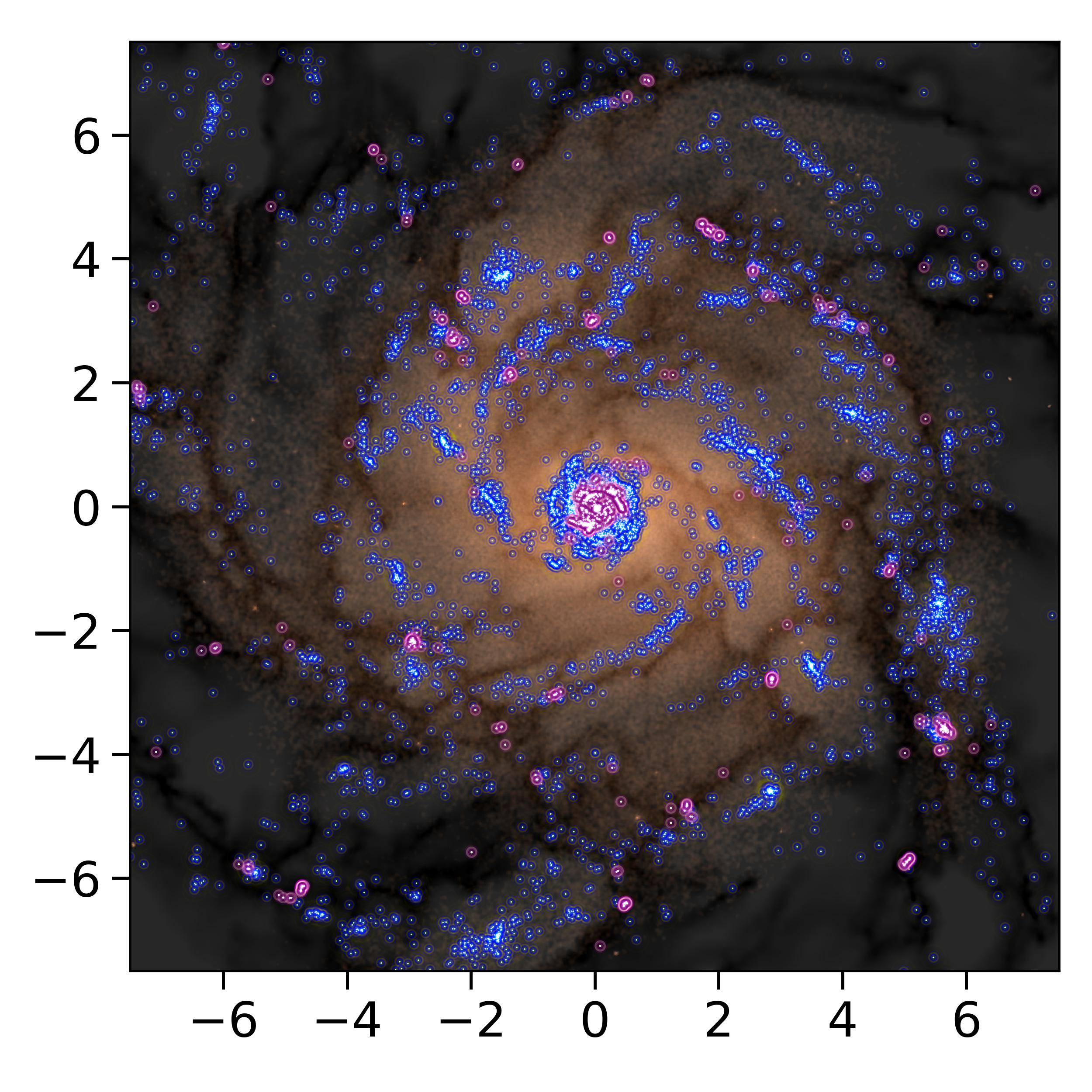}
        \\
        \raisebox{2\height}{\rotatebox{90}{\textbf{2 Gyr}}} &
        \includegraphics[width=0.22\textwidth]{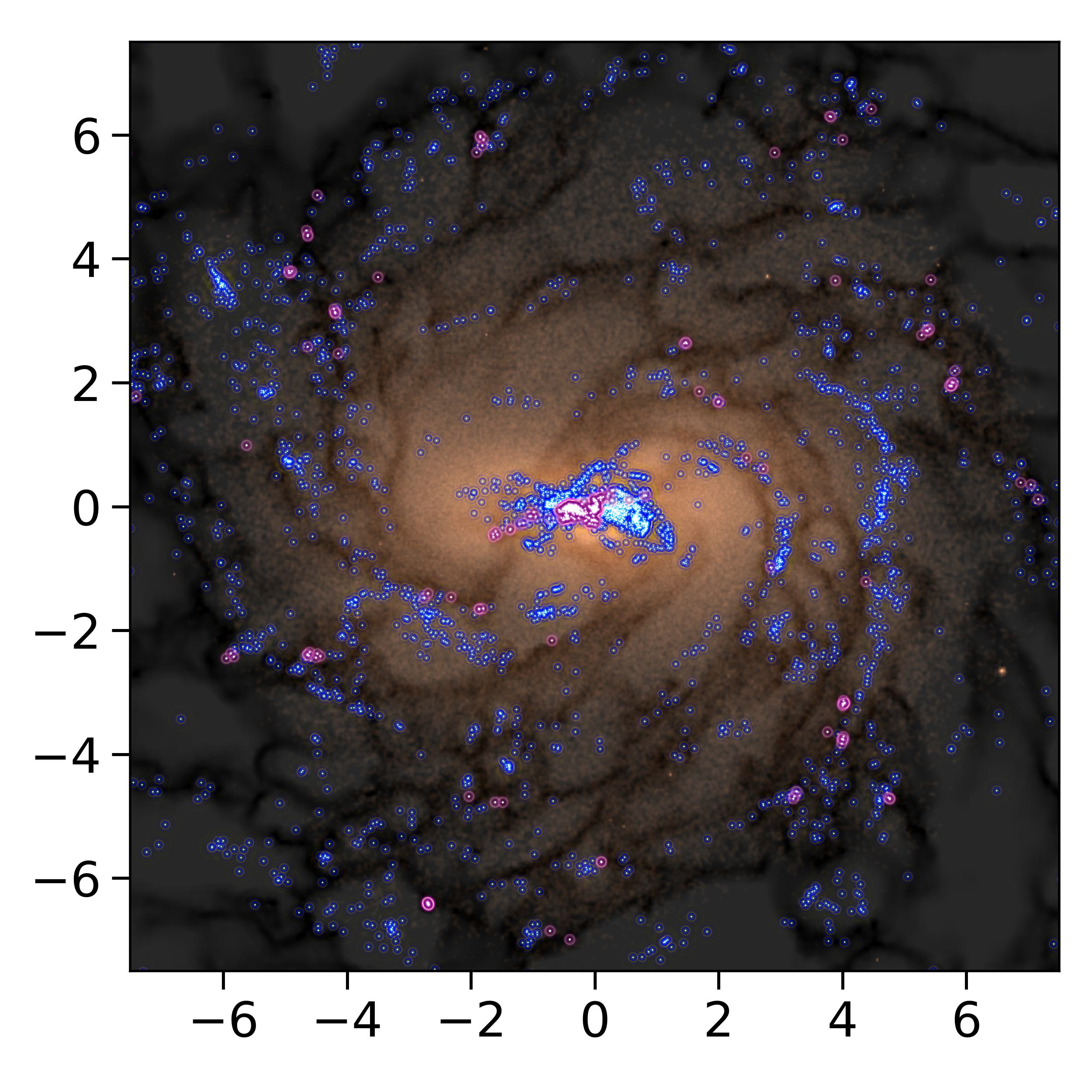} &
        \includegraphics[width=0.22\textwidth]{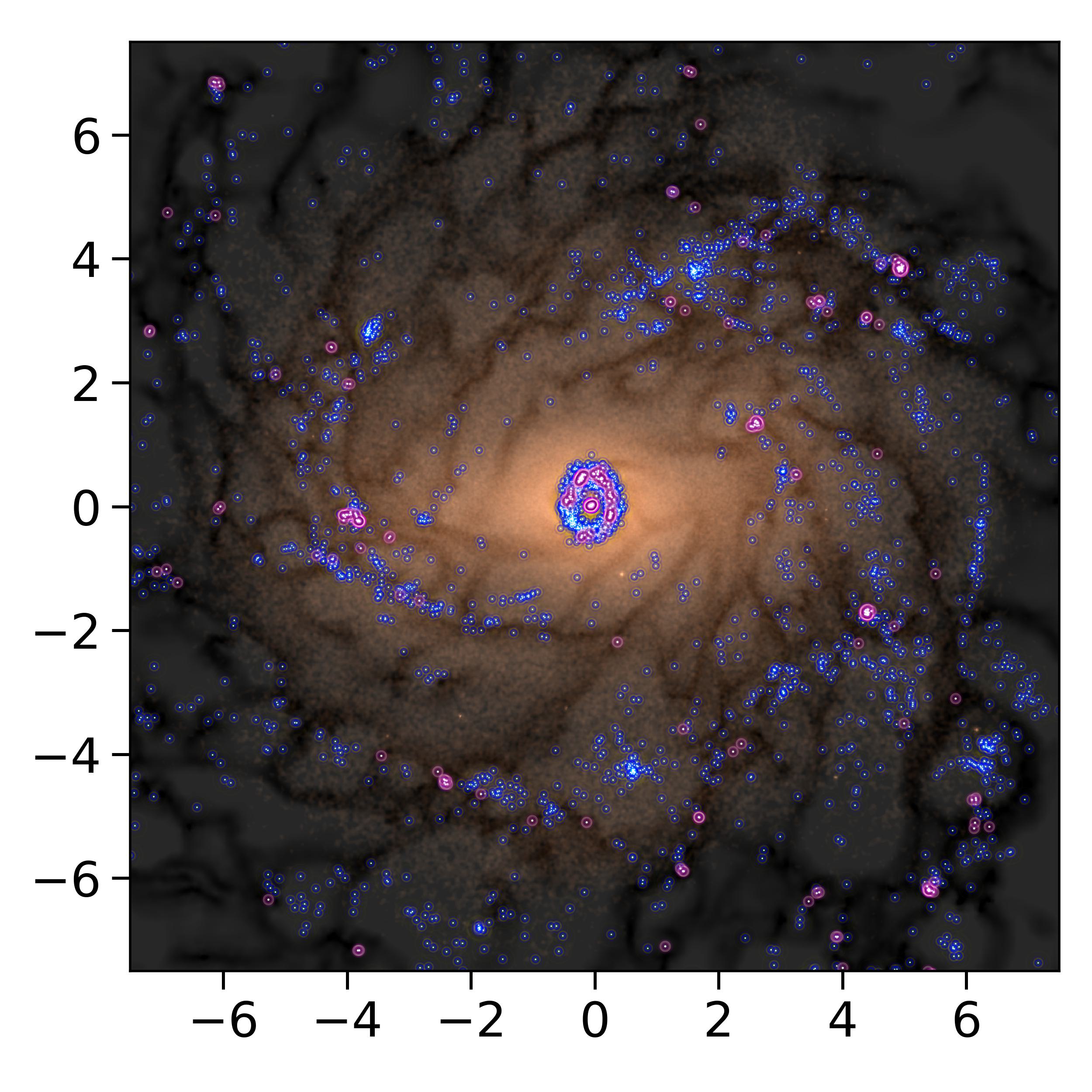} &
        \includegraphics[width=0.22\textwidth]{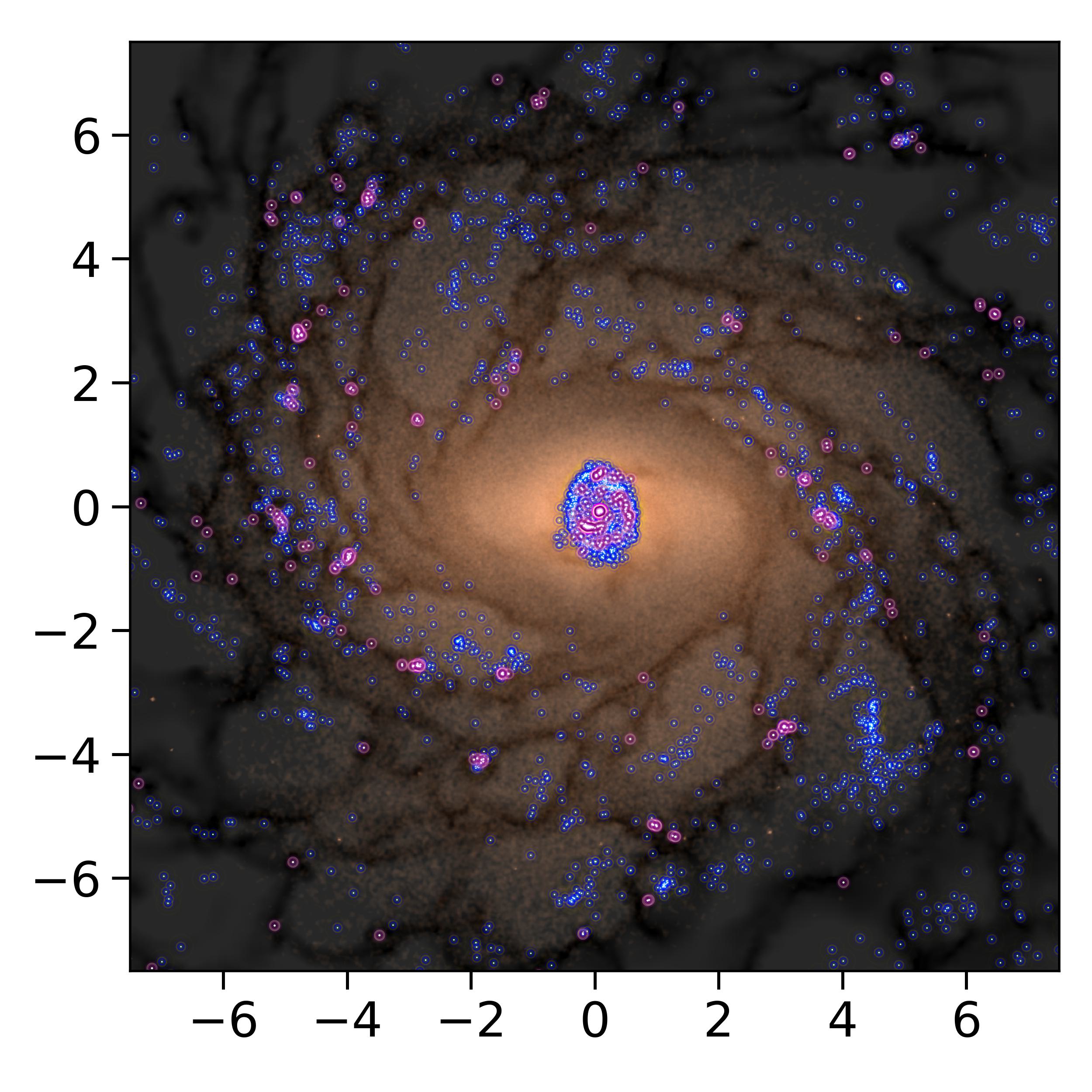}
        \\
        \raisebox{2\height}{\rotatebox{90}{\textbf{3 Gyr}}} &
        \includegraphics[width=0.22\textwidth]{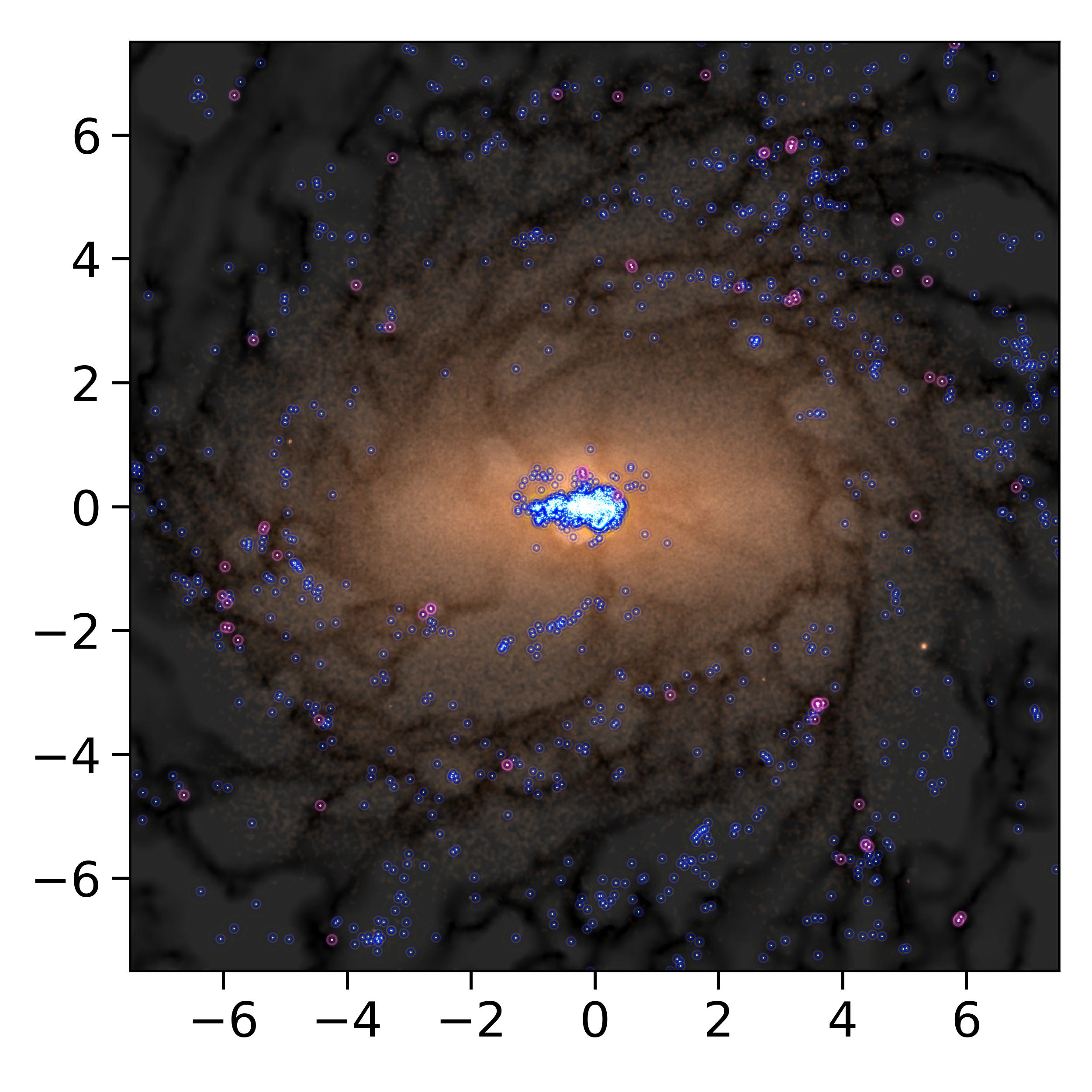} &
        \includegraphics[width=0.22\textwidth]{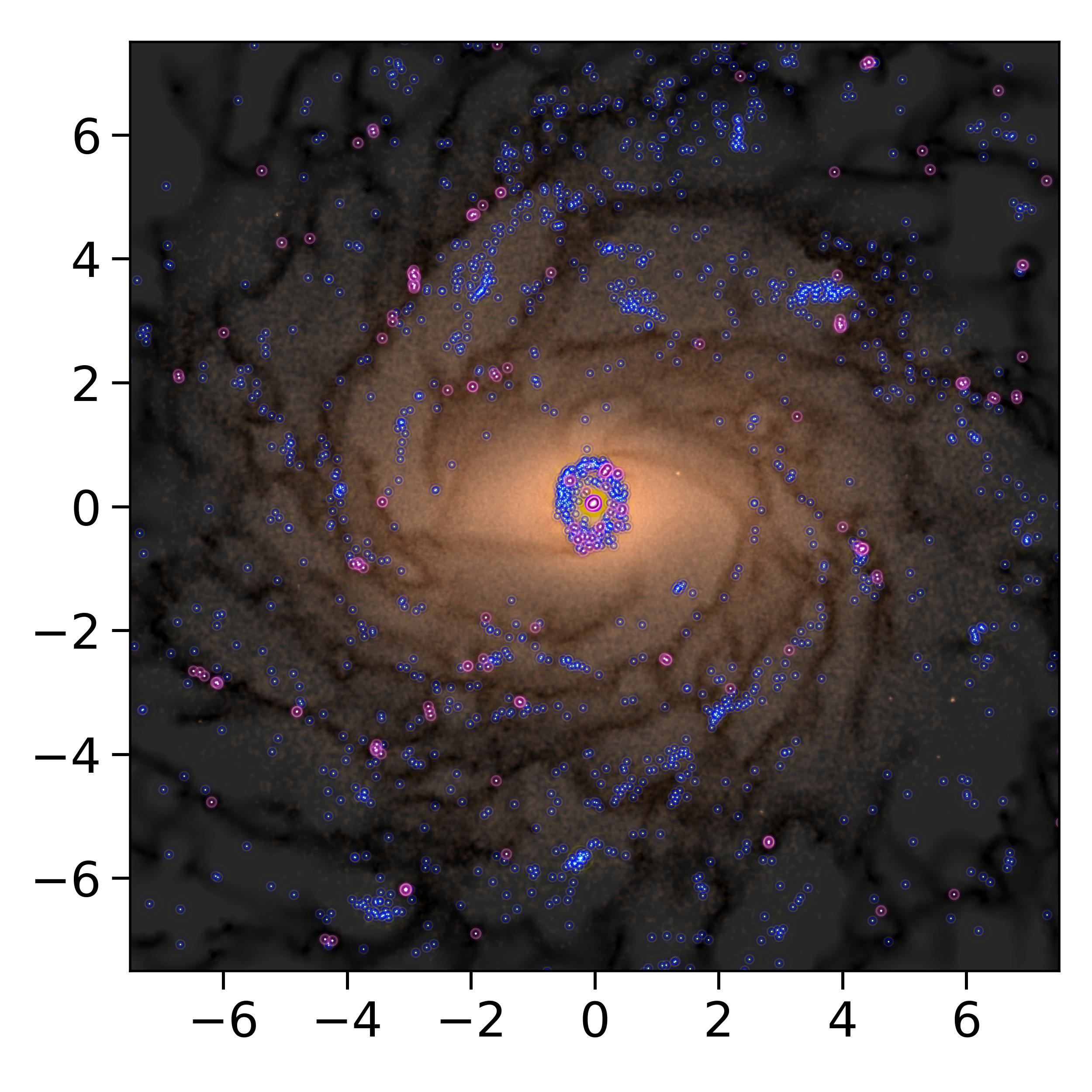} &
        \includegraphics[width=0.22\textwidth]{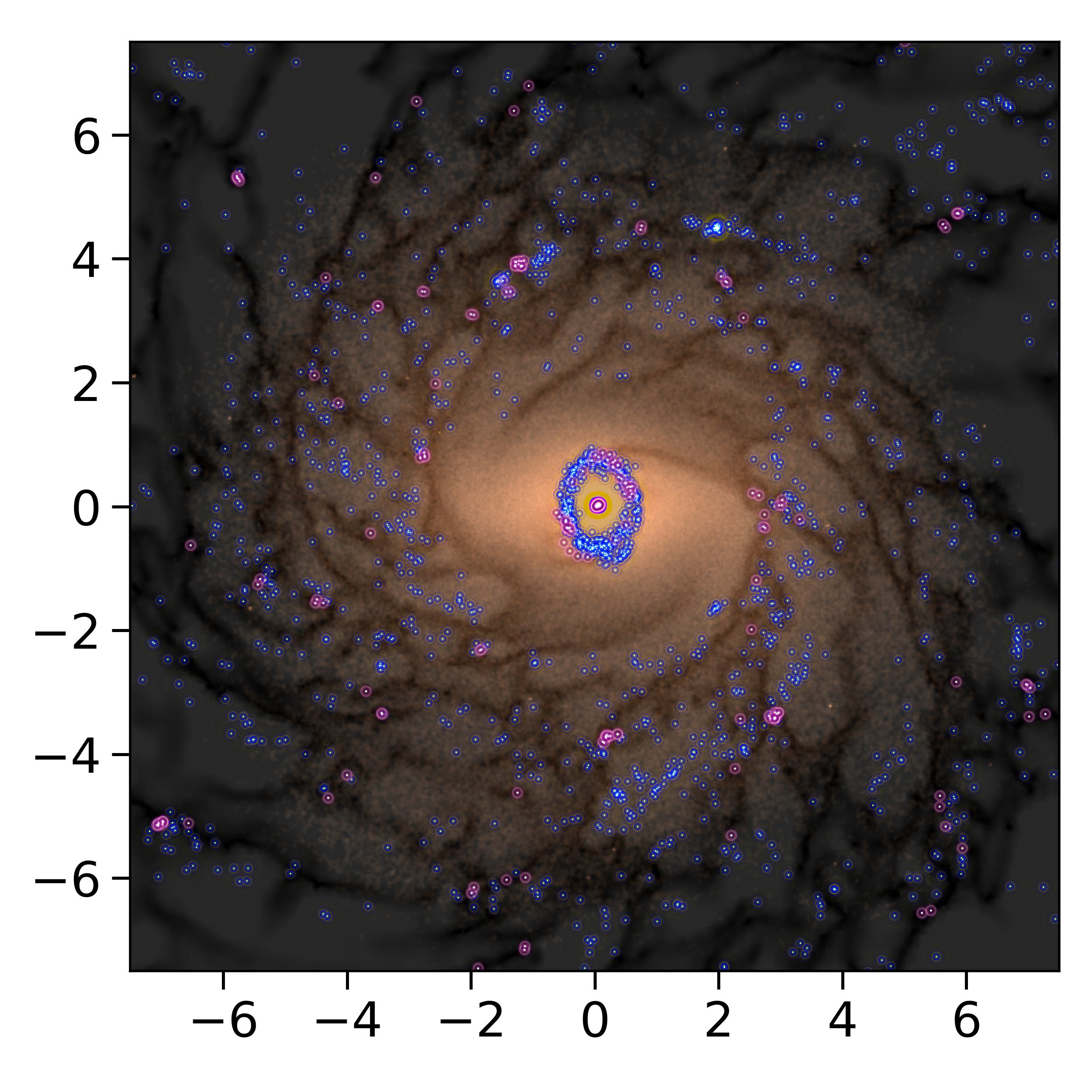}
        \\
        \raisebox{2\height}{\rotatebox{90}{\textbf{4 Gyr}}} &
        \includegraphics[width=0.22\textwidth]{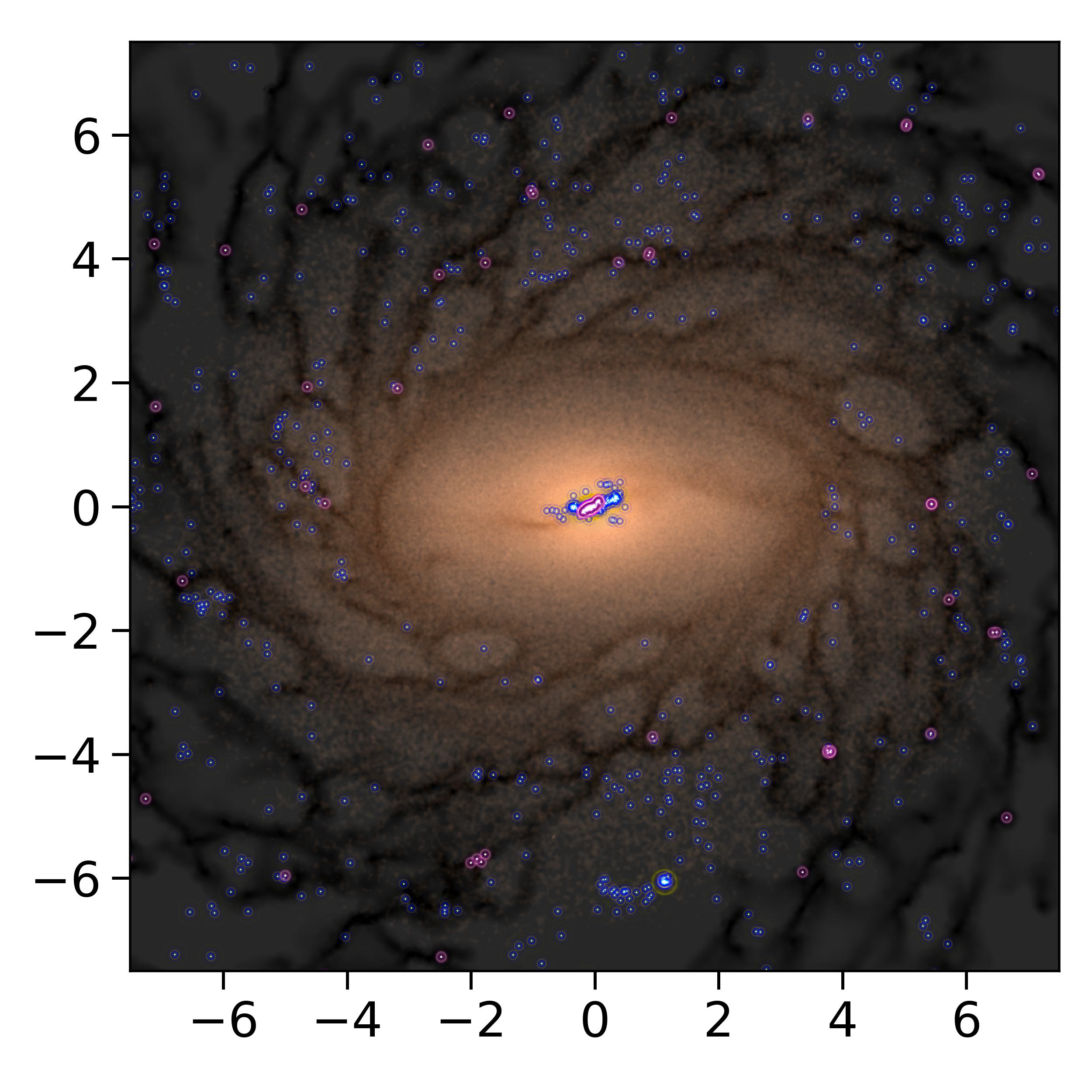} &
        \includegraphics[width=0.22\textwidth]{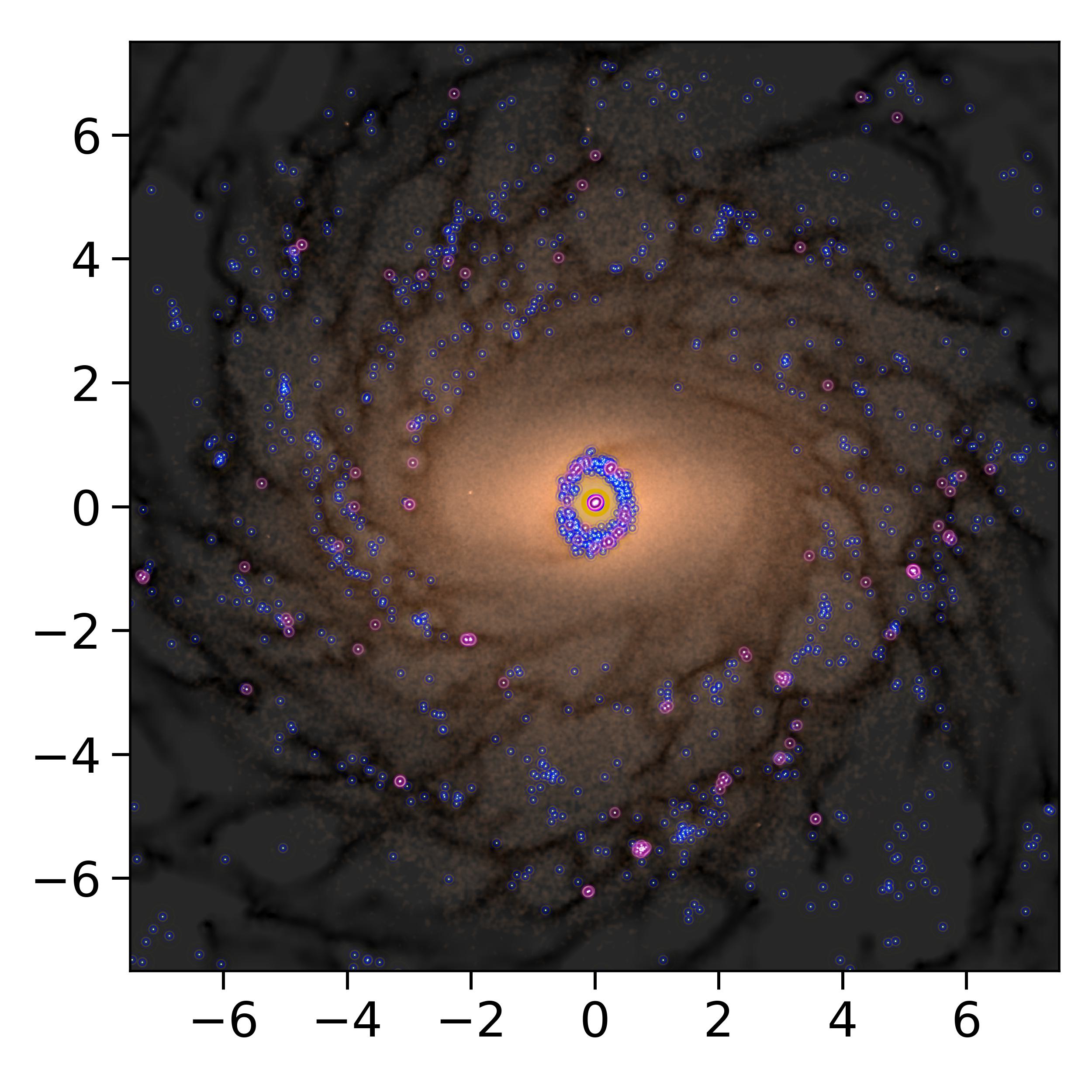} &
        \includegraphics[width=0.22\textwidth]{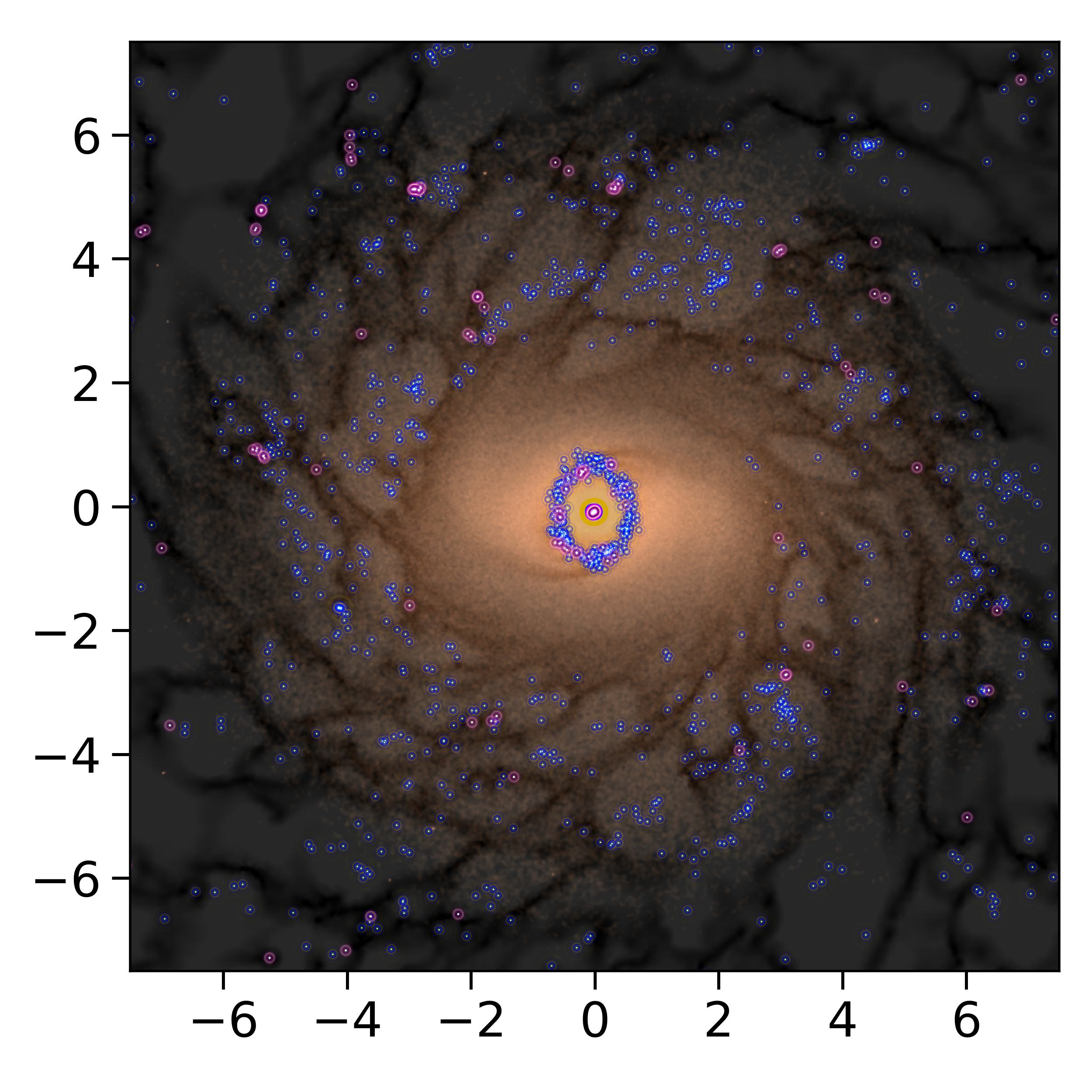}
        \\
        \raisebox{2\height}{\rotatebox{90}{\textbf{5 Gyr}}} &
        \includegraphics[width=0.22\textwidth]{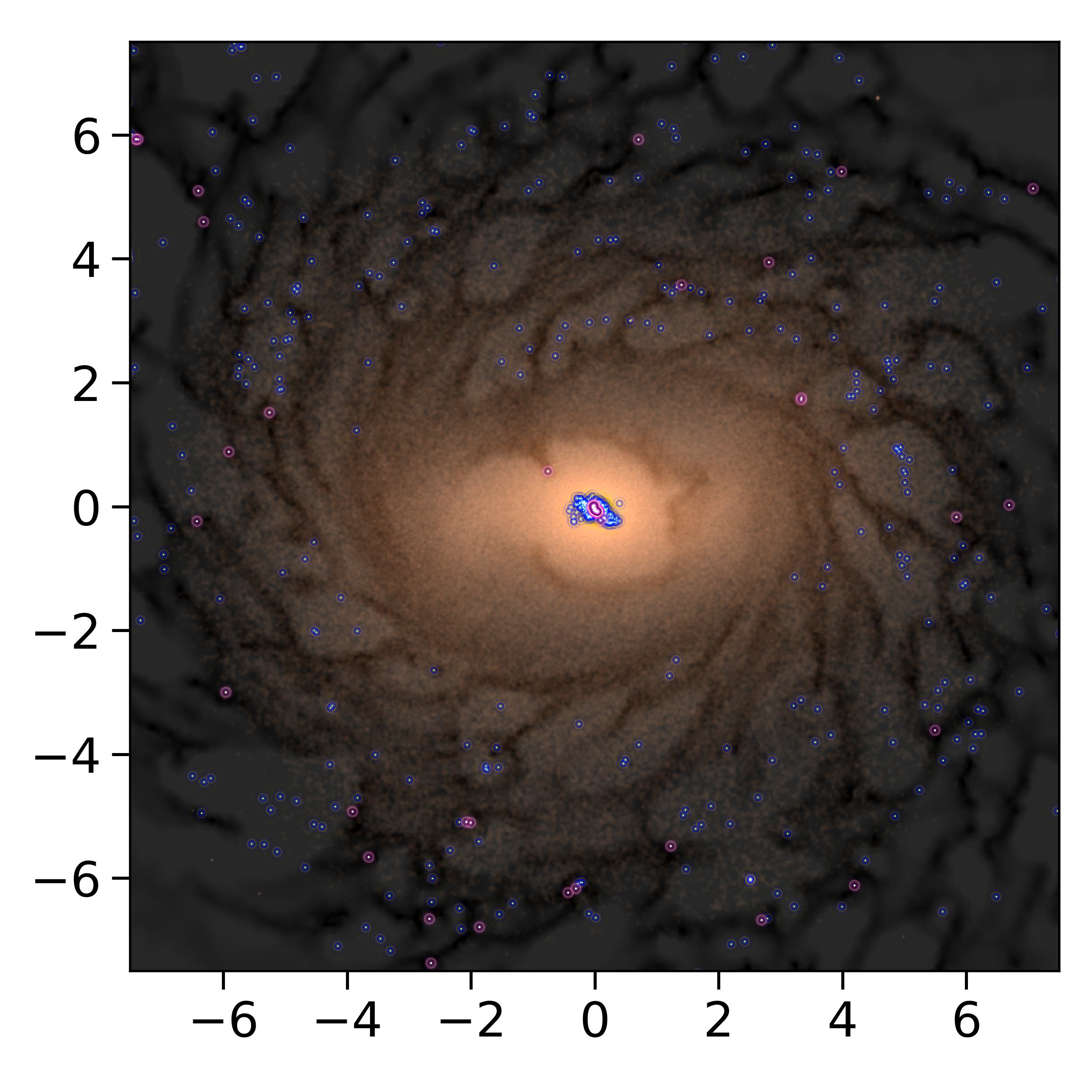} &
        \includegraphics[width=0.22\textwidth]{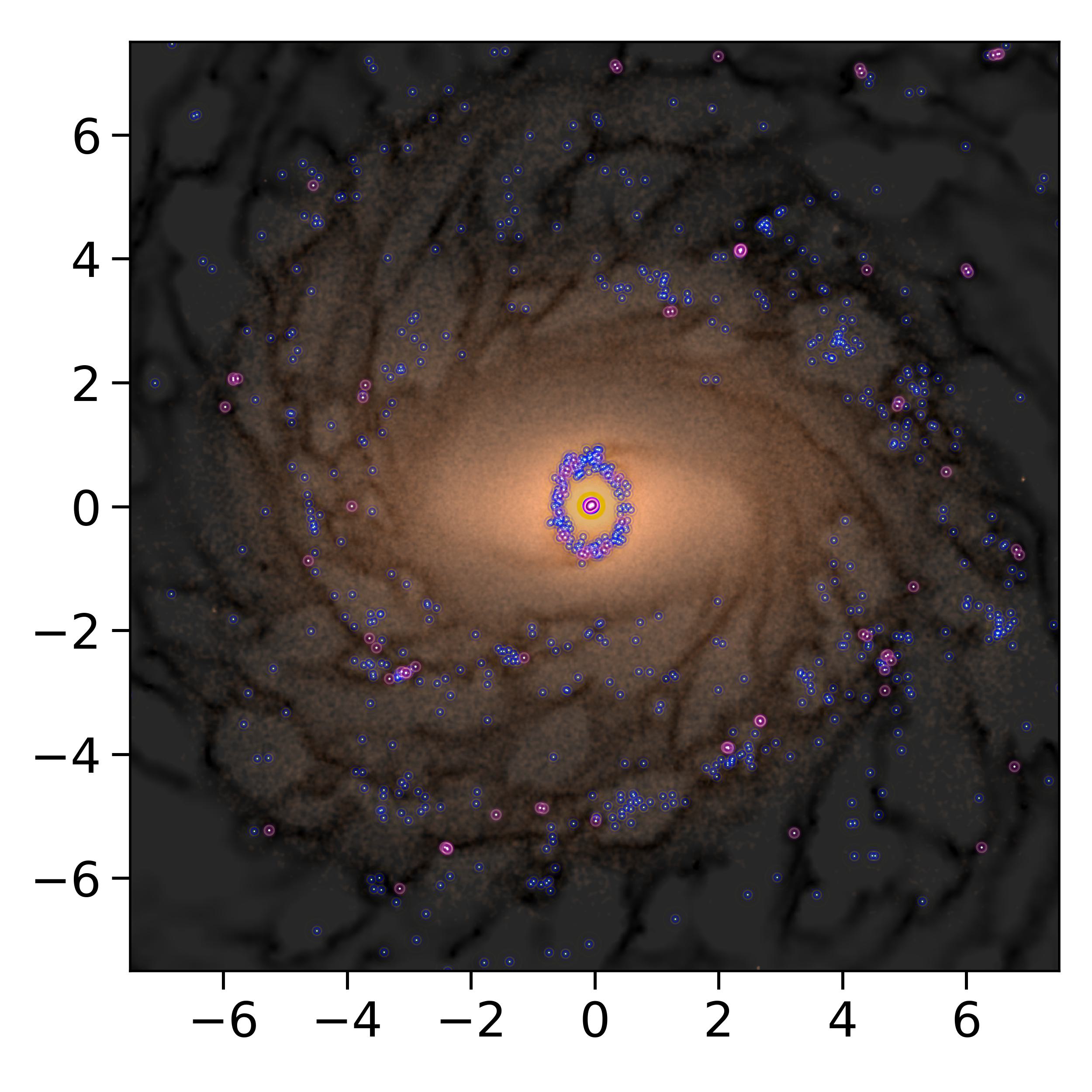} &
        \includegraphics[width=0.22\textwidth]{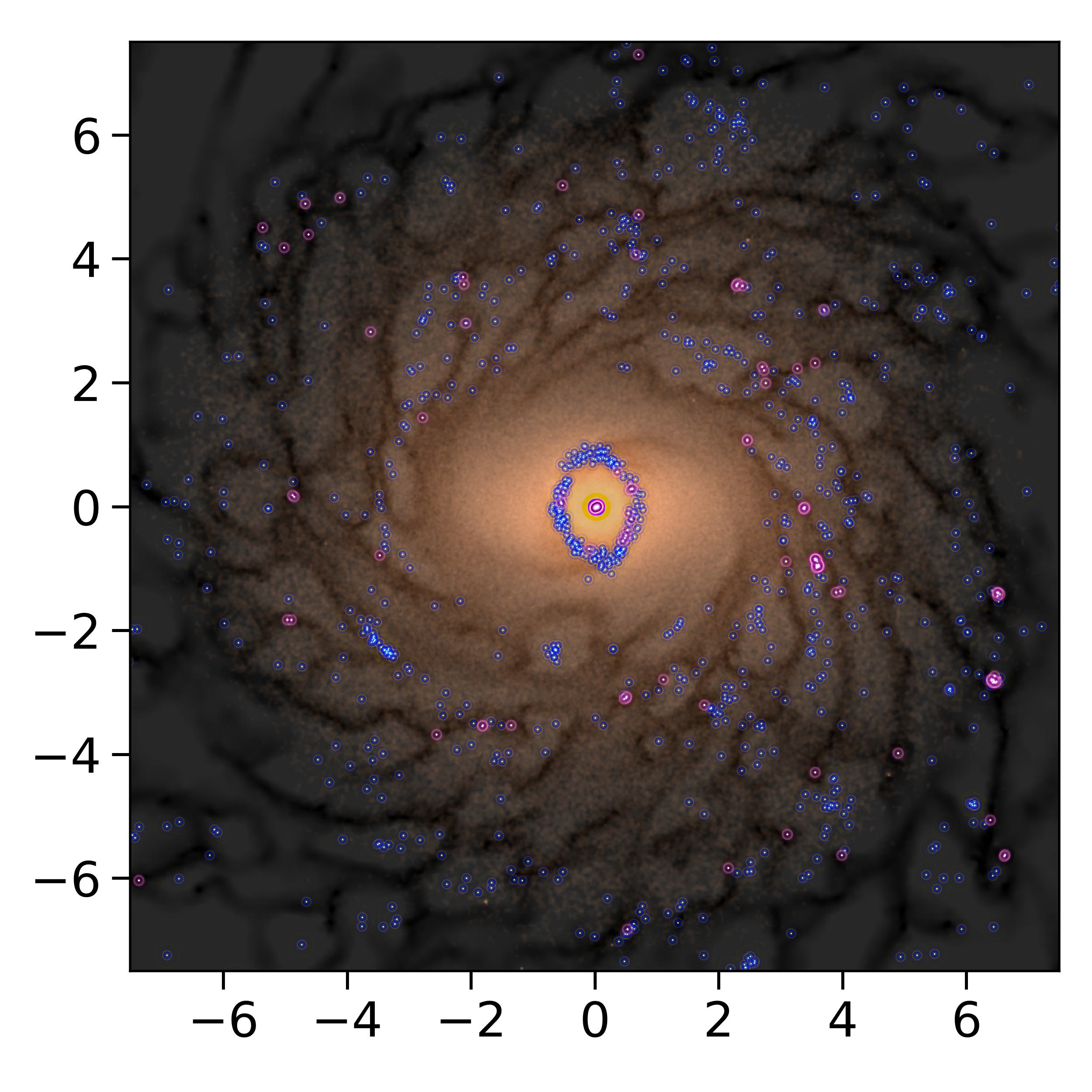}
        \\
    \end{tabular}
    \caption{Face-on projections of the stacked surface density distributions in the X-Y plane within a $15 \times 15$ kpc$^2$ box at 1, 2, 3, 4, and 5 Gyr (top to bottom rows) for models r1c14b00, r1c14b05, and r1c14b10 (left to right columns). Blue and purple points are overlaid to indicate young stars with ages $0.01 < t_\mathrm{age} < 0.1 \, \Gyr$ and $t_\mathrm{age} \leq 0.01 \, \Gyr$, respectively.}
    \label{fig:face_stack}
\end{figure*}

\begin{figure*}[htbp]
    \centering
    \includegraphics[width=0.33\textwidth]{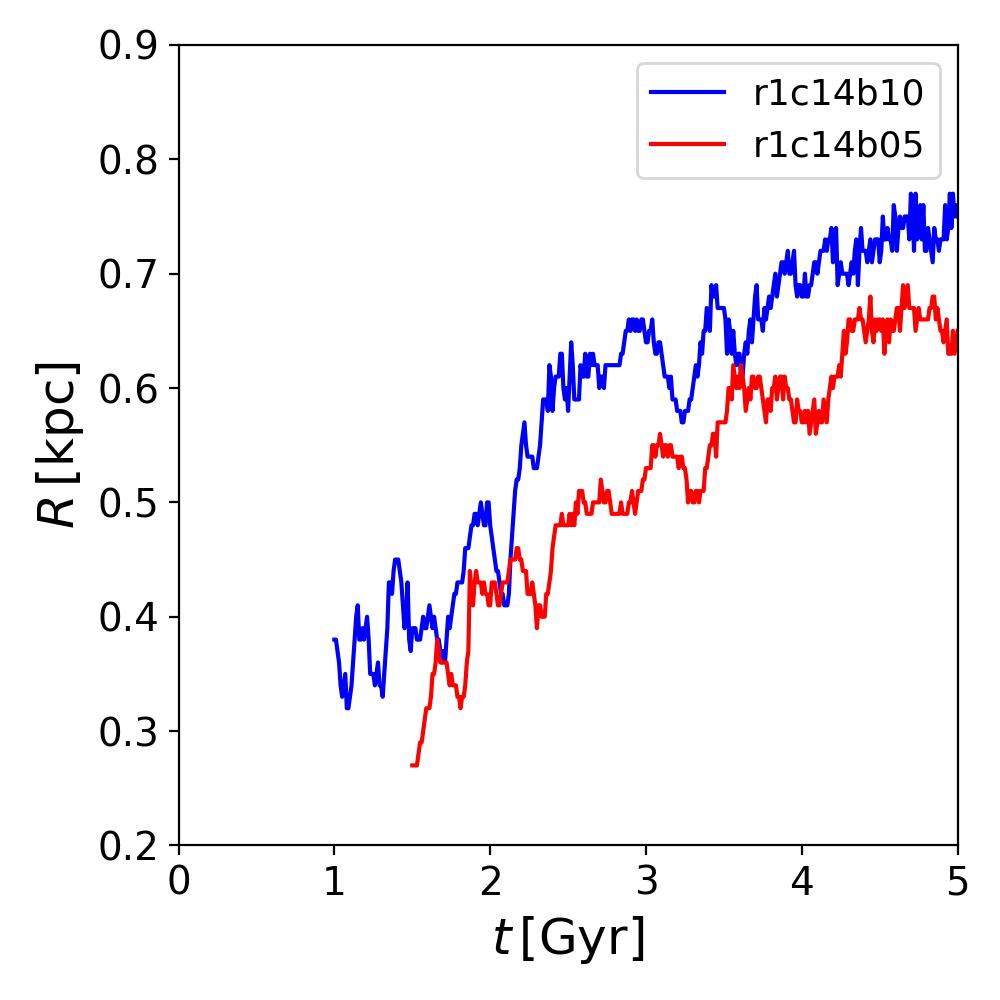}\includegraphics[width=0.33\textwidth]{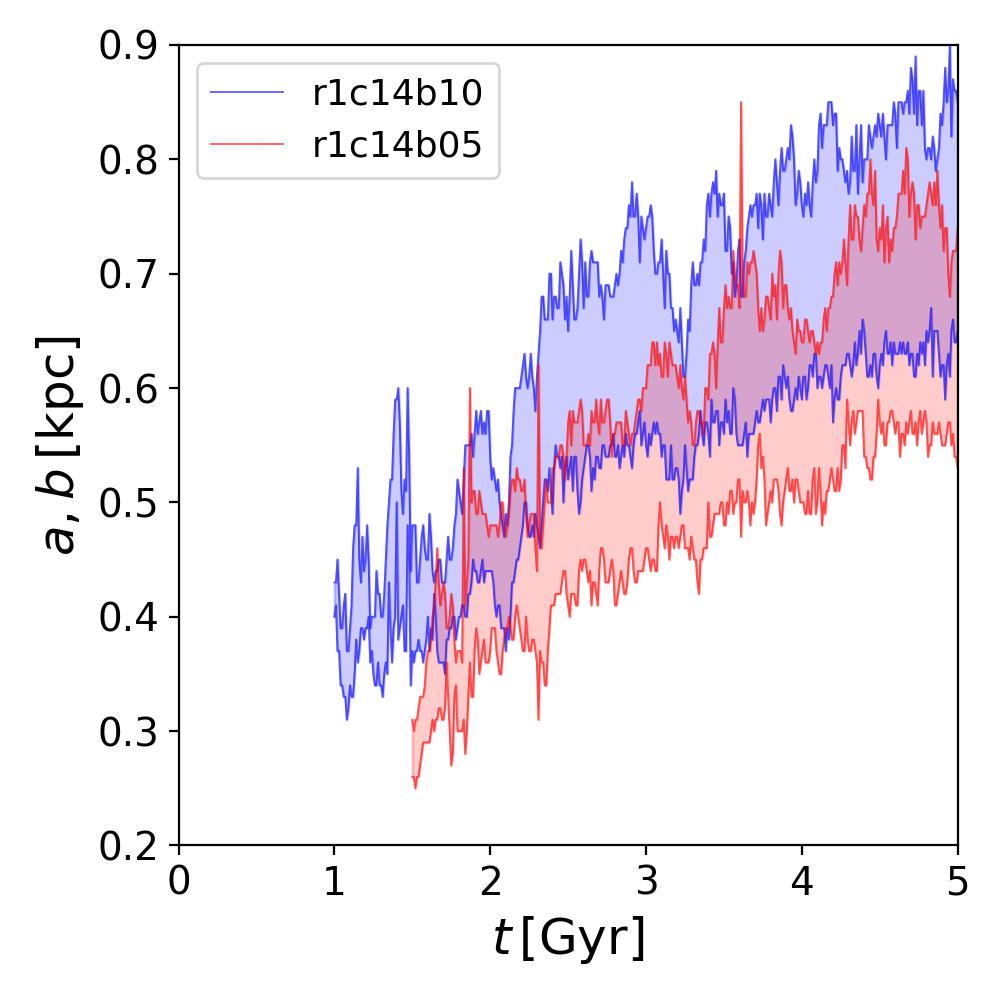}     \includegraphics[width=0.33\textwidth]{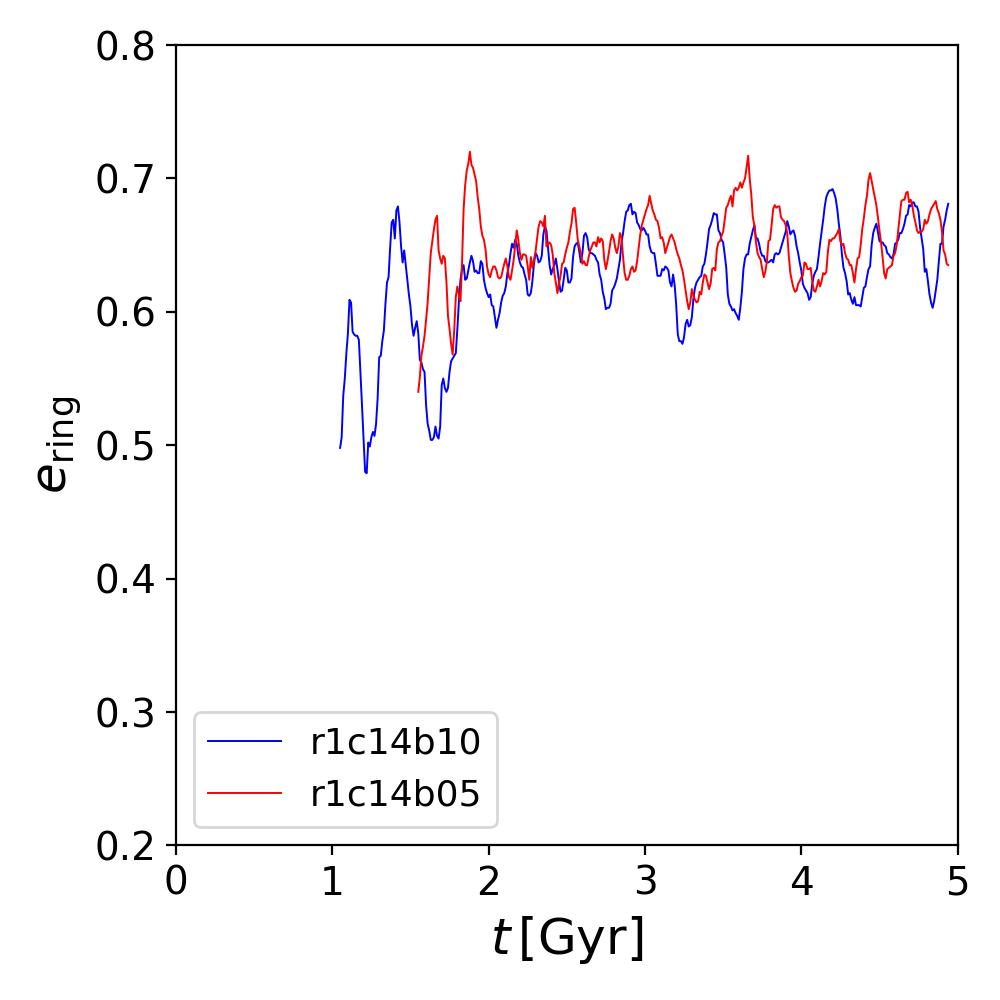}  
    
    \includegraphics[width=0.33\textwidth]{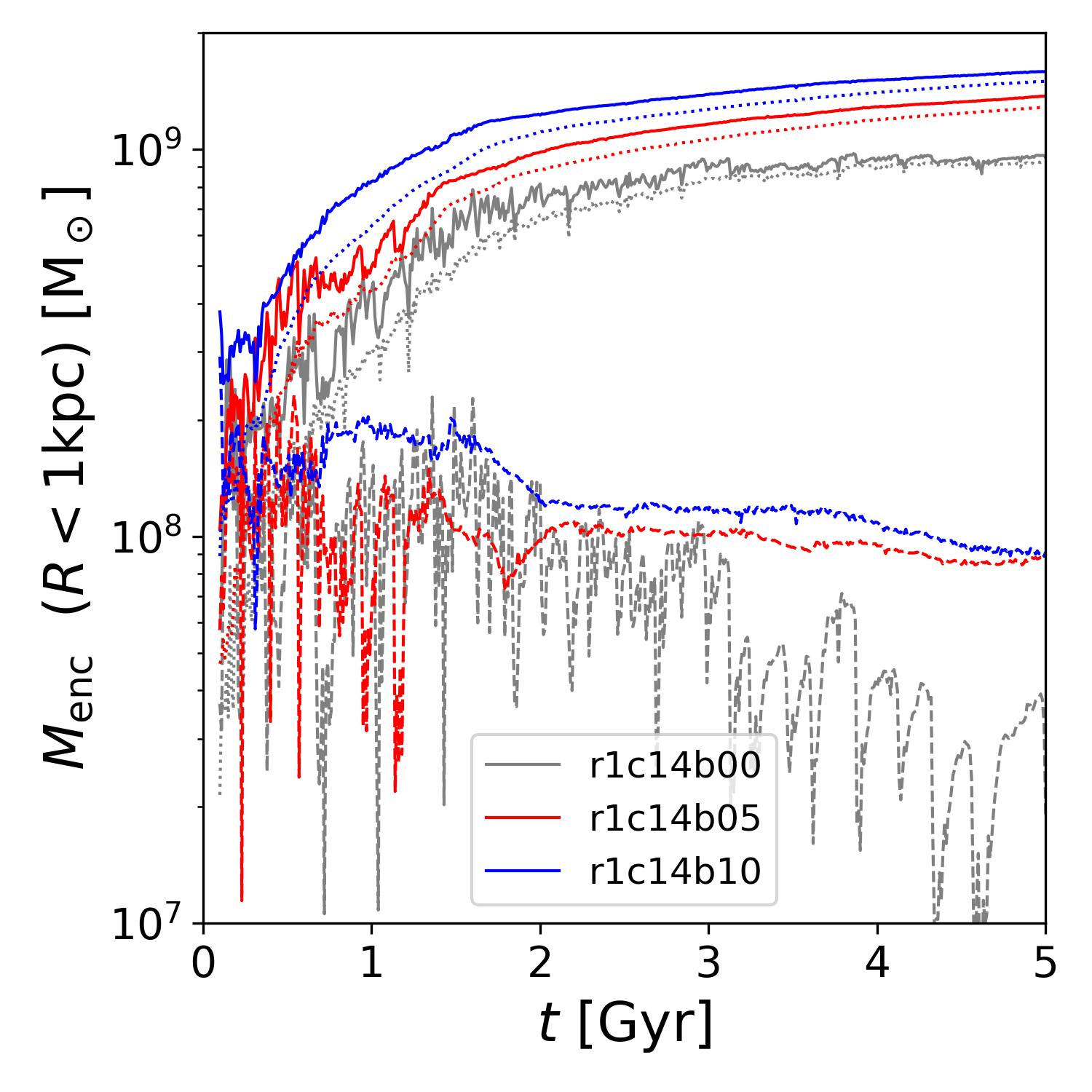}    \includegraphics[width=0.33\textwidth]{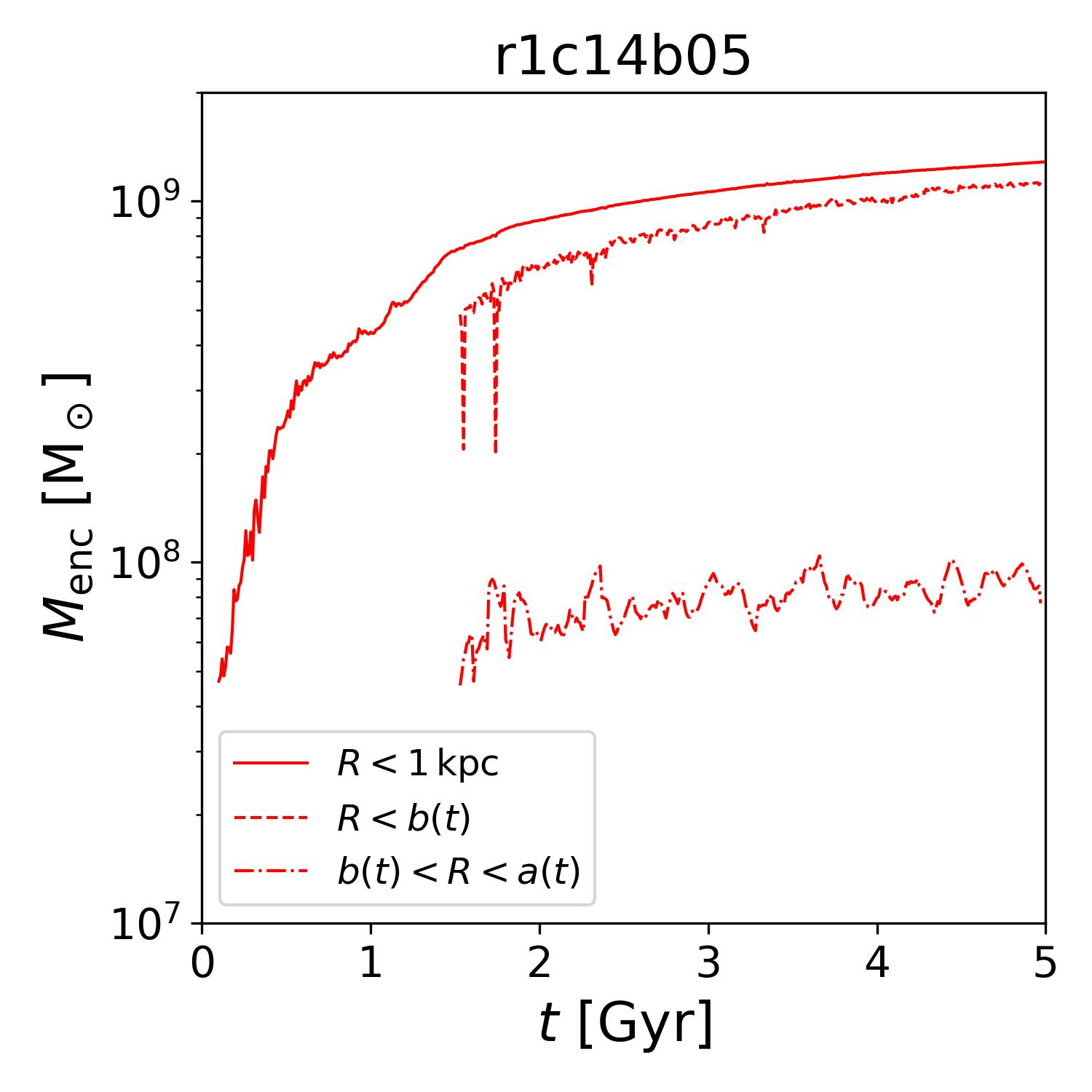}    \includegraphics[width=0.33\textwidth]{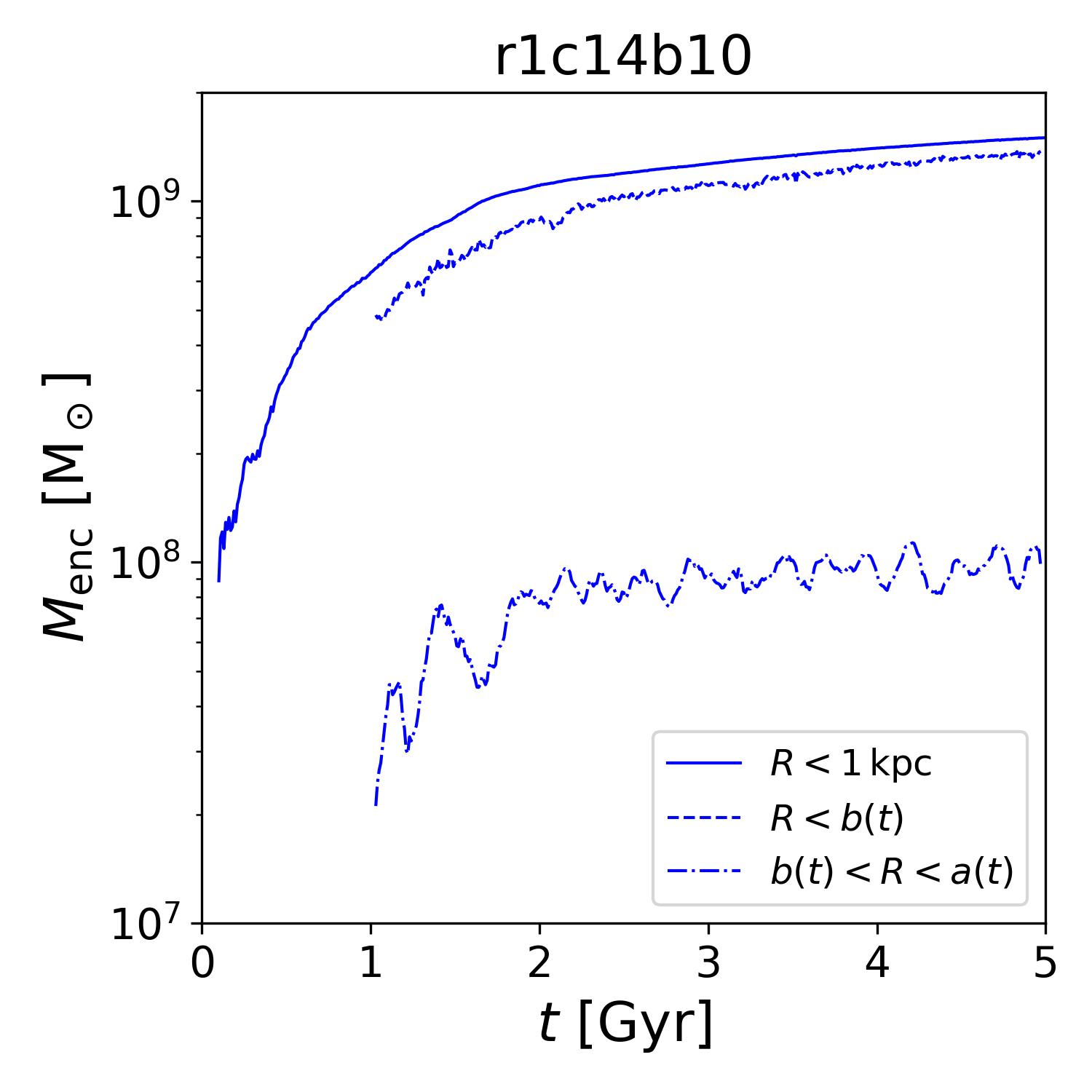}
    
    \caption{Temporal evolution of nuclear ring properties, enclosed masses of specified components within different radial extents. Model names are labeled inside the panel. Gray, red, and blue colors indicate models r1c14b00, r1c14b05, and r1c14b10. Model r1c14b00 does not form a long-lived nuclear ring. The top-left panel shows the ring radius as a function of time, determined by the mass-weighted mean radius of young stars. Using this radius, ellipse fitting is applied to measure the semi-major (upper line) and minor (lower line) axes of the nuclear ring over time in the top-middle panel, while the ellipticity is shown in the top-right panel. The bottom-left panel displays the enclosed mass within 1 kpc for the three models, including newly formed stars (dotted lines), gas (dashed lines), and their combined mass (solid lines). In the bottom-middle and bottom-right panels, the enclosed masses of these new stars between the semi-major and minor axes (nuclear ring) and within the semi-minor axis (nuclear star cluster) are plotted.}
    \label{fig:ring_info}
\end{figure*}

\begin{figure*}[htbp]
    \centering
    \includegraphics[width=0.95\textwidth]{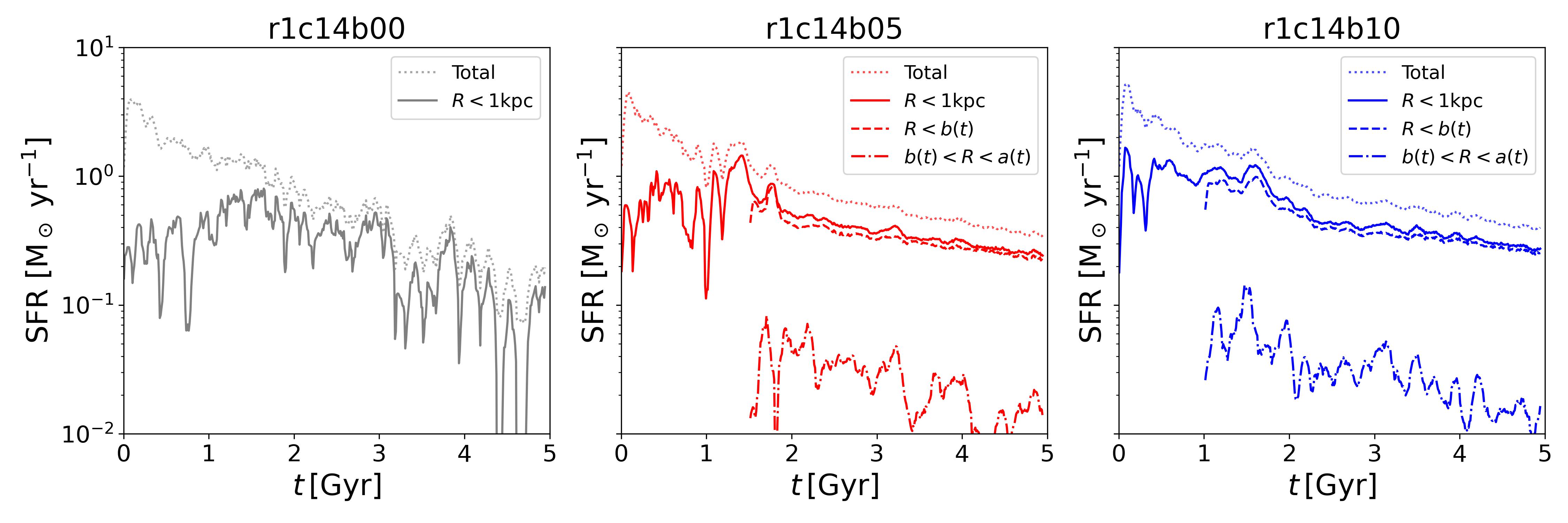}
    \caption{Star formation rates for the three models, both total and within $R < 1$ kpc. The corresponding SFR for each region is labeled with different line styles, where $a$ and $b$ indicate the semi-major and semi-minor axes of the nuclear ring, respectively.}
    \label{fig:sfr}
\end{figure*}

\begin{figure}[htbp]
    \centering
    \includegraphics[width=0.50\textwidth]{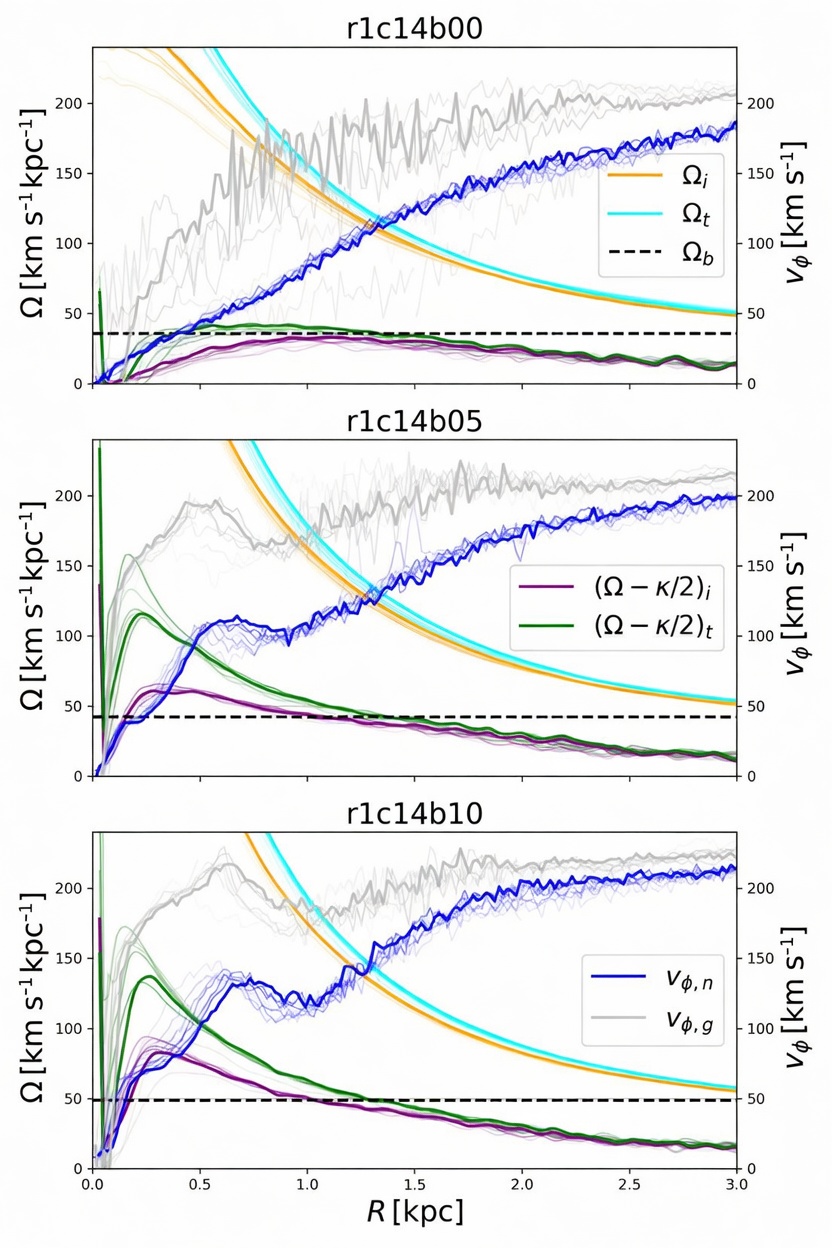}
    \caption{Angular velocity $\Omega$, $\Omega-\kappa/2$ curves, and $v_\phi$ from 1.5 to 5 Gyr in 0.5 Gyr steps, with increasing line transparency for earlier times. The profile at 5 Gyr is displayed with a thick solid line. In the labels, `i' denotes initial stars, `n' denotes new stars, `g' denotes the gas component, and `t' indicates the total stellar component. The bar pattern speed $\Omega_b$ at 5 Gyr is marked as a dashed horizontal line. The bar pattern speeds are 35.8, 42.3, and 48.7 km s$^{-1}$ kpc$^{-1}$, which remains nearly constant in our models (Fig.~\ref{fig:barpattern}), and the corotation radii ($R_{CR}$ where $\Omega_b=\Omega_t$) are 4.4, 3.9, and 3.6 kpc for models r1c14b00, r1c14b05, and r1c14b10, respectively. The time evolution of the ILRs as a function of bar pattern speed is shown in Fig.~\ref{fig:ilr_evolution} and compared to the nuclear ring region. See also Fig.~\ref{fig:radial_vphi} for the radial profiles of mean rotation.}
    \label{fig:resonance}
\end{figure}

\begin{figure*}[htbp]
    \centering
    \includegraphics[width=0.8\textwidth]{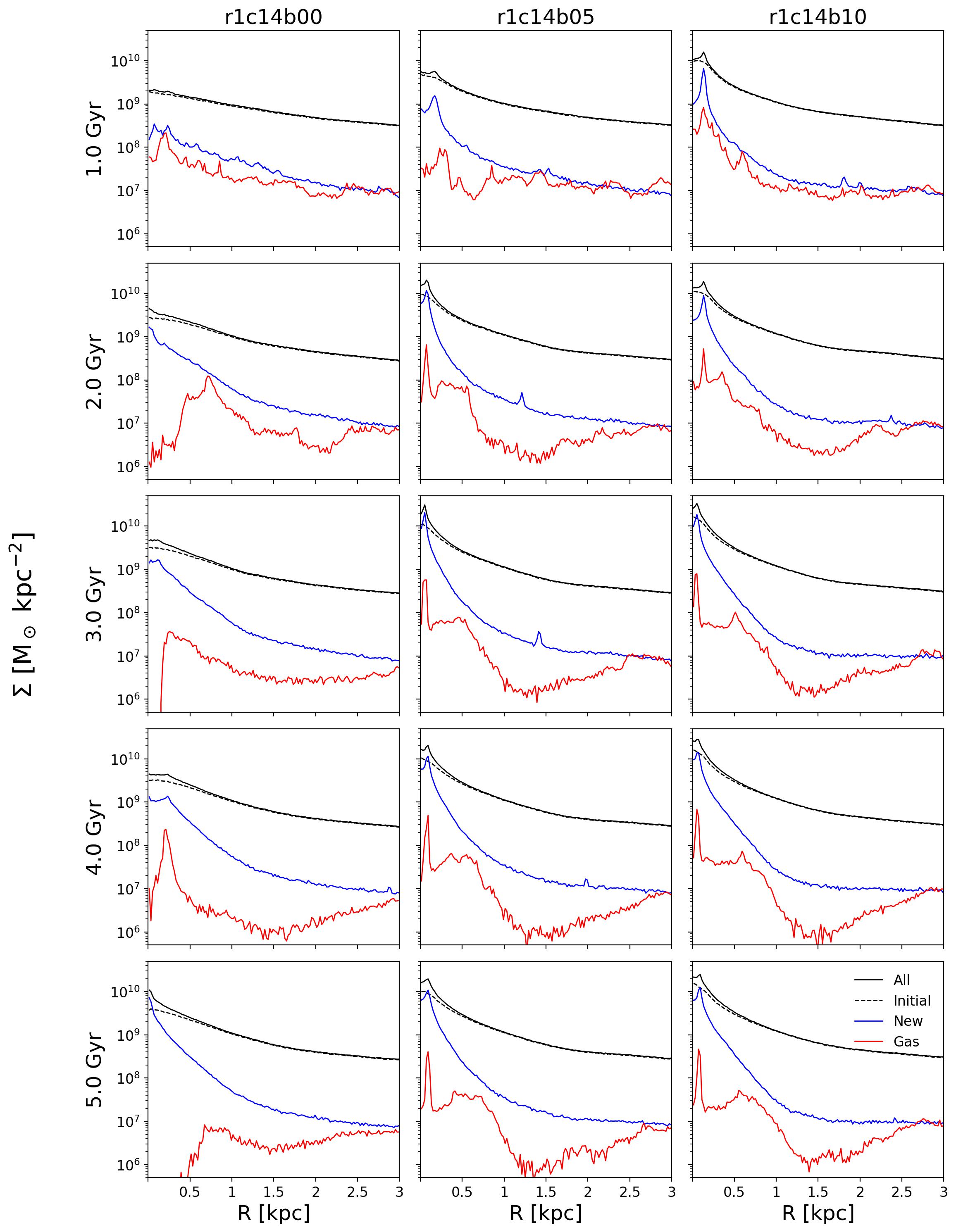}
    \caption{Density profiles of all stars, initial stars, new stars, and gas within 3 kpc for all models at 1, 2, 3, 4, and 5 Gyr. From left to right, the columns correspond to the models r1c14b00, r1c14b05, and r1c14b10. Corresponding evolutionary times and components are labeled.}
    \label{fig:den}
\end{figure*}

\begin{figure*}[htbp]
    \centering
    $\begin{array}{ccc}
        {\fontsize{15pt}{20pt}\selectfont v_{\phi} \, (\mathrm{New \, Stars}) \, [\kms]} & {\fontsize{15pt}{20pt}\selectfont v_{\phi} \, (\mathrm{Initial \, Stars}) \, [\kms]} & {\fontsize{15pt}{20pt}\selectfont v_{\phi} \, (\mathrm{Gas}) \, [\kms]} \\
        \includegraphics[width=0.33\textwidth]{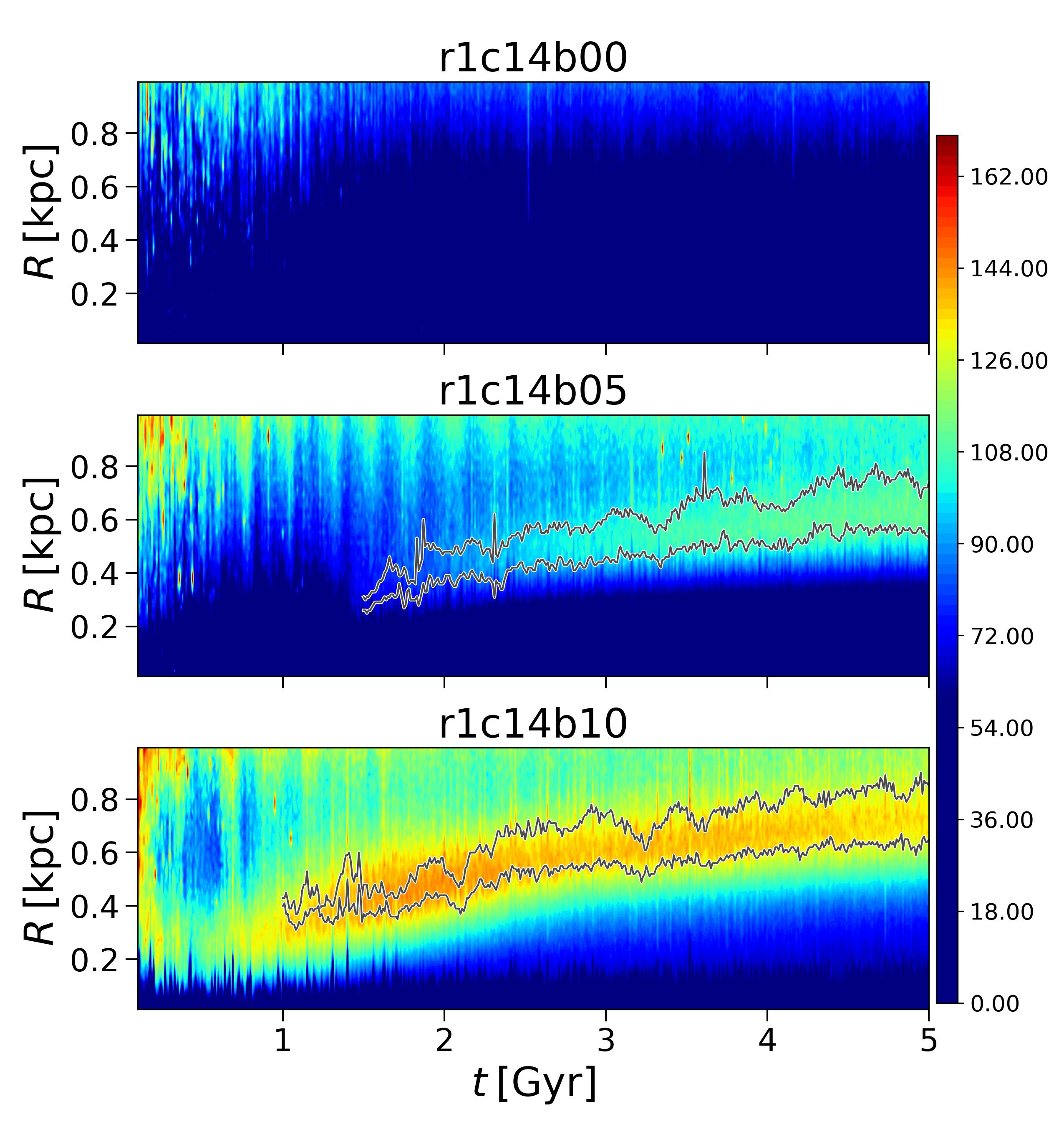} & 
        \includegraphics[width=0.33\textwidth]{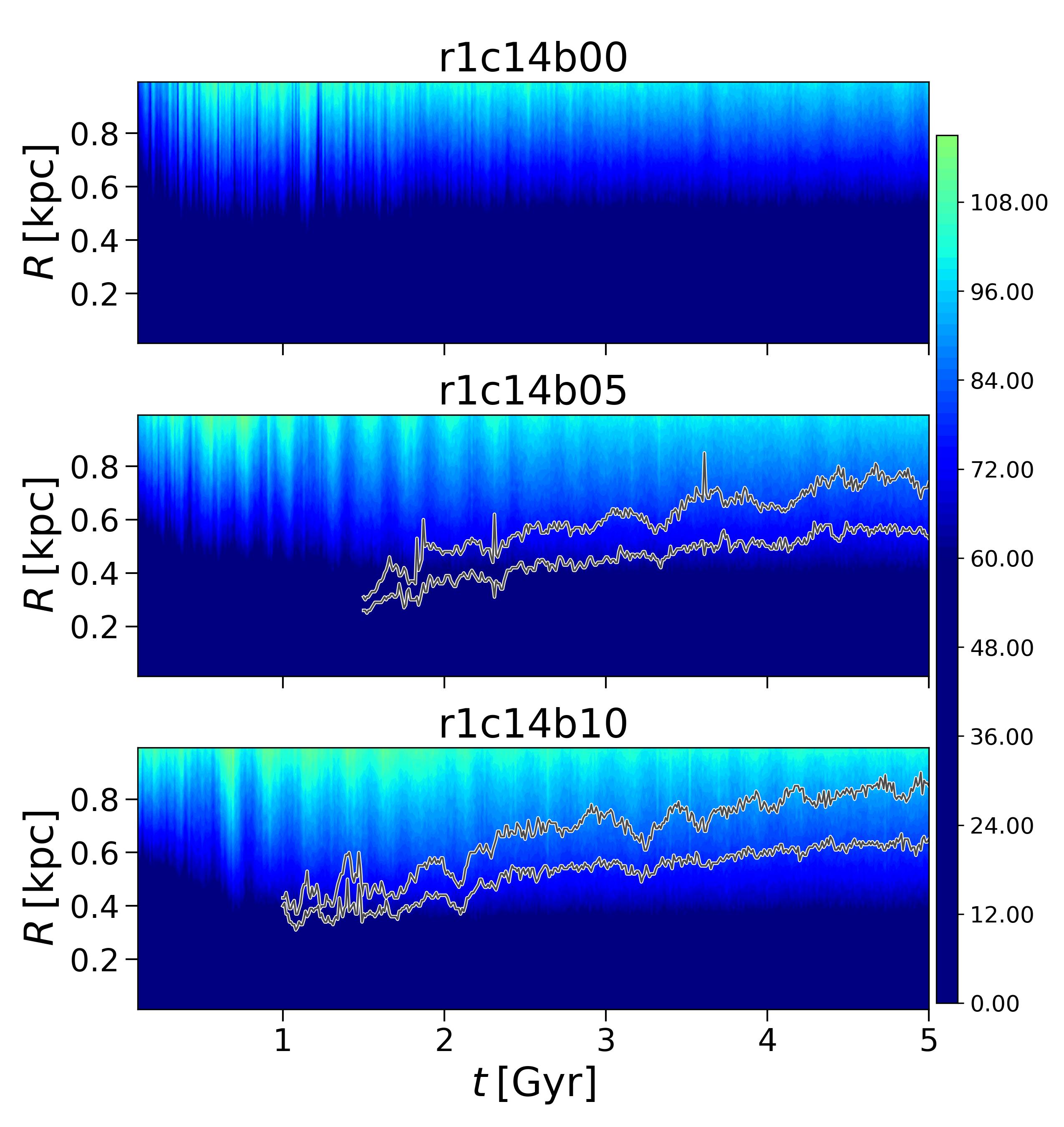}& 
        \includegraphics[width=0.33\textwidth]{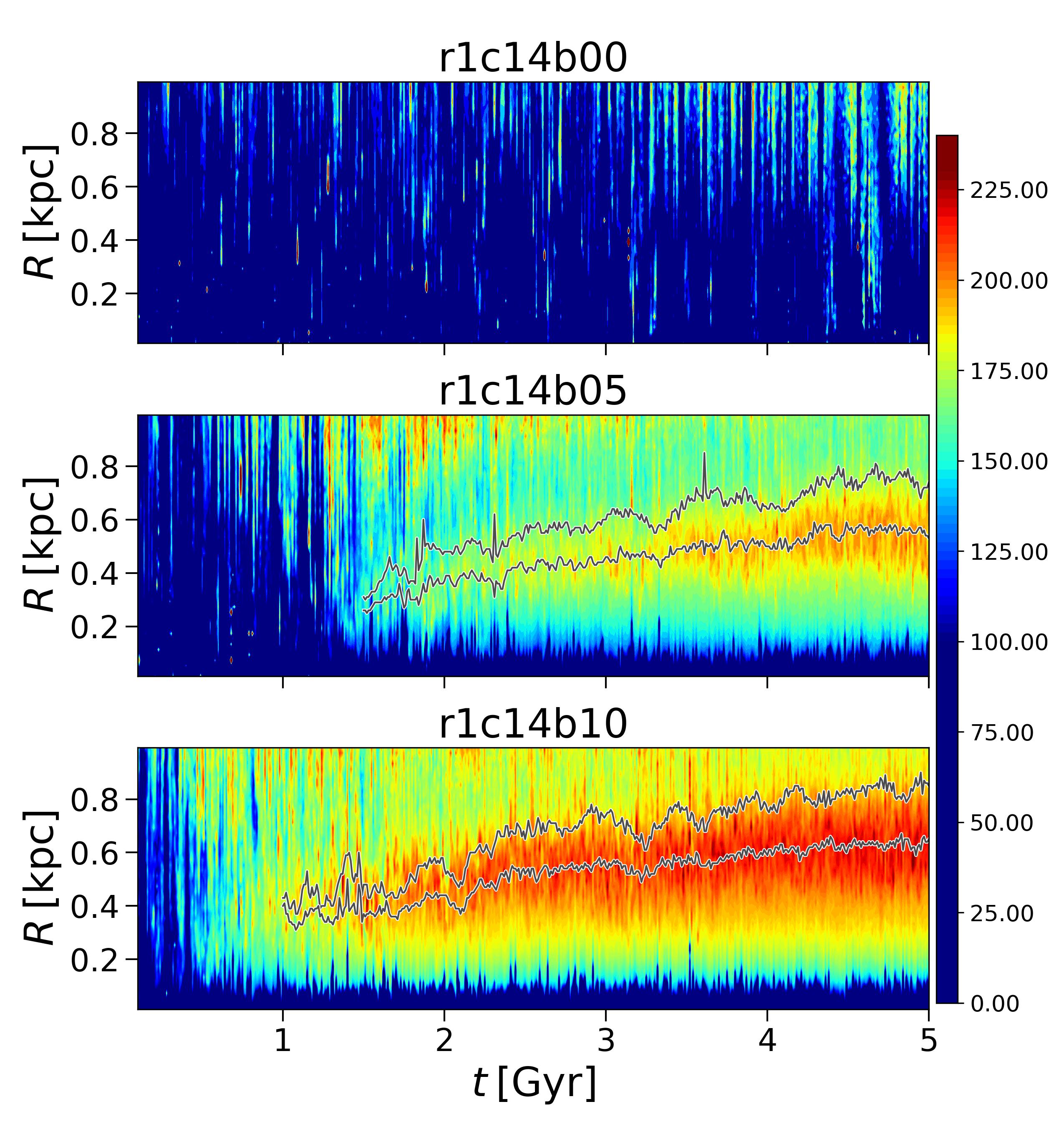}
        
    \end{array}$
    \caption{Distribution maps of the central azimuthal velocity $v_\phi$ for newly formed stars (left column), initial stellar disk particles (middle column), and gas (right column) within 1 kpc for 5 Gyr. To ensure that the same color corresponds to the same value for ease of comparison, the color distribution in the color bar is fixed for each component across all models: 60 to 170 $\kms$ for both new and initial stars, and 100 to 230 $\kms$ for the gas. Overlaid gray lines indicate the semi-major and semi-minor axes of the nuclear ring, as measured in Figure \ref{fig:ring_info}. The radial profiles of $v_\phi$ at 1, 2, 3, 4, and 5 Gyr are also presented in Fig.~\ref{fig:radial_vphi}.}
    \label{fig:vphi}
\end{figure*}

\begin{figure*}[htbp]
    \centering
    $\begin{array}{ccc}
        {\fontsize{20pt}{30pt}\selectfont \sigma_{R} \, (\mathrm{New \, Stars}) \, [\kms]} & {\fontsize{20pt}{30pt}\selectfont \sigma_{\phi} \, (\mathrm{New \, Stars}) \, [\kms]} & {\fontsize{20pt}{30pt}\selectfont \sigma_{z} \, (\mathrm{New \, Stars}) \, [\kms]}\\
        \includegraphics[width=0.33\textwidth]{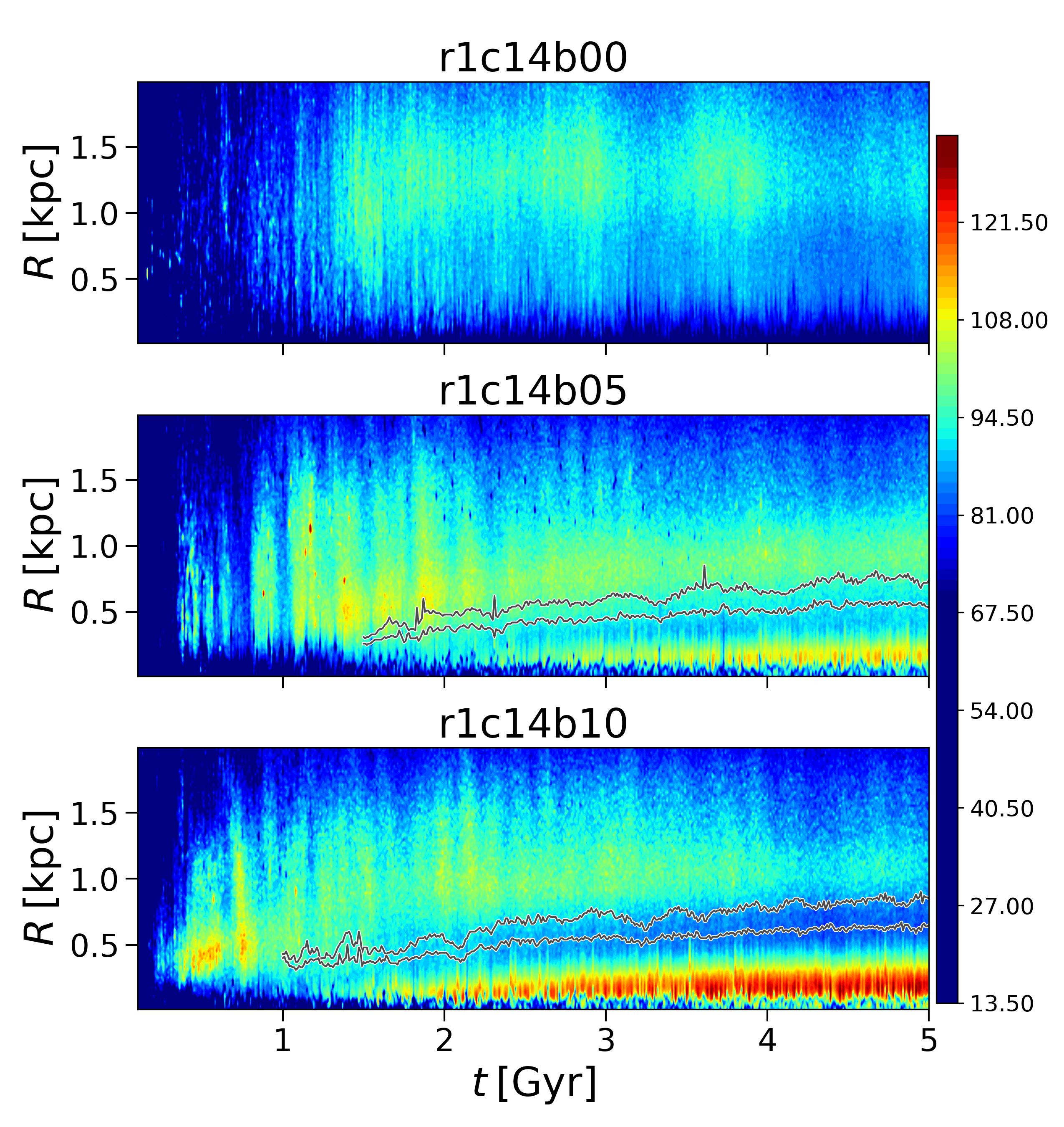} & 
        \includegraphics[width=0.33\textwidth]{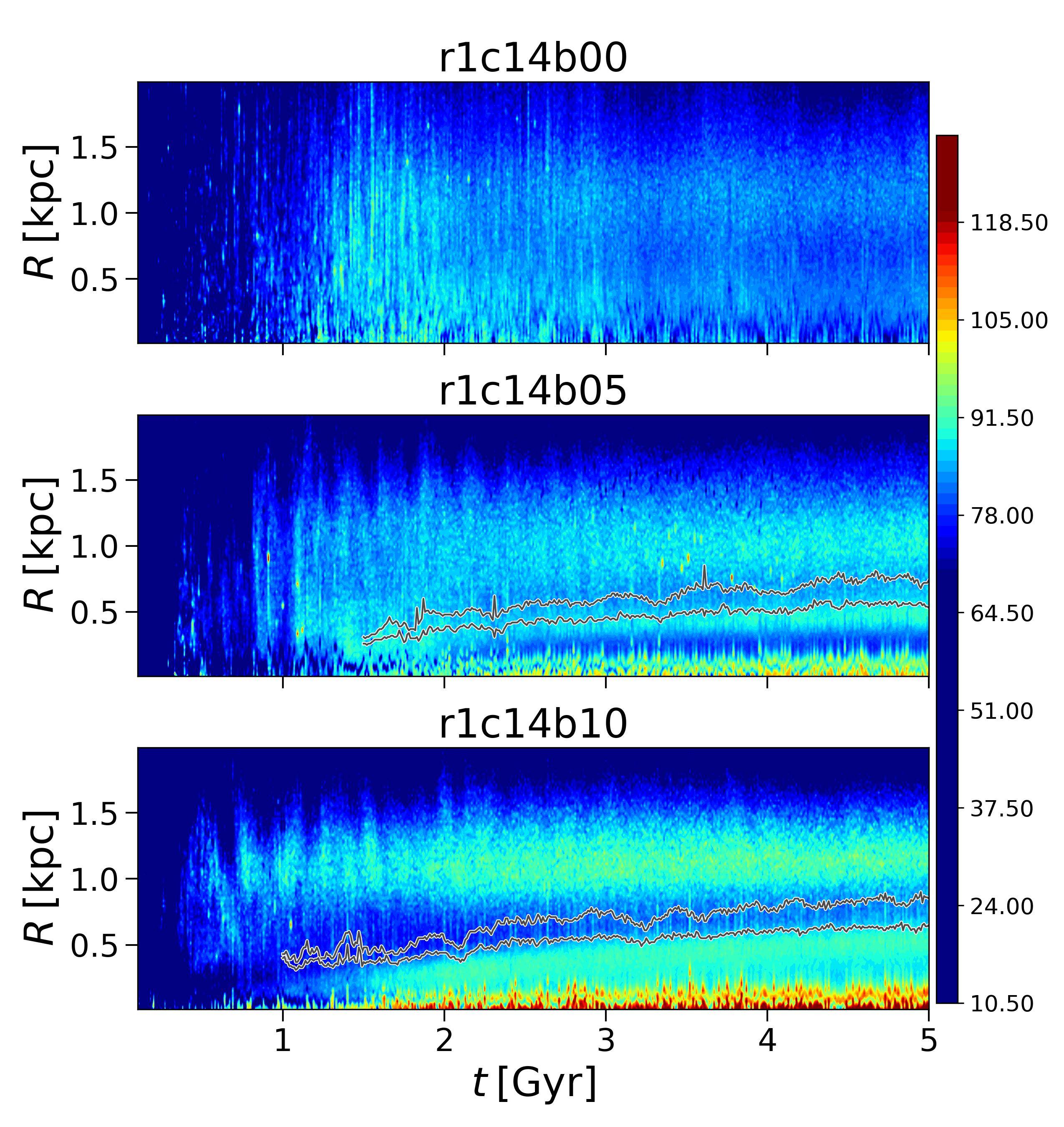}& 
        \includegraphics[width=0.33\textwidth]{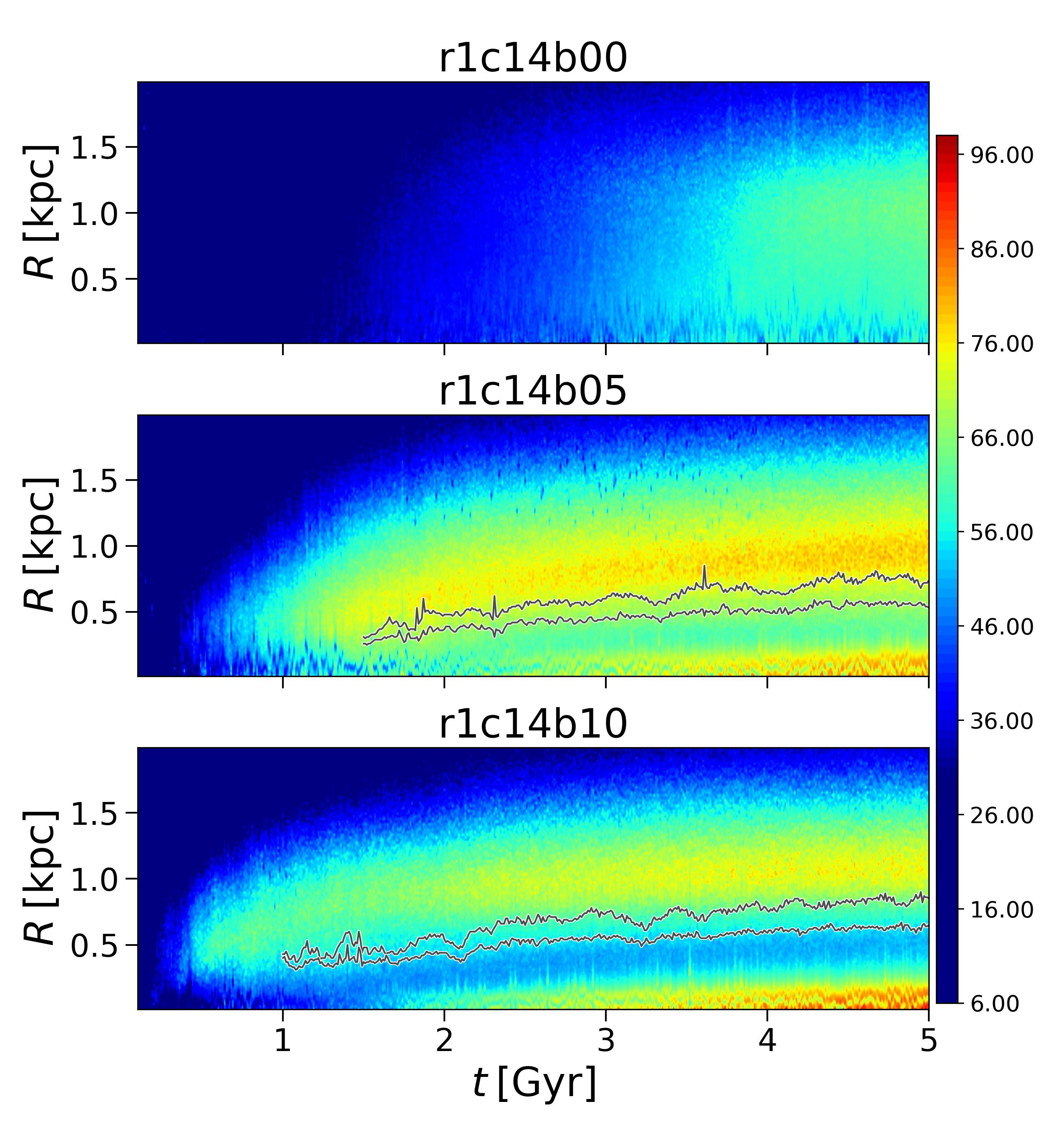}
    \end{array}$
    \caption{Distribution maps of the central velocity dispersions for newly formed stars for 5 Gyr within 1 kpc: $\sigma_R$, $\sigma_\phi$, and $\sigma_z$ from left to right. To ensure that the same color corresponds to the same value for ease of comparison, the color distribution in the color bar is fixed for each component across all models: 70 to 130 $\kms$ for $\sigma_R$, 70 to 120 $\kms$ for $\sigma_\phi$, and 30 to 100 $\kms$ for $\sigma_z$. The radial profiles of these properties at 1, 2, 3, 4, and 5 Gyr are also shown in Fig.~\ref{fig:radial_disp}.
    }
    \label{fig:disp_ns}
\end{figure*}

\begin{figure*}[htbp]
    \centering

    \includegraphics[width=0.8\textwidth]{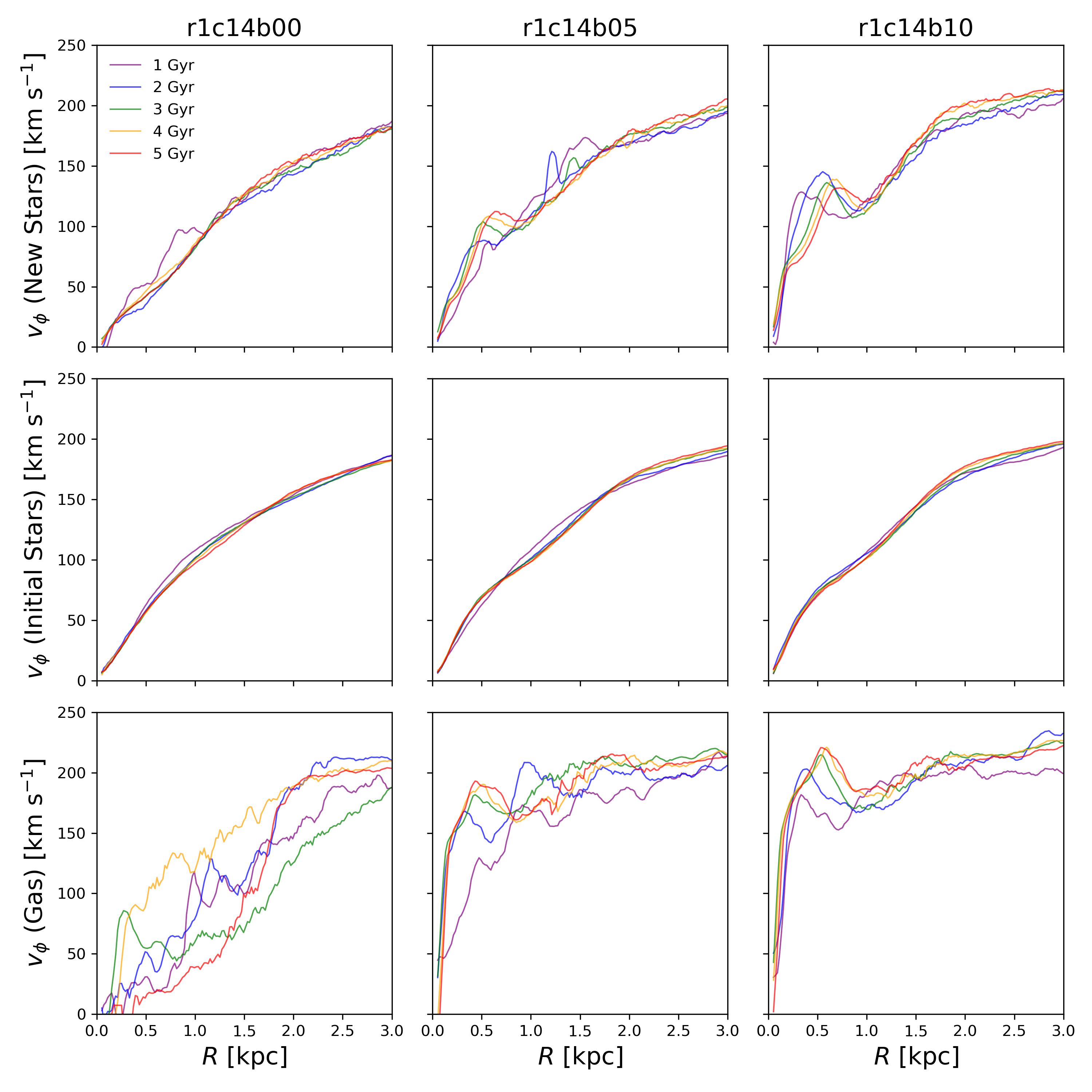}

    \caption{Radial profiles of $v_\phi$ for stars and gas at 1, 2, 3, 4, and 5 Gyr in the three models. Colors corresponding to each time are labeled. The top row shows $v_\phi$ for newly formed stars, the middle row for stars from the initial conditions, and the bottom row for the gas. From left to right, the columns correspond to the models r1c14b00, r1c14b05, and r1c14b10.}
    \label{fig:radial_vphi}
\end{figure*}

\begin{figure*}[htbp]
    \centering    \includegraphics[width=0.8\textwidth]{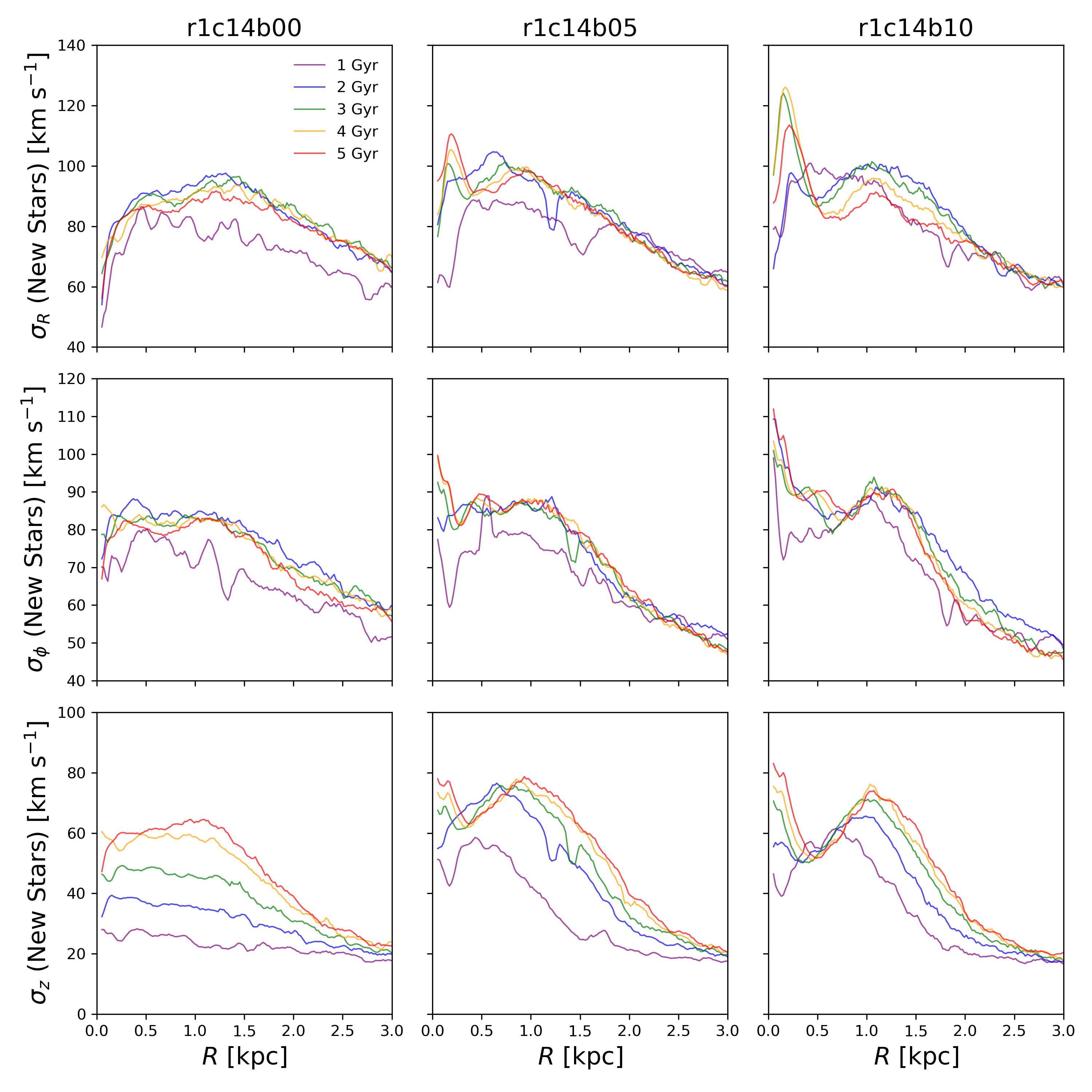}

    \caption{Radial profiles of velocity dispersion for newly formed stars at 1, 2, 3, 4, and 5 Gyr in the three models. Colors corresponding to each time are labeled. The top row shows $\sigma_R$ for newly formed stars, the middle row $\sigma_\phi$, and the bottom row $\sigma_z$. From left to right, the columns correspond to the models r1c14b00, r1c14b05, and r1c14b10.}
    \label{fig:radial_disp}
\end{figure*}

\begin{figure*}[htbp]
    \centering
    \includegraphics[width=0.9\textwidth]{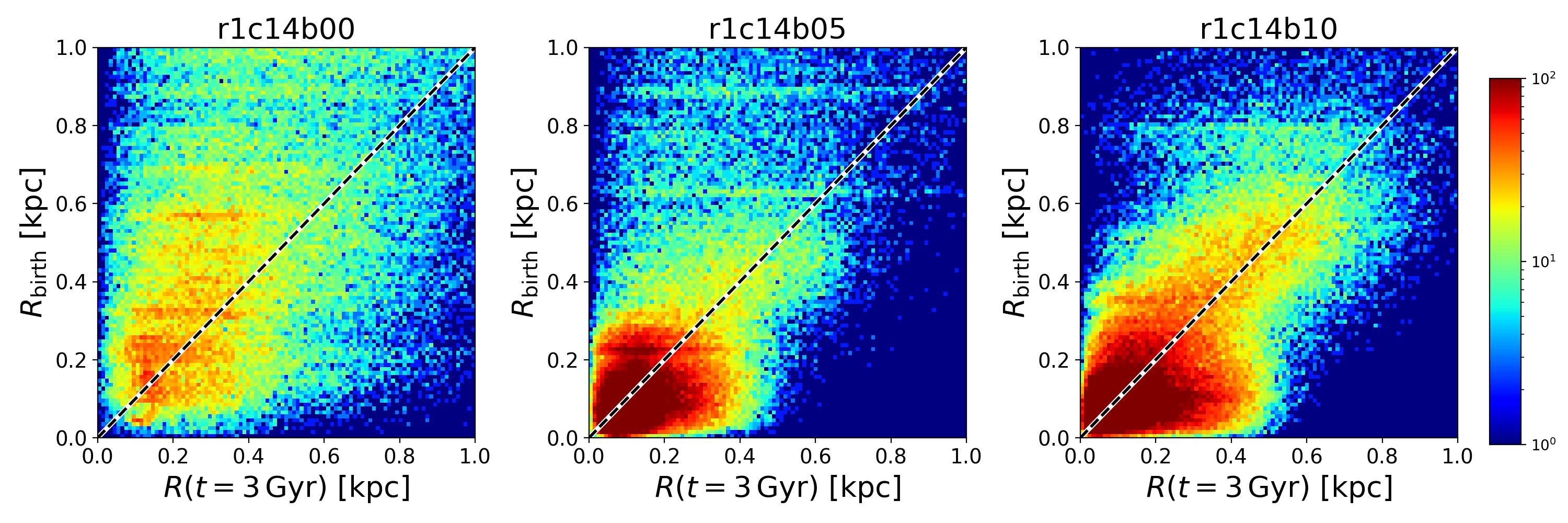}
    
    \includegraphics[width=0.9\textwidth]{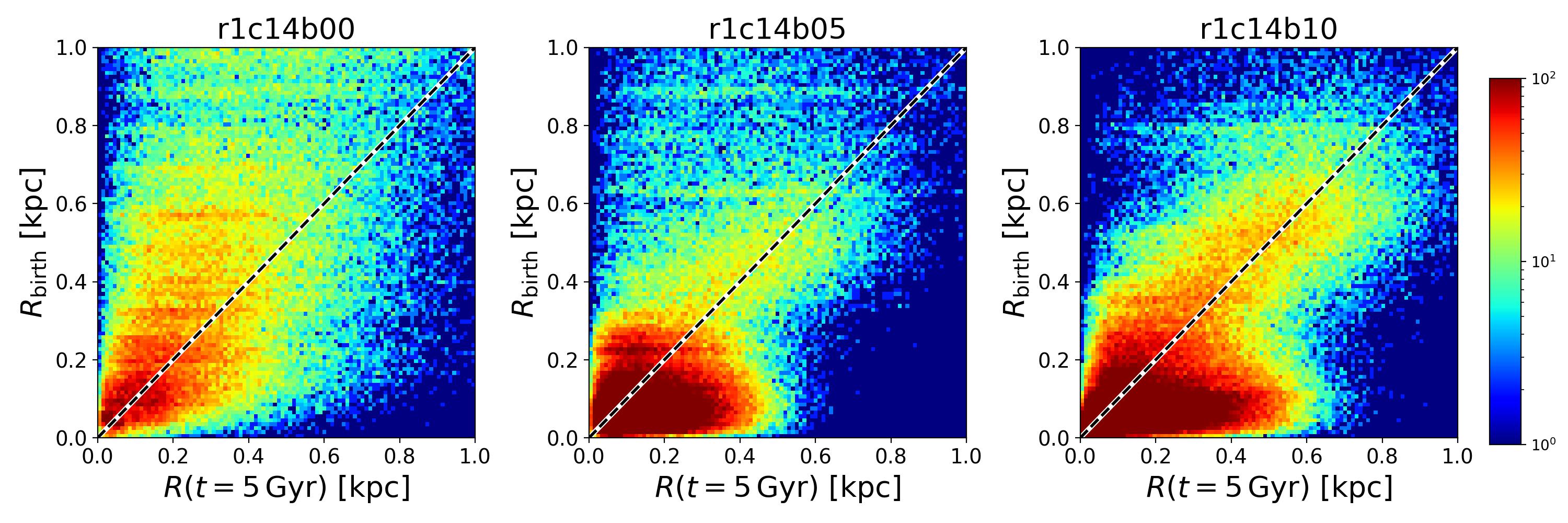}
   
    \caption{Distribution map of final radius versus birth radius for all newly formed stars within 1 kpc at 3 Gyr (top row) and 5 Gyr (bottom row), illustrating radial displacement of stars relative to their birth radii. The black dashed diagonal line indicates the locus where no radial shift has occurred. The colour bar indicates the number of stars per $  0.01 \times 0.01\,\mathrm{kpc^2}  $ bin on a logarithmic scale.}
    \label{fig:migration}
\end{figure*}

\subsection{Bar Formation}
To trace the formation and evolutionary path of bars, we calculate their strength using Fourier analysis:
\begin{eqnarray}
F_{m}(R) &=& \frac{\sum_{j} \mu_{j} \, e^{i m \phi_{j}}}{\sum_{j} \mu_{j}},
\end{eqnarray}
where \(\mu_{j}\) and \(\phi_{j}\) represent the mass and azimuthal angle, respectively, of the \(j\)th particle in an radial annulus of width \(\Delta R = 0.02\,\mathrm{kpc}\), \(m\) is an integer that denotes the multipole order, and we calculate $m=2$ to focus on the bar mode within 8 kpc.

Figure \ref{fig:f2map} illustrates the time evolution of the radial distribution of the Fourier mode $F_2$ ($m=2$) over 5 Gyr. The $F_2$ mode has been computed for the initial stars in the disk component.To ensure that the same color corresponds to the same value for ease of comparison, the color distribution in the color bar is fixed for each component across all models. For example, all values above 0.15 are shown in red.
Soon after the evolution begins, non-axisymmetric structures appear throughout the disk region due to disk instabilities.
These turn into a stable bar structure around 1 Gyr (see also Fig.~\ref{fig:f2radial}). 
Since our stellar disk is dynamically cold and unstable to bar formation, the bar formation epoch is nearly identical in all three models, regardless of bulge mass.
After that, non-axisymmetric structures such as spiral arms diminish, especially in the outer region. 

The inner radial region in Fig.~\ref{fig:f2map} with $F_2>0.15$ (red color) is taken to define the bar region. Its radial extent, however, is not constant, but fluctuates with time. During episodes of bar--spiral mode coupling, the apparent bar length can extend by a few kpc and vary on a dynamical timescale comparable to the bar--spiral beat period \citep{minchev10,quillen11,marques25}. The minimum extent of these fluctuations then provides a better estimate of the underlying bar length \citep{hilmi20}.

Overall, the radial range effectively influenced by the bar with $F_2>0.15$ decreases as the bulge mass increases across the models. For instance, the red region in model r1c14b10 with the most massive bulge covers less than 4 kpc while the bulgeless model r1c14b00 extends beyond 4 kpc. The presence of a massive spheroidal component such as a dark matter halo is known to stabilize the disk against bar instability \citep{athanassoula02a,kwak17}. In our models with the same halo concentration, the classical bulge enhances such stabilizing effects by increasing central mass concentrations of spheroidal components \citep{kataria18, jang23}. Consequently, bars become weaker and shorter with a more massive bulge. A shorter bar that forms under a more massive bulge tends to rotate faster \citep{combes93,athanassoula03,kataria19,jang23}. Thus, model r1c14b10 exhibits the highest bar pattern speed among our models by the end of the evolution (see Table \ref{table:model} and Fig.~\ref{fig:barpattern}).

Figure \ref{fig:f2radial} shows the radial profile of the bar strength taken at representative times. The maximum strength ranges from $  F_2=0.25  $ to 0.30. Once the bars form around 1 Gyr, their peak strength slightly decreases by the end of the simulation. However, the overall shapes of the radial distributions are similar as a function of time. We measure the outer radii where $F_2=0.15$ intersects with the radial profiles of $F_2$ at each time. In the bulgeless model, these radii reach $3.2$ to $4.1\,\kpc$. In contrast, the radii range from $2.1$ to $2.7\,\kpc$ in the model with the most massive bulge. As the bar gradually weakens over time, the intersecting radii decrease in all models.

In addition to the influence of the classical bulge, the gas component in our models contributes to the formation of relatively weaker and shorter stellar bars compared to those in collisionless systems. In gas-free $N$-body simulations, bars grow stronger and their pattern speeds decelerate significantly through angular momentum transfer to the DM halo \citep{athanassoula02b,kwak17,kwak19,jang24,jang25}. However, the gas disk suppresses bar formation by damping disk instabilities and angular momentum transfer, yielding weaker and shorter bars \citep{athanassoula13,seo19,lokas20}. Moreover, gas mitigates not only bar strength but also the impact of vertical buckling instabilities \citep{seo19,lokas20}. In highly turbulent, gas-rich environments, increased gas fractions can even dissolve bars via enhanced dissipation, transforming them into central bulges as star formation-driven kinematic heating disrupts non-axisymmetric features \citep{bland24}. Similar progressive weakening and shortening of bars appear in \cite{beane23}, which employs the SMUGGLE model for their Milky Way-like galaxies with varying gas fractions. As a result, the bars in our models are weaker than in gas-free $N$-body simulations, exhibiting a gradual decline in both length and strength over time.

Another effect of the presence of a gas component is that the bar pattern speed does not decrease. Indeed, the pattern speed remains nearly constant in our models (see Fig.~\ref{fig:barpattern}). Such constant pattern speeds induced by the gas disk have also been reported in previous studies \citep{friedli93,berentzen07,villa09,villa10,athanassoula14,beane23}. According to \cite{beane23}, this constancy arises for the following reasons. First, the positive torque from infalling gas prevents the bar from slowing down rapidly. As a result, the resonant phase space in the halo at a given pattern speed becomes mixed and unable to sustain a negative torque. Second, the ongoing positive torque from the gas increases the pattern speed once more. At these elevated speeds, the halo remains unmixed, allowing it to reimpose a negative torque on the bar. Consequently, the pattern speed stabilizes at a slightly higher value where the gas and halo torques balance. Over time, this leads to a gradual increase or a nearly constant speed in the bar rotation. Higher gas fractions amplify this effect and provide a natural explanation for why most observed bars, including the Milky Way’s, are fast rotators \citep{gerhard11,portail17,barbuy18,bovy19}, although the estimation of the pattern speed depends on the assumed bar length \citep{hilmi20}.

\subsection{Nuclear Disk and Ring Formation}

After the bar formation, we observe the formation of NSRs in the models that include a classical bulge. Figure \ref{fig:face_250} shows face-on images of the superimposed surface density distributions of gas and stars within a 15 kpc box at 2.5 Gyr. Young stars are overlaid: blue points represent those with ages between 0.01 and 0.1 Gyr, while purple points represent stars younger than 0.01 Gyr at the chosen simulation time. The separate surface density distributions of gas and stars, along with stacked images both with and without young stars, are presented in Fig.~\ref{fig:face_stacking} for model r1c14b05. The distribution of young stars (blue and purple) resembles the NSR and traces the gaseous spiral arms beyond the bar, forming a void region of young stars between the bar ends and the NSRs. Interestingly, classical Cepheids, which serve as young stellar tracers \citep{bono05,bono24, nunnari26}, are also found in either the inner disk or outside the bar region in the Milky Way \citep{inno19}.

New stars are born preferentially along the spiral arms of the gaseous disk and in the central gas disk. Stellar feedback is particularly prominent in these regions, which carves out low-density cavities through the combined effects of radiation pressure and supernova explosions in the SMUGGLE model \citep{marinacci19}. Over time, the gas distribution becomes increasingly structured with dense, actively star-forming regions surrounded by lower-density cavities. Such morphology -- which can be clearly observed in Fig.~\ref{fig:face_stacking} -- resembles the gas distributions of galaxies observed by JWST in nearby galaxies \citep{williams24} with visible supernova cavities and outflows.

As the stellar bar rotates, it perturbs the gaseous medium, inducing angular momentum loss and driving inward mass flows that form dust lanes \citep{combes85, shlosman89, athanassoula92, combes93, heller94, kim12a, kim12b,  seo19}. Figure \ref{fig:face_gas} presents the face-on projections of the gas distribution within a 15 kpc box at 1, 2, 3, 4, and 5 Gyr, with model names labeled atop each column. The final column provides a zoomed-in view of model r1c14b10 within 6 kpc, using a different color scale with the color bar at the bottom. At 1 Gyr, in the early stage of the bars, only model r1c14b10 with the most massive classical bulge begins forming a central gas disk. By 2 Gyr, a similar structure emerges in model r1c14b05 with a less massive bulge. These nuclear gas disks form as a combined result of accreted gas via non-axisymmetric features and pre-existing gas in the central region. Over time, these central gas disks become more prominent, while the gas distribution in the bar region fades due to mass inflows.

Since the bulgeless model r1c14b00 lacks an ILR (see Section \ref{sec:dynamics}), it cannot form a long-lived nuclear ring. The relatively strong bar causes the gas to rapidly lose angular momentum and plunge toward the center from the outset, while the ensuing stellar feedback continually disrupts the central region. Using the SMUGGLE model, \cite{li23} showed that accumulated gas can trigger bursty feedback, destroying a nuclear bar and yielding a remnant spheroidal component akin to a small classical bulge \citep{guo20} where the local instability gradually becomes dominant against the disruption \citep{romeo16}. Similarly, \cite{seo19} found that even with weaker feedback, bar-driven gas buildup and the ensuing nascent nuclear disks become disrupted during the early phase of bar evolution, ultimately delaying the formation and growth of nuclear gas disk or nuclear ring. We visually inspected the gas component at every time step (0.01 Gyr) in our bulgeless model. We find that it forms a thin gas bar with morphology similar to that reported in late-type galaxies \citep{combes93} and galaxies without ILRs \citep{athanassoula92}. Despite the gas inflow, our bulgeless model fails to maintain a stable nuclear gas disk until the end of the simulation, as stellar feedback repeatedly disperses the central gas accumulation during the early phases over.

A classical bulge is known to trap gas and form an inner gas reservoir during the early phases of bar formation \citep{combes93}. By increasing the central mass concentration, the bulge steepens the inner rotation curve and promotes the accumulation of bar-driven gas inflows within the ILRs. Additional gas inflows via tidal forcing can enlarge the central gas disk, although its extent stays limited to the ILR \citep{pettitt18}. In the bulge models, the rapid buildup of this central reservoir allows the nuclear gas disk to become sufficiently massive and extended to withstand disruption by the stellar feedback, despite ongoing active star formation. The nuclear gas disks (and rings) grow over time because, as the bar rotates, more gas accretes onto the nuclear regions and enhances their angular momentum \citep{seo19}. Due to star formation and ensuing gas depletion, a gap between its innermost region and outer edge dims, ultimately forming a nuclear gas ring (see the zoomed-in view of model r1c14b10 in Fig. \ref{fig:face_gas}). These morphological features resemble those observed in galaxies such as NGC 1097 (see Fig.~4 in \citealt{kolcu23}).

Figure \ref{fig:face_stack} illustrates the formation of young stars at intervals of 1 Gyr, displaying young stars (blue and purple) overlaid on the face-on projections of the stacked surface density distribution. In the bulgeless model, active star formation occurs along the spirals and at the innermost center. As already shown in the gas distribution (Fig. \ref{fig:face_gas}), this model fails to develop a NSD and NSR for not being able to build up a stable nuclear gas disk. Meanwhile, the bulge models form the nuclear gas disks earlier with the more massive bulge, leading to earlier NSD emergence in model r1c14b10. Also, the more massive bulge produces initially the larger NSD. As seen in the zoomed-in view of the nuclear gas disk (Fig. \ref{fig:face_gas}), gas depletion from active star formation also creates an inner gap in the young stellar distribution between the center and the ring. Consequently, new stars form primarily at the center and along the outer edge of the nuclear gas disk, where bar-driven inflows intersect the nuclear gas ring. By the end of evolution, approximately half the initial gas mass has converted to new stars, making voids around the bar and inside the nuclear ring more prominent. Despite shorter and weaker bars in the bulge models, the final nuclear ring size scales with initial bulge mass. This suggests that, in our isolated models without external gas supply, the nuclear ring size depends more on the initial central gas reservoir set by the bulge than on bar-driven inflows.

Figure \ref{fig:ring_info} presents the time evolution of nuclear ring properties, including size, ellipticity, and enclosed mass. To measure the mean radius of the ring at a given time, we analyze the spatial distribution of young stars formed in the last 0.1 Gyr between $R=0.1$ and 1.5 kpc using annuli with a width of 0.03 kpc. We exclude bins containing fewer than 1\% of the total young stars ($R=0.1$--1.5 kpc) and compute the mass-weighted mean radius as an estimate of the nuclear ring's extent. We then perform an elliptical fit on the distribution of young stars within this radial shell via a direct least-squares method. This yields semi-major ($a$) and semi-minor ($b$) axes, with ellipticity defined as $e_{\rm ring} = \sqrt{1 - (b/a)^2}$. The position angle of semi-major axis remains fixed and perpendicular to the bar's major axis throughout the evolution, and therefore it is not shown. By targeting recently formed young stars, this approach isolates the ring's active star forming structure and reduces contamination from older stellar populations. We select initial times of 1.0 Gyr for model r1c14b05 and 1.5 Gyr for r1c14b10 based on visual inspection of central gas reservoir formation in face-on images every 0.01 Gyr. The results indicate nuclear disk sizes ranging from 0.25 to 0.76 kpc, with larger initial and final sizes for more massive bulges due to the faster buildup of the nuclear gas disk. Note that the models in \cite{seo19} lack a classical bulge, leading to delayed nuclear gas disk and ring formation with a smaller size. Their largest NSR has a size of about $\sim$0.5 kpc over 5 Gyr despite rapid bar growth (see their Fig. 21), while the mean radius of the observed NSDs is about 0.5 kpc in the TIMER survey \citep{gadotti19}. Our NSR ellipticity starts around 0.5 and oscillates between 0.6 and 0.7 owing to the active star formation and ensuing stellar feedback. Such a feedback-driven oscillation is also observed in a local simulation setup \citep{moon22}. Accounting for elongation, the semi-major axis reaches up to 0.9 kpc. This result suggests that the large nuclear disk observed in the barred galaxy (CEERS-4031) at $z=1.5$ \citep{leconte26} may require a massive classical bulge or extended ILRs with ample gas inflow via bar and circumgalactic medium to rapidly trap gas and form a 1 kpc-scale nuclear disk.

The bottom-left panel of Figure \ref{fig:ring_info} illustrates the enclosed mass of new stars (dotted lines), gas (dashed lines), and their combination (solid lines) within 1 kpc in cylindrical coordinates. Gray, red, and blue colors denote models r1c14b00, r1c14b05, and r1c14b10, respectively. Consistent with the early formation of the nuclear gas disk formation observed (Fig. \ref{fig:face_gas}), model r1c14b10 accumulates $2\times10^8\,\Msun$ of gas within 1 kpc around 1 Gyr. Model r1c14b05, with a less massive bulge, reaches about half that amount at 1.5 Gyr. In the bulgeless model, the central enclosed gas mass fluctuates and decreases quickly over time due to the absence of a stable nuclear gas disk against stellar feedback. The other two panels in the bottom row of Fig.~\ref{fig:ring_info} measure enclosed masses between $a$ and $b$ (for the NSR) and within $b$ (for the NSC). These measurements use new stars only. The enclosed mass of initial stars within 1 kpc ranges from 3.5 to $7\times10^9\,\Msun$ in model r1c14b05 and from 5.0 to $8.0\times10^9\,\Msun$ in model r1c14b10, while the total enclosed mass of new stars slows down around $10^9\,\Msun$. The range of these enclosed masses is comparable to the mass of NSDs in \cite{baba20}, in which their bars and NSDs begin to form concurrently. The nuclear disk masses also agree with those derived from the TIMER collaboration \citep{desafreitas25b, gadotti25}. New stars in the NSR reach about $10^8\,\Msun$, whereas the NSC contains substantially more mass, $\sim10^9\,\Msun$. Given the initial stellar mass within 1 kpc, accurate observational mass measurements of nuclear disk and ring require subtracting the background stellar contribution \citep{schultheis19, gadotti20, erwin21, desafreitas23b, fraser24, gadotti26}.

Figure \ref{fig:sfr} shows the total star formation rate (SFR) as a dotted line and regional SFRs as follows. The solid line represents the central 1 kpc, the dashed line indicates the NSC region ($R < b(t)$), and the dash-dotted line denotes the NSR region ($b(t) < R < a(t)$). During the first 1 Gyr before the bar formation, SFRs in all models primarily occur outside 1 kpc and then gradually decline due to the absence of external gas supply. In the bulgeless model r1c14b00, the central SFR remains lower, and the total SFR decreases more rapidly compared to the bulge models that develop actively star-forming NSDs \citep{shlosman89,heller94,friedli95,seo19,baba20}. Repeated disruptions of central gas concentrations in the bulgeless case cause fluctuating and declining central SFRs, which is also shown in their central enclosed mass. The bulge models exhibit similar patterns, with gradually declining yet more elevated SFRs relative to the bulgeless model. Central SFRs become less variable once nuclear gas disks form. After 2 Gyr, when total SFRs fall below 1 $\Msun$yr$^{-1}$, most star formation shifts to the central region. For example, at 5 Gyr in model r1c14b10, the central SFR is 0.3 $\Msun$yr$^{-1}$ while the total is 0.4 $\Msun$yr$^{-1}$. In contrast, the SFRs in the NSRs fluctuate near the nuclear gas disk edge, where bar-driven inflows arrive and stellar feedback acts more effectively on those lower local densities. This fluctuation contributes to varying ellipticity (Fig. \ref{fig:ring_info}). Nevertheless, most star formation takes place in the NSC region inside the NSR's semi-minor axis. Both bulge models display distinct star formation at the very center and NSD edge, with visibly low activity in the gap between NSC and NSR (Fig. \ref{fig:face_stack}). Overall, the SFRs in the NSRs stay below 0.1 $\Msun$yr$^{-1}$, whereas the SFRs in the NSCs range from 1.0 to 0.1 $\Msun$yr$^{-1}$ over time. Nuclear gas rings in 20 nearby galaxies from the PHANGS-ALMA survey have sizes of $\sim0.4^{+0.25}_{-0.15}$ kpc and SFRs of $\sim0.21^{+0.15}_{-0.16}$ $\Msun$yr$^{-1}$ \citep{gleis26}. Interestingly, our models yield comparable NSR sizes and SFRs, though the latter are slightly underestimated due to the gas depletion but still fall within the observational error range. External gas inflows from mergers or the circumgalactic medium could elevate the SFRs and the size of nuclear structures in our models, which could explain large nuclear rings without adopting a massive bulge.

\subsection{Dynamics of Nuclear Structures}\label{sec:dynamics} 
To understand the formation of a NSD and its subsequent evolution into a NSC and NSR, we examine the Lindblad resonances in all models. Figure \ref{fig:resonance} displays the circular velocities, bar pattern speed $  \Omega_b  $, and azimuthal velocities $  v_\phi  $, where the subscript `t', `i', and `g' represent total, initial stars, and gas, respectively. Thick solid lines denote values at 5 Gyr, while transparency increases for lines from 1.5 Gyr to 4.5 Gyr in 0.5 Gyr steps. Lindblad resonances in galactic disks result from interactions between stellar or gaseous orbits and non-axisymmetric perturbations, such as bars. The corotation resonance (CR) arises where the angular frequency $  \Omega  $ equals $  \Omega_b  $, defining the radius at which orbits co-rotate with the bar. The angular frequency $  \Omega(R)  $ is calculated from the gravitational potential $  \Phi  $ as
$\Omega(R) = \sqrt{\frac{1}{R} \frac{\mathrm{d}\Phi}{\mathrm{d}R}}$. This radius decreases with more massive bulges with the higher pattern speeds (see Table \ref{table:model}). ILRs satisfy $\Omega - \kappa/2 = \Omega_b$, while outer LRs (OLRs) satisfy $\Omega + \kappa/2 = \Omega_b$, with the radial epicyclic frequency $\kappa$ given by $\kappa^2 = 4\Omega^2 + Rd(\Omega^2)/dR$. The formation of a bar induces large random motions, reducing the mean rotational velocities of stars $v_{\phi}$ in the bar region \citep{baek25}, but the bar-driven gas inflows continuously add gas and newly formed stars to the nuclear region, leading to an increase in mass and $\Omega_t$ over time. In systems with high central mass concentration, the $\Omega - \kappa/2$ profile exhibits a peak, enabling two ILRs for $\Omega_b$ below the peak \citep{combes85,combes93}. These consist of an inner ILR (iILR) at smaller radii and an outer ILR (oILR) at larger radii. While the bulgeless model r1c14b00 shows initially the absence of ILR, the bulge models display a rising-then-falling $\Omega -\kappa/2$ curve, yielding dual ILRs that intersect $\Omega_b$. The x2 orbits, which are rounded and elongated perpendicular to the bar, exist solely between these two ILRs, and increasing central density enhances ILR effectiveness in this process \citep{combes96,buta96}. In the bulge models, enhanced star formation over time within the NSD elevates the $  (\Omega - \kappa/2)_t  $ curve above $  (\Omega - \kappa/2)_i  $, but also in the innermost regions inside the peak of  $(\Omega - \kappa/2)$ curves, where massive NSCs develop by the end of evolution (Fig. \ref{fig:resonance}). After evolving for a few Gyr, the bulgeless model also possesses an elevated peak in the $(\Omega - \kappa/2)_t$ curve, which intersects its bar pattern speed. However, it is not effective enough to trap gas and form a stable nuclear gas disk against the feedback effects.

Figure \ref{fig:resonance} also shows a peak in $  v_{\phi,g}  $ with values around 200 to 220 km s$^{-1}$ in the bulge models only. The rotational velocity of young stars is lower than that of the interstellar gas, yet higher than that of old stars. Since young stars are formed inheriting the angular momentum of the gas, they possess higher rotational velocities than old stars. Furthermore, because young stars exhibit a smaller velocity dispersion than old stars, they rotate faster than the older population. At the end of the simulated time span, the positions of these velocity  peaks are found approximately at 0.5 and 0.6 kpc in models r1c14b05 and r1c14b10, respectively, which are located inside the peaks of $  v_{\phi,n}$. Thus, a more massive bulge induces a larger and faster rotating nuclear gas disk. As \cite{kim12c} demonstrated, the nuclear ring position is not determined by Lindblad resonances. Indeed, the peaks of $  v_{\phi,g}  $ and $  v_{\phi,n}  $ in our models do not align with those resonances and gradually shift outward over time (see also Fig. \ref{fig:radial_vphi}). Instead, our results indicate that the radius of the nuclear gas disk and subsequent radial growth of NSRs depend on the $\Omega - \kappa/2$ curve and subsequent gas inflow. 
Larger bulge masses steepen the $  \Omega - \kappa/2  $ curve and enable greater gas trapping after the bar formation. Note that our models differ by  bulge mass only, but changing the scale length of a classical bulge can also alter the distribution of the nuclear gas disk \citep{pettitt18}.

Additional gas inflow through the bar enlarges the nuclear gas disks by transferring angular momentum. As shown in \cite{pettitt18}, oILRs may establish the maximum radius to which the nuclear gas disk can expand. Their Dip models with the strongest bulge component and strongest tidal forcing produce the largest nuclear gas disks, where tidal forcing supplies excessive gas to the nuclear region. The radius of their largest nuclear gas disk coincides with the oILR radius, and the dust lanes align with the disk edge (see their Fig. 26). We conjecture that negative torque amplitude decreases as gas approaches the oILR \citep{goldreich79}. This decrease leads to sign reversal and gradual halting of radial inflow. The process ultimately compels gas to settle onto x2 orbits between the iILR and oILR, and the settlement location is determined by the point where torques balance, depending on the bulge strength, pre-existing gas disk, and inflowing gas. 
In fact, the size of our nuclear rings lies between the iILR and the oILR and increases with the oILR, as shown in Figure \ref{fig:ilr_evolution}, and this is consistent with \cite{sormani18,sormani24}.
In future work, we will revisit what determines the radius of x2 gas orbits by examining galaxies with various bar and bulge properties.

The formation of the inner gas disk, which later diverges into a central gas peak and nuclear ring, can be seen in the density profiles of initial stars, new stars, and gas within 3 kpc in Fig.~\ref{fig:den}. Outside the peak of $\Omega - \kappa/2$ ($\sim0.25$ kpc in our bulge models), the centrally accumulated gas initially settles into a rotational disk due to a torque balance where bar perturbations decrease inward. As time progresses, a fraction of the gas crosses the peak of $\Omega - \kappa/2$ through stellar feedback or instabilities (e.g., nuclear spirals in \citealt{kim12c}, \citealt{kolcu23}, \citealt{pastras26}, and \citealt{kolcu26}), moving toward the innermost region. Inside the iILR ($\lesssim0.1$ kpc), torques become positive again and cause gas to pile up at the iILR from the central region. This process of gas trapping and subsequent divergence into two density peaks at the edge and center of the nuclear gas disk appears in the bulge models where the active star formation forms the NSC and NSRs (Fig. \ref{fig:face_stack}). For example, model r1c14b10 initially forms a nuclear gas disk and NSD at 1 Gyr. At 2 Gyr, within 1 kpc, the inner and outer peaks emerge, and the gap between the iILR and nuclear gas ring widens accordingly. Due to active star formation in the innermost region, the iILR shifts inward as the $\Omega - \kappa/2$ curve steepens, causing the gas density peak to move inward from 2 to 5 Gyr. Meanwhile, gas density around the bar effective regions decreases as inflowing to the nuclear gas disk, which enlarges the size and enhance the density peak at its edge where the young stars actively form as the NSRs. Such a process of forming dual peaks in gas density is not observed in the bulgeless model.

To examine radial changes in rotations of the new stars, initial stars, and gas components, we present distribution maps of azimuthal velocities over time within 1 kpc, overlaid with the semi-major and semi-minor axes of the NSRs, in Fig.~\ref{fig:vphi}. To highlight the emergence of high rotation relative to background stars, we define the color bar such that the same color palette maps the velocity range 60–170 km s$^{-1}$ for both new and initial stars, and 100–230 km s$^{-1}$ for gas, ensuring that a given color corresponds to the same velocity value across all models. The bulgeless model displays chaotic and fluctuating $v_\phi$ in gas, with no stable nuclear structures due to the absence of dual ILRs and feedback disrupting central accumulations. This underscores the bulge's role in inducing gas trapping and kinematic coherence in nuclear regions. Indeed, our bulge models exhibit distinct regions with higher rotation, which gradually shift outward due to angular momentum gains from bar-driven inflows and subsequent star formation. Notably, the NSR regions in the gray solid line (radial area between $a$ and $b$ of the NSR) overlap well with these rotation peaks. Meanwhile, the $v_\phi$ distributions for initial stars show no signs of higher rotation, indicating that they do not contribute to nuclear ring formation. This finding supports using nuclear disks for age-dating bars \citep{baba20,sanders24, desafreitas23a, desafreitas23b, desafreitas25a, desafreitas25b} as they are chemodynamically distinct structures from bar/bulge stars \citep{schultheis21,nogueras24}. The $v_\phi$ distributions in the gas components of bulge models confirm that the gas disk resides slightly inside the NSR, as the stellar ring forms from bar-driven inflowing gas at the disk edge. Radial profiles at 1, 2, 3, 4, and 5 Gyr for all models appear in Fig.~\ref{fig:radial_vphi}. Gas velocities fluctuate over time in model r1c14b00, whereas the bulge models quickly develop rotating gas within 1 kpc that gradually rises and peaks around $v_\phi \approx 200$ to 220 km s$^{-1}$, with higher values for the more massive bulge. The NSRs from new stars peak around 110 and 140 km s$^{-1}$ in models r1c14b05 and r1c14b10, respectively. As already shown in the distribution maps of the mean rotations (Fig.~\ref{fig:vphi}), initial stars exhibit no distinct rotation peaks beyond the bar rotations.

Due to the kinematically cold nature of NSDs, many observations report central $\sigma$-drops in galaxies that possess NSDs \citep{kormendy82,emsellem01,marquez03,delorenzo08,gadotti15,erwin15,gadotti19, gadotti20}. These drops indicate that NSDs are rotationally supported with $v/\sigma \sim 1$ to 3 typically. Figure \ref{fig:disp_ns} shows the radial, azimuthal, and vertical velocity dispersions of new stars within 2 kpc in all models. The NSR regions are overlaid as solid gray lines. We identify distinct $\sigma$ regions: the low $\sigma$ region overlaps with the NSR region, while $\sigma$ in the radial, azimuthal, and vertical directions gradually increases in the innermost region where the NSC resides. Due to the presence of a classical bulge and brief NSD formation from the gas reservoir, new stars in the bulge models exhibit higher $\sigma$ than in the bulgeless model. However, the formation of the nuclear ring creates a dynamically cold region between the oILR and the innermost area. After the bar formation, the NSC region begins to widen, and its $\sigma$ gradually increases due to the gas trapping and the formation of a gas density peak. The high gas density drives vigorous star formation, leading to a rapid increase in mass within the compact central region. Consequently, NSCs undergo heating in all directions, but especially more in the radial direction as radial gas inflow forms the peak around the iILR at $\sim0.1 \, \kpc$ (Fig. \ref{fig:den}).

Figure \ref{fig:radial_disp} illustrates the radial profiles of $\sigma_R$, $\sigma_\phi$, and $\sigma_z$ for new stars at 1, 2, 3, 4, and 5 Gyr across all models. Compared to the bulgeless model, the NSCs exhibit distinct high peaks in $\sigma$. Innermost radial heating appears more pronounced in the bulge models with more massive bulges. For example, model r1c14b10 shows $\sigma_R \approx 110$, $\sigma_\phi \approx 110$, and $\sigma_z \approx 80$ at 5 Gyr, though $\sigma_R$ peaks at 0.1--0.2 kpc around the iILR. The low values of $v_\phi \approx 0$ to 70 within this small radial area indicate that the NSC is a pressure-supported system with $v_\phi / \sigma_R \approx 0$ to 0.63. In contrast, the NSR in the same model has $v_\phi \approx 140$ and $\sigma_R \approx 85$ at $R_{\rm ring} = 0.76$ kpc at 5 Gyr (Table \ref{table:model}), yielding $v_\phi / \sigma_R \approx 1.65$. The NSR in model r1c14b05 with a less massive bulge has $v_\phi / \sigma_R \approx 1.22$ at $R_{\rm ring} = 0.64$ kpc at 5 Gyr. The central region of the bulgeless model is pressure-supported with $v_\phi / \sigma_R \approx 0.55$  at $R_{\rm ring} = 0.50$ kpc at 5 Gyr. These results confirms that the NSR is rotationally supported, and a more massive bulge produces a larger and kinematically colder ring.

Figure \ref{fig:migration} illustrates the radial displacement of new stars within 1 kpc by comparing their birth positions to their positions at 3 Gyr and 5 Gyr. The bulgeless model, with less star formation in the 1 kpc region, exhibits mild inward radial displacement. The bulge models display distinct formation of NSCs and NSRs. At 3 Gyr, the region of NSCs shows no visible radial displacement, and most new stars concentrate within 0.2 to 0.3 kpc. At 5 Gyr, stars born within 0.2 kpc move outward as the NSC grows more massive, accompanied by increasing radial heating -- an effect that becomes more pronounced with a more massive bulge (bottom panel of Fig. \ref{fig:radial_disp}). In model r1c14b10, a considerable fraction of NSC stars born within 0.2 kpc migrate radially outward. Specifically, 26.6\% and 8.4\% of the new stars born within 0.2 kpc end up in the 0.2--0.4 kpc and 0.4--0.6 kpc annuli, respectively. In model r1c14b05, 3.4\% of the newly born stars within 0.2 kpc move to the 0.4--0.6 kpc annulus. These radial shifts also appear in the $\sigma_R$ of new stars, where the central peak shifts slightly outward from 3 to 5 Gyr (top panel of Fig. \ref{fig:radial_disp}). This outcome suggests that careful kinematic selection is required to distinguish NSC and NSR components, as the outward-migrating NSC may contaminate the former radial extent of the NSR with its massive and extended structure. Unlike NSCs, NSRs show no significant radial shift and remain primarily along the diagonal line in Fig.~\ref{fig:migration}.
In a follow-up study, we will employ higher-resolution models to trace the ages and radial shift of the NSRs, together with their interactions with NSC heating, which will be essential for age-dating the bar using nuclear structures.

\section{Discussion and Summary}\label{sec:discussion}
We investigate the formation and evolution of nuclear structures, such as nuclear disk, NSC, and NSR in isolated bar-forming galaxies. We construct three Milky Way-mass galaxy models by varying the bulge mass only, while keeping the disk and halo structure identical. To examine the detailed ISM substructures, our galaxy models are evolved for 5 Gyr by using the \textsc{SMUGGLE} ISM and stellar feedback model  \citep{marinacci19}. Without adopting a fixed bar potential, our isolated galaxies with a classical bulge form a bar and nuclear disk and ring with a NSC at the center.

Our self-consistent simulations demonstrate the pivotal role of a classical bulge in modulating bar formation and the emergence of nuclear structures in isolated Milky Way-mass galaxies. In the presence of a bulge, bars develop weaker and shorter, consistent with the stabilizing influence of central mass concentrations on disk instabilities \citep{combes93,athanassoula02a}. The bulge augments the spheroidal component's effect, akin to a concentrated dark matter halo, by increasing the central density and suppressing non-axisymmetric growth \citep{kataria18, jang23}. This leads to reduced bar strengths and lengths. The gas disk further tempers bar evolution \citep{seo19}, yielding nearly constant pattern speeds through torque equilibrium between infalling gas and the halo \citep{beane23}. Such dynamics prevents significant bar deceleration observed in collisionless $N$-body simulations \citep{athanassoula02b}, aligning with the prevalence of fast bars in observations. However, despite the weaker bar formation, our bulge models possess a nuclear ring at the end of simulation.

The formation of NSDs and NSRs hinges on the bulge's ability to trap gas within the ILR during early bar phases, as evidenced by the absence of stable nuclear structures in our bulgeless model. 
Without the ILR, a thin gas bar forms that funnels gas directly toward the center \citep{athanassoula92, combes93, seo19}, while stellar feedback disrupts the central gas accumulations. This delays or even prevents NSD development despite ongoing gas inflows \citep{seo19,li23}, depending on the strength of the feedback. We note that a massive classical bulge is not always necessary for nuclear disk formation. In fact, many barred galaxies lacking large spheroidal components are observed to host NSDs in the TIMER survey \citep{bittner20}. We suggest that the time required to form a nuclear disk and to reach the sizes observed in those galaxies becomes longer without a central spheroidal component. This implies that the bars in such galaxies would need to be relatively older if the systems have undergone isolated evolution. In the early phase of bar formation in the bulgeless model, a small spheroidal stellar component forms in the central region \citep{combes93,guo20}, gradually steepening the $\Omega-\kappa/2$ curve and eventually creating the ILRs. However, due to the stronger feedback, the formation of a stable nuclear gas disk is continuously disrupted over the 5 Gyr of evolution. The galaxy model in \cite{seo19} also lacks a classical bulge but adopts weaker feedback, leading to delayed nuclear disk formation and a smaller final size. Over the 5 Gyr of evolution, the largest NSR in \cite{seo19} reaches only $\sim$0.5 kpc, which may not be sufficient to explain the larger NSRs or NSDs observed in the TIMER survey \citep{gadotti19} unless the simulation is evolved for a longer period or includes external gas inflows \citep{pettitt18}. In addition to the effects of feedback and external gas inflows, these scenarios should be revisited by including additional mechanisms such as turbulence and magnetic fields, which can suppress star formation \citep{moon23}, thereby allowing the inflowing gas to remain dynamically colder for longer and potentially enabling the formation of a small nuclear gas disk even in the absence of a classical bulge.

In contrast, bulge models rapidly form nuclear gas reservoirs, with more massive bulges yielding earlier and more extended nuclear disks, evolving into NSRs of 0.64--0.76 kpc by 5 Gyr. This process involves bar-driven inflows along dust lanes, accumulating gas inside the ILR where torques balance \citep{athanassoula92, combes93}. Subsequent star formation depletes the inner disk, creating gaps and bifurcating into a central NSC and peripheral NSR, with young stars tracing active sites at the center and ring edge. This result agrees with the suggestion by \cite{nogueras23} that, based on the kinematic and metallicity gradients of the Milky Way, nuclear structures might have diverged from the same structure. NSDs are observed across a wide range of galaxy types \citep{schultheis25}, and our models naturally explain their presence in galaxies hosting central spheroidal components. As suggested in \cite{gadotti26}, a proper estimation of the classical bulge component is essential for understanding the initial conditions that give rise to the observed variety of NSD sizes.

Our NSR sizes and ellipticities (0.6--0.7, oscillating due to feedback) compare favorably to TIMER survey observations (mean NSD radius $\sim$0.5 kpc in \citealt{gadotti20}) and PHANGS-ALMA nuclear rings ($\sim$0.4 kpc in \citealt{gleis26}). However, the isolated setup limits the gas supply, resulting in declining SFRs (central 0.1--1 $\Msun/\rm yr$, NSR $<0.1$ $\Msun/\rm yr$) below, but still compatible with, the observed values of \cite{gleis26} ($\sim0.21^{+0.15}_{-0.16}$ $\Msun/\rm yr$). Also, additional gas inflow can grow the size of the nuclear gas disk larger by transferring angular momentum over time. In fact, \cite{erwin21} presented contrasting cases of two barred galaxies that possess different nuclear structures, as they presumably have undergone different evolutionary paths for the gaseous medium in field and cluster environments. Future SMUGGLE-Ring projects with various galaxy properties, gas fractions, tidal effects, and merger scenario will probe NSD diversity and interesting outliers such as the JWST-observed high-z galaxy (CEERS-4031) with a 1 kpc-scale nuclear disk \citep{leconte26}.

Dynamically, the NSDs in our bulge models exhibit kinematic decoupling, with rotationally supported NSRs ($v_{\phi}/\sigma_R \approx 1.2-1.7$) and pressure-supported NSCs ($v_{\phi}/\sigma_R \approx 0-0.7$), reflecting central heating from vigorous star formation and radial inflows. The $\sigma$-drops in NSRs indicate cold kinematics, consistent with observations of central velocity dispersion decreases in NSD-hosting galaxies \citep{erwin15,gadotti19,gadotti20, schultheis21}. Radial profiles show peaks in gas and new star rotations (200-220 km s$^{-1}$ and 110-140 km s$^{-1}$, respectively) shifting outward, driven by angular momentum gains and following size growth. The $\Omega - \kappa/2$ curve's steepening from the presence of a bulge and NSC buildup enables dual ILRs, trapping gas between iILR and oILR, although the formation of the NSR has nothing to do with the ILR \citep{kim12c}. Radial displacement relative to birth radii over time is insignificant in NSRs but prominent in NSCs possibly due to radial heating and/or radial migration, although further investigation at higher resolution will be required to separately quantify the effects of radial heating and migration. This suggests accurate kinematic separation is needed for future component analysis \citep{schultheis25}. These kinematics affirm NSDs as younger, distinct populations for bar age-dating \citep{baba20, desafreitas23a, desafreitas23b,desafreitas25a,desafreitas25b}. Finding correlations among the properties of NSDs (e.g., the mass and size of the NSC and NSR), the bar, and the classical bulge will allow us to better understand the evolutionary history of barred galaxies, including our Milky Way.

Our simulations support the inside-out formation scenario for NSDs and NSRs. The bulge establishes a nuclear gas disk that expands via bar-driven gas inflows, and the star formation proceeds on the edge of the NSRs. This is evident from the radial offset between the velocity peaks of the nuclear gas disks and the NSRs, together with the outward shift of these peaks over time. In accordance with our results, the TIMER survey interprets nuclear rings as the expanding, star-forming outer edges of continuous and expanding nuclear disks \citep{bittner20}. This interpretation implies that the disk builds from a compact central reservoir outward. The survey provides further evidence for the inside-out formation of NSDs by demonstrating radial gradients where stellar ages decrease and metallicities decline with increasing radius from the center to the nuclear ring. Evidence for the inside-out scenario is also found in the Milky Way through chemical analysis \citep[see][]{ryde25, schultheis25b}. In future work, we plan to explore the chemodynamics and ages of nuclear structures under various initial conditions to provide detailed references for understanding the nuclear disks and potential bar-forming epochs of the Milky Way and other galaxies.

The Milky Way's NSD is notably compact with a radius of $  \sim 0.1\,{\rm kpc}$  \citep[][and references therein]{schultheis25}. This radius is significantly smaller than the NSRs in our bulge models (0.25--0.76 kpc). This difference implies that the Milky Way's classical bulge or ILR may not be as dominant in promoting early, extended nuclear gas reservoirs as in simulations with more massive bulges. The formation scenario of the Milky Way's NSD can depend on the bar age and bulge mass. For instance, if the bar is old and the mass of the classical bulge was negligible at the time of bar formation, it would have evolved over a long period of secular evolution to build a sufficient central mass concentration for NSD formation \citep{seo19}. Alternatively, if the bar is young, a considerable mass of central spheroidal component should pre-exist. Recent observational studies suggest that the Milky Way may indeed host a small, early-formed classical bulge \citep{nepal26}, and that the Galactic bar may be relatively young \citep{nepal24}. This component could trap bar-driven infaling gas and settle it into a rotational structure. These scenarios should be considered alongside external gas injection, since a history of gas depletion could limit outward growth. For instance, our models adopt a fixed gas fraction of $  M_g/M_s \approx 9.1\%  $, which is comparable to the observed gas fractions in Milky Way-mass galaxies \citep{sage93, leroy09, saintonge11, papovich16}. However, this value might also correspond to the gas fraction present at an intermediate stage of bar evolution, rather than to the initial gas fraction. In the young-bar scenario, a slight increase in the initial gas fraction would be sufficient to explain the observations and would not produce significant differences from our current results. In the old-bar scenario, either a substantially higher initial gas fraction or additional external gas accretion would be required. A higher initial gas fraction would lead to a weaker bar \citep{seo19} and would take longer to form owing to less efficient angular-momentum exchange between the bar and the DM halo \citep{beane23}. As a result, even with more gas available in the disk, the bar would drive gas toward the center less efficiently. This would result in reduced central star formation activity while making the nuclear disk to become larger owing to the shorter bar length \citep{seo19}.

Kinematically, the Milky Way's NSD has $v/\sigma \approx 1.4$ \citep{sormani22,shahzamanian22}. This value aligns closely with the range found for our NSRs with $v_\phi/\sigma_R \approx$1.2--1.7. Both indicate rotationally supported, dynamically cool systems. Our measurements refer to the peaks of the azimuthal velocity profiles at 5 Gyr. These values evolve with time and also vary radially because of the presence of NSCs. At 5 Gyr, our peak rotations reach 110--140 km s$^{-1}$ for new stars and 200--220 km s$^{-1}$ for gas. These exceed the Milky Way NSD value of $v_{\phi} \approx 100$ km s$^{-1}$. The difference likely arises from steeper inner rotation curves in our models, which result from more massive bulges and higher central densities. At earlier times, our model with the less massive bulge shows $v_{\phi}\approx 100$ km s$^{-1}$ at 3 Gyr, while it forms the NSD around 1.5 Gyr. This suggests that the rotation of the Milky Way NSD within the range of values in our models, presumably at early evolutionary stages. Lastly, our pressure-supported NSCs, with $v_{\phi}/\sigma < 0.7$, resemble the kinematically hot and massive core of the Milky Way's embedded the NSC.
We will revisit the formation and evolutionary pathways of nuclear structures at higher resolution and discuss the effects of numerical resolution, which influences not only bar properties \citep{kwak26} but also the details of the ISM and feedback processes \citep{deng24}.
In future, high-resolution simulations of the SMUGGLE-Ring project, we will be able to interpret the origin of NSCs and their connection with NSDs as a function of galaxy mass and infall history of star cluster \citep{fahrion22,fahrion24}, as well as the detailed role of nuclear spirals, which alter the in-situ growth of NSCs \citep{kolcu23,kolcu26}.

Our findings are summarized as below:
\begin{enumerate}
\item
Nuclear disk size scales with the mass of the classical bulge when the disk mass is fixed, despite the formation of weaker bars in our bulge models.

\item 
The presence of a classical bulge is not always necessary for nuclear disk formation, but its gravitational influence relative to stellar feedback regulates the timing and size of the nuclear disk during the isolated evolution of barred galaxies.

\item
NSRs are kinematically colder, exhibiting rotationally supported structures with $v_{\phi}/\sigma_R > 1.2-1.7$ and $\sigma$-drops, while NSCs are hotter toward the central region with $v_{\phi}/\sigma_R < 0.7$.
\item
NSCs are more massive ($\sim10^9$ M$_{\odot}$) than nuclear rings ($\sim10^8$ M$_{\odot}$). Beyond a certain density threshold of NSCs, they undergo radial displacement due to heating.
\item
NSCs and NSRs diverge from the initial nuclear disk through star formation-driven depletion and feedback, creating inner gaps and bifurcating the structure.
\item
The radius of the NSRs, estimated from the mass distribution and rotation peak, lies slightly outside the nuclear gas disk, as bar-driven inflowing gas triggers star formation primarily at the edge of the nuclear gas disk.
\item
Our results support the inside-out formation scenario, where NSCs and rings diverge from a compact nuclear disk and follow kinematically distinct evolutionary paths, with outward shift of density and velocity peaks driven by bar inflows. We will revisit the evolutionary path of nuclear structures by including external factors such as mergers in future work.
\end{enumerate}

\begin{acknowledgements}
We thank the referee for their constructive comments, which have helped us improve our manuscript. We gratefully acknowledge the computing time made available for the SMUGGLE-Ring project on the high-performance computer "Lise" at the NHR Center NHR@ZIB. This center is jointly supported by the Federal Ministry of Education and Research and the state governments participating in the NHR (www.nhr-verein.de/unsere-partner). SK appreciates Woo-Young Seo and In Sung Jang for the stimulating discussions.
The work of W-T.K.\ was supported by the grant of the National Research Foundation of Korea (RS-2025-00517264). IM acknowledges support by the Deutsche Forschungsgemeinschaft under the grant MI 2009/2-1. HL is supported by the National Key R\&D Program of China No. 2023YFB3002502, the National Natural Science Foundation of China under No. 12373006 and 12533004, and the China Manned Space Program with grant No. CMS-CSST-2025-A10.
\end{acknowledgements}

\bibliographystyle{aa} 
\bibliography{ref}

@ARTICLE{gadotti09,
       author = {{Gadotti}, Dimitri A.},
        title = "{Structural properties of pseudo-bulges, classical bulges and elliptical galaxies: a Sloan Digital Sky Survey perspective}",
      journal = {\mnras},
         year = 2009,
        month = mar,
       volume = {393},
       number = {4},
        pages = {1531-1552},
          doi = {10.1111/j.1365-2966.2008.14257.x},
archivePrefix = {arXiv},
       eprint = {0810.1953},
 primaryClass = {astro-ph},
       adsurl = {https://ui.adsabs.harvard.edu/abs/2009MNRAS.393.1531G}
}

@ARTICLE{nunnari26,
       author = {{Nunnari}, A. and {D'Orazi}, V. and {Fiorentino}, G. and {Braga}, V.~F. and {Bono}, G. and {Fabrizio}, M. and {J{\"o}nsson}, H. and {Kudritzki}, R.-P. and {da Silva}, R. and {Bergemann}, M. and {Poggio}, E. and {Otto}, J.~M. and {Baeza-Villagra}, K. and {Bragaglia}, A. and {Ceci}, G. and {Dall'Ora}, M. and {Inno}, L. and {Lardo}, C. and {Matsunaga}, N. and {Monelli}, M. and {S{\'a}nchez-Benavente}, M. and {Sneden}, C. and {Tantalo}, M. and {Th{\'e}v{\'e}nin}, F. and {Kovtyukh}, V. and {Di Criscienzo}, M. and {B{\"o}cek Topcu}, G.},
        title = "{Classical Cepheids in the Galactic thin disk: I. Abundance gradients via non-local thermodynamic equilibrium spectral analysis}",
      journal = {\aap},
         year = 2026,
        month = mar,
       volume = {708},
          eid = {A17},
        pages = {A17},
          doi = {10.1051/0004-6361/202558288},
archivePrefix = {arXiv},
       eprint = {2511.22491},
 primaryClass = {astro-ph.GA},
       adsurl = {https://ui.adsabs.harvard.edu/abs/2026A&A...708A..17N}
}

@ARTICLE{romeo16,
       author = {{Romeo}, Alessandro B. and {Fathi}, Kambiz},
        title = "{What powers the starburst activity of NGC 1068? Star-driven gravitational instabilities caught in the act}",
      journal = {\mnras},
         year = 2016,
        month = aug,
       volume = {460},
       number = {3},
        pages = {2360-2367},
          doi = {10.1093/mnras/stw1147},
archivePrefix = {arXiv},
       eprint = {1602.03049},
 primaryClass = {astro-ph.GA},
       adsurl = {https://ui.adsabs.harvard.edu/abs/2016MNRAS.460.2360R}
}

@ARTICLE{gadotti26,
       author = {{Gadotti}, Dimitri A.},
        title = "{Robust galaxy image decompositions with differential evolution optimization and the problem of classical bulges in and beyond the nearby Universe}",
      journal = {\mnras},
         year = 2026,
        month = feb,
       volume = {545},
       number = {4},
          eid = {staf2072},
        pages = {staf2072},
          doi = {10.1093/mnras/staf2072},
archivePrefix = {arXiv},
       eprint = {2511.13823},
 primaryClass = {astro-ph.GA},
       adsurl = {https://ui.adsabs.harvard.edu/abs/2026MNRAS.545f2072G}
}

@ARTICLE{kolcu26,
       author = {{Kolcu}, Tutku and {Maciejewski}, Witold and {Erwin}, Peter and {Gadotti}, Dimitri A. and {Fragkoudi}, Francesca and {Coelho}, Paula R.~T. and {Debattista}, Victor P. and {de Lorenzo-C{\'a}ceres}, Adriana and {de S{\'a}-Freitas}, Camila and {S{\'a}nchez-Bl{\'a}zquez}, Patricia},
        title = "{Composite Bulges -- V. Detecting signatures of gas inflows in IFU data: The MUSE view of ionised gas kinematics in nearby galaxies}",
      journal = {arXiv e-prints},
         year = 2026,
        month = jan,
          eid = {arXiv:2601.02534},
        pages = {arXiv:2601.02534},
          doi = {10.48550/arXiv.2601.02534},
archivePrefix = {arXiv},
       eprint = {2601.02534},
 primaryClass = {astro-ph.GA},
       adsurl = {https://ui.adsabs.harvard.edu/abs/2026arXiv260102534K}
}

@ARTICLE{kolcu23,
       author = {{Kolcu}, Tutku and {Maciejewski}, Witold and {Gadotti}, Dimitri A. and {Fragkoudi}, Francesca and {Erwin}, Peter and {S{\'a}nchez-Bl{\'a}zquez}, Patricia and {Neumann}, Justus and {Van de Ven}, Glenn and {de S{\'a}-Freitas}, Camila and {Longmore}, Steven and {Debattista}, Victor P.},
        title = "{Composite bulges - IV. Detecting signatures of gas inflows in the IFU data: the MUSE view of ionized gas kinematics in NGC 1097}",
      journal = {\mnras},
         year = 2023,
        month = sep,
       volume = {524},
       number = {1},
        pages = {207-223},
          doi = {10.1093/mnras/stad1896},
archivePrefix = {arXiv},
       eprint = {2306.11091},
 primaryClass = {astro-ph.GA},
       adsurl = {https://ui.adsabs.harvard.edu/abs/2023MNRAS.524..207K}
}

@ARTICLE{jang25,
       author = {{Jang}, Dajeong and {Kim}, Woong-Tae and {Lee}, Yun Hee},
        title = "{Conditions for Bar Formation in Bulgeless Disk Galaxies}",
      journal = {\apj},
         year = 2025,
        month = nov,
       volume = {993},
       number = {2},
          eid = {236},
        pages = {236},
          doi = {10.3847/1538-4357/ae0302},
archivePrefix = {arXiv},
       eprint = {2509.07353},
 primaryClass = {astro-ph.GA},
       adsurl = {https://ui.adsabs.harvard.edu/abs/2025ApJ...993..236J}
}

@ARTICLE{baek25,
       author = {{Baek}, Seungwon and {Kim}, Woong-Tae and {Jang}, Dajeong and {Kim}, Taehyun},
        title = "{Dissecting Bar-induced Stellar Kinematics in Disk Galaxies: The Bisymmetric Model and Rotation Curve Modifications}",
      journal = {\apj},
         year = 2025,
        month = sep,
       volume = {990},
       number = {2},
          eid = {184},
        pages = {184},
          doi = {10.3847/1538-4357/adf213},
archivePrefix = {arXiv},
       eprint = {2507.15204},
 primaryClass = {astro-ph.GA},
       adsurl = {https://ui.adsabs.harvard.edu/abs/2025ApJ...990..184B}
}

@ARTICLE{kwak26,
       author = {{Kwak}, S. and {Minchev}, I. and {Steinmetz}, M. and {Yi}, S.~K.},
        title = "{Effects of Resolution and Local Stability on Galactic Disks: 2. Halo Resolution and Softening on Bar Formation}",
      journal = {arXiv e-prints},
         year = 2026,
        month = mar,
          eid = {arXiv:2603.02494},
        pages = {arXiv:2603.02494},
          doi = {10.48550/arXiv.2603.02494},
archivePrefix = {arXiv},
       eprint = {2603.02494},
 primaryClass = {astro-ph.GA},
       adsurl = {https://ui.adsabs.harvard.edu/abs/2026arXiv260302494K}
}

@ARTICLE{kwak19,
       author = {{Kwak}, SungWon and {Kim}, Woong-Tae and {Rey}, Soo-Chang and {Quinn}, Thomas R.},
        title = "{Origin of Nonaxisymmetric Features of Virgo Cluster Early-type Dwarf Galaxies. II. Tidal Effects on Disk Features and Stability}",
      journal = {\apj},
         year = 2019,
        month = dec,
       volume = {887},
       number = {2},
          eid = {139},
        pages = {139},
          doi = {10.3847/1538-4357/ab5716},
archivePrefix = {arXiv},
       eprint = {1911.05094},
 primaryClass = {astro-ph.GA},
       adsurl = {https://ui.adsabs.harvard.edu/abs/2019ApJ...887..139K}
}

@ARTICLE{marinova07,
       author = {{Marinova}, Irina and {Jogee}, Shardha},
        title = "{Characterizing Bars at z \raisebox{-0.5ex}\textasciitilde 0 in the Optical and NIR: Implications for the Evolution of Barred Disks with Redshift}",
      journal = {\apj},
         year = 2007,
        month = apr,
       volume = {659},
       number = {2},
        pages = {1176-1197},
          doi = {10.1086/512355},
archivePrefix = {arXiv},
       eprint = {astro-ph/0608039},
 primaryClass = {astro-ph},
       adsurl = {https://ui.adsabs.harvard.edu/abs/2007ApJ...659.1176M}
}

@ARTICLE{menendez07,
       author = {{Men{\'e}ndez-Delmestre}, Kar{\'\i}n and {Sheth}, Kartik and {Schinnerer}, Eva and {Jarrett}, Thomas H. and {Scoville}, Nick Z.},
        title = "{A Near-Infrared Study of 2MASS Bars in Local Galaxies: An Anchor for High-Redshift Studies}",
      journal = {\apj},
         year = 2007,
        month = mar,
       volume = {657},
       number = {2},
        pages = {790-804},
          doi = {10.1086/511025},
archivePrefix = {arXiv},
       eprint = {astro-ph/0611540},
 primaryClass = {astro-ph},
       adsurl = {https://ui.adsabs.harvard.edu/abs/2007ApJ...657..790M}
}

@ARTICLE{whyte02,
       author = {{Whyte}, L.~F. and {Abraham}, R.~G. and {Merrifield}, M.~R. and {Eskridge}, P.~B. and {Frogel}, J.~A. and {Pogge}, R.~W.},
        title = "{Morphological classification of the OSU Bright Spiral Galaxy Survey}",
      journal = {\mnras},
         year = 2002,
        month = nov,
       volume = {336},
       number = {4},
        pages = {1281-1286},
          doi = {10.1046/j.1365-8711.2002.05879.x},
archivePrefix = {arXiv},
       eprint = {astro-ph/0207461},
 primaryClass = {astro-ph},
       adsurl = {https://ui.adsabs.harvard.edu/abs/2002MNRAS.336.1281W}
}

@ARTICLE{eskridge00,
       author = {{Eskridge}, Paul B. and {Frogel}, Jay A. and {Pogge}, Richard W. and {Quillen}, Alice C. and {Davies}, Roger L. and {DePoy}, D.~L. and {Houdashelt}, Mark L. and {Kuchinski}, Leslie E. and {Ram{\'\i}rez}, Solange V. and {Sellgren}, K. and {Terndrup}, Donald M. and {Tiede}, Glenn P.},
        title = "{The Frequency of Barred Spiral Galaxies in the Near-Infrared}",
      journal = {\aj},
         year = 2000,
        month = feb,
       volume = {119},
       number = {2},
        pages = {536-544},
          doi = {10.1086/301203},
archivePrefix = {arXiv},
       eprint = {astro-ph/9910479},
 primaryClass = {astro-ph},
       adsurl = {https://ui.adsabs.harvard.edu/abs/2000AJ....119..536E}
}

@ARTICLE{bittner20,
       author = {{Bittner}, Adrian and {S{\'a}nchez-Bl{\'a}zquez}, Patricia and {Gadotti}, Dimitri A. and {Neumann}, Justus and {Fragkoudi}, Francesca and {Coelho}, Paula and {de Lorenzo-C{\'a}ceres}, Adriana and {Falc{\'o}n-Barroso}, Jes{\'u}s and {Kim}, Taehyun and {Leaman}, Ryan and {Mart{\'\i}n-Navarro}, Ignacio and {M{\'e}ndez-Abreu}, Jairo and {P{\'e}rez}, Isabel and {Querejeta}, Miguel and {Seidel}, Marja K. and {van de Ven}, Glenn},
        title = "{Inside-out formation of nuclear discs and the absence of old central spheroids in barred galaxies of the TIMER survey}",
      journal = {\aap},
         year = 2020,
        month = nov,
       volume = {643},
          eid = {A65},
        pages = {A65},
          doi = {10.1051/0004-6361/202038450},
archivePrefix = {arXiv},
       eprint = {2009.01856},
 primaryClass = {astro-ph.GA},
       adsurl = {https://ui.adsabs.harvard.edu/abs/2020A&A...643A..65B}
}

@ARTICLE{delorenzo08,
       author = {{de Lorenzo-C{\'a}ceres}, A. and {Falc{\'o}n-Barroso}, J. and {Vazdekis}, A. and {Mart{\'\i}nez-Valpuesta}, I.},
        title = "{Stellar Kinematics in Double-Barred Galaxies: The {\ensuremath{\sigma}}-Hollows}",
      journal = {\apjl},
         year = 2008,
        month = sep,
       volume = {684},
       number = {2},
        pages = {L83},
          doi = {10.1086/592145},
archivePrefix = {arXiv},
       eprint = {0808.0517},
 primaryClass = {astro-ph},
       adsurl = {https://ui.adsabs.harvard.edu/abs/2008ApJ...684L..83D}
}

@ARTICLE{emsellem01,
       author = {{Emsellem}, E. and {Greusard}, D. and {Combes}, F. and {Friedli}, D. and {Leon}, S. and {P{\'e}contal}, E. and {Wozniak}, H.},
        title = "{Dynamics of embedded bars and the connection with AGN. I. ISAAC/VLT stellar kinematics}",
      journal = {\aap},
         year = 2001,
        month = mar,
       volume = {368},
        pages = {52-63},
          doi = {10.1051/0004-6361:20000523},
archivePrefix = {arXiv},
       eprint = {astro-ph/0012480},
 primaryClass = {astro-ph},
       adsurl = {https://ui.adsabs.harvard.edu/abs/2001A&A...368...52E}
}

@ARTICLE{kormendy82,
       author = {{Kormendy}, J.},
        title = "{Rotation of the bulge components of barred galaxies.}",
      journal = {\apj},
         year = 1982,
        month = jun,
       volume = {257},
        pages = {75-88},
          doi = {10.1086/159964},
       adsurl = {https://ui.adsabs.harvard.edu/abs/1982ApJ...257...75K}
}

@ARTICLE{bosch98,
       author = {{van den Bosch}, Frank C. and {Jaffe}, Walter and {van der Marel}, Roeland P.},
        title = "{Nuclear stellar discs in early-type galaxies - I. HST and WHT observations}",
      journal = {\mnras},
         year = 1998,
        month = feb,
       volume = {293},
       number = {4},
        pages = {343-363},
          doi = {10.1046/j.1365-8711.1998.01069.x},
archivePrefix = {arXiv},
       eprint = {astro-ph/9708027},
 primaryClass = {astro-ph},
       adsurl = {https://ui.adsabs.harvard.edu/abs/1998MNRAS.293..343V}
}

@ARTICLE{barbuy18,
       author = {{Barbuy}, Beatriz and {Chiappini}, Cristina and {Gerhard}, Ortwin},
        title = "{Chemodynamical History of the Galactic Bulge}",
      journal = {\araa},
         year = 2018,
        month = sep,
       volume = {56},
        pages = {223-276},
          doi = {10.1146/annurev-astro-081817-051826},
archivePrefix = {arXiv},
       eprint = {1805.01142},
 primaryClass = {astro-ph.GA},
       adsurl = {https://ui.adsabs.harvard.edu/abs/2018ARA&A..56..223B}
}

@ARTICLE{portail17,
       author = {{Portail}, Matthieu and {Gerhard}, Ortwin and {Wegg}, Christopher and {Ness}, Melissa},
        title = "{Dynamical modelling of the galactic bulge and bar: the Milky Way's pattern speed, stellar and dark matter mass distribution}",
      journal = {\mnras},
         year = 2017,
        month = feb,
       volume = {465},
       number = {2},
        pages = {1621-1644},
          doi = {10.1093/mnras/stw2819},
archivePrefix = {arXiv},
       eprint = {1608.07954},
 primaryClass = {astro-ph.GA},
       adsurl = {https://ui.adsabs.harvard.edu/abs/2017MNRAS.465.1621P}
}

@ARTICLE{gerhard11,
       author = {{Gerhard}, O.},
        title = "{Pattern speeds in the Milky Way.}",
      journal = {Memorie della Societa Astronomica Italiana Supplementi},
         year = 2011,
        month = jan,
       volume = {18},
        pages = {185},
          doi = {10.48550/arXiv.1003.2489},
archivePrefix = {arXiv},
       eprint = {1003.2489},
 primaryClass = {astro-ph.GA},
       adsurl = {https://ui.adsabs.harvard.edu/abs/2011MSAIS..18..185G}
}

@ARTICLE{moon22,
       author = {{Moon}, Sanghyuk and {Kim}, Woong-Tae and {Kim}, Chang-Goo and {Ostriker}, Eve C.},
        title = "{Effects of Varying Mass Inflows on Star Formation in Nuclear Rings of Barred Galaxies}",
      journal = {\apj},
         year = 2022,
        month = jan,
       volume = {925},
       number = {1},
          eid = {99},
        pages = {99},
          doi = {10.3847/1538-4357/ac3a7b},
archivePrefix = {arXiv},
       eprint = {2110.14882},
 primaryClass = {astro-ph.GA},
       adsurl = {https://ui.adsabs.harvard.edu/abs/2022ApJ...925...99M}
}

@ARTICLE{kim12a,
       author = {{Kim}, Woong-Tae and {Seo}, Woo-Young and {Stone}, James M. and {Yoon}, Doosoo and {Teuben}, Peter J.},
        title = "{Central Regions of Barred Galaxies: Two-dimensional Non-self-gravitating Hydrodynamic Simulations}",
      journal = {\apj},
         year = 2012,
        month = mar,
       volume = {747},
       number = {1},
          eid = {60},
        pages = {60},
          doi = {10.1088/0004-637X/747/1/60},
archivePrefix = {arXiv},
       eprint = {1112.6055},
 primaryClass = {astro-ph.GA},
       adsurl = {https://ui.adsabs.harvard.edu/abs/2012ApJ...747...60K}
}

@ARTICLE{kim12b,
       author = {{Kim}, Woong-Tae and {Stone}, James M.},
        title = "{Two-dimensional Magnetohydrodynamic Simulations of Barred Galaxies}",
      journal = {\apj},
         year = 2012,
        month = jun,
       volume = {751},
       number = {2},
          eid = {124},
        pages = {124},
          doi = {10.1088/0004-637X/751/2/124},
archivePrefix = {arXiv},
       eprint = {1204.0073},
 primaryClass = {astro-ph.CO},
       adsurl = {https://ui.adsabs.harvard.edu/abs/2012ApJ...751..124K}
}

@ARTICLE{kim12c,
       author = {{Kim}, Woong-Tae and {Seo}, Woo-Young and {Kim}, Yonghwi},
        title = "{Gaseous Structures in Barred Galaxies: Effects of the Bar Strength}",
      journal = {\apj},
         year = 2012,
        month = oct,
       volume = {758},
       number = {1},
          eid = {14},
        pages = {14},
          doi = {10.1088/0004-637X/758/1/14},
archivePrefix = {arXiv},
       eprint = {1208.1821},
 primaryClass = {astro-ph.CO},
       adsurl = {https://ui.adsabs.harvard.edu/abs/2012ApJ...758...14K}
}

@ARTICLE{goldreich79,
       author = {{Goldreich}, P. and {Tremaine}, S.},
        title = "{The excitation of density waves at the Lindblad and corotation resonances by an external potential.}",
      journal = {\apj},
         year = 1979,
        month = nov,
       volume = {233},
        pages = {857-871},
          doi = {10.1086/157448},
       adsurl = {https://ui.adsabs.harvard.edu/abs/1979ApJ...233..857G}
}

@ARTICLE{friedli95,
       author = {{Friedli}, D. and {Benz}, W.},
        title = "{Secular evolution of isolated barred galaxies. II. Coupling between stars and interstellar medium via star formation.}",
      journal = {\aap},
         year = 1995,
        month = sep,
       volume = {301},
        pages = {649},
       adsurl = {https://ui.adsabs.harvard.edu/abs/1995A&A...301..649F}
}

@ARTICLE{gadotti15,
       author = {{Gadotti}, Dimitri A. and {Seidel}, Marja K. and {S{\'a}nchez-Bl{\'a}zquez}, Patricia and {Falc{\'o}n-Barroso}, Jesus and {Husemann}, Bernd and {Coelho}, Paula and {P{\'e}rez}, Isabel},
        title = "{MUSE tells the story of NGC 4371: The dawning of secular evolution}",
      journal = {\aap},
         year = 2015,
        month = dec,
       volume = {584},
          eid = {A90},
        pages = {A90},
          doi = {10.1051/0004-6361/201526677},
archivePrefix = {arXiv},
       eprint = {1509.00032},
 primaryClass = {astro-ph.GA},
       adsurl = {https://ui.adsabs.harvard.edu/abs/2015A&A...584A..90G}
}

@ARTICLE{gadotti19,
       author = {{Gadotti}, Dimitri A. and {S{\'a}nchez-Bl{\'a}zquez}, Patricia and {Falc{\'o}n-Barroso}, Jes{\'u}s and {Husemann}, Bernd and {Seidel}, Marja K. and {P{\'e}rez}, Isabel and {de Lorenzo-C{\'a}ceres}, Adriana and {Martinez-Valpuesta}, Inma and {Fragkoudi}, Francesca and {Leung}, Gigi and {van de Ven}, Glenn and {Leaman}, Ryan and {Coelho}, Paula and {Martig}, Marie and {Kim}, Taehyun and {Neumann}, Justus and {Querejeta}, Miguel},
        title = "{Time Inference with MUSE in Extragalactic Rings (TIMER): properties of the survey and high-level data products}",
      journal = {\mnras},
         year = 2019,
        month = jan,
       volume = {482},
       number = {1},
        pages = {506-529},
          doi = {10.1093/mnras/sty2666},
archivePrefix = {arXiv},
       eprint = {1810.01425},
 primaryClass = {astro-ph.GA},
       adsurl = {https://ui.adsabs.harvard.edu/abs/2019MNRAS.482..506G}
}

@ARTICLE{baba20,
       author = {{Baba}, Junichi and {Kawata}, Daisuke},
        title = "{Age dating the Galactic bar with the nuclear stellar disc}",
      journal = {\mnras},
         year = 2020,
        month = mar,
       volume = {492},
       number = {3},
        pages = {4500-4511},
          doi = {10.1093/mnras/staa140},
archivePrefix = {arXiv},
       eprint = {1909.07548},
 primaryClass = {astro-ph.GA},
       adsurl = {https://ui.adsabs.harvard.edu/abs/2020MNRAS.492.4500B}
}

@ARTICLE{gadotti20,
       author = {{Gadotti}, Dimitri A. and {Bittner}, Adrian and {Falc{\'o}n-Barroso}, Jes{\'u}s and {M{\'e}ndez-Abreu}, Jairo and {Kim}, Taehyun and {Fragkoudi}, Francesca and {de Lorenzo-C{\'a}ceres}, Adriana and {Leaman}, Ryan and {Neumann}, Justus and {Querejeta}, Miguel and {S{\'a}nchez-Bl{\'a}zquez}, Patricia and {Martig}, Marie and {Mart{\'\i}n-Navarro}, Ignacio and {P{\'e}rez}, Isabel and {Seidel}, Marja K. and {van de Ven}, Glenn},
        title = "{Kinematic signatures of nuclear discs and bar-driven secular evolution in nearby galaxies of the MUSE TIMER project}",
      journal = {\aap},
         year = 2020,
        month = nov,
       volume = {643},
          eid = {A14},
        pages = {A14},
          doi = {10.1051/0004-6361/202038448},
archivePrefix = {arXiv},
       eprint = {2009.01852},
 primaryClass = {astro-ph.GA},
       adsurl = {https://ui.adsabs.harvard.edu/abs/2020A&A...643A..14G}
}

@ARTICLE{shahzamanian22,
       author = {{Shahzamanian}, B. and {Sch{\"o}del}, R. and {Nogueras-Lara}, F. and {Mart{\'\i}nez-Arranz}, A. and {Sormani}, M.~C. and {Gallego-Calvente}, A.~T. and {Gallego-Cano}, E. and {Alburai}, A.},
        title = "{A proper motion catalogue for the Milky Way's nuclear stellar disc}",
      journal = {\aap},
         year = 2022,
        month = jun,
       volume = {662},
          eid = {A11},
        pages = {A11},
          doi = {10.1051/0004-6361/202142687},
archivePrefix = {arXiv},
       eprint = {2108.11847},
 primaryClass = {astro-ph.GA},
       adsurl = {https://ui.adsabs.harvard.edu/abs/2022A&A...662A..11S}
}

@ARTICLE{fraser24,
       author = {{Fraser-McKelvie}, A. and {van de Sande}, J. and {Gadotti}, D.~A. and {Emsellem}, E. and {Brown}, T. and {Fisher}, D.~B. and {Martig}, M. and {Bureau}, M. and {Gerhard}, O. and {Battisti}, A.~J. and {Bland-Hawthorn}, J. and {Boecker}, A. and {Catinella}, B. and {Combes}, F. and {Cortese}, L. and {Croom}, S.~M. and {Davis}, T.~A. and {Falc{\'o}n-Barroso}, J. and {Fragkoudi}, F. and {Freeman}, K.~C. and {Hayden}, M.~R. and {McDermid}, R. and {Mazzilli Ciraulo}, B. and {Mendel}, J.~T. and {Pinna}, F. and {Poci}, A. and {Rutherford}, T.~H. and {de S{\'a}-Freitas}, C. and {Silva-Lima}, L.~A. and {Valenzuela}, L.~M. and {van de Ven}, G. and {Wang}, Z. and {Watts}, A.~B.},
        title = "{The GECKOS survey: Identifying kinematic sub-structures in edge-on galaxies}",
      journal = {\aap},
         year = 2025,
        month = aug,
       volume = {700},
          eid = {A237},
        pages = {A237},
          doi = {10.1051/0004-6361/202452891},
archivePrefix = {arXiv},
       eprint = {2411.03430},
 primaryClass = {astro-ph.GA},
       adsurl = {https://ui.adsabs.harvard.edu/abs/2025A&A...700A.237F}
}

@ARTICLE{desafreitas25b,
       author = {{de S{\'a}-Freitas}, Camila and {Gadotti}, Dimitri A. and {Fragkoudi}, Francesca and {Coelho}, Paula and {de Lorenzo-C{\'a}ceres}, Adriana and {Falc{\'o}n-Barroso}, Jes{\'u}s and {S{\'a}nchez-Bl{\'a}zquez}, Patricia and {Kim}, Taehyun and {Mendez-Abreu}, Jairo and {Neumann}, Justus and {Querejeta}, Miguel and {van de Ven}, Glenn},
        title = "{Bar ages derived for the first time in nearby galaxies: Insights into secular evolution from the TIMER sample (Corrigendum)}",
      journal = {\aap},
         year = 2025,
        month = dec,
       volume = {704},
          eid = {C1},
        pages = {C1},
          doi = {10.1051/0004-6361/202557798e},
       adsurl = {https://ui.adsabs.harvard.edu/abs/2025A&A...704C...1D}
}

@ARTICLE{desafreitas25a,
       author = {{de S{\'a}-Freitas}, Camila and {Gadotti}, Dimitri A. and {Fragkoudi}, Francesca and {Coelho}, Paula and {de Lorenzo-C{\'a}ceres}, Adriana and {Falc{\'o}n-Barroso}, Jes{\'u}s and {S{\'a}nchez-Bl{\'a}zquez}, Patricia and {Kim}, Taehyun and {Mendez-Abreu}, Jairo and {Neumann}, Justus and {Querejeta}, Miguel and {van de Ven}, Glenn},
        title = "{Bar ages derived for the first time in nearby galaxies: Insights into secular evolution from the TIMER sample}",
      journal = {\aap},
         year = 2025,
        month = jun,
       volume = {698},
          eid = {A5},
        pages = {A5},
          doi = {10.1051/0004-6361/202453367},
archivePrefix = {arXiv},
       eprint = {2503.20864},
 primaryClass = {astro-ph.GA},
       adsurl = {https://ui.adsabs.harvard.edu/abs/2025A&A...698A...5D}
}

@ARTICLE{desafreitas23a,
       author = {{de S{\'a}-Freitas}, Camila and {Fragkoudi}, Francesca and {Gadotti}, Dimitri A. and {Falc{\'o}n-Barroso}, Jes{\'u}s and {Bittner}, Adrian and {S{\'a}nchez-Bl{\'a}zquez}, Patricia and {van de Ven}, Glenn and {Bieri}, Rebekka and {Coccato}, Lodovico and {Coelho}, Paula and {Fahrion}, Katja and {Gon{\c{c}}alves}, Geraldo and {Kim}, Taehyun and {de Lorenzo-C{\'a}ceres}, Adriana and {Martig}, Marie and {Mart{\'\i}n-Navarro}, Ignacio and {Mendez-Abreu}, Jairo and {Neumann}, Justus and {Querejeta}, Miguel},
        title = "{A new method for age-dating the formation of bars in disc galaxies. The TIMER view on NGC1433's old bar and the inside-out growth of its nuclear disc}",
      journal = {\aap},
         year = 2023,
        month = mar,
       volume = {671},
          eid = {A8},
        pages = {A8},
          doi = {10.1051/0004-6361/202244667},
archivePrefix = {arXiv},
       eprint = {2211.07670},
 primaryClass = {astro-ph.GA},
       adsurl = {https://ui.adsabs.harvard.edu/abs/2023A&A...671A...8D}
}

@ARTICLE{desafreitas23b,
       author = {{de S{\'a}-Freitas}, Camila and {Gadotti}, Dimitri A. and {Fragkoudi}, Francesca and {Coccato}, Lodovico and {Coelho}, Paula and {de Lorenzo-C{\'a}ceres}, Adriana and {Falc{\'o}n-Barroso}, Jes{\'u}s and {Kolcu}, Tutku and {Mart{\'\i}n-Navarro}, Ignacio and {Mendez-Abreu}, Jairo and {Neumann}, Justus and {Blazquez}, Patricia Sanchez and {Querejeta}, Miguel and {van de Ven}, Glenn},
        title = "{Disc galaxies are still settling. Discovery of the smallest nuclear discs and their young stellar bars}",
      journal = {\aap},
         year = 2023,
        month = oct,
       volume = {678},
          eid = {A202},
        pages = {A202},
          doi = {10.1051/0004-6361/202347028},
archivePrefix = {arXiv},
       eprint = {2308.04482},
 primaryClass = {astro-ph.GA},
       adsurl = {https://ui.adsabs.harvard.edu/abs/2023A&A...678A.202D}
}

@ARTICLE{fahrion24,
       author = {{Fahrion}, Katja and {B{\"o}ker}, Torsten and {Perna}, Michele and {Beck}, Tracy L. and {Maiolino}, Roberto and {Arribas}, Santiago and {Bunker}, Andrew J. and {Charlot}, Stephane and {Ceci}, Matteo and {Cresci}, Giovanni and {De Marchi}, Guido and {L{\"u}tzgendorf}, Nora and {Ulivi}, Lorenzo},
        title = "{Growing a nuclear star cluster from star formation and cluster mergers: The JWST NIRSpec view of NGC 4654}",
      journal = {\aap},
         year = 2024,
        month = jul,
       volume = {687},
          eid = {A83},
        pages = {A83},
          doi = {10.1051/0004-6361/202449629},
archivePrefix = {arXiv},
       eprint = {2404.08910},
 primaryClass = {astro-ph.GA},
       adsurl = {https://ui.adsabs.harvard.edu/abs/2024A&A...687A..83F}
}

@ARTICLE{fahrion22,
       author = {{Fahrion}, Katja and {Leaman}, Ryan and {Lyubenova}, Mariya and {van de Ven}, Glenn},
        title = "{Disentangling the formation mechanisms of nuclear star clusters}",
      journal = {\aap},
         year = 2022,
        month = feb,
       volume = {658},
          eid = {A172},
        pages = {A172},
          doi = {10.1051/0004-6361/202039778},
archivePrefix = {arXiv},
       eprint = {2112.05610},
 primaryClass = {astro-ph.GA},
       adsurl = {https://ui.adsabs.harvard.edu/abs/2022A&A...658A.172F}
}

@ARTICLE{heller94,
       author = {{Heller}, Clayton H. and {Shlosman}, Isaac},
        title = "{Fueling Nuclear Activity in Disk Galaxies: Starbursts and Monsters}",
      journal = {\apj},
         year = 1994,
        month = mar,
       volume = {424},
        pages = {84},
          doi = {10.1086/173874},
       adsurl = {https://ui.adsabs.harvard.edu/abs/1994ApJ...424...84H}
}

@ARTICLE{shlosman89,
       author = {{Shlosman}, Isaac and {Frank}, Juhan and {Begelman}, Mitchell C.},
        title = "{Bars within bars: a mechanism for fuelling active galactic nuclei}",
      journal = {\nat},
         year = 1989,
        month = mar,
       volume = {338},
       number = {6210},
        pages = {45-47},
          doi = {10.1038/338045a0},
       adsurl = {https://ui.adsabs.harvard.edu/abs/1989Natur.338...45S}
}

@ARTICLE{buta96,
       author = {{Buta}, R. and {Combes}, F.},
        title = "{Galactic Rings}",
      journal = {\fcp},
         year = 1996,
        month = jan,
       volume = {17},
        pages = {95-281},
       adsurl = {https://ui.adsabs.harvard.edu/abs/1996FCPh...17...95B}
}

@INPROCEEDINGS{combes96,
       author = {{Combes}, F.},
        title = "{Ring and Lens Formation}",
    booktitle = {IAU Colloquium 157: Barred Galaxies},
         year = 1996,
       editor = {{Buta}, R. and {Crocker}, D.~A. and {Elmegreen}, B.~G.},
       series = {Astronomical Society of the Pacific Conference Series},
       volume = {91},
        month = jan,
        pages = {286},
       adsurl = {https://ui.adsabs.harvard.edu/abs/1996ASPC...91..286C}
}

@ARTICLE{combes85,
       author = {{Combes}, F. and {Gerin}, M.},
        title = "{Spiral structure of molecular clouds in response to bar forcing: a particle simulation.}",
      journal = {\aap},
         year = 1985,
        month = sep,
       volume = {150},
        pages = {327-338},
       adsurl = {https://ui.adsabs.harvard.edu/abs/1985A&A...150..327C}
}

@ARTICLE{guo20,
       author = {{Guo}, Minghao and {Du}, Min and {Ho}, Luis C. and {Debattista}, Victor P. and {Zhao}, Dongyao},
        title = "{A New Channel of Bulge Formation via the Destruction of Short Bars}",
      journal = {\apj},
         year = 2020,
        month = jan,
       volume = {888},
       number = {2},
          eid = {65},
        pages = {65},
          doi = {10.3847/1538-4357/ab584a},
archivePrefix = {arXiv},
       eprint = {1911.07002},
 primaryClass = {astro-ph.GA},
       adsurl = {https://ui.adsabs.harvard.edu/abs/2020ApJ...888...65G}
}

@ARTICLE{athanassoula92,
       author = {{Athanassoula}, E.},
        title = "{The existence and shapes of dust lanes in galactic bars.}",
      journal = {\mnras},
         year = 1992,
        month = nov,
       volume = {259},
        pages = {345-364},
          doi = {10.1093/mnras/259.2.345},
       adsurl = {https://ui.adsabs.harvard.edu/abs/1992MNRAS.259..345A}
}

@ARTICLE{sormani24,
       author = {{Sormani}, Mattia C. and {Sobacchi}, Emanuele and {Sanders}, Jason L.},
        title = "{Nuclear rings are the inner edge of a gap around the Lindblad Resonance}",
      journal = {\mnras},
         year = 2024,
        month = mar,
       volume = {528},
       number = {4},
        pages = {5742-5762},
          doi = {10.1093/mnras/stae082},
archivePrefix = {arXiv},
       eprint = {2309.14093},
 primaryClass = {astro-ph.GA},
       adsurl = {https://ui.adsabs.harvard.edu/abs/2024MNRAS.528.5742S}
}

@ARTICLE{sormani18,
       author = {{Sormani}, Mattia C. and {Sobacchi}, Emanuele and {Fragkoudi}, Francesca and {Ridley}, Matthew and {Tre{\ss}}, Robin G. and {Glover}, Simon C.~O. and {Klessen}, Ralf S.},
        title = "{A dynamical mechanism for the origin of nuclear rings}",
      journal = {\mnras},
         year = 2018,
        month = nov,
       volume = {481},
       number = {1},
        pages = {2-19},
          doi = {10.1093/mnras/sty2246},
archivePrefix = {arXiv},
       eprint = {1805.07969},
 primaryClass = {astro-ph.GA},
       adsurl = {https://ui.adsabs.harvard.edu/abs/2018MNRAS.481....2S}
}

@ARTICLE{sormani22,
       author = {{Sormani}, Mattia C. and {Sanders}, Jason L. and {Fritz}, Tobias K. and {Smith}, Leigh C. and {Gerhard}, Ortwin and {Sch{\"o}del}, Rainer and {Magorrian}, John and {Neumayer}, Nadine and {Nogueras-Lara}, Francisco and {Feldmeier-Krause}, Anja and {Mastrobuono-Battisti}, Alessandra and {Schultheis}, Mathias and {Shahzamanian}, Banafsheh and {Vasiliev}, Eugene and {Klessen}, Ralf S. and {Lucas}, Philip and {Minniti}, Dante},
        title = "{Self-consistent modelling of the Milky Way's nuclear stellar disc}",
      journal = {\mnras},
         year = 2022,
        month = may,
       volume = {512},
       number = {2},
        pages = {1857-1884},
          doi = {10.1093/mnras/stac639},
archivePrefix = {arXiv},
       eprint = {2111.12713},
 primaryClass = {astro-ph.GA},
       adsurl = {https://ui.adsabs.harvard.edu/abs/2022MNRAS.512.1857S}
}

@ARTICLE{schultheis19,
       author = {{Schultheis}, M. and {Rich}, R.~M. and {Origlia}, L. and {Ryde}, N. and {Nandakumar}, G. and {Thorsbro}, B. and {Neumayer}, N.},
        title = "{The inner two degrees of the Milky Way. Evidence of a chemical difference between the Galactic Center and the surrounding inner bulge stellar populations}",
      journal = {\aap},
         year = 2019,
        month = jul,
       volume = {627},
          eid = {A152},
        pages = {A152},
          doi = {10.1051/0004-6361/201935772},
archivePrefix = {arXiv},
       eprint = {1906.07985},
 primaryClass = {astro-ph.GA},
       adsurl = {https://ui.adsabs.harvard.edu/abs/2019A&A...627A.152S}
}

@ARTICLE{pettitt18,
       author = {{Pettitt}, Alex R. and {Wadsley}, J.~W.},
        title = "{Bars and spirals in tidal interactions with an ensemble of galaxy mass models}",
      journal = {\mnras},
         year = 2018,
        month = mar,
       volume = {474},
       number = {4},
        pages = {5645-5671},
          doi = {10.1093/mnras/stx3129},
archivePrefix = {arXiv},
       eprint = {1712.00882},
 primaryClass = {astro-ph.GA},
       adsurl = {https://ui.adsabs.harvard.edu/abs/2018MNRAS.474.5645P}
}

@ARTICLE{ryde25,
       author = {{Ryde}, N. and {Nandakumar}, G. and {Albarrac{\'\i}n}, R. and {Schultheis}, M. and {Rojas-Arriagada}, A. and {Zoccali}, M.},
        title = "{Chemical abundances in the Milky Way's nuclear stellar disc}",
      journal = {\aap},
         year = 2025,
        month = jul,
       volume = {699},
          eid = {A176},
        pages = {A176},
          doi = {10.1051/0004-6361/202554791},
archivePrefix = {arXiv},
       eprint = {2505.15924},
 primaryClass = {astro-ph.GA},
       adsurl = {https://ui.adsabs.harvard.edu/abs/2025A&A...699A.176R}
}

@ARTICLE{marquez03,
       author = {{M{\'a}rquez}, I. and {Masegosa}, J. and {Durret}, F. and {Gonz{\'a}lez Delgado}, R.~M. and {Moles}, M. and {Maza}, J. and {P{\'e}rez}, E. and {Roth}, M.},
        title = "{The detection of stellar velocity dispersion drops in the central regions of five isolated Seyfert spirals}",
      journal = {\aap},
         year = 2003,
        month = oct,
       volume = {409},
        pages = {459-467},
          doi = {10.1051/0004-6361:20031059},
archivePrefix = {arXiv},
       eprint = {astro-ph/0306497},
 primaryClass = {astro-ph},
       adsurl = {https://ui.adsabs.harvard.edu/abs/2003A&A...409..459M}
}

@ARTICLE{erwin15,
       author = {{Erwin}, Peter and {Saglia}, Roberto P. and {Fabricius}, Maximilian and {Thomas}, Jens and {Nowak}, Nina and {Rusli}, Stephanie and {Bender}, Ralf and {Vega Beltr{\'a}n}, Juan Carlos and {Beckman}, John E.},
        title = "{Composite bulges: the coexistence of classical bulges and discy pseudo-bulges in S0 and spiral galaxies}",
      journal = {\mnras},
         year = 2015,
        month = feb,
       volume = {446},
       number = {4},
        pages = {4039-4077},
          doi = {10.1093/mnras/stu2376},
archivePrefix = {arXiv},
       eprint = {1411.2599},
 primaryClass = {astro-ph.GA},
       adsurl = {https://ui.adsabs.harvard.edu/abs/2015MNRAS.446.4039E}
}

@ARTICLE{erwin21,
       author = {{Erwin}, Peter and {Seth}, Anil and {Debattista}, Victor P. and {Seidel}, Marja and {Mehrgan}, Kianusch and {Thomas}, Jens and {Saglia}, Roberto and {de Lorenzo-C{\'a}ceres}, Adriana and {Maciejewski}, Witold and {Fabricius}, Maximilian and {M{\'e}ndez-Abreu}, Jairo and {Hopp}, Ulrich and {Kluge}, Matthias and {Beckman}, John E. and {Bender}, Ralf and {Drory}, Niv and {Fisher}, Deanne},
        title = "{Composite bulges - II. Classical bulges and nuclear discs in barred galaxies: the contrasting cases of NGC 4608 and NGC 4643}",
      journal = {\mnras},
         year = 2021,
        month = apr,
       volume = {502},
       number = {2},
        pages = {2446-2473},
          doi = {10.1093/mnras/stab126},
archivePrefix = {arXiv},
       eprint = {2101.05321},
 primaryClass = {astro-ph.GA},
       adsurl = {https://ui.adsabs.harvard.edu/abs/2021MNRAS.502.2446E}
}

@ARTICLE{bono05,
       author = {{Bono}, G. and {Marconi}, M. and {Cassisi}, S. and {Caputo}, F. and {Gieren}, W. and {Pietrzynski}, G.},
        title = "{Classical Cepheid Pulsation Models. X. The Period-Age Relation}",
      journal = {\apj},
         year = 2005,
        month = mar,
       volume = {621},
       number = {2},
        pages = {966-977},
          doi = {10.1086/427744},
archivePrefix = {arXiv},
       eprint = {astro-ph/0411756},
 primaryClass = {astro-ph},
       adsurl = {https://ui.adsabs.harvard.edu/abs/2005ApJ...621..966B}
}

@ARTICLE{bono24,
       author = {{Bono}, G. and {Braga}, V.~F. and {Pietrinferni}, A.},
        title = "{Cepheids as distance indicators and stellar tracers}",
      journal = {\aapr},
         year = 2024,
        month = apr,
       volume = {32},
       number = {1},
          eid = {4},
        pages = {4},
          doi = {10.1007/s00159-024-00153-0},
archivePrefix = {arXiv},
       eprint = {2405.04893},
 primaryClass = {astro-ph.SR},
       adsurl = {https://ui.adsabs.harvard.edu/abs/2024A&ARv..32....4B}
}

@ARTICLE{inno19,
       author = {{Inno}, L. and {Urbaneja}, M.~A. and {Matsunaga}, N. and {Bono}, G. and {Nonino}, M. and {Debattista}, V.~P. and {Sormani}, M.~C. and {Bergemann}, M. and {da Silva}, R. and {Lemasle}, B. and {Romaniello}, M. and {Rix}, H.-W.},
        title = "{First metallicity determination from near-infrared spectra for five obscured Cepheids discovered in the inner disc}",
      journal = {\mnras},
         year = 2019,
        month = jan,
       volume = {482},
       number = {1},
        pages = {83-97},
          doi = {10.1093/mnras/sty2661},
archivePrefix = {arXiv},
       eprint = {1805.03212},
 primaryClass = {astro-ph.GA},
       adsurl = {https://ui.adsabs.harvard.edu/abs/2019MNRAS.482...83I}
}

@ARTICLE{williams24,
       author = {{Williams}, Thomas G. and {Lee}, Janice C. and {Larson}, Kirsten L. and {Leroy}, Adam K. and {Sandstrom}, Karin and {Schinnerer}, Eva and {Thilker}, David A. and {Belfiore}, Francesco and {Egorov}, Oleg V. and {Rosolowsky}, Erik and {Sutter}, Jessica and {DePasquale}, Joseph and {Pagan}, Alyssa and {Berger}, Travis A. and {Anand}, Gagandeep S. and {Barnes}, Ashley T. and {Bigiel}, Frank and {Boquien}, M{\'e}d{\'e}ric and {Cao}, Yixian and {Chastenet}, J{\'e}r{\'e}my and {Chevance}, M{\'e}lanie and {Chown}, Ryan and {Dale}, Daniel A. and {Deger}, Sinan and {Eibensteiner}, Cosima and {Emsellem}, Eric and {Faesi}, Christopher M. and {Glover}, Simon C.~O. and {Grasha}, Kathryn and {Hannon}, Stephen and {Hassani}, Hamid and {Henshaw}, Jonathan D. and {Jim{\'e}nez-Donaire}, Mar{\'\i}a J. and {Kim}, Jaeyeon and {Klessen}, Ralf S. and {Koch}, Eric W. and {Li}, Jing and {Liu}, Daizhong and {Meidt}, Sharon E. and {M{\'e}ndez-Delgado}, J. Eduardo and {Murphy}, Eric J. and {Neumann}, Justus and {Neumann}, Lukas and {Neumayer}, Nadine and {Oakes}, Elias K. and {Pathak}, Debosmita and {Pety}, J{\'e}r{\^o}me and {Pinna}, Francesca and {Querejeta}, Miguel and {Ramambason}, Lise and {Romanelli}, Andrea and {Sormani}, Mattia C. and {Stuber}, Sophia K. and {Sun}, Jiayi and {Teng}, Yu-Hsuan and {Usero}, Antonio and {Watkins}, Elizabeth J. and {Weinbeck}, Tony D.},
        title = "{PHANGS-JWST: Data-processing Pipeline and First Full Public Data Release}",
      journal = {\apjs},
         year = 2024,
        month = jul,
       volume = {273},
       number = {1},
          eid = {13},
        pages = {13},
          doi = {10.3847/1538-4365/ad4be5},
archivePrefix = {arXiv},
       eprint = {2401.15142},
 primaryClass = {astro-ph.GA},
       adsurl = {https://ui.adsabs.harvard.edu/abs/2024ApJS..273...13W}
}

@ARTICLE{combes93,
       author = {{Combes}, F. and {Elmegreen}, B.~G.},
        title = "{Bars in early- and late-type galaxies.}",
      journal = {\aap},
         year = 1993,
        month = apr,
       volume = {271},
        pages = {391-401},
       adsurl = {https://ui.adsabs.harvard.edu/abs/1993A&A...271..391C}
}

@ARTICLE{lokas20,
       author = {{{\L}okas}, Ewa L.},
        title = "{The effect of warm gas on the buckling instability in galactic bars}",
      journal = {\aap},
         year = 2020,
        month = feb,
       volume = {634},
          eid = {A122},
        pages = {A122},
          doi = {10.1051/0004-6361/201937165},
archivePrefix = {arXiv},
       eprint = {2001.07041},
 primaryClass = {astro-ph.GA},
       adsurl = {https://ui.adsabs.harvard.edu/abs/2020A&A...634A.122L}
}

@ARTICLE{bovy19,
       author = {{Bovy}, Jo and {Leung}, Henry W. and {Hunt}, Jason A.~S. and {Mackereth}, J. Ted and {Garc{\'\i}a-Hern{\'a}ndez}, Domingo A. and {Roman-Lopes}, Alexandre},
        title = "{Life in the fast lane: a direct view of the dynamics, formation, and evolution of the Milky Way's bar}",
      journal = {\mnras},
         year = 2019,
        month = dec,
       volume = {490},
       number = {4},
        pages = {4740-4747},
          doi = {10.1093/mnras/stz2891},
archivePrefix = {arXiv},
       eprint = {1905.11404},
 primaryClass = {astro-ph.GA},
       adsurl = {https://ui.adsabs.harvard.edu/abs/2019MNRAS.490.4740B}
}

@ARTICLE{jang24,
       author = {{Jang}, Dajeong and {Kim}, Woong-Tae},
        title = "{Effects of Halo Spin on the Formation and Evolution of Bars in Disk Galaxies}",
      journal = {\apj},
         year = 2024,
        month = aug,
       volume = {971},
       number = {1},
          eid = {67},
        pages = {67},
          doi = {10.3847/1538-4357/ad54b9},
archivePrefix = {arXiv},
       eprint = {2406.08823},
 primaryClass = {astro-ph.GA},
       adsurl = {https://ui.adsabs.harvard.edu/abs/2024ApJ...971...67J}
}

@ARTICLE{athanassoula14,
       author = {{Athanassoula}, E.},
        title = "{Bar slowdown and the distribution of dark matter in barred galaxies}",
      journal = {\mnras},
         year = 2014,
        month = feb,
       volume = {438},
       number = {1},
        pages = {L81-L85},
          doi = {10.1093/mnrasl/slt163},
archivePrefix = {arXiv},
       eprint = {1312.1690},
 primaryClass = {astro-ph.GA},
       adsurl = {https://ui.adsabs.harvard.edu/abs/2014MNRAS.438L..81A}
}

@ARTICLE{villa10,
       author = {{Villa-Vargas}, Jorge and {Shlosman}, Isaac and {Heller}, Clayton},
        title = "{Dark Matter Halos and Evolution of Bars in Disk Galaxies: Varying Gas Fraction and Gas Spatial Resolution}",
      journal = {\apj},
         year = 2010,
        month = aug,
       volume = {719},
       number = {2},
        pages = {1470-1480},
          doi = {10.1088/0004-637X/719/2/1470},
archivePrefix = {arXiv},
       eprint = {1004.4899},
 primaryClass = {astro-ph.CO},
       adsurl = {https://ui.adsabs.harvard.edu/abs/2010ApJ...719.1470V}
}

@ARTICLE{villa09,
       author = {{Villa-Vargas}, Jorge and {Shlosman}, Isaac and {Heller}, Clayton},
        title = "{Dark Matter Halos and Evolution of Bars in Disk Galaxies: Collisionless Models Revisited}",
      journal = {\apj},
         year = 2009,
        month = dec,
       volume = {707},
       number = {1},
        pages = {218-232},
          doi = {10.1088/0004-637X/707/1/218},
archivePrefix = {arXiv},
       eprint = {0904.0646},
 primaryClass = {astro-ph.CO},
       adsurl = {https://ui.adsabs.harvard.edu/abs/2009ApJ...707..218V}
}

@ARTICLE{berentzen07,
       author = {{Berentzen}, Ingo and {Shlosman}, Isaac and {Martinez-Valpuesta}, Inma and {Heller}, Clayton H.},
        title = "{Gas Feedback on Stellar Bar Evolution}",
      journal = {\apj},
         year = 2007,
        month = sep,
       volume = {666},
       number = {1},
        pages = {189-200},
          doi = {10.1086/520531},
archivePrefix = {arXiv},
       eprint = {astro-ph/0703028},
 primaryClass = {astro-ph},
       adsurl = {https://ui.adsabs.harvard.edu/abs/2007ApJ...666..189B}
}

@ARTICLE{friedli93,
       author = {{Friedli}, D. and {Benz}, W.},
        title = "{Secular evolution of isolated barred galaxies. I. Gravitational coupling between stellar bars and interstellar medium.}",
      journal = {\aap},
         year = 1993,
        month = feb,
       volume = {268},
        pages = {65-85},
       adsurl = {https://ui.adsabs.harvard.edu/abs/1993A&A...268...65F}
}

@ARTICLE{spitoni26,
       author = {{Spitoni}, E. and {Schultheis}, M. and {Matteucci}, F. and {Ryde}, N. and {Cescutti}, G. and {Saro}, A. and {Sormani}, M.~C. and {Thorsbro}, B.},
        title = "{The LEGARE Project: I. Chemical evolution model of the Nuclear Stellar Disc in a Bayesian framework}",
      journal = {\aap},
         year = 2026,
        month = mar,
       volume = {707},
          eid = {A202},
        pages = {A202},
          doi = {10.1051/0004-6361/202558155},
archivePrefix = {arXiv},
       eprint = {2601.21032},
 primaryClass = {astro-ph.GA},
       adsurl = {https://ui.adsabs.harvard.edu/abs/2026A&A...707A.202S}
}

@ARTICLE{leconte26,
       author = {{Le Conte}, Zoe A. and {Gadotti}, Dimitri A. and {Harvey}, Thomas and {Ferreira}, Leonardo and {Conselice}, Christopher J. and {Kim}, Taehyun and {de S{\'a}-Freitas}, Camila and {Fragkoudi}, Francesca and {Neumann}, Justus and {Athanassoula}, E.},
        title = "{A nuclear disc at Cosmic Noon: evidence of early bar-driven galaxy evolution}",
      journal = {arXiv e-prints},
         year = 2026,
        month = jan,
          eid = {arXiv:2601.18871},
        pages = {arXiv:2601.18871},
archivePrefix = {arXiv},
       eprint = {2601.18871},
 primaryClass = {astro-ph.GA},
       adsurl = {https://ui.adsabs.harvard.edu/abs/2026arXiv260118871L}
}

@ARTICLE{neumayer20,
       author = {{Neumayer}, Nadine and {Seth}, Anil and {B{\"o}ker}, Torsten},
        title = "{Nuclear star clusters}",
      journal = {\aapr},
         year = 2020,
        month = jul,
       volume = {28},
       number = {1},
          eid = {4},
        pages = {4},
          doi = {10.1007/s00159-020-00125-0},
archivePrefix = {arXiv},
       eprint = {2001.03626},
 primaryClass = {astro-ph.GA},
       adsurl = {https://ui.adsabs.harvard.edu/abs/2020A&ARv..28....4N}
}

@ARTICLE{bland24,
       author = {{Bland-Hawthorn}, Joss and {Tepper-Garcia}, Thor and {Agertz}, Oscar and {Federrath}, Christoph},
        title = "{Turbulent Gas-rich Disks at High Redshift: Bars and Bulges in a Radial Shear Flow}",
      journal = {\apj},
         year = 2024,
        month = jun,
       volume = {968},
       number = {2},
          eid = {86},
        pages = {86},
          doi = {10.3847/1538-4357/ad4118},
archivePrefix = {arXiv},
       eprint = {2402.06060},
 primaryClass = {astro-ph.GA},
       adsurl = {https://ui.adsabs.harvard.edu/abs/2024ApJ...968...86B}
}

@ARTICLE{athanassoula02a,
       author = {{Athanassoula}, E. and {Misiriotis}, A.},
        title = "{Morphology, photometry and kinematics of N -body bars - I. Three models with different halo central concentrations}",
      journal = {\mnras},
         year = 2002,
        month = feb,
       volume = {330},
       number = {1},
        pages = {35-52},
          doi = {10.1046/j.1365-8711.2002.05028.x},
archivePrefix = {arXiv},
       eprint = {astro-ph/0111449},
 primaryClass = {astro-ph},
       adsurl = {https://ui.adsabs.harvard.edu/abs/2002MNRAS.330...35A}
}

@ARTICLE{kataria19,
       author = {{Kataria}, Sandeep Kumar and {Das}, Mousumi},
        title = "{The Effect of Bulge Mass on Bar Pattern Speed in Disk Galaxies}",
      journal = {\apj},
         year = 2019,
        month = nov,
       volume = {886},
       number = {1},
          eid = {43},
        pages = {43},
          doi = {10.3847/1538-4357/ab48f7},
archivePrefix = {arXiv},
       eprint = {1910.03967},
 primaryClass = {astro-ph.GA},
       adsurl = {https://ui.adsabs.harvard.edu/abs/2019ApJ...886...43K}
}

@ARTICLE{marques25,
       author = {{Marques}, L. and {Minchev}, I. and {Ratcliffe}, B. and {Khoperskov}, S. and {Steinmetz}, M. and {Wenger}, T.~V. and {Buck}, T. and {Martig}, M. and {Kordopatis}, G. and {Schultheis}, M. and {Zucker}, D.~B.},
        title = "{Bar-spiral interaction induces radial migration and star formation bursts}",
      journal = {\aap},
         year = 2025,
        month = sep,
       volume = {701},
          eid = {A88},
        pages = {A88},
          doi = {10.1051/0004-6361/202554020},
archivePrefix = {arXiv},
       eprint = {2502.02651},
 primaryClass = {astro-ph.GA},
       adsurl = {https://ui.adsabs.harvard.edu/abs/2025A&A...701A..88M}
}

@ARTICLE{hilmi20,
       author = {{Hilmi}, T. and {Minchev}, I. and {Buck}, T. and {Martig}, M. and {Quillen}, A.~C. and {Monari}, G. and {Famaey}, B. and {de Jong}, R.~S. and {Laporte}, C.~F.~P. and {Read}, J. and {Sanders}, J.~L. and {Steinmetz}, M. and {Wegg}, C.},
        title = "{Fluctuations in galactic bar parameters due to bar-spiral interaction}",
      journal = {\mnras},
         year = 2020,
        month = sep,
       volume = {497},
       number = {1},
        pages = {933-955},
          doi = {10.1093/mnras/staa1934},
archivePrefix = {arXiv},
       eprint = {2003.05457},
 primaryClass = {astro-ph.GA},
       adsurl = {https://ui.adsabs.harvard.edu/abs/2020MNRAS.497..933H}
}

@ARTICLE{quillen11,
       author = {{Quillen}, Alice C. and {Dougherty}, Jamie and {Bagley}, Micaela B. and {Minchev}, Ivan and {Comparetta}, Justin},
        title = "{Structure in phase space associated with spiral and bar density waves in an N-body hybrid galactic disc}",
      journal = {\mnras},
         year = 2011,
        month = oct,
       volume = {417},
       number = {1},
        pages = {762-784},
          doi = {10.1111/j.1365-2966.2011.19349.x},
archivePrefix = {arXiv},
       eprint = {1010.5745},
 primaryClass = {astro-ph.GA},
       adsurl = {https://ui.adsabs.harvard.edu/abs/2011MNRAS.417..762Q}
}

@ARTICLE{minchev10,
       author = {{Minchev}, I. and {Famaey}, B.},
        title = "{A New Mechanism for Radial Migration in Galactic Disks: Spiral-Bar Resonance Overlap}",
      journal = {\apj},
         year = 2010,
        month = oct,
       volume = {722},
       number = {1},
        pages = {112-121},
          doi = {10.1088/0004-637X/722/1/112},
archivePrefix = {arXiv},
       eprint = {0911.1794},
 primaryClass = {astro-ph.GA},
       adsurl = {https://ui.adsabs.harvard.edu/abs/2010ApJ...722..112M}
}

@ARTICLE{schultheis25b,
       author = {{Schultheis}, M. and {Serrano}, L. and {Thorsbro}, B. and {Nogueras-Lara}, F. and {Feldmeier-Krause}, A. and {Nandakumar}, G. and {Fiteni}, K. and {Sormani}, M.~C. and {Ryde}, N.},
        title = "{Chemical analysis of the Milky Way's nuclear star cluster: Evidence for a metallicity gradient}",
      journal = {\aap},
         year = 2026,
        month = jan,
       volume = {705},
          eid = {A235},
        pages = {A235},
          doi = {10.1051/0004-6361/202557723},
archivePrefix = {arXiv},
       eprint = {2512.08387},
 primaryClass = {astro-ph.GA},
       adsurl = {https://ui.adsabs.harvard.edu/abs/2026A&A...705A.235S}
}

@ARTICLE{schultheis21,
       author = {{Schultheis}, M. and {Fritz}, T.~K. and {Nandakumar}, G. and {Rojas-Arriagada}, A. and {Nogueras-Lara}, F. and {Feldmeier-Krause}, A. and {Gerhard}, O. and {Neumayer}, N. and {Patrick}, L.~R. and {Prieto}, M.~A. and {Sch{\"o}del}, R. and {Mastrobuono-Battisti}, A. and {Sormani}, M.~C.},
        title = "{The nuclear stellar disc of the Milky Way: A dynamically cool and metal-rich component possibly formed from the central molecular zone}",
      journal = {\aap},
         year = 2021,
        month = jun,
       volume = {650},
          eid = {A191},
        pages = {A191},
          doi = {10.1051/0004-6361/202140499},
archivePrefix = {arXiv},
       eprint = {2104.10439},
 primaryClass = {astro-ph.GA},
       adsurl = {https://ui.adsabs.harvard.edu/abs/2021A&A...650A.191S}
}

@ARTICLE{nogueras24,
       author = {{Nogueras-Lara}, F. and {Nieuwmunster}, N. and {Schultheis}, M. and {Sormani}, M.~C. and {Fragkoudi}, F. and {Thorsbro}, B. and {Rich}, R.~M. and {Ryde}, N. and {Sanders}, J.~L. and {Smith}, L.~C.},
        title = "{Metallicity-dependent kinematics and orbits in the Milky Way's nuclear stellar disc}",
      journal = {\aap},
         year = 2024,
        month = oct,
       volume = {690},
          eid = {A313},
        pages = {A313},
          doi = {10.1051/0004-6361/202450946},
archivePrefix = {arXiv},
       eprint = {2409.15279},
 primaryClass = {astro-ph.GA},
       adsurl = {https://ui.adsabs.harvard.edu/abs/2024A&A...690A.313N}
}

@ARTICLE{nogueras23,
       author = {{Nogueras-Lara}, F. and {Feldmeier-Krause}, A. and {Sch{\"o}del}, R. and {Sormani}, M.~C. and {de Lorenzo-C{\'a}ceres}, A. and {Mastrobuono-Battisti}, A. and {Schultheis}, M. and {Neumayer}, N. and {Rich}, R.~M. and {Nieuwmunster}, N.},
        title = "{Smooth kinematic and metallicity gradients reveal that the Milky Way's nuclear star cluster and disc might be part of the same structure}",
      journal = {\aap},
         year = 2023,
        month = dec,
       volume = {680},
          eid = {A75},
        pages = {A75},
          doi = {10.1051/0004-6361/202347421},
archivePrefix = {arXiv},
       eprint = {2309.07219},
 primaryClass = {astro-ph.GA},
       adsurl = {https://ui.adsabs.harvard.edu/abs/2023A&A...680A..75N}
}

@ARTICLE{kataria18,
       author = {{Kataria}, Sandeep Kumar and {Das}, Mousumi},
        title = "{A study of the effect of bulges on bar formation in disc galaxies}",
      journal = {\mnras},
         year = 2018,
        month = apr,
       volume = {475},
       number = {2},
        pages = {1653-1664},
          doi = {10.1093/mnras/stx3279},
       adsurl = {https://ui.adsabs.harvard.edu/abs/2018MNRAS.475.1653K}
}

@ARTICLE{jang23,
       author = {{Jang}, Dajeong and {Kim}, Woong-Tae},
        title = "{Effects of the Central Mass Concentration on Bar Formation in Disk Galaxies}",
      journal = {\apj},
         year = 2023,
        month = jan,
       volume = {942},
       number = {2},
          eid = {106},
        pages = {106},
          doi = {10.3847/1538-4357/aca7bc},
archivePrefix = {arXiv},
       eprint = {2211.16816},
 primaryClass = {astro-ph.GA},
       adsurl = {https://ui.adsabs.harvard.edu/abs/2023ApJ...942..106J}
}

@ARTICLE{gleis26,
       author = {{Gleis}, Damian R. and {Stuber}, Sophia K. and {Schinnerer}, Eva and {Neumann}, Justus and {Meidt}, Sharon E. and {Querejeta}, Miguel and {Emsellem}, Eric and {Leroy}, Adam K. and {Barnes}, Ashley T. and {Bigiel}, Frank and {Burton}, Charlie and {Chevance}, M{\'e}lanie and {Dale}, Daniel A. and {Grasha}, Kathryn and {Klessen}, Ralf S. and {Levy}, Rebecca C. and {Neumann}, Lukas and {Pan}, Hsi-An and {Ruiz-Garc{\'\i}a}, Marina and {Sormani}, Mattia C. and {Sun}, Jiayi and {Teng}, Yu-Hsuan and {Williams}, Thomas G.},
        title = "{Molecular gas and star formation in central rings across nearby galaxies}",
      journal = {\aap},
         year = 2026,
        month = mar,
       volume = {707},
          eid = {A121},
        pages = {A121},
          doi = {10.1051/0004-6361/202557430},
archivePrefix = {arXiv},
       eprint = {2601.11127},
 primaryClass = {astro-ph.GA},
       adsurl = {https://ui.adsabs.harvard.edu/abs/2026A&A...707A.121G}
}

@ARTICLE{kwak25,
       author = {{Kwak}, SungWon and {Minchev}, Ivan and {Pfrommer}, Christoph and {Steinmetz}, Matthias and {Yi}, Sukyoung K.},
        title = "{Effects of Resolution and Local Stability on Galactic Disks: I. Multiple Spiral Mode Formation via Swing Amplification}",
      journal = {arXiv e-prints},
         year = 2025,
        month = nov,
          eid = {arXiv:2511.21805},
        pages = {arXiv:2511.21805},
          doi = {10.48550/arXiv.2511.21805},
archivePrefix = {arXiv},
       eprint = {2511.21805},
 primaryClass = {astro-ph.GA},
       adsurl = {https://ui.adsabs.harvard.edu/abs/2025arXiv251121805K}
}

@ARTICLE{papovich16,
       author = {{Papovich}, C. and {Labb{\'e}}, I. and {Glazebrook}, K. and {Quadri}, R. and {Bekiaris}, G. and {Dickinson}, M. and {Finkelstein}, S.~L. and {Fisher}, D. and {Inami}, H. and {Livermore}, R.~C. and {Spitler}, L. and {Straatman}, C. and {Tran}, K.-V.},
        title = "{Large molecular gas reservoirs in ancestors of Milky Way-mass galaxies nine billion years ago}",
      journal = {Nature Astronomy},
         year = 2016,
        month = dec,
       volume = {1},
          eid = {0003},
        pages = {0003},
          doi = {10.1038/s41550-016-0003},
archivePrefix = {arXiv},
       eprint = {1610.05313},
 primaryClass = {astro-ph.GA},
       adsurl = {https://ui.adsabs.harvard.edu/abs/2016NatAs...1E...3P}
}

@ARTICLE{saintonge11,
       author = {{Saintonge}, Am{\'e}lie and {Kauffmann}, Guinevere and {Kramer}, Carsten and {Tacconi}, Linda J. and {Buchbender}, Christof and {Catinella}, Barbara and {Fabello}, Silvia and {Graci{\'a}-Carpio}, Javier and {Wang}, Jing and {Cortese}, Luca and {Fu}, Jian and {Genzel}, Reinhard and {Giovanelli}, Riccardo and {Guo}, Qi and {Haynes}, Martha P. and {Heckman}, Timothy M. and {Krumholz}, Mark R. and {Lemonias}, Jenna and {Li}, Cheng and {Moran}, Sean and {Rodriguez-Fernandez}, Nemesio and {Schiminovich}, David and {Schuster}, Karl and {Sievers}, Albrecht},
        title = "{COLD GASS, an IRAM legacy survey of molecular gas in massive galaxies - I. Relations between H$_{2}$, H I, stellar content and structural properties}",
      journal = {\mnras},
         year = 2011,
        month = jul,
       volume = {415},
       number = {1},
        pages = {32-60},
          doi = {10.1111/j.1365-2966.2011.18677.x},
archivePrefix = {arXiv},
       eprint = {1103.1642},
 primaryClass = {astro-ph.CO},
       adsurl = {https://ui.adsabs.harvard.edu/abs/2011MNRAS.415...32S}
}

@ARTICLE{leroy09,
       author = {{Leroy}, Adam K. and {Walter}, Fabian and {Bigiel}, Frank and {Usero}, Antonio and {Weiss}, Axel and {Brinks}, Elias and {de Blok}, W.~J.~G. and {Kennicutt}, Robert C. and {Schuster}, Karl-Friedrich and {Kramer}, Carsten and {Wiesemeyer}, H.~W. and {Roussel}, H{\'e}l{\`e}ne},
        title = "{Heracles: The HERA CO Line Extragalactic Survey}",
      journal = {\aj},
         year = 2009,
        month = jun,
       volume = {137},
       number = {6},
        pages = {4670-4696},
          doi = {10.1088/0004-6256/137/6/4670},
archivePrefix = {arXiv},
       eprint = {0905.4742},
 primaryClass = {astro-ph.CO},
       adsurl = {https://ui.adsabs.harvard.edu/abs/2009AJ....137.4670L}
}

@ARTICLE{sage93,
       author = {{Sage}, L.~J.},
        title = "{Molecular gas in nearby galaxies. I. CO observations of a distance-limited sample.}",
      journal = {\aap},
         year = 1993,
        month = may,
       volume = {272},
        pages = {123-136},
       adsurl = {https://ui.adsabs.harvard.edu/abs/1993A&A...272..123S}
}

@ARTICLE{guiglion15,
       author = {{Guiglion}, G. and {Recio-Blanco}, A. and {de Laverny}, P. and {Kordopatis}, G. and {Hill}, V. and {Mikolaitis}, {\v{S}}. and {Minchev}, I. and {Chiappini}, C. and {Wyse}, R.~F.~G. and {Gilmore}, G. and {Randich}, S. and {Feltzing}, S. and {Bensby}, T. and {Flaccomio}, E. and {Koposov}, S.~E. and {Pancino}, E. and {Bayo}, A. and {Costado}, M.~T. and {Franciosini}, E. and {Hourihane}, A. and {Jofr{\'e}}, P. and {Lardo}, C. and {Lewis}, J. and {Lind}, K. and {Magrini}, L. and {Morbidelli}, L. and {Sacco}, G.~G. and {Ruchti}, G. and {Worley}, C.~C. and {Zaggia}, S.},
        title = "{The Gaia-ESO Survey: New constraints on the Galactic disc velocity dispersion and its chemical dependencies}",
      journal = {\aap},
         year = 2015,
        month = nov,
       volume = {583},
          eid = {A91},
        pages = {A91},
          doi = {10.1051/0004-6361/201525883},
archivePrefix = {arXiv},
       eprint = {1509.05271},
 primaryClass = {astro-ph.GA},
       adsurl = {https://ui.adsabs.harvard.edu/abs/2015A&A...583A..91G}
}

@ARTICLE{sharma14,
       author = {{Sharma}, S. and {Bland-Hawthorn}, J. and {Binney}, J. and {Freeman}, K.~C. and {Steinmetz}, M. and {Boeche}, C. and {Bienaym{\'e}}, O. and {Gibson}, B.~K. and {Gilmore}, G.~F. and {Grebel}, E.~K. and {Helmi}, A. and {Kordopatis}, G. and {Munari}, U. and {Navarro}, J.~F. and {Parker}, Q.~A. and {Reid}, W.~A. and {Seabroke}, G.~M. and {Siebert}, A. and {Watson}, F. and {Williams}, M.~E.~K. and {Wyse}, R.~F.~G. and {Zwitter}, T.},
        title = "{Kinematic Modeling of the Milky Way Using the RAVE and GCS Stellar Surveys}",
      journal = {\apj},
         year = 2014,
        month = sep,
       volume = {793},
       number = {1},
          eid = {51},
        pages = {51},
          doi = {10.1088/0004-637X/793/1/51},
archivePrefix = {arXiv},
       eprint = {1405.7435},
 primaryClass = {astro-ph.GA},
       adsurl = {https://ui.adsabs.harvard.edu/abs/2014ApJ...793...51S}
}

@ARTICLE{hernquist90,
       author = {{Hernquist}, Lars},
        title = "{An Analytical Model for Spherical Galaxies and Bulges}",
      journal = {\apj},
         year = 1990,
        month = jun,
       volume = {356},
        pages = {359},
          doi = {10.1086/168845},
       adsurl = {https://ui.adsabs.harvard.edu/abs/1990ApJ...356..359H}
}

@ARTICLE{springel05,
       author = {{Springel}, Volker and {Di Matteo}, Tiziana and {Hernquist}, Lars},
        title = "{Modelling feedback from stars and black holes in galaxy mergers}",
      journal = {\mnras},
         year = 2005,
        month = aug,
       volume = {361},
       number = {3},
        pages = {776-794},
          doi = {10.1111/j.1365-2966.2005.09238.x},
archivePrefix = {arXiv},
       eprint = {astro-ph/0411108},
 primaryClass = {astro-ph},
       adsurl = {https://ui.adsabs.harvard.edu/abs/2005MNRAS.361..776S}
}

@ARTICLE{bose19,
       author = {{Bose}, Sownak and {Eisenstein}, Daniel J. and {Hernquist}, Lars and {Pillepich}, Annalisa and {Nelson}, Dylan and {Marinacci}, Federico and {Springel}, Volker and {Vogelsberger}, Mark},
        title = "{Revealing the galaxy-halo connection in IllustrisTNG}",
      journal = {\mnras},
         year = 2019,
        month = dec,
       volume = {490},
       number = {4},
        pages = {5693-5711},
          doi = {10.1093/mnras/stz2546},
archivePrefix = {arXiv},
       eprint = {1905.08799},
 primaryClass = {astro-ph.CO},
       adsurl = {https://ui.adsabs.harvard.edu/abs/2019MNRAS.490.5693B}
}

@ARTICLE{springel08,
       author = {{Springel}, V. and {Wang}, J. and {Vogelsberger}, M. and {Ludlow}, A. and {Jenkins}, A. and {Helmi}, A. and {Navarro}, J.~F. and {Frenk}, C.~S. and {White}, S.~D.~M.},
        title = "{The Aquarius Project: the subhaloes of galactic haloes}",
      journal = {\mnras},
         year = 2008,
        month = dec,
       volume = {391},
       number = {4},
        pages = {1685-1711},
          doi = {10.1111/j.1365-2966.2008.14066.x},
archivePrefix = {arXiv},
       eprint = {0809.0898},
 primaryClass = {astro-ph},
       adsurl = {https://ui.adsabs.harvard.edu/abs/2008MNRAS.391.1685S}
}

@ARTICLE{mo98,
       author = {{Mo}, H.~J. and {Mao}, Shude and {White}, Simon D.~M.},
        title = "{The formation of galactic discs}",
      journal = {\mnras},
         year = 1998,
        month = apr,
       volume = {295},
       number = {2},
        pages = {319-336},
          doi = {10.1046/j.1365-8711.1998.01227.x},
archivePrefix = {arXiv},
       eprint = {astro-ph/9707093},
 primaryClass = {astro-ph},
       adsurl = {https://ui.adsabs.harvard.edu/abs/1998MNRAS.295..319M}
}

@ARTICLE{deng24,
       author = {{Deng}, Yunwei and {Li}, Hui and {Kannan}, Rahul and {Smith}, Aaron and {Vogelsberger}, Mark and {Bryan}, Greg L.},
        title = "{Simulating ionization feedback from young massive stars: impact of numerical resolution}",
      journal = {\mnras},
         year = 2024,
        month = jan,
       volume = {527},
       number = {1},
        pages = {478-500},
          doi = {10.1093/mnras/stad3202},
archivePrefix = {arXiv},
       eprint = {2309.15900},
 primaryClass = {astro-ph.GA},
       adsurl = {https://ui.adsabs.harvard.edu/abs/2024MNRAS.527..478D}
}

@ARTICLE{wang25,
       author = {{Wang}, Wenwen and {Zhou}, Zhimin},
        title = "{Identification and Analysis of Galactic Bars in DESI Legacy Imaging Surveys}",
      journal = {\apj},
         year = 2025,
        month = apr,
       volume = {982},
       number = {2},
          eid = {129},
        pages = {129},
          doi = {10.3847/1538-4357/adbcac},
archivePrefix = {arXiv},
       eprint = {2503.22075},
 primaryClass = {astro-ph.GA},
       adsurl = {https://ui.adsabs.harvard.edu/abs/2025ApJ...982..129W}
}

@ARTICLE{sivasankaran26,
       author = {{Sivasankaran}, Aneesh and {Blecha}, Laura and {Torrey}, Paul and {Kelley}, Luke Zoltan and {Bhowmick}, Aklant and {Vogelsberger}, Mark and {Hernquist}, Lars and {Marinacci}, Federico and {Sales}, Laura V.},
        title = "{AGN feedback in merging galaxies with a SMUGGLE multiphase ISM}",
      journal = {\mnras},
         year = 2026,
        month = jan,
       volume = {545},
       number = {3},
          eid = {staf2044},
        pages = {staf2044},
          doi = {10.1093/mnras/staf2044},
archivePrefix = {arXiv},
       eprint = {2511.13024},
 primaryClass = {astro-ph.GA},
       adsurl = {https://ui.adsabs.harvard.edu/abs/2026MNRAS.545f2044S}
}

@ARTICLE{zhangz25,
       author = {{Zhang}, Zhijie and {Zhang}, Xiaoxia and {Li}, Hui and {Fang}, Taotao and {Luo}, Yang and {Marinacci}, Federico and {Sales}, Laura V. and {Torrey}, Paul and {Vogelsberger}, Mark and {Yu}, Qingzheng and {Yuan}, Feng},
        title = "{Tracing the Origins of Hot Halo Gas in Milky Way─type Galaxies with SMUGGLE}",
      journal = {\apj},
         year = 2025,
        month = oct,
       volume = {991},
       number = {2},
          eid = {170},
        pages = {170},
          doi = {10.3847/1538-4357/ae019f},
archivePrefix = {arXiv},
       eprint = {2508.21576},
 primaryClass = {astro-ph.GA},
       adsurl = {https://ui.adsabs.harvard.edu/abs/2025ApJ...991..170Z}
}

@ARTICLE{sivasankaran25,
       author = {{Sivasankaran}, Aneesh and {Blecha}, Laura and {Torrey}, Paul and {Kelley}, Luke Zoltan and {Bhowmick}, Aklant and {Vogelsberger}, Mark and {Hernquist}, Lars and {Marinacci}, Federico and {Sales}, Laura V.},
        title = "{AGN feedback in isolated galaxies with a SMUGGLE multiphase ISM}",
      journal = {\mnras},
         year = 2025,
        month = feb,
       volume = {537},
       number = {2},
        pages = {817-830},
          doi = {10.1093/mnras/staf062},
archivePrefix = {arXiv},
       eprint = {2402.15240},
 primaryClass = {astro-ph.GA},
       adsurl = {https://ui.adsabs.harvard.edu/abs/2025MNRAS.537..817S}
}

@ARTICLE{zhange24,
       author = {{Zhang}, Eric and {Sales}, Laura V. and {Marinacci}, Federico and {Torrey}, Paul and {Vogelsberger}, Mark and {Springel}, Volker and {Li}, Hui and {Pakmor}, R{\"u}diger and {Gutcke}, Thales A.},
        title = "{Bursty Star Formation in Dwarfs is Sensitive to Numerical Choices in Supernova Feedback Models}",
      journal = {\apj},
         year = 2024,
        month = nov,
       volume = {975},
       number = {2},
          eid = {229},
        pages = {229},
          doi = {10.3847/1538-4357/ad7f57},
archivePrefix = {arXiv},
       eprint = {2406.10338},
 primaryClass = {astro-ph.GA},
       adsurl = {https://ui.adsabs.harvard.edu/abs/2024ApJ...975..229Z}
}

@ARTICLE{li24,
       author = {{Li}, Chengzhe and {Li}, Hui and {Cui}, Wei and {Marinacci}, Federico and {Sales}, Laura V. and {Vogelsberger}, Mark and {Torrey}, Paul},
        title = "{Evolution and distribution of superbubbles in simulated Milky Way-like galaxies}",
      journal = {\mnras},
         year = 2024,
        month = apr,
       volume = {529},
       number = {4},
        pages = {4073-4086},
          doi = {10.1093/mnras/stae797},
archivePrefix = {arXiv},
       eprint = {2403.12135},
 primaryClass = {astro-ph.GA},
       adsurl = {https://ui.adsabs.harvard.edu/abs/2024MNRAS.529.4073L}
}

@ARTICLE{zhangz24,
       author = {{Zhang}, Zhijie and {Zhang}, Xiaoxia and {Li}, Hui and {Fang}, Taotao and {Yu}, Qingzheng and {Luo}, Yang and {Marinacci}, Federico and {Sales}, Laura V. and {Torrey}, Paul and {Vogelsberger}, Mark},
        title = "{Low- and High-velocity O VI in Milky Way-like Galaxies: The Role of Stellar Feedback}",
      journal = {\apj},
         year = 2024,
        month = feb,
       volume = {962},
       number = {1},
          eid = {15},
        pages = {15},
          doi = {10.3847/1538-4357/ad10a4},
archivePrefix = {arXiv},
       eprint = {2311.16354},
 primaryClass = {astro-ph.GA},
       adsurl = {https://ui.adsabs.harvard.edu/abs/2024ApJ...962...15Z}
}

@ARTICLE{barbani23,
       author = {{Barbani}, Filippo and {Pascale}, Raffaele and {Marinacci}, Federico and {Sales}, Laura V. and {Vogelsberger}, Mark and {Torrey}, Paul and {Li}, Hui},
        title = "{Galactic coronae in Milky Way-like galaxies: the role of stellar feedback in gas accretion}",
      journal = {\mnras},
         year = 2023,
        month = sep,
       volume = {524},
       number = {3},
        pages = {4091-4108},
          doi = {10.1093/mnras/stad2152},
archivePrefix = {arXiv},
       eprint = {2306.11791},
 primaryClass = {astro-ph.GA},
       adsurl = {https://ui.adsabs.harvard.edu/abs/2023MNRAS.524.4091B}
}

@ARTICLE{narayanan23,
       author = {{Narayanan}, Desika and {Smith}, J.-D.~T. and {Hensley}, Brandon S. and {Li}, Qi and {Hu}, Chia-Yu and {Sandstrom}, Karin and {Torrey}, Paul and {Vogelsberger}, Mark and {Marinacci}, Federico and {Sales}, Laura V.},
        title = "{A Framework for Modeling Polycyclic Aromatic Hydrocarbon Emission in Galaxy Evolution Simulations}",
      journal = {\apj},
         year = 2023,
        month = jul,
       volume = {951},
       number = {2},
          eid = {100},
        pages = {100},
          doi = {10.3847/1538-4357/accf8d},
archivePrefix = {arXiv},
       eprint = {2301.07136},
 primaryClass = {astro-ph.GA},
       adsurl = {https://ui.adsabs.harvard.edu/abs/2023ApJ...951..100N}
}

@ARTICLE{sivasankaran22,
       author = {{Sivasankaran}, Aneesh and {Blecha}, Laura and {Torrey}, Paul and {Kelley}, Luke Zoltan and {Bhowmick}, Aklant and {Vogelsberger}, Mark and {Losacco}, Rachel and {Weinberger}, Rainer and {Hernquist}, Lars and {Marinacci}, Federico and {Sales}, Laura V. and {Qi}, Jia},
        title = "{Simulations of black hole fueling in isolated and merging galaxies with an explicit, multiphase ISM}",
      journal = {\mnras},
         year = 2022,
        month = dec,
       volume = {517},
       number = {4},
        pages = {4752-4767},
          doi = {10.1093/mnras/stac2759},
archivePrefix = {arXiv},
       eprint = {2203.14985},
 primaryClass = {astro-ph.GA},
       adsurl = {https://ui.adsabs.harvard.edu/abs/2022MNRAS.517.4752S}
}

@ARTICLE{sanders24,
       author = {{Sanders}, Jason L. and {Kawata}, Daisuke and {Matsunaga}, Noriyuki and {Sormani}, Mattia C. and {Smith}, Leigh C. and {Minniti}, Dante and {Gerhard}, Ortwin},
        title = "{The epoch of the Milky Way's bar formation: dynamical modelling of Mira variables in the nuclear stellar disc}",
      journal = {\mnras},
         year = 2024,
        month = may,
       volume = {530},
       number = {3},
        pages = {2972-2993},
          doi = {10.1093/mnras/stae711},
archivePrefix = {arXiv},
       eprint = {2311.00035},
 primaryClass = {astro-ph.GA},
       adsurl = {https://ui.adsabs.harvard.edu/abs/2024MNRAS.530.2972S}
}

@ARTICLE{smith22,
       author = {{Smith}, Aaron and {Kannan}, Rahul and {Tacchella}, Sandro and {Vogelsberger}, Mark and {Hernquist}, Lars and {Marinacci}, Federico and {Sales}, Laura V. and {Torrey}, Paul and {Li}, Hui and {Yeh}, Jessica Y.-C. and {Qi}, Jia},
        title = "{The physics of Lyman-{\ensuremath{\alpha}} escape from disc-like galaxies}",
      journal = {\mnras},
         year = 2022,
        month = nov,
       volume = {517},
       number = {1},
        pages = {1-27},
          doi = {10.1093/mnras/stac2641},
archivePrefix = {arXiv},
       eprint = {2111.13721},
 primaryClass = {astro-ph.GA},
       adsurl = {https://ui.adsabs.harvard.edu/abs/2022MNRAS.517....1S}
}

@ARTICLE{li22,
       author = {{Li}, Hui and {Vogelsberger}, Mark and {Bryan}, Greg L. and {Marinacci}, Federico and {Sales}, Laura V. and {Torrey}, Paul},
        title = "{Formation and evolution of young massive clusters in galaxy mergers: the SMUGGLE view}",
      journal = {\mnras},
         year = 2022,
        month = jul,
       volume = {514},
       number = {1},
        pages = {265-279},
          doi = {10.1093/mnras/stac1136},
archivePrefix = {arXiv},
       eprint = {2109.10356},
 primaryClass = {astro-ph.GA},
       adsurl = {https://ui.adsabs.harvard.edu/abs/2022MNRAS.514..265L}
}

@ARTICLE{burger22,
       author = {{Burger}, Jan D. and {Zavala}, Jes{\'u}s and {Sales}, Laura V. and {Vogelsberger}, Mark and {Marinacci}, Federico and {Torrey}, Paul},
        title = "{Degeneracies between self-interacting dark matter and supernova feedback as cusp-core transformation mechanisms}",
      journal = {\mnras},
         year = 2022,
        month = jul,
       volume = {513},
       number = {3},
        pages = {3458-3481},
          doi = {10.1093/mnras/stac994},
archivePrefix = {arXiv},
       eprint = {2108.07358},
 primaryClass = {astro-ph.GA},
       adsurl = {https://ui.adsabs.harvard.edu/abs/2022MNRAS.513.3458B}
}

@ARTICLE{li20,
       author = {{Li}, Hui and {Vogelsberger}, Mark and {Marinacci}, Federico and {Sales}, Laura V. and {Torrey}, Paul},
        title = "{The effects of subgrid models on the properties of giant molecular clouds in galaxy formation simulations}",
      journal = {\mnras},
         year = 2020,
        month = dec,
       volume = {499},
       number = {4},
        pages = {5862-5872},
          doi = {10.1093/mnras/staa3122},
archivePrefix = {arXiv},
       eprint = {2001.07214},
 primaryClass = {astro-ph.GA},
       adsurl = {https://ui.adsabs.harvard.edu/abs/2020MNRAS.499.5862L}
}

@ARTICLE{tacchella22,
       author = {{Tacchella}, Sandro and {Smith}, Aaron and {Kannan}, Rahul and {Marinacci}, Federico and {Hernquist}, Lars and {Vogelsberger}, Mark and {Torrey}, Paul and {Sales}, Laura and {Li}, Hui},
        title = "{H {\ensuremath{\alpha}} emission in local galaxies: star formation, time variability, and the diffuse ionized gas}",
      journal = {\mnras},
         year = 2022,
        month = jun,
       volume = {513},
       number = {2},
        pages = {2904-2929},
          doi = {10.1093/mnras/stac818},
archivePrefix = {arXiv},
       eprint = {2112.00027},
 primaryClass = {astro-ph.GA},
       adsurl = {https://ui.adsabs.harvard.edu/abs/2022MNRAS.513.2904T}
}

@ARTICLE{kannan20,
       author = {{Kannan}, Rahul and {Marinacci}, Federico and {Vogelsberger}, Mark and {Sales}, Laura V. and {Torrey}, Paul and {Springel}, Volker and {Hernquist}, Lars},
        title = "{Simulating the interstellar medium of galaxies with radiative transfer, non-equilibrium thermochemistry, and dust}",
      journal = {\mnras},
         year = 2020,
        month = dec,
       volume = {499},
       number = {4},
        pages = {5732-5748},
          doi = {10.1093/mnras/staa3249},
archivePrefix = {arXiv},
       eprint = {1910.14041},
 primaryClass = {astro-ph.GA},
       adsurl = {https://ui.adsabs.harvard.edu/abs/2020MNRAS.499.5732K}
}

@ARTICLE{pastras26,
       author = {{Pastras}, S. and {Patsis}, P.~A. and {Athanassoula}, E.},
        title = "{Morphologies arising from the gas flow in the innermost kiloparsec of barred galaxy models}",
      journal = {\aap},
         year = 2026,
        month = feb,
       volume = {707},
          eid = {A65},
        pages = {A65},
          doi = {10.1051/0004-6361/202556733},
archivePrefix = {arXiv},
       eprint = {2601.04306},
 primaryClass = {astro-ph.GA},
       adsurl = {https://ui.adsabs.harvard.edu/abs/2026A&A...707A..65P}
}

@ARTICLE{schultheis25,
       author = {{Schultheis}, Mathias and {Sormani}, Mattia C. and {Gadotti}, Dimitri A.},
        title = "{Nuclear stellar discs}",
      journal = {\aapr},
         year = 2025,
        month = nov,
       volume = {33},
       number = {1},
          eid = {7},
        pages = {7},
          doi = {10.1007/s00159-025-00163-6},
archivePrefix = {arXiv},
       eprint = {2509.04562},
 primaryClass = {astro-ph.GA},
       adsurl = {https://ui.adsabs.harvard.edu/abs/2025A&ARv..33....7S}
}

@ARTICLE{kim18,
       author = {{Kim}, Chang-Goo and {Ostriker}, Eve C.},
        title = "{Numerical Simulations of Multiphase Winds and Fountains from Star-forming Galactic Disks. I. Solar Neighborhood TIGRESS Model}",
      journal = {\apj},
         year = 2018,
        month = feb,
       volume = {853},
       number = {2},
          eid = {173},
        pages = {173},
          doi = {10.3847/1538-4357/aaa5ff},
archivePrefix = {arXiv},
       eprint = {1801.03952},
 primaryClass = {astro-ph.GA},
       adsurl = {https://ui.adsabs.harvard.edu/abs/2018ApJ...853..173K}
}

@ARTICLE{li17,
       author = {{Li}, Miao and {Bryan}, Greg L. and {Ostriker}, Jeremiah P.},
        title = "{Quantifying Supernovae-driven Multiphase Galactic Outflows}",
      journal = {\apj},
         year = 2017,
        month = jun,
       volume = {841},
       number = {2},
          eid = {101},
        pages = {101},
          doi = {10.3847/1538-4357/aa7263},
archivePrefix = {arXiv},
       eprint = {1610.08971},
 primaryClass = {astro-ph.GA},
       adsurl = {https://ui.adsabs.harvard.edu/abs/2017ApJ...841..101L}
}

@ARTICLE{matzner02,
       author = {{Matzner}, Christopher D.},
        title = "{On the Role of Massive Stars in the Support and Destruction of Giant Molecular Clouds}",
      journal = {\apj},
         year = 2002,
        month = feb,
       volume = {566},
       number = {1},
        pages = {302-314},
          doi = {10.1086/338030},
archivePrefix = {arXiv},
       eprint = {astro-ph/0110278},
 primaryClass = {astro-ph},
       adsurl = {https://ui.adsabs.harvard.edu/abs/2002ApJ...566..302M}
}

@ARTICLE{agertz13,
       author = {{Agertz}, Oscar and {Kravtsov}, Andrey V. and {Leitner}, Samuel N. and {Gnedin}, Nickolay Y.},
        title = "{Toward a Complete Accounting of Energy and Momentum from Stellar Feedback in Galaxy Formation Simulations}",
      journal = {\apj},
         year = 2013,
        month = jun,
       volume = {770},
       number = {1},
          eid = {25},
        pages = {25},
          doi = {10.1088/0004-637X/770/1/25},
archivePrefix = {arXiv},
       eprint = {1210.4957},
 primaryClass = {astro-ph.CO},
       adsurl = {https://ui.adsabs.harvard.edu/abs/2013ApJ...770...25A}
}

@BOOK{rybicki86,
       author = {{Rybicki}, George B. and {Lightman}, Alan P.},
        title = "{Radiative Processes in Astrophysics}",
         year = 1986,
       adsurl = {https://ui.adsabs.harvard.edu/abs/1986rpa..book.....R}
}

@ARTICLE{weaver77,
       author = {{Weaver}, R. and {McCray}, R. and {Castor}, J. and {Shapiro}, P. and {Moore}, R.},
        title = "{Interstellar bubbles. II. Structure and evolution.}",
      journal = {\apj},
         year = 1977,
        month = dec,
       volume = {218},
        pages = {377-395},
          doi = {10.1086/155692},
       adsurl = {https://ui.adsabs.harvard.edu/abs/1977ApJ...218..377W}
}

@ARTICLE{maclow88,
       author = {{Mac Low}, Mordecai-Mark and {McCray}, Richard},
        title = "{Superbubbles in Disk Galaxies}",
      journal = {\apj},
         year = 1988,
        month = jan,
       volume = {324},
        pages = {776},
          doi = {10.1086/165936},
       adsurl = {https://ui.adsabs.harvard.edu/abs/1988ApJ...324..776M}
}

@ARTICLE{cioffi88,
       author = {{Cioffi}, Denis F. and {McKee}, Christopher F. and {Bertschinger}, Edmund},
        title = "{Dynamics of Radiative Supernova Remnants}",
      journal = {\apj},
         year = 1988,
        month = nov,
       volume = {334},
        pages = {252},
          doi = {10.1086/166834},
       adsurl = {https://ui.adsabs.harvard.edu/abs/1988ApJ...334..252C}
}

@ARTICLE{martizzi15,
       author = {{Martizzi}, Davide and {Faucher-Gigu{\`e}re}, Claude-Andr{\'e} and {Quataert}, Eliot},
        title = "{Supernova feedback in an inhomogeneous interstellar medium}",
      journal = {\mnras},
         year = 2015,
        month = jun,
       volume = {450},
       number = {1},
        pages = {504-522},
          doi = {10.1093/mnras/stv562},
archivePrefix = {arXiv},
       eprint = {1409.4425},
 primaryClass = {astro-ph.GA},
       adsurl = {https://ui.adsabs.harvard.edu/abs/2015MNRAS.450..504M}
}

@ARTICLE{hopkins18a,
       author = {{Hopkins}, Philip F. and {Wetzel}, Andrew and {Kere{\v{s}}}, Du{\v{s}}an and {Faucher-Gigu{\`e}re}, Claude-Andr{\'e} and {Quataert}, Eliot and {Boylan-Kolchin}, Michael and {Murray}, Norman and {Hayward}, Christopher C. and {El-Badry}, Kareem},
        title = "{How to model supernovae in simulations of star and galaxy formation}",
      journal = {\mnras},
         year = 2018,
        month = jun,
       volume = {477},
       number = {2},
        pages = {1578-1603},
          doi = {10.1093/mnras/sty674},
archivePrefix = {arXiv},
       eprint = {1707.07010},
 primaryClass = {astro-ph.GA},
       adsurl = {https://ui.adsabs.harvard.edu/abs/2018MNRAS.477.1578H}
}

@ARTICLE{springel03,
       author = {{Springel}, Volker and {Hernquist}, Lars},
        title = "{Cosmological smoothed particle hydrodynamics simulations: a hybrid multiphase model for star formation}",
      journal = {\mnras},
         year = 2003,
        month = feb,
       volume = {339},
       number = {2},
        pages = {289-311},
          doi = {10.1046/j.1365-8711.2003.06206.x},
archivePrefix = {arXiv},
       eprint = {astro-ph/0206393},
 primaryClass = {astro-ph},
       adsurl = {https://ui.adsabs.harvard.edu/abs/2003MNRAS.339..289S}
}

@ARTICLE{chabrier01,
       author = {{Chabrier}, Gilles},
        title = "{The Galactic Disk Mass Budget. I. Stellar Mass Function and Density}",
      journal = {\apj},
         year = 2001,
        month = jun,
       volume = {554},
       number = {2},
        pages = {1274-1281},
          doi = {10.1086/321401},
archivePrefix = {arXiv},
       eprint = {astro-ph/0107018},
 primaryClass = {astro-ph},
       adsurl = {https://ui.adsabs.harvard.edu/abs/2001ApJ...554.1274C}
}

@ARTICLE{bigiel08,
       author = {{Bigiel}, F. and {Leroy}, A. and {Walter}, F. and {Brinks}, E. and {de Blok}, W.~J.~G. and {Madore}, B. and {Thornley}, M.~D.},
        title = "{The Star Formation Law in Nearby Galaxies on Sub-Kpc Scales}",
      journal = {\aj},
         year = 2008,
        month = dec,
       volume = {136},
       number = {6},
        pages = {2846-2871},
          doi = {10.1088/0004-6256/136/6/2846},
archivePrefix = {arXiv},
       eprint = {0810.2541},
 primaryClass = {astro-ph},
       adsurl = {https://ui.adsabs.harvard.edu/abs/2008AJ....136.2846B}
}

@ARTICLE{krumholz09,
       author = {{Krumholz}, Mark R. and {Matzner}, Christopher D.},
        title = "{The Dynamics of Radiation-pressure-dominated H II Regions}",
      journal = {\apj},
         year = 2009,
        month = oct,
       volume = {703},
       number = {2},
        pages = {1352-1362},
          doi = {10.1088/0004-637X/703/2/1352},
archivePrefix = {arXiv},
       eprint = {0906.4343},
 primaryClass = {astro-ph.SR},
       adsurl = {https://ui.adsabs.harvard.edu/abs/2009ApJ...703.1352K}
}

@ARTICLE{krumholz11,
       author = {{Krumholz}, Mark R. and {Gnedin}, Nickolay Y.},
        title = "{A Comparison of Methods for Determining the Molecular Content of Model Galaxies}",
      journal = {\apj},
         year = 2011,
        month = mar,
       volume = {729},
       number = {1},
          eid = {36},
        pages = {36},
          doi = {10.1088/0004-637X/729/1/36},
archivePrefix = {arXiv},
       eprint = {1011.4065},
 primaryClass = {astro-ph.CO},
       adsurl = {https://ui.adsabs.harvard.edu/abs/2011ApJ...729...36K}
}

@ARTICLE{mckee10,
       author = {{McKee}, Christopher F. and {Krumholz}, Mark R.},
        title = "{The Atomic-to-Molecular Transition in Galaxies. III. A New Method for Determining the Molecular Content of Primordial and Dusty Clouds}",
      journal = {\apj},
         year = 2010,
        month = jan,
       volume = {709},
       number = {1},
        pages = {308-320},
          doi = {10.1088/0004-637X/709/1/308},
archivePrefix = {arXiv},
       eprint = {0908.0330},
 primaryClass = {astro-ph.GA},
       adsurl = {https://ui.adsabs.harvard.edu/abs/2010ApJ...709..308M}
}

@ARTICLE{krumholz07,
       author = {{Krumholz}, Mark R. and {Tan}, Jonathan C.},
        title = "{Slow Star Formation in Dense Gas: Evidence and Implications}",
      journal = {\apj},
         year = 2007,
        month = jan,
       volume = {654},
       number = {1},
        pages = {304-315},
          doi = {10.1086/509101},
archivePrefix = {arXiv},
       eprint = {astro-ph/0606277},
 primaryClass = {astro-ph},
       adsurl = {https://ui.adsabs.harvard.edu/abs/2007ApJ...654..304K}
}

@ARTICLE{semenov17,
       author = {{Semenov}, Vadim A. and {Kravtsov}, Andrey V. and {Gnedin}, Nickolay Y.},
        title = "{The Physical Origin of Long Gas Depletion Times in Galaxies}",
      journal = {\apj},
         year = 2017,
        month = aug,
       volume = {845},
       number = {2},
          eid = {133},
        pages = {133},
          doi = {10.3847/1538-4357/aa8096},
archivePrefix = {arXiv},
       eprint = {1704.04239},
 primaryClass = {astro-ph.GA},
       adsurl = {https://ui.adsabs.harvard.edu/abs/2017ApJ...845..133S}
}

@ARTICLE{ferriere01,
       author = {{Ferri{\`e}re}, Katia M.},
        title = "{The interstellar environment of our galaxy}",
      journal = {Reviews of Modern Physics},
         year = 2001,
        month = oct,
       volume = {73},
       number = {4},
        pages = {1031-1066},
          doi = {10.1103/RevModPhys.73.1031},
archivePrefix = {arXiv},
       eprint = {astro-ph/0106359},
 primaryClass = {astro-ph},
       adsurl = {https://ui.adsabs.harvard.edu/abs/2001RvMP...73.1031F}
}

@ARTICLE{wolfire95,
       author = {{Wolfire}, M.~G. and {Hollenbach}, D. and {McKee}, C.~F. and {Tielens}, A.~G.~G.~M. and {Bakes}, E.~L.~O.},
        title = "{The Neutral Atomic Phases of the Interstellar Medium}",
      journal = {\apj},
         year = 1995,
        month = apr,
       volume = {443},
        pages = {152},
          doi = {10.1086/175510},
       adsurl = {https://ui.adsabs.harvard.edu/abs/1995ApJ...443..152W}
}

@ARTICLE{field69,
       author = {{Field}, G.~B. and {Goldsmith}, D.~W. and {Habing}, H.~J.},
        title = "{Cosmic-Ray Heating of the Interstellar Gas}",
      journal = {\apjl},
         year = 1969,
        month = mar,
       volume = {155},
        pages = {L149},
          doi = {10.1086/180324},
       adsurl = {https://ui.adsabs.harvard.edu/abs/1969ApJ...155L.149F}
}

@ARTICLE{wolfire03,
       author = {{Wolfire}, Mark G. and {McKee}, Christopher F. and {Hollenbach}, David and {Tielens}, A.~G.~G.~M.},
        title = "{Neutral Atomic Phases of the Interstellar Medium in the Galaxy}",
      journal = {\apj},
         year = 2003,
        month = apr,
       volume = {587},
       number = {1},
        pages = {278-311},
          doi = {10.1086/368016},
archivePrefix = {arXiv},
       eprint = {astro-ph/0207098},
 primaryClass = {astro-ph},
       adsurl = {https://ui.adsabs.harvard.edu/abs/2003ApJ...587..278W}
}

@ARTICLE{guo08,
       author = {{Guo}, Fulai and {Oh}, S. Peng},
        title = "{Feedback heating by cosmic rays in clusters of galaxies}",
      journal = {\mnras},
         year = 2008,
        month = feb,
       volume = {384},
       number = {1},
        pages = {251-266},
          doi = {10.1111/j.1365-2966.2007.12692.x},
archivePrefix = {arXiv},
       eprint = {0706.1274},
 primaryClass = {astro-ph},
       adsurl = {https://ui.adsabs.harvard.edu/abs/2008MNRAS.384..251G}
}

@ARTICLE{rahmati13,
       author = {{Rahmati}, Alireza and {Pawlik}, Andreas H. and {Rai{\v{c}}evi{\'c}}, Milan and {Schaye}, Joop},
        title = "{On the evolution of the H I column density distribution in cosmological simulations}",
      journal = {\mnras},
         year = 2013,
        month = apr,
       volume = {430},
       number = {3},
        pages = {2427-2445},
          doi = {10.1093/mnras/stt066},
archivePrefix = {arXiv},
       eprint = {1210.7808},
 primaryClass = {astro-ph.CO},
       adsurl = {https://ui.adsabs.harvard.edu/abs/2013MNRAS.430.2427R}
}

@ARTICLE{hopkins18b,
       author = {{Hopkins}, Philip F. and {Wetzel}, Andrew and {Kere{\v{s}}}, Du{\v{s}}an and {Faucher-Gigu{\`e}re}, Claude-Andr{\'e} and {Quataert}, Eliot and {Boylan-Kolchin}, Michael and {Murray}, Norman and {Hayward}, Christopher C. and {Garrison-Kimmel}, Shea and {Hummels}, Cameron and {Feldmann}, Robert and {Torrey}, Paul and {Ma}, Xiangcheng and {Angl{\'e}s-Alc{\'a}zar}, Daniel and {Su}, Kung-Yi and {Orr}, Matthew and {Schmitz}, Denise and {Escala}, Ivanna and {Sanderson}, Robyn and {Grudi{\'c}}, Michael Y. and {Hafen}, Zachary and {Kim}, Ji-Hoon and {Fitts}, Alex and {Bullock}, James S. and {Wheeler}, Coral and {Chan}, T.~K. and {Elbert}, Oliver D. and {Narayanan}, Desika},
        title = "{FIRE-2 simulations: physics versus numerics in galaxy formation}",
      journal = {\mnras},
         year = 2018,
        month = oct,
       volume = {480},
       number = {1},
        pages = {800-863},
          doi = {10.1093/mnras/sty1690},
archivePrefix = {arXiv},
       eprint = {1702.06148},
 primaryClass = {astro-ph.GA},
       adsurl = {https://ui.adsabs.harvard.edu/abs/2018MNRAS.480..800H}
}

@ARTICLE{vogelsberger20,
       author = {{Vogelsberger}, Mark and {Marinacci}, Federico and {Torrey}, Paul and {Puchwein}, Ewald},
        title = "{Cosmological simulations of galaxy formation}",
      journal = {Nature Reviews Physics},
         year = 2020,
        month = jan,
       volume = {2},
       number = {1},
        pages = {42-66},
          doi = {10.1038/s42254-019-0127-2},
archivePrefix = {arXiv},
       eprint = {1909.07976},
 primaryClass = {astro-ph.GA},
       adsurl = {https://ui.adsabs.harvard.edu/abs/2020NatRP...2...42V}
}

@ARTICLE{vogelsberger14a,
       author = {{Vogelsberger}, M. and {Genel}, S. and {Springel}, V. and {Torrey}, P. and {Sijacki}, D. and {Xu}, D. and {Snyder}, G. and {Bird}, S. and {Nelson}, D. and {Hernquist}, L.},
        title = "{Properties of galaxies reproduced by a hydrodynamic simulation}",
      journal = {\nat},
         year = 2014,
        month = may,
       volume = {509},
       number = {7499},
        pages = {177-182},
          doi = {10.1038/nature13316},
archivePrefix = {arXiv},
       eprint = {1405.1418},
 primaryClass = {astro-ph.CO},
       adsurl = {https://ui.adsabs.harvard.edu/abs/2014Natur.509..177V}
}

@ARTICLE{vogelsberger14b,
       author = {{Vogelsberger}, Mark and {Genel}, Shy and {Springel}, Volker and {Torrey}, Paul and {Sijacki}, Debora and {Xu}, Dandan and {Snyder}, Greg and {Nelson}, Dylan and {Hernquist}, Lars},
        title = "{Introducing the Illustris Project: simulating the coevolution of dark and visible matter in the Universe}",
      journal = {\mnras},
         year = 2014,
        month = oct,
       volume = {444},
       number = {2},
        pages = {1518-1547},
          doi = {10.1093/mnras/stu1536},
archivePrefix = {arXiv},
       eprint = {1405.2921},
 primaryClass = {astro-ph.CO},
       adsurl = {https://ui.adsabs.harvard.edu/abs/2014MNRAS.444.1518V}
}

@ARTICLE{vogelsberger13,
       author = {{Vogelsberger}, Mark and {Genel}, Shy and {Sijacki}, Debora and {Torrey}, Paul and {Springel}, Volker and {Hernquist}, Lars},
        title = "{A model for cosmological simulations of galaxy formation physics}",
      journal = {\mnras},
         year = 2013,
        month = dec,
       volume = {436},
       number = {4},
        pages = {3031-3067},
          doi = {10.1093/mnras/stt1789},
archivePrefix = {arXiv},
       eprint = {1305.2913},
 primaryClass = {astro-ph.CO},
       adsurl = {https://ui.adsabs.harvard.edu/abs/2013MNRAS.436.3031V}
}

@ARTICLE{ferland98,
       author = {{Ferland}, G.~J. and {Korista}, K.~T. and {Verner}, D.~A. and {Ferguson}, J.~W. and {Kingdon}, J.~B. and {Verner}, E.~M.},
        title = "{CLOUDY 90: Numerical Simulation of Plasmas and Their Spectra}",
      journal = {\pasp},
         year = 1998,
        month = jul,
       volume = {110},
       number = {749},
        pages = {761-778},
          doi = {10.1086/316190},
       adsurl = {https://ui.adsabs.harvard.edu/abs/1998PASP..110..761F}
}

@ARTICLE{faucher09,
       author = {{Faucher-Gigu{\`e}re}, Claude-Andr{\'e} and {Lidz}, Adam and {Zaldarriaga}, Matias and {Hernquist}, Lars},
        title = "{A New Calculation of the Ionizing Background Spectrum and the Effects of He II Reionization}",
      journal = {\apj},
         year = 2009,
        month = oct,
       volume = {703},
       number = {2},
        pages = {1416-1443},
          doi = {10.1088/0004-637X/703/2/1416},
archivePrefix = {arXiv},
       eprint = {0901.4554},
 primaryClass = {astro-ph.CO},
       adsurl = {https://ui.adsabs.harvard.edu/abs/2009ApJ...703.1416F}
}

@ARTICLE{ikeuchi86,
       author = {{Ikeuchi}, S. and {Ostriker}, J.~P.},
        title = "{Evolution of the Intergalactic Medium: What Happened during the Epoch Z = 3--10?}",
      journal = {\apj},
         year = 1986,
        month = feb,
       volume = {301},
        pages = {522},
          doi = {10.1086/163921},
       adsurl = {https://ui.adsabs.harvard.edu/abs/1986ApJ...301..522I}
}

@ARTICLE{katz96,
       author = {{Katz}, Neal and {Weinberg}, David H. and {Hernquist}, Lars},
        title = "{Cosmological Simulations with TreeSPH}",
      journal = {\apjs},
         year = 1996,
        month = jul,
       volume = {105},
        pages = {19},
          doi = {10.1086/192305},
archivePrefix = {arXiv},
       eprint = {astro-ph/9509107},
 primaryClass = {astro-ph},
       adsurl = {https://ui.adsabs.harvard.edu/abs/1996ApJS..105...19K}
}

@ARTICLE{barnes86,
       author = {{Barnes}, Josh and {Hut}, Piet},
        title = "{A hierarchical O(N log N) force-calculation algorithm}",
      journal = {\nat},
         year = 1986,
        month = dec,
       volume = {324},
       number = {6096},
        pages = {446-449},
          doi = {10.1038/324446a0},
       adsurl = {https://ui.adsabs.harvard.edu/abs/1986Natur.324..446B}
}

@ARTICLE{pakmor16,
       author = {{Pakmor}, R{\"u}diger and {Springel}, Volker and {Bauer}, Andreas and {Mocz}, Philip and {Munoz}, Diego J. and {Ohlmann}, Sebastian T. and {Schaal}, Kevin and {Zhu}, Chenchong},
        title = "{Improving the convergence properties of the moving-mesh code AREPO}",
      journal = {\mnras},
         year = 2016,
        month = jan,
       volume = {455},
       number = {1},
        pages = {1134-1143},
          doi = {10.1093/mnras/stv2380},
archivePrefix = {arXiv},
       eprint = {1503.00562},
 primaryClass = {astro-ph.GA},
       adsurl = {https://ui.adsabs.harvard.edu/abs/2016MNRAS.455.1134P}
}

\begin{appendix} 
\section{Supplementary Figures}\label{appendix:one}

\begin{figure}[htbp]
    \centering
    \includegraphics[width=0.5\textwidth]{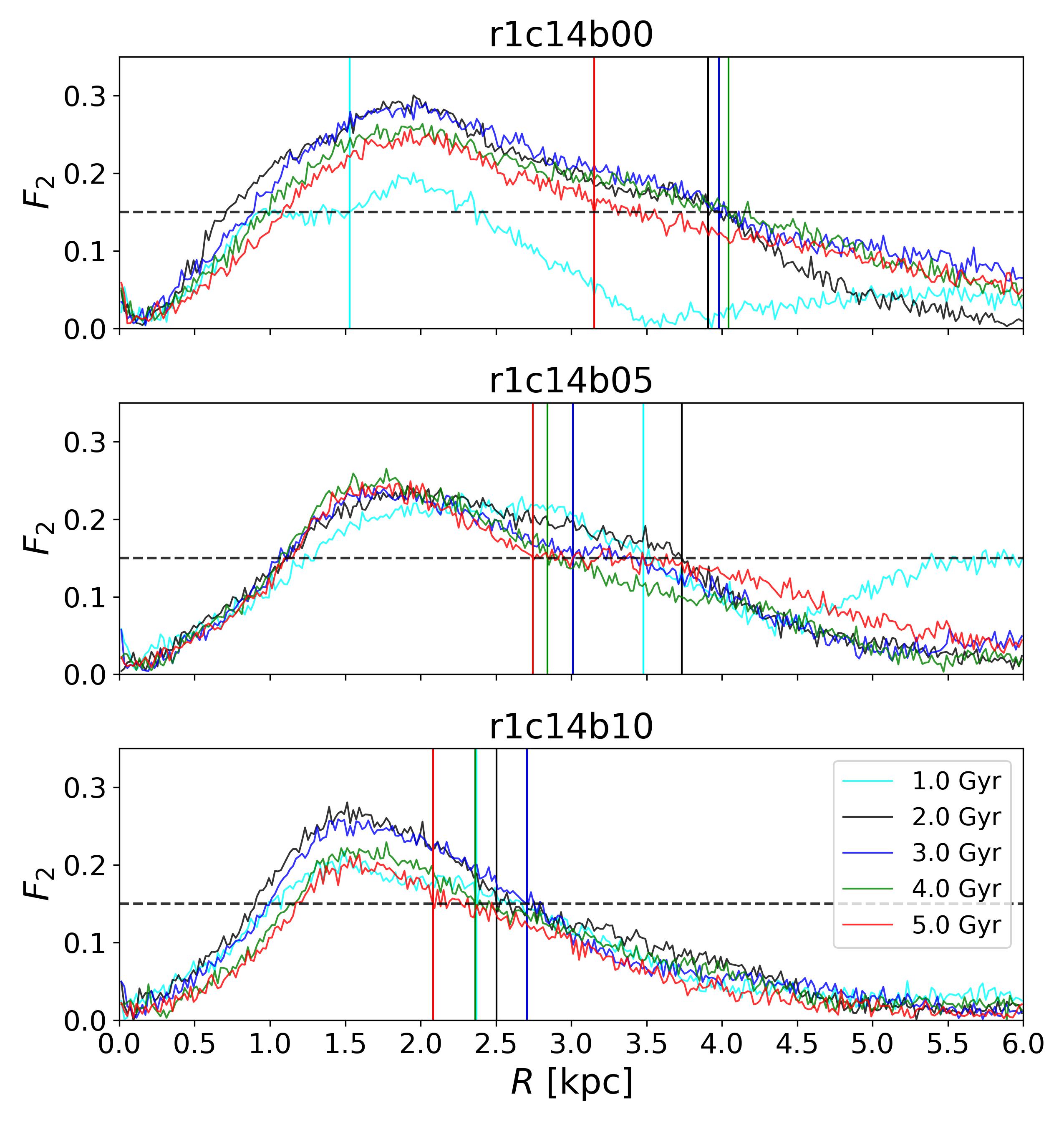}

    \caption{Radial distributions of the Fourier mode $m=2$ for all models at 2, 3, 4, and 5 Gyr within 6 kpc. Model names and colors corresponding to each time are labeled. The horizontal line indicates $F_2 = 0.15$, while the vertical lines mark the outermost intersections with this line to estimate the radial extent of the bar's influence.}
    \label{fig:f2radial}
\end{figure}

\begin{figure}[htbp]
    \centering
    \includegraphics[width=0.5\textwidth]{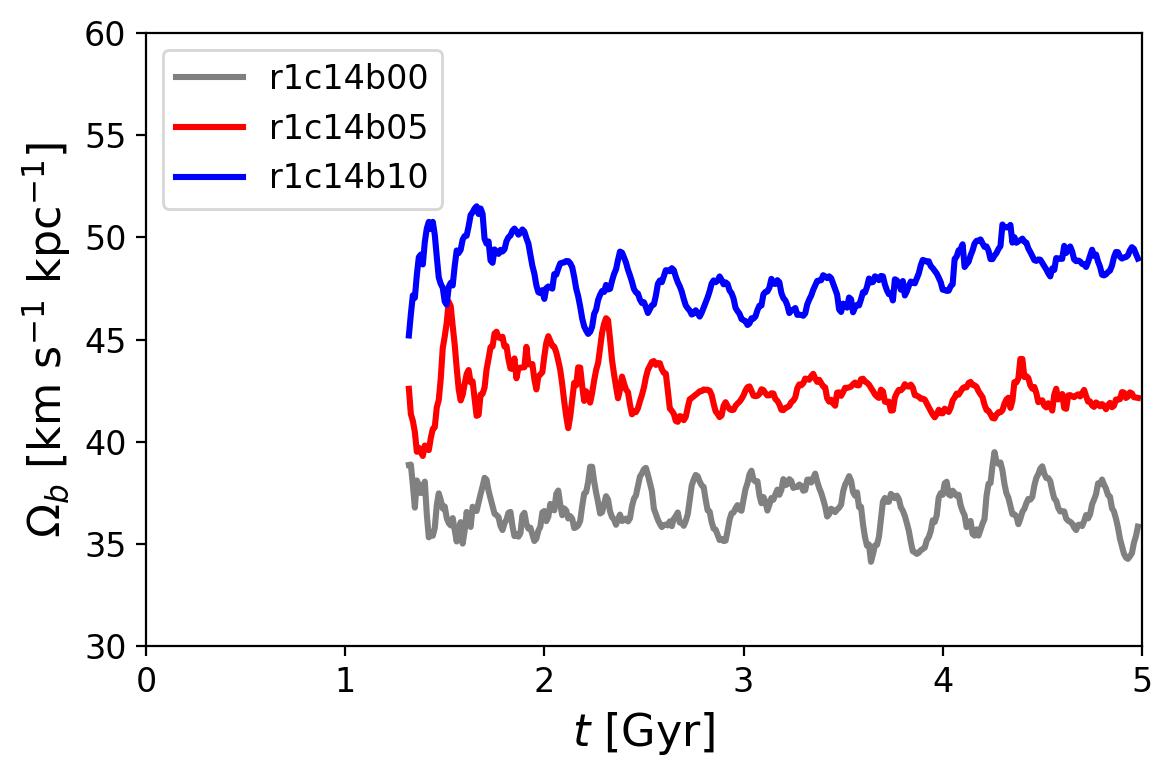}
    \caption{Bar pattern speed evolution of all models between 1.3 and 5 Gyr. The pattern speeds are derived from the mean phase angle of the Fourier mode $m=2$ averaged over the radial range 1.0--2.0 kpc. The time derivative of this phase is converted to bar pattern speed in units of km s$^{-1}$kpc$^{-1}$.}
    \label{fig:barpattern}
\end{figure}

\begin{figure}[htbp]
    \centering
    \includegraphics[width=0.30\textwidth]{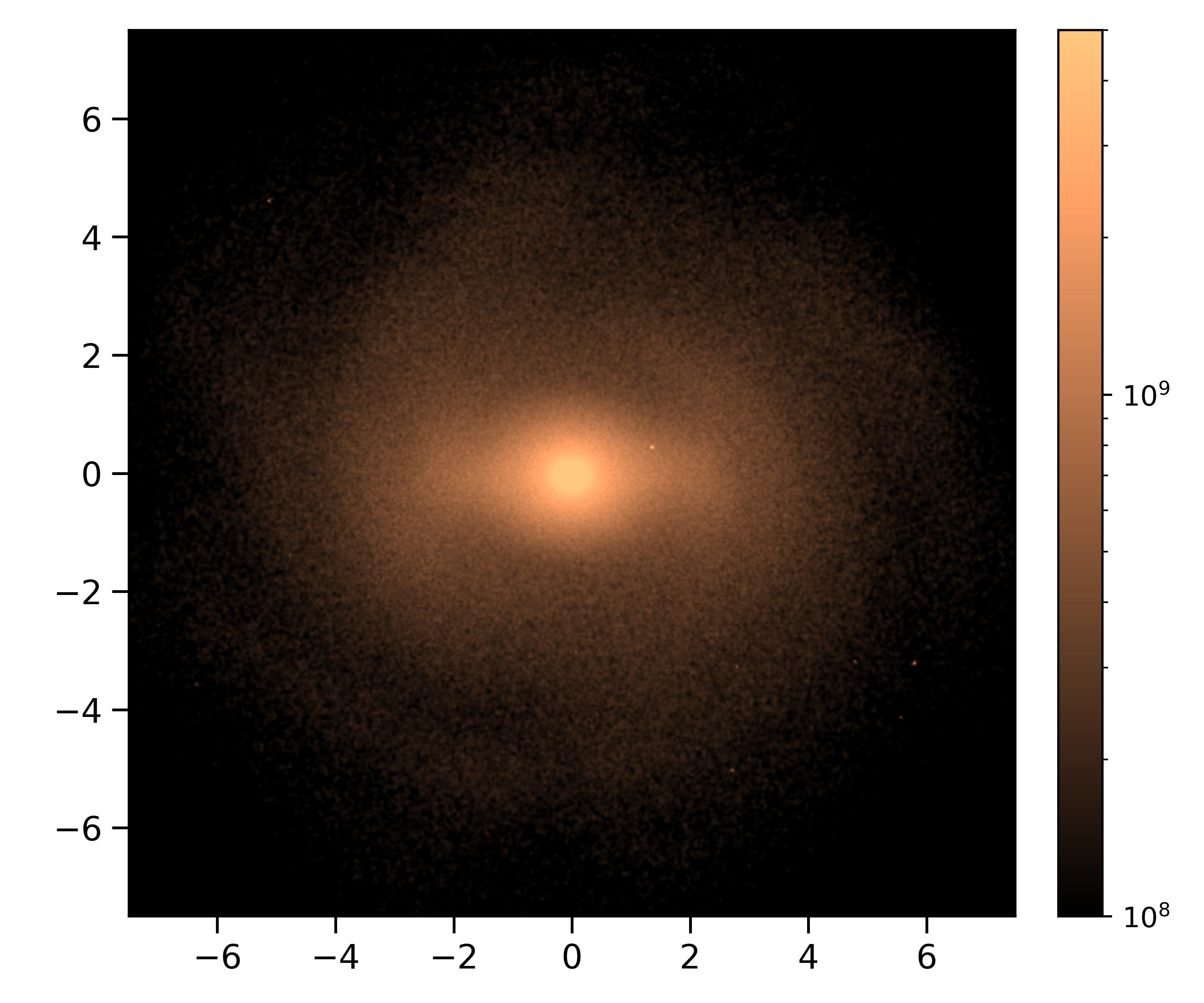}
    
    \includegraphics[width=0.30\textwidth]{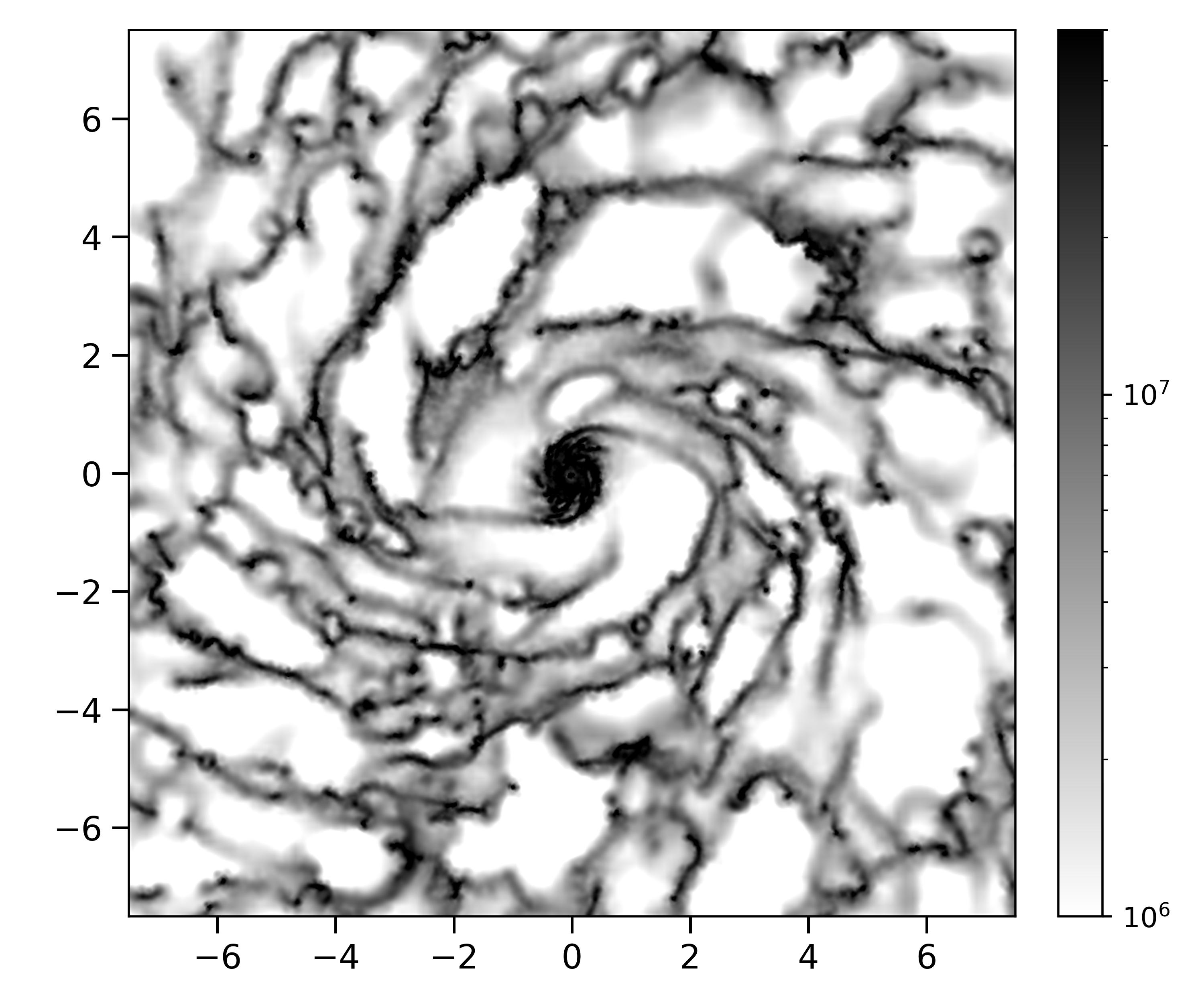}

    \includegraphics[width=0.30\textwidth]{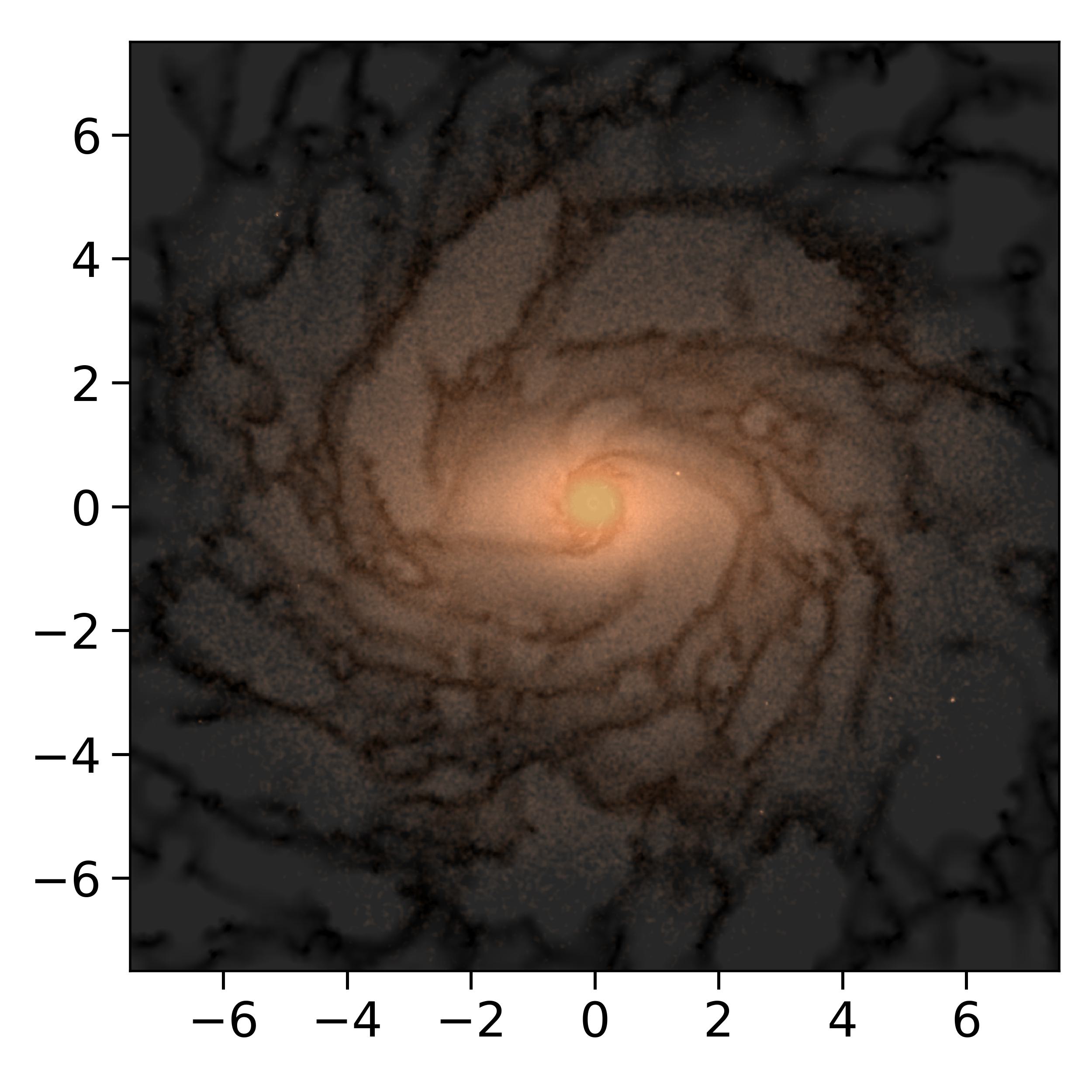}
    
    \includegraphics[width=0.30\textwidth]{fig/stack_ring2_r3c14b05md60dp14g10_300.jpg}
    \caption{Face-on projections of the surface density distributions of stars, gas, and their stacked image in the X-Y plane within a $15 \times 15$ kpc$^2$ box at 3.0 Gyr for model r1c14b05. The color bars indicate units of $M_\sun$ kpc$^{-2}$. The surface density images of stars and gas are superimposed with partial transparency to produce the third panel, over which blue and purple points are overlaid in the bottom panel to indicate young stars with ages $0.01 < t_\mathrm{age} < 0.1 \, \Gyr$ and $t_\mathrm{age} \leq 0.01 \, \Gyr$, respectively.}
    \label{fig:face_stacking}
\end{figure}

\begin{figure}[htbp]
    \centering
    \includegraphics[width=0.5\textwidth]{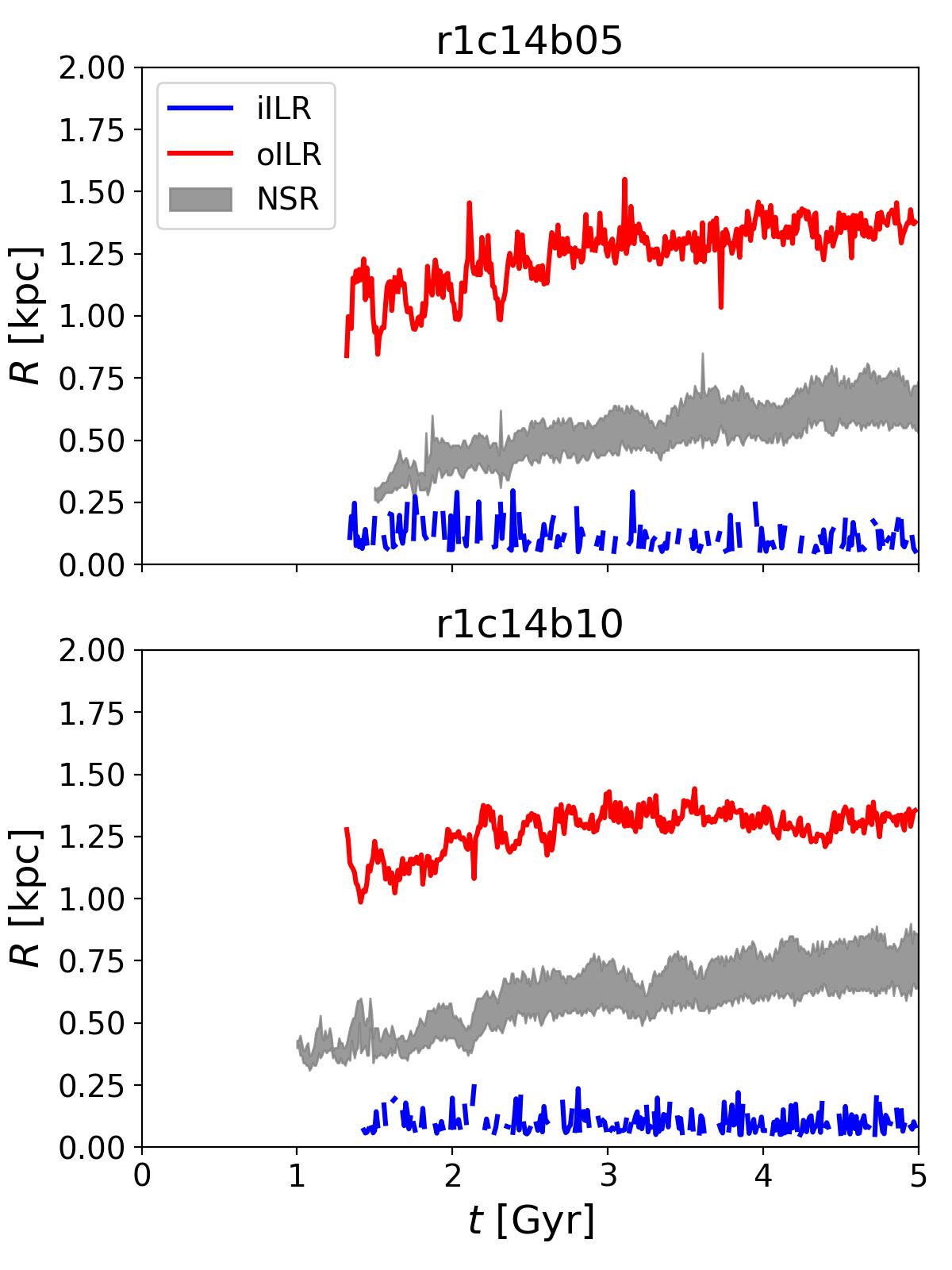}
    \caption{Time evolution of inner inner Lindblad resonance (iILR) and outer inner Lindblad resonance (oILR) for model r1c14b05 and r1c14b10, which form double ILRs and a nuclear stellar ring (NSR). The bar pattern speed in Fig.~\ref{fig:barpattern} is used for the calculation. The gray shaded region indicates the radial extent of NSR, bounded by its semi-major and semi-minor axes. }
    \label{fig:ilr_evolution}
\end{figure}

\end{appendix}

\end{document}